\documentclass[11pt]{article}
\usepackage{amssymb}
\usepackage{amsmath}
\usepackage{amsfonts}
\usepackage{sw20elba}
\usepackage[bottom]{footmisc}
\usepackage{geometry}
\usepackage[onehalfspacing]{setspace}
\usepackage[authoryear]{natbib}
\usepackage{hyperref}
\usepackage{hyphenat}
\usepackage{portland}
\usepackage{lscape}
\usepackage{scalefnt}
\usepackage{graphicx}

\newtheorem{theorem}{Theorem}

\newtheorem{corollary}{Corollary}

\newtheorem{lemma}{Lemma}

\allowdisplaybreaks
\input{tcilatex}
\begin{document}

\title{Threshold Regression with Nonparametric Sample Splitting\thanks{%
We are grateful to Xiaohong Chen, Jonathan Dingel, Bo
Honor\'{e}, Sokbae Lee, Yuan Liao, Francesca Molinari, Ingmar Prucha, Myung
Seo, Ping Yu, and participants at numerous seminar/conference presentations
for very helpful comments. Financial supports from the Appleby-Mosher grant
and the CUSE grant are highly appreciated.}}
\author{\textsc{Yoonseok Lee}\thanks{\textit{Address}: Department of
Economics and Center for Policy Research, Syracuse University, 426 Eggers
Hall, Syracuse, NY 13244. \textit{E-mail}: \texttt{ylee41@maxwell.syr.edu}} 
\\
%EndAName
Syracuse University \and \textsc{Yulong Wang}\thanks{\textit{Address}:
Department of Economics and Center for Policy Research, Syracuse University,
127 Eggers Hall, Syracuse, NY 13244. \textit{E-mail}: \texttt{%
ywang402@maxwell.syr.edu}} \\
%EndAName
Syracuse University}
\date{January 2021}
\maketitle

\begin{abstract}
\noindent This paper develops a threshold regression model where an unknown
relationship between two variables nonparametrically determines the
threshold. We allow the observations to be cross-sectionally dependent so
that the model can be applied to determine an unknown spatial border for
sample splitting over a random field. We derive the uniform rate of
convergence and the nonstandard limiting distribution of the nonparametric
threshold estimator. We also obtain the root-n consistency and the
asymptotic normality of the regression coefficient estimator. Our model has
broad empirical relevance as illustrated by estimating the tipping point in
social segregation problems as a function of demographic characteristics;
and determining metropolitan area boundaries using nighttime light intensity
collected from satellite imagery. We find that the new empirical results are
substantially different from those in the existing studies.

\vspace{0.15in}


\end{abstract}

\thispagestyle{empty}\setcounter{page}{0}\newpage

\setcounter{page}{1}\renewcommand{\baselinestretch}{1.5}

\section{Introduction}

Sample splitting and threshold regression models have spawned a vast
literature in econometrics and statistics. Existing studies typically
specify the sample splitting criteria in a parametric way as whether a
single random variable or a linear combination of variables crosses some
unknown threshold. See, for example, \cite{Hansen00a}, \cite{Caner04}, \cite%
{SeoLinton07}, \cite{LeeSeoShin11}, \cite{LiLing12}, \cite{Yu12}, \cite%
{LLSS18}, \cite{Hidalgo19}, and \cite{Yu19}. In this paper, we study a novel
extension to consider a \textit{nonparametric} sample splitting model. Such
an extension leads to new theoretical results and substantially generalizes
the empirical applicability of threshold models.

Specifically, we consider a model given by 
\begin{equation}
y_{i}=x_{i}^{\top }\beta _{0}+x_{i}^{\top }\delta _{0}\mathbf{1}\left[
q_{i}\leq \gamma _{0}\left( s_{i}\right) \right] +u_{i}  \label{model}
\end{equation}%
for $i=1,\ldots ,n$, where $\mathbf{1}\left[ \cdot \right] $ is the binary
indicator. In this model, the marginal effect of $x_{i}$ to $y_{i}$ can be
different across $i$ as $(\beta _{0}+\delta _{0})$ or $\beta _{0}$ depending
on whether $q_{i}\leq \gamma _{0}\left( s_{i}\right) $ or not. The threshold
function $\gamma _{0}(\cdot )$ is unknown, and the main parameters of
interest are $\beta _{0}$, $\delta _{0}$, and $\gamma _{0}(\cdot )$. The
novel feature of this model is that the sample splitting is determined by an
unknown relationship between two variables $q_{i}$ and $s_{i}$, and their
relationship is characterized by the nonparametric threshold function $%
\gamma _{0}(\cdot )$. In contrast, the classical threshold regression models
assume $\gamma _{0}\left( \cdot \right) $ to be a constant or a linear
index. Our new specification can cover interesting cases that have not been
studied. For example, we can consider the threshold to be heterogeneous and
specific to each observation $i$ if we see $\gamma _{0}\left( s_{i}\right)
=\gamma _{0i}$; or the threshold to be determined by the direction of some
moment condition $\gamma _{0}(s_{i})=\mathbb{E}[q_{i}|s_{i}]$. Apparently,
when $\gamma _{0}(s)=\gamma _{0}$ or $\gamma _{0}(s)=\gamma _{0}s$ for some
parameter $\gamma _{0}$ and $s\neq 0$, it reduces to the standard threshold
regression model.

The new model is motivated by the following two applications: estimating
potentially heterogeneous thresholds in public economics and determining
spatial sample splitting in urban economics. The first one is about the
tipping point model proposed by \cite{Schelling71}, who analyzes the
phenomenon that a neighborhood's white population substantially decreases
once the minority share exceeds a certain threshold, called the tipping
point. \cite{Card08} empirically estimate the tipping point model by
considering the constant threshold regression, $y_{i}=\beta _{10}+\delta
_{10}\mathbf{1}\left[ q_{i}\leq \gamma _{0}\right] +x_{2i}^{\top }\beta
_{20}+u_{i}$, where $y_{i}$ is the white population change in a decade and $%
q_{i}$ is the initial minority share in the $i$th tract. The parameters $%
\delta _{10}$ and $\gamma _{0}$ denote the change size and the threshold,
respectively. In Section VII of \cite{Card08}, however, they find that the
tipping point $\gamma _{0}$ varies depending on the attitudes of white
residents toward the minority. This finding raises the concern on the
constant threshold model and motivates us to study the more general model (%
\ref{model}) by specifying the tipping point $\gamma _{0}$ as a
nonparametric function of local demographic characteristics. We estimate
such a tipping function in Section \ref{Section tipping}.

For the second application, we use the model (\ref{model}) to define
metropolitan area boundaries, which is a fundamental problem in urban
economics. Recently, many studies propose to use nighttime light intensity
collected from satellite imagery to define the metropolitan area. They set
an \textit{ad hoc} level of light intensity as a threshold and categorize a
pixel in the satellite imagery as a part of the metropolitan area if the
light intensity of that pixel is higher than the threshold. See, for
example, \cite{Rozenfeld11}, \cite{Henderson12}, \cite{Dingel19}, and \cite%
{Vogel19}. In contrast, the model (\ref{model}) can provide a data-driven
guidance of choosing the intensity threshold from the econometric
perspective, if we let $y_{i}$ as the light intensity in the $i$th pixel and 
$(q_{i},s_{i})$ as the location information of that pixel (more precisely,
the coordinate of a point on a rotated map as described in Section \ref%
{Section contour}). In Section \ref{Section boundary}, we estimate the
metropolitan area of Dallas, Texas, especially its development from 1995 to
2010, and find substantially different results from the conventional
approaches. To the best of our knowledge, this is the first study to
nonparametrically determine the metropolitan area using a threshold model.

We develop a two-step estimation procedure of (\ref{model}), where we
estimate $\gamma _{0}\left( \cdot \right) $ by the local constant least
squares. Under the shrinking threshold asymptotics as in \cite{Bai97b}, \cite%
{Bai98}, and \cite{Hansen00a}, we show that the nonparametric estimator $%
\widehat{\gamma }(\cdot )$ is uniformly consistent and has a highly
nonstandard limiting distribution. Based on such distribution, we develop a
pointwise specification test of $\gamma _{0}(s)$ for any given $s$, which
enables us to construct a confidence interval by inverting the test.
Besides, the parametric part $(\widehat{\beta }^{\top },\widehat{\delta }%
^{\top })^{\top }$ is shown to satisfy the root-$n$ asymptotic normality.

We highlight some novel technical features of the new estimator as follows.
First, since the nonparametric function $\gamma _{0}\left( \cdot \right) $
is inside the indicator function, technical proofs of the asymptotic results
are non-standard.\ In particular, we establish the uniform rate of
convergence of $\widehat{\gamma }\left( \cdot \right) $, which involves
substantially more complicated derivations than the standard (constant)
threshold regression model. Second, we find that, unlike the standard kernel
estimator, $\widehat{\gamma }(\cdot )$ is asymptotically unbiased even if
the optimal bandwidth is used. Also, when the change size $\delta _{0}$
shrinks very slowly, the optimal rate of convergence of $\widehat{\gamma }%
(\cdot )$ becomes close to the root-$n$ rate. In the standard kernel
regression, such a fast rate of convergence can be obtained when the unknown
function is infinitely differentiable, while we only require the
second-order differentiability of $\gamma _{0}\left( \cdot \right) $. Third,
to limit the effect of estimating $\gamma _{0}\left( \cdot \right) $ to $(%
\widehat{\beta }^{\top },\widehat{\delta }^{\top })^{\top }$, we propose to
use the observations that are sufficiently away from the estimated threshold
in the second-step parametric estimation. The choice of this distance is
obtained by the uniform convergence rate of $\widehat{\gamma }\left( \cdot
\right) $. Fourth, we let the variables be cross-sectionally dependent by
considering the strong-mixing random field as in \cite{Conley99} and \cite%
{Conley07}. This generalization allows us to study nonparametric sample
splitting of spatial observations. For instance, if we let $(q_{i},s_{i})$
correspond to the geographical location (i.e., latitude and longitude on the
map), then the threshold $\mathbf{1}\left[ q_{i}\leq \gamma _{0}\left(
s_{i}\right) \right] $ identifies the unknown border yielding a
two-dimensional sample splitting. In more general contexts, the model can be
applied to identify social or economic segregation over interacting agents.
Finally, noting that $\mathbf{1}\left[ q_{i}\leq \gamma _{0}\left(
s_{i}\right) \right] $ can be considered as the special case of $\mathbf{1}%
\left[ g_{0}\left( q_{i},s_{i}\right) \leq 0\right] $ when $g_{0}$ is
monotonically increasing in $q_{i}$, we discuss how to extend the proposed
method to such a more general case that leads to a threshold contour model.

The rest of the paper is organized as follows. Section \ref{Section
estimation} sets up the model, establishes the identification, and defines
the estimator. Section \ref{Section asymptotics} derives the asymptotic
properties of the estimators and develops a likelihood ratio test of the
threshold function. Section \ref{Section contour} describes how to extend
the main model to estimate a threshold contour. Section \ref{Section
simulation} studies small sample properties of the proposed statistics by
Monte Carlo simulations. Section \ref{Section empirics} applies the new
method to estimate the tipping point function and to determine metropolitan
areas. Section \ref{Section conclusion} concludes this paper with some
remarks. The main proofs are in the Appendix, and all the omitted proofs are
collected in the supplementary material.

We use the following notations. Let $\rightarrow _{p}$ denote convergence in
probability, $\rightarrow _{d}$ convergence in distribution, and $%
\Rightarrow $ weak convergence of the underlying probability measure as $%
n\rightarrow \infty $. Let $\left\lfloor r\right\rfloor $ denote the biggest
integer smaller than or equal to $r$, $\mathbf{1}[E]$ the indicator function
of a generic event $E$, and $\left\Vert A\right\Vert $ the Euclidean norm of
a vector or matrix $A$. For any set $B$, let $|B|$ as the cardinality of $B$.

\section{Model Setup\label{Section estimation}}

We assume spatial processes located on an evenly spaced lattice $\Lambda
\subset 
%TCIMACRO{\U{211d} }%
%BeginExpansion
\mathbb{R}
%EndExpansion
^{2}$, following \cite{Conley99}, \cite{Conley07}, and \cite%
{CarbonFrancqTran2007}.\footnote{%
It can be extended to an unevenly spaced lattice as in \cite{Bolthausen82}
and \cite{Jenish09} with substantially more complicated notations.\ (cf.\
footnote 9 in \cite{Conley99}).} We consider the threshold regression model
given by (\ref{model}), which is 
\begin{equation*}
y_{i}=x_{i}^{\top }\beta _{0}+x_{i}^{\top }\delta _{0}\mathbf{1}\left[
q_{i}\leq \gamma _{0}\left( s_{i}\right) \right] +u_{i}\text{,}
\end{equation*}%
where the observations $\{(y_{i},x_{i}^{\top },q_{i},s_{i})^{\top }\in 
\mathbb{R}^{1+\dim (x)+1+1};i\in \Lambda _{n}\}$ are a triangular array of
real random variables defined on some probability space with $\Lambda _{n}$
being a fixed sequence of finite subsets of $\Lambda $. In this setup, the
cardinality of $\Lambda _{n}$, $n=|\Lambda _{n}|$, is the sample size and $%
\sum_{i\in \Lambda _{n}}$ denotes the summation of all observations. For
readability, we postpone the regularity conditions on $\Lambda _{n}$ in
Assumption A later. The threshold function $\gamma _{0}:\mathbb{R\rightarrow
R}$ as well as the regression coefficients $\theta _{0}=(\beta _{0}^{\top
},\delta _{0}^{\top })^{\top }\in \mathbb{R}^{2\dim (x)}$ are unknown, and
they are the parameters of interest.\footnote{%
The main results of this paper can be extended to consider multi-dimensional 
$s_{i}$ using multivariate kernels. However, we only consider the scalar
case for the expositional simplicity. Furthermore, the results are readily
generalized to the case where only a subset of parameters differ between
regimes.} Since we consider a shrinking threshold effect, the parameter $%
\delta _{0}$ is to depend on the sample size $n$ as in Assumption A-(ii)
below; hence $\delta _{0}$ and $\theta _{0}$ should be written as $\delta
_{n0}$ and $\theta _{n0}$, respectively. However, we write $\delta _{0}$ and 
$\theta _{0}$ for simplicity. We let $\mathcal{Q}\subset 
%TCIMACRO{\U{211d} }%
%BeginExpansion
\mathbb{R}
%EndExpansion
$ and $\mathcal{S}\subset 
%TCIMACRO{\U{211d} }%
%BeginExpansion
\mathbb{R}
%EndExpansion
$ denote the supports of $q_{i}$ and $s_{i}$, respectively. Suppose the
space of $\gamma _{0}\left( s\right) $ for any $s$ is a compact set $\Gamma
\subset 
%TCIMACRO{\U{211d} }%
%BeginExpansion
\mathbb{R}
%EndExpansion
$.

First, we establish the identification, which requires the following
conditions.

\paragraph{Assumption ID}

\begin{description}
\item \textit{(i) }$\mathbb{E}\left[ u_{i}|x_{i},q_{i},s_{i}\right] =0$%
\textit{\ almost surely.}

\item \textit{(ii) }$\mathbb{E}\left[ x_{i}x_{i}^{\top }\right] >\mathbb{E}%
\left[ x_{i}x_{i}^{\top }\mathbf{1}\left[ q_{i}\leq \gamma \right] \right]
>0 $\textit{\ for any }$\gamma \in \Gamma $.

\item \textit{(iii) For any }$s\in \mathcal{S}$, \textit{there exists }$%
\varepsilon (s)>0$\textit{\ such that }$\varepsilon (s)<\mathbb{P}\left(
q_{i}\leq \gamma _{0}(s_{i})|s_{i}=s\right) <1-\varepsilon (s)$\textit{\ and 
}$\delta _{0}^{\top }\mathbb{E}\left[ x_{i}x_{i}^{\top }|q_{i}=q,s_{i}=s%
\right] \delta _{0}>0$ \textit{for all} $\left( q,s\right) \in \mathcal{Q}%
\times \mathcal{S}$.

\item \textit{(iv) }$q_{i}$\textit{\ is continuously distributed with its
conditional density }$f(q|s)$\textit{\ satisfying that }$%
0<C_{1}<f(q|s)<C_{2}<\infty $\textit{\ for all} $\left( q,s\right) \in
\Gamma \times \mathcal{S}$\textit{\ and some constants }$C_{1}$ and $C_{2}$.
\end{description}

\medskip

Assumption ID is mild. The condition (i) excludes endogeneity, and (ii) is
the full rank condition to identify the global parameters $\beta _{0}$ and $%
\delta _{0}$. The conditions (ii) and (iii) require that the location of the
threshold is not on the boundary of the support of $q_{i}$ for any $s\in 
\mathcal{S}$, which is inevitable for identification and has been commonly
assumed in the existing threshold literature (e.g., \cite{Hansen00a}). If $%
\gamma _{0}(s)$ reaches the boundary of $q_{i}$ for some $s\in \mathcal{S}$,
then no observation can be generated from one side of the threshold function
at this $s$, and identification is failed. The second condition in (iii)
assumes the coefficient change exists (i.e., $\delta _{0}\neq 0$). Note that
it does not require $\mathbb{E}\left[ x_{i}x_{i}^{\top }|q_{i}=q,s_{i}=s%
\right] $ to be of full rank, and hence $q_{i}$ or $s_{i}$ can be one of the
elements of $x_{i}$ (e.g., the threshold autoregressive model by \cite%
{Tong83}) or a linear combination of $x_{i}$. The condition (iv) requires
the conditional density of $q_{i}$ given any $s_{i}$ is positive and bounded
in $\Gamma $.

Under Assumption ID, the following theorem establishes the identification of
the semiparametric threshold regression model (\ref{model}).

\begin{theorem}
\label{Thm id}Under Assumption ID, the parameters $(\beta _{0}^{\top
},\delta _{0}^{\top })^{\top }$ are the unique minimizer of $\mathbb{E}%
[(y_{i}-x_{i}^{\top }\beta -x_{i}^{\top }\delta \mathbf{1}\left[ q_{i}\leq
\gamma \right] )^{2}]$ for any $\gamma \in \Gamma $, and the threshold
function $\gamma _{0}\left( s\right) $ is the unique minimizer of $\mathbb{E}%
[(y_{i}-x_{i}^{\top }\beta _{0}-x_{i}^{\top }\delta _{0}\mathbf{1}\left[
q_{i}\leq \gamma (s_{i})\right] )^{2}|s_{i}=s]$ for each given $s\in 
\mathcal{S}$.
\end{theorem}

Given identification, we proceed to estimate this semiparametric model in
two steps. First, for given $s\in \mathcal{S}$, we fix $\gamma
_{0}(s)=\gamma $ and obtain $\widehat{\beta }\left( \gamma ;s\right) $ and $%
\widehat{\delta }\left( \gamma ;s\right) $ by the local constant least
squares conditional on $\gamma $:%
\begin{equation}
(\widehat{\beta }\left( \gamma ;s\right) ^{\intercal },\widehat{\delta }%
\left( \gamma ;s\right) ^{\intercal })^{\intercal }=\arg \min_{\beta ,\delta
}Q_{n}\left( \beta ,\delta ,\gamma ;s\right) \text{,}  \label{first}
\end{equation}%
where 
\begin{equation}
Q_{n}\left( \beta ,\delta ,\gamma ;s\right) =\dsum_{i\in \Lambda
_{n}}K\left( \frac{s_{i}-s}{b_{n}}\right) \left( y_{i}-x_{i}^{\top }\beta
-x_{i}^{\top }\delta \mathbf{1}\left[ q_{i}\leq \gamma \right] \right) ^{2}
\label{sse}
\end{equation}%
for some kernel function $K\left( \cdot \right) $ and a bandwidth parameter $%
b_{n}$. Then $\gamma _{0}(s)$ is estimated by%
\begin{equation}
\widehat{\gamma }\left( s\right) =\arg \min_{\gamma \in \Gamma
_{n}}Q_{n}\left( \gamma ;s\right)  \label{g-hat0}
\end{equation}%
for given $s$, where $\Gamma _{n}=\Gamma \cap \{q_{1},\ldots ,q_{n}\}$ and $%
Q_{n}\left( \gamma ;s\right) $ is the concentrated sum of squares defined as 
\begin{equation}
Q_{n}\left( \gamma ;s\right) =Q_{n}\left( \widehat{\beta }\left( \gamma
;s\right) ,\widehat{\delta }\left( \gamma ;s\right) ,\gamma ;s\right) \text{.%
}  \label{reg}
\end{equation}%
To avoid any additional technical complexity, we focus on estimation of $%
\gamma _{0}(s)$ at $s\in \mathcal{S}_{0}\subset \mathcal{S}$ for some
compact interior subset $\mathcal{S}_{0}$ of the support, say the middle
70\% quantiles. Note that, given $s$, the nonparametric estimator $\widehat{%
\gamma }\left( s\right) $ can be seen as a local version of the standard
(constant) threshold regression estimator. Therefore, the computation of (%
\ref{g-hat0}) requires one-dimensional grid search of the threshold for only 
$n$ times as in the standard threshold regression estimation.

In the second step, we estimate the parametric components $\beta _{0}$ and $%
\delta _{0}$. To minimize any potential effects from the first step
estimation, we estimate $\beta _{0}$ and $\delta _{0}^{\ast }=\beta
_{0}+\delta _{0}$ using the observations that are sufficiently away from the
estimated threshold. This is implemented by considering%
\begin{eqnarray}
\widehat{\beta } &=&\arg \min_{\beta }\dsum_{i\in \Lambda _{n}}\left(
y_{i}-x_{i}^{\top }\beta \right) ^{2}\mathbf{1}\left[ q_{i}>\widehat{\gamma }%
\left( s_{i}\right) +\pi _{n}\right] \mathbf{1}[s_{i}\in \mathcal{S}_{0}]%
\text{,}  \label{para-b} \\
\widehat{\delta }^{\ast } &=&\arg \min_{\delta ^{\ast }}\dsum_{i\in \Lambda
_{n}}\left( y_{i}-x_{i}^{\top }\delta ^{\ast }\right) ^{2}\mathbf{1}\left[
q_{i}<\widehat{\gamma }\left( s_{i}\right) -\pi _{n}\right] \mathbf{1}%
[s_{i}\in \mathcal{S}_{0}]  \label{para-d}
\end{eqnarray}%
for some constant $\pi _{n}>0$ satisfying $\pi _{n}\rightarrow 0$ as $%
n\rightarrow \infty $, which is defined later. The change size $\delta $ can
be estimated as $\widehat{\delta }=\widehat{\delta }^{\ast }-\widehat{\beta }
$.

For the asymptotic behavior of the threshold estimator, the existing
literature typically assumes martingale difference arrays (e.g., \cite%
{Hansen00a} and \cite{LLSS18}) or random samples (e.g., \cite{Yu12} and \cite%
{Yu19}). In this paper, we allow for cross-sectional dependence by
considering spatial $\alpha $-mixing processes as in \cite{Bolthausen82} and 
\cite{Conley99}. More precisely, for any indices (or locations) $i,j\in
\Lambda $, we define the metric $\lambda \left( i,j\right) =\max_{1\leq \ell
\leq \dim \left( \Lambda \right) }\left\vert i_{\ell }-j_{\ell }\right\vert $
and the corresponding norm $\max_{1\leq \ell \leq \dim \left( \Lambda
\right) }\left\vert i_{\ell }\right\vert $, where $i_{\ell }$ denotes the $%
\ell $th component of $i$. The distance of any two subsets $\Lambda
_{1},\Lambda _{2}\subset \Lambda $ is defined as $\lambda (\Lambda
_{1},\Lambda _{2})=\inf \{\lambda (i,j):i\in \Lambda _{1},j\in \Lambda
_{2}\} $. We let $\mathcal{F}_{\Lambda }$ be the $\sigma $-algebra generated
by a random sequence $\left( x_{i}^{\top },q_{i},s_{i},u_{i}\right) ^{\top }$
for $i\in \Lambda $ and define the spatial $\alpha $-mixing coefficient as%
\begin{equation}
\alpha _{k,l}(m)=\sup \left\{ \left\vert \mathbb{P}\left( A\cap B\right) -%
\mathbb{P}\left( A\right) \mathbb{P}\left( B\right) \right\vert :A\in 
\mathcal{F}_{\Lambda _{1}}\text{, }B\in \mathcal{F}_{\Lambda _{2}}\text{, }%
\lambda \left( \Lambda _{1},\Lambda _{2}\right) \geq m\right\} \text{,}
\label{mix}
\end{equation}%
where $\left\vert \Lambda _{1}\right\vert \leq k$ and $\left\vert \Lambda
_{2}\right\vert \leq l$. Without loss of generality, we assume $\alpha
_{k,l}(0)=1$ and $\alpha _{k,l}(m)$ is monotonically decreasing in $m$ for
all $k$ and $l$.

The following conditions are imposed for deriving the asymptotic properties
of our two-step estimator. Let $f\left( q,s\right) $\ be the joint density
function of $(q_{i},s_{i})$ and%
\begin{eqnarray}
D\left( q,s\right) &=&\mathbb{E}[x_{i}x_{i}^{\top }|\left(
q_{i},s_{i}\right) =\left( q,s\right) ]\text{,}  \label{D} \\
V\left( q,s\right) &=&\mathbb{E}[x_{i}x_{i}^{\top }u_{i}^{2}|\left(
q_{i},s_{i}\right) =\left( q,s\right) ]\text{.}  \label{V}
\end{eqnarray}

\paragraph{Assumption A}

\begin{description}
\item \textit{(i) The lattice }$\Lambda _{n}\subset 
%TCIMACRO{\U{211d} }%
%BeginExpansion
\mathbb{R}
%EndExpansion
^{2}$\textit{\ is infinitely countable; all the elements in }$\Lambda _{n}$%
\textit{\ are located at distances at least }$\lambda _{0}>1$\textit{\ from
each other (i.e., for any }$i,j\in \Lambda _{n}$, $\lambda \left( i,j\right)
\geq \lambda _{0}$\textit{); and }$\lim_{n\rightarrow \infty }\left\vert
\partial \Lambda _{n}\right\vert /n=0$\textit{, where }$\partial \Lambda
_{n}=\{i\in \Lambda _{n}:\exists j\not\in \Lambda _{n}$\textit{\ with }$%
\lambda (i,j)=1\}$\textit{.}

\item \textit{(ii) }$\delta _{0}=c_{0}n^{-\epsilon }$\textit{\ for some }$%
c_{0}\neq 0$\textit{\ and }$\epsilon \in (0,1/2)$; $\left( c_{0}^{\top
},\beta _{0}^{\top }\right) ^{\top }$\textit{\ belongs to some compact
subset of }$\mathbb{R}^{2\dim (x)}$\textit{.}

\item \textit{(iii) }$\left( x_{i}^{\top },q_{i},s_{i},u_{i}\right) ^{\top }$%
\textit{\ is strictly stationary and }$\alpha $\textit{-mixing with the
mixing coefficient }$\alpha _{k,l}(m)$\textit{\ defined in (\ref{mix}),
which satisfies that for all }$k$ and $l$,\textit{\ }$\alpha _{k,l}(m)\leq
C_{1}\exp (-C_{2}m)$\textit{\ for some positive constants }$C_{1}$\textit{\
and }$C_{2}$\textit{.}

\item \textit{(iv) }$0<\mathbb{E}\left[ u_{i}^{2}|x_{i},q_{i},s_{i}\right]
<\infty $\textit{\ almost surely.}

\item \textit{(v) Uniformly in }$\left( q,s\right) $\textit{, there exist
some finite constants }$\varphi >0$ and $C>0$\textit{\ such that }$\mathbb{E}%
[||x_{i}x_{i}^{\top }||^{4+\varphi }|(q_{i},s_{i})=(q,s)]<C$\textit{\ and }$%
\mathbb{E}[||x_{i}u_{i}||^{4+\varphi }|(q_{i},s_{i})=(q,s)]<C$.

\item \textit{(vi) }$\gamma _{0}:\mathcal{S}\mapsto \Gamma $\textit{\ is a
twice continuously differentiable function with bounded derivatives.}

\item \textit{(vii) }$D\left( q,s\right) $\textit{, }$V\left( q,s\right) $%
\textit{, and }$f\left( q,s\right) $\textit{\ are uniformly bounded in }$%
(q,s)$\textit{, continuous in }$q$\textit{, and twice continuously
differentiable in }$s$ \textit{with bounded derivatives. For any }$i,j\in
\Lambda _{n}$,\textit{\ the joint density of }$(q_{i},q_{j},s_{i},s_{j})^{%
\intercal }$\textit{\ is uniformly bounded above by some constant }$C<\infty 
$\textit{\ and continuously differentiable in all components.}

\item \textit{(viii) }$c_{0}^{\top }D\left( \gamma _{0}(s),s\right) c_{0}>0$%
\textit{, }$c_{0}^{\top }V\left( \gamma _{0}(s),s\right) c_{0}>0$\textit{,
and }$f\left( \gamma _{0}(s),s\right) >0$ \textit{for all }$s\in \mathcal{S}$%
\textit{.}

\item \textit{(ix) As }$n\rightarrow \infty $\textit{, }$b_{n}\rightarrow 0$%
\textit{, }$n^{1-2\epsilon }b_{n}/\log n\rightarrow \infty $\textit{, }$\log
n/(nb_{n}^{2})\rightarrow 0$, \textit{and }$nb_{n}^{(2+2\varphi )/(2+\varphi
)}\rightarrow \infty $\textit{\ for some }$\varphi >0$ \textit{given in (v).}

\item \textit{(x) }$K\left( \cdot \right) $\textit{\ is a positive
second-order kernel, which is Lipschitz, symmetric around zero, and
nonincreasing on }$%
%TCIMACRO{\U{211d} }%
%BeginExpansion
\mathbb{R}
%EndExpansion
^{+}$ \textit{and satisfies }$\int K\left( v\right) dv=1$\textit{, }$\int
v^{2}K\left( v\right) dv<\infty $.
\end{description}

\medskip

We provide some discussions about Assumption A. First, we assume that $q_{i}$
and $s_{i}$ are continuous random variables as in the example in Section \ref%
{Section tipping}. This setup also covers the two-dimensional
\textquotedblleft spatial structural break\textquotedblright\ model as a
special case as in the example in Section \ref{Section boundary}. For the
latter case, we denote $n_{1}$ and $n_{2}$ as the numbers of rows
(latitudes) and columns (longitudes) in the grid of pixels, and normalize $q$
and $s$ in the way that $q\in \{1/n_{1},2/n_{1},\ldots ,1\}$ and $s\in
\{1/n_{2},2/n_{2},\ldots ,1\}$. Under similar (and even weaker) regularity
conditions as Assumption A, we can show that the asymptotic results in the
following sections extend to this case once we treat $(q_{i},s_{i})^{\top }$
as independently and uniformly distributed random variables over $\left[ 0,1%
\right] ^{2}$. Such similarity is also found in the standard structural
break and the threshold regression models (e.g.,\ Proposition 5 in \cite%
{Bai98} and Theorem 1 in \cite{Hansen00a}).

Second, Assumption A is mild and common in the existing literature. In
particular, the condition (i) is the same as in \cite{Bolthausen82} to
define the latent random field. Note that $\lambda _{0}$ can be any strictly
positive value, and hence we can impose $\lambda _{0}>1$ without loss of
generality. The condition (ii) adopts the widely used shrinking change size
setup as in \cite{Bai97b}, \cite{Bai98}, and \cite{Hansen00a} to obtain a
simple limiting distribution. In contrast, a constant change size (when $%
\epsilon =0$) leads to a complicated asymptotic distribution of the
threshold estimator, which depends on nuisance parameters (e.g., \cite%
{Chan93}). The condition (iii) is required to establish the maximal
inequality and uniform convergence in a spatially dependent random field. We
impose a stronger condition than \cite{Jenish09} to obtain the maximal
inequality uniformly over $\gamma $ and $s$. We could weaken this condition
such that $\alpha _{k,l}(m)$ decays at a polynomial rate (e.g., $\alpha
_{k,l}(m)\leq Cm^{-r}$ for some $r>8$ and a constant $C$ as in \cite%
{CarbonFrancqTran2007}) if we impose higher moment restrictions in the
condition (v). However, this exponential decay rate simplifies the technical
proofs. The conditions (iv) to (viii) are similar to Assumption 1 of \cite%
{Hansen00a}. The condition (ix) imposes restrictions on the bandwidth $b_{n}$%
, which now depends on $\epsilon $ and $\varphi $. The condition (x) holds
for many commonly used kernel functions including the Gaussian and the
uniform kernels.

Third, we assume $\gamma _{0}\left( \cdot \right) $ to be a function from $%
\mathcal{S}$ to $\Gamma $ in Assumption A-(vi), which is not necessarily
one-to-one. For this reason, sample splitting based on $\mathbf{1}\left[
q_{i}\leq \gamma _{0}\left( s_{i}\right) \right] $ can be different from
that based on $\mathbf{1}\left[ s_{i}\geq \breve{\gamma}_{0}\left(
q_{i}\right) \right] $ for some function $\breve{\gamma}_{0}\left( \cdot
\right) $. Instead of restricting $\gamma _{0}\left( \cdot \right) $ to be
one-to-one in this paper, we presume that one knows which variables should
be respectively assigned as $q_{i}$ and $s_{i}$ from the context.
Alternatively, we can consider a function $g_{0}\left( q,s\right) $ such
that $g_{0}$ is monotonically increasing in $q$ for any $s$. Then, $\mathbf{1%
}\left[ q_{i}\leq \gamma _{0}\left( s_{i}\right) \right] $ is viewed as a
special case of $\mathbf{1}\left[ g_{0}\left( q_{i},s_{i}\right) \leq 0%
\right] $ by inverting $g_{0}\left( \cdot ,s\right) =q^{\ast }$, where $%
g_{0}\left( q^{\ast },s\right) =0$. We discuss such extension to identify a
threshold contour in Section \ref{Section contour}.

\section{Asymptotic Results\label{Section asymptotics}}

We first obtain the asymptotic properties of $\widehat{\gamma }\left(
s\right) $. The following theorem derives the pointwise consistency and the
pointwise rate of convergence of $\widehat{\gamma }\left( s\right) $ at the
interior points of $\mathcal{S}$.

\begin{theorem}
\label{p-roc}For a given $s\in \mathcal{S}_{0}$, under Assumptions ID and A, 
$\widehat{\gamma }\left( s\right) \rightarrow _{p}\gamma _{0}\left( s\right) 
$ as $n\rightarrow \infty $. Furthermore, 
\begin{equation}
\widehat{\gamma }\left( s\right) -\gamma _{0}\left( s\right) =O_{p}\left( 
\frac{1}{n^{1-2\epsilon }b_{n}}\right)  \label{proc}
\end{equation}%
provided that $n^{1-2\epsilon }b_{n}^{2}$ does not diverge.
\end{theorem}

The pointwise rate of convergence of $\widehat{\gamma }\left( s\right) $
depends on two parameters, $\epsilon $ and $b_{n}$. It is decreasing in $%
\epsilon $ like the parametric (constant) threshold case: a larger $\epsilon 
$ reduces the threshold effect $\delta _{0}=c_{0}n^{-\epsilon }$ and hence
decreases the effective sampling information on the threshold. Since we
estimate $\gamma _{0}(\cdot )$ using the kernel estimation method, the rate
of convergence depends on the bandwidth $b_{n}$ as well. As in the standard
kernel estimator case, a smaller bandwidth decreases the effective local
sample size, which reduces the precision of the estimator $\widehat{\gamma }%
\left( s\right) $. Therefore, in order to have a sufficiently fast rate of
convergence, we need to choose $b_{n}$ large enough when the threshold
effect $\delta _{0}$ is expected to be small (i.e., when $\epsilon $ is
close to $1/2$).

Unlike the standard kernel estimator, there seems no bias-variance trade-off
in $\widehat{\gamma }\left( s\right) $ in (\ref{proc}), implying that we
could improve the rate of convergence by choosing a larger bandwidth $b_{n}$%
. However, as we can find in Theorem \ref{g-an} below, $b_{n}$ cannot be
chosen too large to result in $n^{1-2\epsilon }b_{n}^{2}\rightarrow \infty $%
, under which $n^{1-2\epsilon }b_{n}(\widehat{\gamma }\left( s\right)
-\gamma _{0}\left( s\right) )$ is no longer $O_{p}(1)$. Therefore, we can
obtain the optimal bandwidth using the restriction that $n^{1-2\epsilon
}b_{n}^{2}$ does not diverge.

Under this restriction, we find the optimal bandwidth as $b_{n}^{\ast
}=n^{-(1-2\epsilon )/2}c^{\ast }$ for some constant $0<c^{\ast }<\infty $,
which yields the optimal pointwise rate of convergence of $\widehat{\gamma }%
\left( s\right) $ as $n^{-(1-2\epsilon )/2}$. However, such a bandwidth
choice is not feasible because of the unknown constant $c^{\ast }$ and the
nuisance parameter $\epsilon $ that are not estimable. In practice, we
suggest cross validation as we implement in Section \ref{Section empirics},
although its statistical properties need to be studied further. Note that,
when the change size $\delta _{0}$ shrinks very slowly with $n$ (i.e., $%
\epsilon $ is close to $0$), the optimal rate of convergence of $\widehat{%
\gamma }(\cdot )$ is close to $n^{-1/2}$. This $\sqrt{n}$-rate is obtained
in the standard kernel regression if the unknown function is infinitely
differentiable, while we only require the second-order differentiability of $%
\gamma _{0}\left( \cdot \right) $.

The next theorem derives the limiting distribution of $\widehat{\gamma }%
\left( s\right) $. We let $W(\cdot )$ be a two-sided Brownian motion defined
as in \cite{Hansen00a}:%
\begin{equation}
W(r)=W_{1}(-r)\mathbf{1}\left[ r<0\right] +W_{2}(r)\mathbf{1}\left[ r>0%
\right] \text{,}  \label{2BM}
\end{equation}%
where $W_{1}(\cdot )$ and $W_{2}(\cdot )$ are independent standard Brownian
motions on $[0,\infty )$.

\begin{theorem}
\label{g-an}Under Assumptions ID and A, for a given $s\in \mathcal{S}_{0}$,
if $n^{1-2\epsilon }b_{n}^{2}\rightarrow \varrho \in (0,\infty )$,%
\begin{equation}
n^{1-2\epsilon }b_{n}\left( \widehat{\gamma }\left( s\right) -\gamma
_{0}\left( s\right) \right) \rightarrow _{d}\xi \left( s\right) \arg
\max_{r\in 
%TCIMACRO{\U{211d} }%
%BeginExpansion
\mathbb{R}
%EndExpansion
}\left( W\left( r\right) +\mu \left( r,\varrho ;s\right) \right)
\label{a-normal}
\end{equation}%
as $n\rightarrow \infty $, where%
\begin{eqnarray*}
\mu \left( r,\varrho ;s\right) &=&-\left\vert r\right\vert \psi _{0}\left(
r,\varrho ;s\right) +\frac{\varrho |\dot{\gamma}_{0}(s)|}{\xi (s)}\psi
_{1}\left( r,\varrho ;s\right) \text{,} \\
\psi _{j}\left( r,\varrho ;s\right) &=&\int_{0}^{|r|\xi (s)/\left( \varrho |%
\dot{\gamma}_{0}(s)|\right) }t^{j}K\left( t\right) dt\ \ \ \text{for }j=0,1%
\text{,} \\
\xi \left( s\right) &=&\frac{\kappa _{2}c_{0}^{\top }V\left( \gamma
_{0}\left( s\right) ,s\right) c_{0}}{\left( c_{0}^{\top }D\left( \gamma
_{0}\left( s\right) ,s\right) c_{0}\right) ^{2}f\left( \gamma _{0}\left(
s\right) ,s\right) }
\end{eqnarray*}%
with $\kappa _{2}=\int K(v)^{2}dv$ and $\dot{\gamma}_{0}\left( s\right) $ is
the first derivative of $\gamma _{0}$ at $s$. Furthermore, 
\begin{equation*}
\mathbb{E}\left[ \arg \max_{r\in 
%TCIMACRO{\U{211d} }%
%BeginExpansion
\mathbb{R}
%EndExpansion
}\left( W\left( r\right) +\mu \left( r,\varrho ;s\right) \right) \right] =0.
\end{equation*}
\end{theorem}

The drift term $\mu \left( r,\varrho ;s\right) $ in (\ref{a-normal}) depends
on the constant $0<\varrho <\infty $, which is the limit of $n^{1-2\epsilon
}b_{n}^{2}=(n^{1-2\epsilon }b_{n})b_{n}$, and $|\dot{\gamma}_{0}(s)|$, the
steepness of $\gamma _{0}(\cdot )$ at $s$. Interestingly, it resembles the
typical $O(b_{n})$ boundary bias of the standard local constant estimator.
However, this non-zero drift term is not because of the typical boundary
effect but because of the inequality restriction inside the indicator
function, $\mathbf{1}\left[ q_{i}\leq \gamma _{0}\left( s_{i}\right) \right] 
$, which characterizes the sample splitting.

It is important to note that having this non-zero drift term in the limiting
expression does not mean that the limiting distribution of $\widehat{\gamma }%
\left( s\right) $ has a non-zero mean, even when we use the optimal
bandwidth $b_{n}^{\ast }=O(n^{-(1-2\epsilon )/2})$ satisfying $%
n^{1-2\epsilon }b_{n}^{\ast 2}\rightarrow \varrho \in (0,\infty )$. This is
mainly because the drift function $\mu \left( r,\varrho ;s\right) $ is
symmetric about zero and hence the limiting random variable $\arg \max_{r\in 
%TCIMACRO{\U{211d} }%
%BeginExpansion
\mathbb{R}
%EndExpansion
}\left( W\left( r\right) +\mu \left( r,\varrho ;s\right) \right) $ is mean
zero. In general, we can show that the random variable $\arg \max_{r\in 
%TCIMACRO{\U{211d} }%
%BeginExpansion
\mathbb{R}
%EndExpansion
}\left( W\left( r\right) +\mu \left( r,\varrho ;s\right) \right) $ always
has mean zero if $\mu \left( r,\varrho ;s\right) $ is a non-random function
that is symmetric about zero and monotonically decreasing fast enough. This
result might be of independent research interest and is summarized in Lemma %
\ref{lemma-drift} in the Appendix. Figure \ref{fig drift} depicts the drift
function $\mu \left( r,\varrho ;s\right) $ for various kernels when $\xi
(s)/\left( \varrho |\dot{\gamma}_{0}(s)|\right) =1$.

\begin{figure}[h]  
\centering
    \caption{Drift function $\mu \left( r,\varrho ;s\right) $ for different kernels (color online)}
    \includegraphics[width=0.6\textwidth]{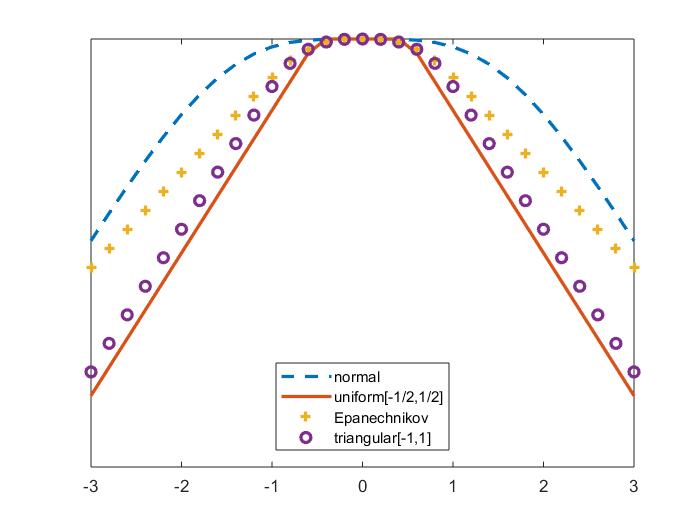}
    
\label{fig drift}
\end{figure}%
%EndExpansion

Since the limiting distribution in (\ref{a-normal}) depends on unknown
components, like $\varrho $ and $\dot{\gamma}_{0}(s)$, it is hard to use
this result for further inference. We instead suggest undersmoothing for
practical use. More precisely, if we suppose $n^{1-2\epsilon
}b_{n}^{2}\rightarrow 0$ as $n\rightarrow \infty $, then the limiting
distribution in (\ref{a-normal}) simplifies to\footnote{%
We let $\psi _{0}\left( r,0;s\right) =\int_{0}^{\infty }K\left( t\right)
dt=1/2$ and $\psi _{1}\left( r,0;s\right) =\int_{0}^{\infty }tK\left(
t\right) dt<\infty $.} 
\begin{equation}
n^{1-2\epsilon }b_{n}\left( \widehat{\gamma }\left( s\right) -\gamma
_{0}\left( s\right) \right) \rightarrow _{d}\xi \left( s\right) \arg
\max_{r\in 
%TCIMACRO{\U{211d} }%
%BeginExpansion
\mathbb{R}
%EndExpansion
}\left( W\left( r\right) -\frac{\left\vert r\right\vert }{2}\right)
\label{an0}
\end{equation}%
as $n\rightarrow \infty $, which appears the same as in the parametric case
in \cite{Hansen00a} except for the scaling factor $n^{1-2\epsilon }b_{n}$.
The distribution of $\arg \max_{r\in 
%TCIMACRO{\U{211d} }%
%BeginExpansion
\mathbb{R}
%EndExpansion
}\left( W\left( r\right) -\left\vert r\right\vert /2\right) $ is known
(e.g., \cite{Bhattacharya76} and \cite{Bai97b}), which is also described in
Hansen (2000, p.581). The $\xi \left( s\right) $ term determines the scale
of the distribution at given $s$ in the way that it increases in the
conditional variance $\mathbb{E}[u_{i}^{2}|x_{i},q_{i},s_{i}]$ and decreases
in the size of the threshold constant $c_{0}$ and the density of $%
(q_{i},s_{i})$ near the threshold.

Even when $n^{1-2\epsilon }b_{n}^{2}\rightarrow 0$ as $n\rightarrow \infty $%
, the asymptotic distribution in (\ref{an0}) still depends on the unknown
parameter $\epsilon $ (or equivalently $c_{0}$) in $\xi \left( s\right) $
that is not estimable. Thus, this result cannot be directly used for
inference of $\gamma _{0}\left( s\right) $. Alternatively, given any $s\in 
\mathcal{S}_{0}$, we can consider a pointwise likelihood ratio test
statistic for 
\begin{equation}
H_{0}:\gamma _{0}\left( s\right) =\gamma _{\ast }\left( s\right) \text{ \ \
against \ \ }H_{1}:\gamma _{0}\left( s\right) \neq \gamma _{\ast }\left(
s\right) ,  \label{hypo}
\end{equation}%
\ which is given as 
\begin{equation}
LR_{n}(s)=\dsum_{i\in \Lambda _{n}}K\left( \frac{s_{i}-s}{b_{n}}\right) 
\frac{Q_{n}\left( \gamma _{\ast }\left( s\right) ,s\right) -Q_{n}\left( 
\widehat{\gamma }\left( s\right) ,s\right) }{Q_{n}\left( \widehat{\gamma }%
\left( s\right) ,s\right) }\text{.}  \label{LR}
\end{equation}%
The following corollary obtains the limiting null distribution of this test
statistic that is free of nuisance parameters. By inverting the likelihood
ratio statistic, we can form a pointwise confidence interval for $\gamma
_{0}\left( s\right) $.

\begin{corollary}
\label{g-test}Suppose $n^{1-2\epsilon }b_{n}^{2}\rightarrow 0$ as $%
n\rightarrow \infty $. Under the same condition in Theorem \ref{g-an}, for
any fixed $s\in \mathcal{S}_{0}$, the test statistic in (\ref{LR}) satisfies 
\begin{equation}
LR_{n}(s)\rightarrow _{d}\xi _{LR}\left( s\right) \max_{r\in 
%TCIMACRO{\U{211d} }%
%BeginExpansion
\mathbb{R}
%EndExpansion
}\left( 2W\left( r\right) -\left\vert r\right\vert \right)  \label{limLR}
\end{equation}%
as $n\rightarrow \infty $ under the null hypothesis (\ref{hypo}), where 
\begin{equation*}
\xi _{LR}\left( s\right) =\frac{\kappa _{2}c_{0}^{\top }V\left( \gamma
_{0}\left( s\right) ,s\right) c_{0}}{\sigma ^{2}(s)c_{0}^{\top }D\left(
\gamma _{0}\left( s\right) ,s\right) c_{0}}
\end{equation*}%
with $\sigma ^{2}(s)=\mathbb{E}\left[ u_{i}^{2}|s_{i}=s\right] $ and $\kappa
_{2}=\int K(v)^{2}dv$.
\end{corollary}

When $\mathbb{E}[u_{i}^{2}|x_{i},q_{i},s_{i}]=\mathbb{E}[u_{i}^{2}|s_{i}]$,
which is the case of local conditional homoskedasticity, the scale parameter 
$\xi _{LR}\left( s\right) $ is simplified as $\kappa _{2}$, and hence the
limiting null distribution of $LR_{n}(s)$ becomes free of nuisance
parameters and the same for all $s\in \mathcal{S}_{0}$. Though this limiting
distribution is still nonstandard, the critical values in this case can be
simulated using the same method as Hansen (2000, p.582) with the scale
adjusted by $\kappa _{2}$. More precisely, since the distribution function
of $\zeta =\max_{r\in 
%TCIMACRO{\U{211d} }%
%BeginExpansion
\mathbb{R}
%EndExpansion
}\left( 2W\left( r\right) -\left\vert r\right\vert \right) $ is given as $%
\mathbb{P}(\zeta \leq z)=(1-\exp (-z/2))^{2}\mathbf{1}\left[ z\geq 0\right] $%
, the distribution function of $\zeta ^{\ast }=\kappa _{2}\zeta $ is $%
\mathbb{P}(\zeta ^{\ast }\leq z)=(1-\exp (-z/2\kappa _{2}))^{2}\mathbf{1}%
\left[ z\geq 0\right] $, where $\zeta ^{\ast }$ is the limiting random
variable of $LR_{n}(s)$ given in (\ref{limLR}) under the local conditional
homoskedasticity. By inverting it, we can obtain the critical values given a
choice of $K(\cdot )$. For instance, the critical values for the Gaussian
kernel is reported in Table \ref{tbl cv}, where $\kappa _{2}=(2\sqrt{\pi }%
)^{-1}\simeq 0.2821$ in this case.

%TCIMACRO{%
%\TeXButton{B}{\begin{table}[tbp] 
%\centering}}%
%BeginExpansion
\begin{table}[tbp] 
\centering%
%EndExpansion
\caption{Simulated Critical Values of the LR Test (Gaussian Kernel)}\label%
{tbl cv}

\bigskip

\begin{tabular}{crrrrrrrr}
\hline\hline
$\mathbb{P}(\zeta ^{\ast }>cv)$ &  & {\small 0.800} & {\small 0.850} & 
{\small 0.900} & {\small 0.925} & {\small 0.950} & {\small 0.975} & {\small %
0.990} \\ \hline
$cv$ &  & {\small 1.268} & {\small 1.439} & {\small 1.675} & {\small 1.842}
& {\small 2.074} & {\small 2.469} & {\small 2.988} \\ \hline
\end{tabular}

\bigskip

\raggedright{\footnotesize Note: $\zeta ^{\ast }$ is the limiting
distribution of $LR_{n}(s)$ under the local conditional homoskedasticity.
The Gaussian kernel is used.}\bigskip 
%TCIMACRO{\TeXButton{E}{\end{table}}}%
%BeginExpansion
\end{table}%
%EndExpansion

In general, we can estimate $\xi _{LR}\left( s\right) $ by%
\begin{equation*}
\widehat{\xi }_{LR}\left( s\right) =\frac{\kappa _{2}\widehat{\delta }^{\top
}\widehat{V}\left( \widehat{\gamma }\left( s\right) ,s\right) \widehat{%
\delta }}{\widehat{\sigma }^{2}(s)\widehat{\delta }^{\top }\widehat{D}\left( 
\widehat{\gamma }\left( s\right) ,s\right) \widehat{\delta }},
\end{equation*}%
where $\widehat{\delta }$ is from (\ref{para-b}) and (\ref{para-d}), and $%
\widehat{\sigma }^{2}(s)$, $\widehat{D}\left( \widehat{\gamma }\left(
s\right) ,s\right) $, and $\widehat{V}\left( \widehat{\gamma }\left(
s\right) ,s\right) $ are the standard Nadaraya-Watson estimators. In
particular, we let $\widehat{\sigma }^{2}(s)=\sum_{i\in \Lambda _{n}}\omega
_{1i}(s)\widehat{u}_{i}^{2}$ with $\widehat{u}_{i}=y_{i}-x_{i}^{\top }%
\widehat{\beta }-x_{i}^{\top }\widehat{\delta }\mathbf{1}\left[ q_{i}\leq 
\widehat{\gamma }\left( s_{i}\right) \right] $, 
\begin{equation*}
\widehat{D}\left( \widehat{\gamma }\left( s\right) ,s\right) =\dsum_{i\in
\Lambda _{n}}\omega _{2i}(s)x_{i}x_{i}^{\top }\text{, \ and }\ \widehat{V}%
\left( \widehat{\gamma }\left( s\right) ,s\right) =\dsum_{i\in \Lambda
_{n}}\omega _{2i}(s)x_{i}x_{i}^{\top }\widehat{u}_{i}^{2}\text{,}
\end{equation*}%
where%
\begin{equation*}
\omega _{1i}(s)=\frac{K\left( (s_{i}-s)/b_{n}\right) }{\sum_{j\in \Lambda
_{n}}K\left( (s_{j}-s)/b_{n}\right) }\text{ \ and \ }\omega _{2i}(s)=\frac{%
\mathbb{K}\left( (q_{i}-\widehat{\gamma }\left( s\right) )/b_{n}^{\prime
},(s_{i}-s)/b_{n}^{\prime \prime }\right) }{\sum_{j\in \Lambda _{n}}\mathbb{K%
}\left( (q_{j}-\widehat{\gamma }\left( s\right) )/b_{n}^{\prime
},(s_{j}-s)/b_{n}^{\prime \prime }\right) }
\end{equation*}%
for some bivariate kernel function $\mathbb{K}(\cdot ,\cdot )$ and bandwidth
parameters $(b_{n}^{\prime },b_{n}^{\prime \prime })$.

Finally, we show the $\sqrt{n}$-consistency of the semiparametric estimators 
$\widehat{\beta }$ and $\widehat{\delta }^{\ast }$ in (\ref{para-b}) and (%
\ref{para-d}). For this purpose, we first obtain the uniform rate of
convergence of $\widehat{\gamma }\left( s\right) $.

\begin{theorem}
\label{u-roc}Under Assumptions ID and A, 
\begin{equation*}
\sup_{s\in \mathcal{S}_{0}}\left\vert \widehat{\gamma }\left( s\right)
-\gamma _{0}\left( s\right) \right\vert =O_{p}\left( \frac{\log n}{%
n^{1-2\epsilon }b_{n}}\right)
\end{equation*}%
provided that $n^{1-2\epsilon }b_{n}^{2}$ does not diverge.
\end{theorem}

\noindent Apparently, the uniform consistency of $\widehat{\gamma }\left(
s\right) $ follows when $\log n/(n^{1-2\epsilon }b_{n})\rightarrow 0$ as $%
n\rightarrow \infty $. Based on this uniform convergence, the following
theorem derives the joint limiting distribution of $\widehat{\beta }$ and $%
\widehat{\delta }^{\ast }$. We let $\widehat{\theta }^{\ast }=(\widehat{%
\beta }^{\top },\widehat{\delta }^{\ast \top })^{\top }$ and $\theta
_{0}^{\ast }=(\beta _{0}^{\top },\delta _{0}^{\ast \top })^{\top }$.

\begin{theorem}
\label{bd}Suppose the conditions in Theorem \ref{u-roc} hold. If we let $\pi
_{n}>0$ such that $\pi _{n}\rightarrow 0$ and $\{\log n/(n^{1-2\epsilon
}b_{n})\}/\pi _{n}\rightarrow 0$ as $n\rightarrow \infty $, we have 
\begin{equation}
\sqrt{n}\left( \widehat{\theta }^{\ast }-\theta _{0}^{\ast }\right)
\rightarrow _{d}\mathcal{N}\left( 0,\Sigma _{X}^{\ast -1}\Omega ^{\ast
}\Sigma _{X}^{\ast -1}\right)  \label{th-an*}
\end{equation}%
as $n\rightarrow \infty $, where%
\begin{equation*}
\Sigma _{X}^{\ast }=\left[ 
\begin{array}{cc}
\mathbb{E}\left[ x_{i}x_{i}^{\top }\mathbf{1}_{i}^{+}\right] & 0 \\ 
0 & \mathbb{E}\left[ x_{i}x_{i}^{\top }\mathbf{1}_{i}^{-}\right]%
\end{array}%
\right] \text{ \ and \ }\Omega ^{\ast }=\lim_{n\rightarrow \infty }\frac{1}{n%
}Var\left[ 
\begin{array}{c}
\sum_{i\in \Lambda _{n}}x_{i}u_{i}\mathbf{1}_{i}^{+} \\ 
\sum_{i\in \Lambda _{n}}x_{i}u_{i}\mathbf{1}_{i}^{-}%
\end{array}%
\right]
\end{equation*}%
with $\mathbf{1}_{i}^{+}=\mathbf{1[}q_{i}>\gamma _{0}(s_{i})]\mathbf{1}%
[s_{i}\in \mathcal{S}_{0}]$ and $\mathbf{1}_{i}^{-}=\mathbf{1[}q_{i}<\gamma
_{0}(s_{i})]\mathbf{1}[s_{i}\in \mathcal{S}_{0}]$.
\end{theorem}

For the second-step estimator $\widehat{\theta }^{\ast }$, we use (\ref%
{para-b}) and (\ref{para-d}), instead of the conventional plug-in
estimation, say $\arg \min_{\beta ,\delta }\sum_{i\in \Lambda
_{n}}(y_{i}-x_{i}^{\top }\beta -x_{i}^{\top }\delta \mathbf{1}[q_{i}\leq 
\widehat{\gamma }\left( s_{i}\right) ])^{2}\mathbf{1}[s_{i}\in \mathcal{S}%
_{0}]$. The reason is that the first-step nonparametric estimator $\widehat{%
\gamma }(\cdot )$ may not be asymptotically orthogonal to the second step.
Unlike the standard semiparametric literature (e.g.,\ Assumption N(c) in 
\cite{Andrews94a}), the asymptotic effect of $\widehat{\gamma }\left(
s\right) $ to the second-step estimation is not easily derived due to the
discontinuity. The new estimation idea above, however, only uses the
observations that are little affected by the estimation error in the first
step to achieve asymptotic orthogonality. As we verify in Lemma \ref{bias1}
in the Appendix, this is done by choosing a large enough $\pi _{n}$ in (\ref%
{para-b}) and (\ref{para-d}) such that the observations that are included in
the second step are outside the uniform convergence bound of $\left\vert 
\widehat{\gamma }\left( s\right) -\gamma _{0}\left( s\right) \right\vert $.
Thanks to the threshold regression structure, we can estimate the parameters
on each side of the threshold even using these subsamples. Meanwhile, we
also want $\pi _{n}\rightarrow 0$ fast enough to include more observations.
By doing so, though we lose some efficiency in finite samples, we can derive
the asymptotic normality of $\widehat{\theta }=(\widehat{\beta }^{\top },%
\widehat{\delta }^{\top })^{\top }$ that has zero mean and achieves the same
asymptotic variance as if $\gamma _{0}(\cdot )$ was known.

By the delta method, Theorem \ref{bd} readily yields the limiting
distribution of $\widehat{\theta }=(\widehat{\beta }^{\top },\widehat{\delta 
}^{\top })^{\top }$ as 
\begin{equation}
\sqrt{n}\left( \widehat{\theta }-\theta _{0}\right) \rightarrow _{d}\mathcal{%
N}\left( 0,\Sigma _{X}^{-1}\Omega \Sigma _{X}^{-1}\right) \text{ \ as }%
n\rightarrow \infty \text{,}  \label{th-an}
\end{equation}%
where 
\begin{equation*}
\Sigma _{X}=\mathbb{E}\left[ z_{i}z_{i}^{\top }\mathbf{1}\left[ s_{i}\in 
\mathcal{S}_{0}\right] \right] \text{ \ and \ }\Omega =\lim_{n\rightarrow
\infty }\frac{1}{n}Var\left[ \sum_{i\in \Lambda _{n}}z_{i}u_{i}\mathbf{1}%
\left[ s_{i}\in \mathcal{S}_{0}\right] \right]
\end{equation*}%
with $z_{i}=[x_{i}^{\top },x_{i}^{\top }\mathbf{1}\left[ q_{i}\leq \gamma
_{0}\left( s_{i}\right) \right] ]^{\top }$. The asymptotic variance
expressions in (\ref{th-an*}) and (\ref{th-an}) allow for cross-sectional
dependence as they have the long-run variance (LRV) forms $\Omega ^{\ast }$
and $\Omega $. They can be consistently estimated by the robust estimator
developed by \cite{Conley07} using $\widehat{u}_{i}=(y_{i}-x_{i}^{\top }%
\widehat{\beta }-x_{i}^{\top }\widehat{\delta }\mathbf{1}[q_{i}\leq \widehat{%
\gamma }\left( s_{i}\right) ])\mathbf{1}[s_{i}\in \mathcal{S}_{0}]$. The
terms $\Sigma _{X}^{\ast }$ and $\Sigma _{X}$ can be estimated by their
sample analogues.

\section{Threshold Contour\label{Section contour}}

When we consider sample splitting over a two-dimensional space (i.e., $q_{i}$
and $s_{i}$ respectively correspond to the latitude and longitude on the
map), the threshold model (\ref{model}) can be generalized to estimate a
nonparametric contour threshold model:%
\begin{equation}
y_{i}=x_{i}^{\top }\beta _{0}+x_{i}^{\top }\delta _{0}\mathbf{1}\left[
g_{0}\left( q_{i},s_{i}\right) \leq 0\right] +u_{i}\text{,}  \label{model2}
\end{equation}%
where the unknown function\ $g_{0}:\mathcal{Q}\times \mathcal{S}\mapsto 
\mathbb{R}$ determines the threshold contour on a random field that yields
sample splitting. An interesting example includes identifying an unknown
closed boundary over the map, such as a city boundary, and an area of a
disease outbreak or airborne pollution. In social science, it can identify a
group boundary or a region in which the agents share common demographic,
political, or economic characteristics.

To relate this generalized form to the original threshold model (\ref{model}%
), we suppose there exists a known center at $\left( q_{i}^{\ast
},s_{i}^{\ast }\right) $ such that $g_{0}\left( q_{i}^{\ast },s_{i}^{\ast
}\right) <0$. Without loss of generality, we can normalize $\left(
q_{i}^{\ast },s_{i}^{\ast }\right) $ to be $\left( 0,0\right) $ and
re-center the original location variables $(q_{i},s_{i})$ accordingly. In
addition, we define the radius distance $l_{i}$ and angle $a_{i}^{\circ }$
of the $i$th observation relative to the origin as 
\begin{eqnarray*}
l_{i} &=&(q_{i}^{2}+s_{i}^{2})^{1/2}\text{,} \\
a_{i}^{\circ } &=&\bar{a}_{i}^{\circ }\mathbf{I}_{i}+\left( 180^{\circ }-%
\bar{a}_{i}^{\circ }\right) \mathbf{II}_{i}+\left( 180^{\circ }+\bar{a}%
_{i}^{\circ }\right) \mathbf{III}_{i}+\left( 360^{\circ }-\bar{a}_{i}^{\circ
}\right) \mathbf{IV}_{i}\text{,}
\end{eqnarray*}%
where $\bar{a}_{i}^{\circ }=\arctan \left( \left\vert q_{i}/s_{i}\right\vert
\right) $, and each of $(\mathbf{I}_{i},\mathbf{II}_{i},\mathbf{III}_{i},%
\mathbf{IV}_{i})$ respectively denotes the indicator that the $i$th
observation locates in the first, second, third, and forth quadrant.

We suppose that there is only one threshold at any angle and the threshold
contour is star-shaped. For each chosen angle $a^{\circ }\in \lbrack
0^{\circ },360^{\circ })$, we rotate the original coordinate
counterclockwise and implement the least squares estimation (\ref{reg}) only
using the observations in the first two quadrants after rotation. It will
ensure that the threshold mapping after rotation is a well-defined function.

\begin{figure}[tbp]  
\centering%
\caption{Illustration of rotation (color online)}\label{fig contour}
\includegraphics[width=1\textwidth]{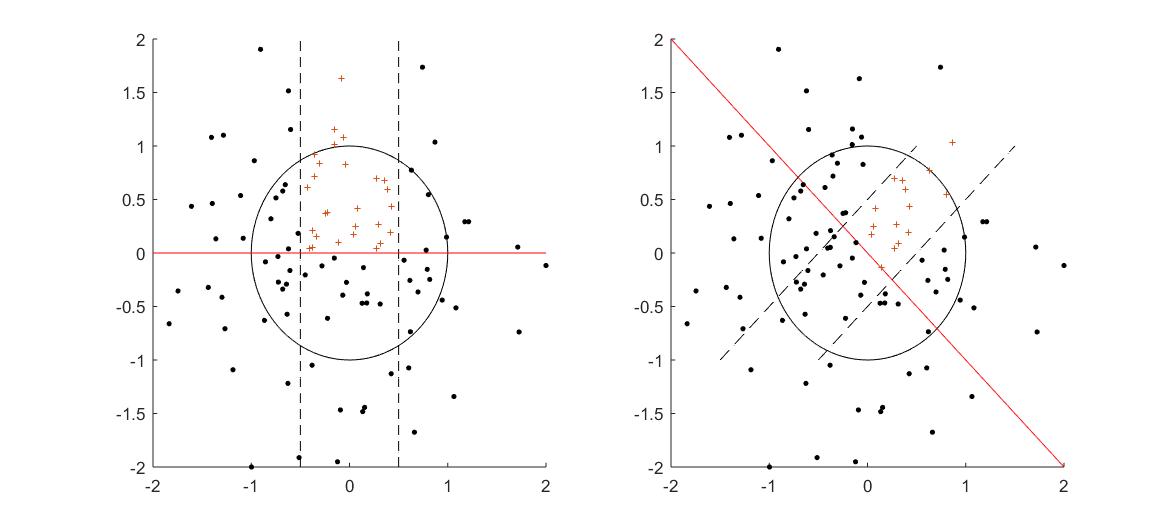}

\end{figure}%

In particular, the angle relative to the origin is $a_{i}^{\circ }-a^{\circ
} $ after rotating the coordinate by $a^{\circ }$ degrees counterclockwise,
and the new location (after the rotation) is given as $(q_{i}\left( a^{\circ
}\right) ,s_{i}\left( a^{\circ }\right) )$, where%
\begin{equation*}
\left( 
\begin{array}{c}
q_{i}\left( a^{\circ }\right) \\ 
s_{i}\left( a^{\circ }\right)%
\end{array}%
\right) =\left( 
\begin{array}{c}
q_{i}\cos \left( a^{\circ }\right) -s_{i}\sin \left( a^{\circ }\right) \\ 
s_{i}\cos \left( a^{\circ }\right) +q_{i}\sin \left( a^{\circ }\right)%
\end{array}%
\right) \text{.}
\end{equation*}%
After this rotation, we estimate the following nonparametric threshold model:%
\begin{equation}
y_{i}=x_{i}^{\top }\beta _{0}+x_{i}^{\top }\delta _{0}\mathbf{1}\left[
q_{i}\left( a^{\circ }\right) \leq \gamma _{a^{\circ }}\left( s_{i}\left(
a^{\circ }\right) \right) \right] +u_{i}  \label{model3}
\end{equation}%
using only the observations $i$ satisfying $q_{i}\left( a^{\circ }\right)
\geq 0$ and in the neighborhood of $s_{i}\left( a^{\circ }\right) =0$, where 
$\gamma _{a^{\circ }}\left( \cdot \right) $ is the unknown threshold curve
as in the original model (\ref{model}) on the $a^{\circ }$-degree-rotated
coordinate plane. Such reparametrization guarantees that $\gamma _{a^{\circ
}}\left( \cdot \right) $ is always positive and it is estimated at the
origin. Figure \ref{fig contour} illustrates the idea of such rotation and
pointwise estimation over a bounded support so that only the red cross
points are included for estimation at different angles. Thus, the estimation
and inference procedures developed in the previous sections are directly
applicable, though we expect some efficiency loss as we only use the
subsample with $q_{i}\left( a^{\circ }\right) \geq 0$ at each $a^{\circ }$.

\section{Monte Carlo Experiments\label{Section simulation}}

We examine the small sample performance of the semiparametric threshold
regression estimator by Monte Carlo simulations. We generate $n$ draws from 
\begin{equation}
y_{i}=x_{i}^{\top }\beta _{0}+x_{i}^{\top }\delta _{0}\mathbf{1}\left[
q_{i}\leq \gamma _{0}\left( s_{i}\right) \right] +u_{i}\text{,}
\label{sim-model}
\end{equation}%
where $x_{i}=(1,x_{2i})^{\top }$ and $x_{2i}\in \mathbb{R}$. We let $\beta
_{0}=(\beta _{10},\beta _{20})^{\top }=0\iota _{2}$ and consider three
different values of $\delta _{0}=(\delta _{10},\delta _{20})^{\top }=\delta
\iota _{2}$ with $\delta =1,2,3,4$, where $\iota _{2}=(1,1)^{\top }$. For
the threshold function, we let $\gamma _{0}\left( s\right) =\sin (s)/2$. We
consider the cross-sectional dependence structure in $\left(
x_{2i},q_{i},s_{i},u_{i}\right) ^{\top }$ as follows:%
\begin{equation}
\left\{ 
\begin{array}{l}
\left( q_{i},s_{i}\right) ^{\top }\sim iid\mathcal{N}\left( 0,I_{2}\right) 
\text{;} \\ 
x_{2i}|\left( q_{i},s_{i}\right) \sim iid\mathcal{N}\left( 0,(1+\rho \left(
s_{i}^{2}+q_{i}^{2}\right) )^{-1}\right) \text{;} \\ 
\underline{\mathbf{u}}|\{(x_{i},q_{i},s_{i})\}_{i=1}^{n}\sim \mathcal{N}%
\left( 0,\Sigma \right) \text{,}%
\end{array}%
\right.  \label{DGP corr}
\end{equation}%
where $\underline{\mathbf{u}}=(u_{1},\ldots ,u_{n})^{\top }$. The $(i,j)$th
element of $\Sigma $ is $\Sigma _{ij}=\rho ^{\lfloor \ell _{ij}n\rfloor }%
\mathbf{1}[\ell _{ij}<m/n]$, where $\ell _{ij}=\{\left( s_{i}-s_{j}\right)
^{2}+\left( q_{i}-q_{j}\right) ^{2}\}^{1/2}$ is the $L^{2}$-distance between
the $i$th and $j$th observations. The diagonal elements of $\Sigma $ are
normalized as $\Sigma _{ii}=1$. This $m$-dependent setup follows from the
Monte Carlo experiment in \cite{Conley07} in the sense that each unit can be
cross-sectionally correlated with at most $2m^{2}$ observations. Within the $%
m$ distance, the dependence decays at a polynomial rate as indicated by $%
\rho ^{\lfloor \ell _{ij}n\rfloor }$. The parameter $\rho $ describes the
strength of cross-sectional dependence in the way that a larger $\rho $
leads to stronger dependence relative to the unit standard deviation. In
particular, we consider the cases with $\rho =0$ (i.e., i.i.d.
observations), $0.5$, and $1$. We consider the sample size $n=100$, $200$,
and $500$, and set $\mathcal{S}_{0}$ to include the middle 70\% observations
of $s_{i}$.

%TCIMACRO{%
%\TeXButton{B}{\begin{table}[tbp] 
%\centering}}%
%BeginExpansion
\begin{table}[tbp] 
\centering%
%EndExpansion
\caption{Rej. Prob. of the LR Test with i.i.d. Data}\label{tbl r1}

\bigskip

\begin{tabular}{cccccccccccccccc}
\hline\hline
&  & \multicolumn{4}{c}{${\small s=0.0}$} &  & \multicolumn{4}{c}{${\small %
s=0.5}$} &  & \multicolumn{4}{c}{${\small s=1.0}$} \\ 
\cline{3-6}\cline{8-11}\cline{13-16}
${\small n}$ & \multicolumn{1}{l}{${\small \delta =}$} & {\small 1} & 
{\small 2} & {\small 3} & {\small 4} &  & {\small 1} & {\small 2} & {\small 3%
} & {\small 4} &  & {\small 1} & {\small 2} & {\small 3} & {\small 4} \\ 
\hline
\multicolumn{1}{r}{\small 100} &  & {\small 0.16} & {\small 0.09} & {\small %
0.06} & {\small 0.08} &  & {\small 0.19} & {\small 0.10} & {\small 0.09} & 
{\small 0.08} &  & {\small 0.25} & {\small 0.18} & {\small 0.16} & {\small %
0.12} \\ 
\multicolumn{1}{r}{\small 200} &  & {\small 0.09} & {\small 0.06} & {\small %
0.06} & {\small 0.07} &  & {\small 0.12} & {\small 0.06} & {\small 0.04} & 
{\small 0.06} &  & {\small 0.18} & {\small 0.09} & {\small 0.08} & {\small %
0.06} \\ 
\multicolumn{1}{r}{\small 500} &  & {\small 0.08} & {\small 0.05} & {\small %
0.05} & {\small 0.06} &  & {\small 0.08} & {\small 0.04} & {\small 0.04} & 
{\small 0.05} &  & {\small 0.09} & {\small 0.04} & {\small 0.04} & {\small %
0.03} \\ \hline
\end{tabular}

\bigskip

\raggedright {\footnotesize Note: Entries are rejection probabilities of the
LR test (\ref{LR}) when data are generated from (\ref{sim-model}) with $%
\gamma _{0}\left( s\right) =\sin (s)/2$. The dependence structure is given
in (\ref{DGP corr})\ with $\rho =0$.\ The significance level\ is $5\%$\ and
the results are based on 1000 simulations.}%
%TCIMACRO{\TeXButton{E}{\end{table}}}%
%BeginExpansion
\end{table}%
%EndExpansion

%TCIMACRO{%
%\TeXButton{B}{\begin{table}[tbp] 
%\centering}}%
%BeginExpansion
\begin{table}[tbp] 
\centering%
%EndExpansion
\caption{Rej. Prob. of the LR Test with Cross-sectionally Correlated Data}%
\label{tbl r3}

\bigskip

\begin{tabular}{cccccccccccccccc}
\hline\hline
&  & \multicolumn{4}{c}{${\small s=0.0}$} &  & \multicolumn{4}{c}{${\small %
s=0.5}$} &  & \multicolumn{4}{c}{${\small s=1.0}$} \\ 
\cline{3-6}\cline{8-11}\cline{13-16}
${\small n}$ & \multicolumn{1}{l}{${\small \delta =}$} & {\small 1} & 
{\small 2} & {\small 3} & {\small 4} &  & {\small 1} & {\small 2} & {\small 3%
} & {\small 4} &  & {\small 1} & {\small 2} & {\small 3} & {\small 4} \\ 
\hline
\multicolumn{1}{r}{\small 100} &  & {\small 0.18} & {\small 0.09} & {\small %
0.07} & {\small 0.08} &  & {\small 0.21} & {\small 0.11} & {\small 0.10} & 
{\small 0.06} &  & {\small 0.28} & {\small 0.20} & {\small 0.17} & {\small %
0.13} \\ 
\multicolumn{1}{r}{\small 200} &  & {\small 0.12} & {\small 0.06} & {\small %
0.06} & {\small 0.06} &  & {\small 0.13} & {\small 0.08} & {\small 0.06} & 
{\small 0.05} &  & {\small 0.20} & {\small 0.37} & {\small 0.09} & {\small %
0.06} \\ 
\multicolumn{1}{r}{\small 500} &  & {\small 0.08} & {\small 0.04} & {\small %
0.06} & {\small 0.06} &  & {\small 0.06} & {\small 0.06} & {\small 0.04} & 
{\small 0.05} &  & {\small 0.13} & {\small 0.08} & {\small 0.05} & {\small %
0.02} \\ \hline
\end{tabular}

\bigskip

\raggedright {\footnotesize Note: Entries are rejection probabilities of the
LR test (\ref{LR}) when data are generated from (\ref{sim-model}) with $%
\gamma _{0}\left( s\right) =\sin (s)/2$. The dependence structure is given
in (\ref{DGP corr})\ with $\rho =1$\ and $m=10$.\ The significance level\ is 
$5\%$\ and the results are based on $1000$ simulations.}%
%TCIMACRO{\TeXButton{E}{\end{table}}}%
%BeginExpansion
\end{table}%
%EndExpansion

First, Tables \ref{tbl r1} and \ref{tbl r3} report the small sample
rejection probabilities of the LR test in (\ref{LR}) for $H_{0}:\gamma
_{0}(s)=\sin (s)/2$ against $H_{1}:\gamma _{0}(s)\neq \sin (s)/2$ at the 5\%
nominal level at three different locations $s=0$, $0.5$, and $1$. In
particular, Table \ref{tbl r1} examines the case with no cross-sectional
dependence ($\rho =0$), while Table \ref{tbl r3} examines the case with
cross-sectional dependence whose dependence decays slowly with $\rho =1$ and 
$m=10$. For the bandwidth parameter, we normalize $s_{i}$ and $q_{i}$ to
have zero mean and unit standard deviation, and choose $b_{n}=0.5n^{-1/2}$
in the main regression. This choice is for undersmoothing so that $%
n^{1-2\epsilon }b_{n}^{2}=n^{-2\epsilon }\rightarrow 0$. To estimate $%
D\left( \gamma _{0}\left( s\right) ,s\right) $ and $V\left( \gamma
_{0}\left( s\right) ,s\right) $, we use the rule-of-thumb bandwidths from
the standard kernel regression satisfying $b_{n}^{\prime }=O(n^{-1/5})$ and $%
b_{n}^{\prime \prime }=O(n^{-1/6})$. All the results are based on $1000$
simulations. In general, the test for $\gamma _{0}$ performs better as (i)
the sample size gets larger; (ii) the coefficient change gets more
significant; (iii) the cross-sectional dependence gets weaker; and (iv) the
target gets closer to the mid-support of $s$. When $\delta _{0}$ and $n$ are
large, the LR test is conservative, which is also found in the classical
threshold regression (e.g., \cite{Hansen00a}).

%TCIMACRO{%
%\TeXButton{B}{\begin{table}[tbp] 
%\centering}}%
%BeginExpansion
\begin{table}[tbp] 
\centering%
%EndExpansion
\caption{Coverage Prob. of the Plug-in Confidence Interval}\label{tbl bd
plugin}

\bigskip

\begin{tabular}{lccccccccccccccc}
\hline\hline
&  & \multicolumn{4}{c}{$\beta _{20}$} &  & \multicolumn{4}{c}{$\beta
_{20}+\delta _{20}$} &  & \multicolumn{4}{c}{$\delta _{20}$} \\ 
\cline{3-6}\cline{8-11}\cline{13-16}
\multicolumn{1}{c}{${\small n}$} & \multicolumn{1}{l}{${\small \delta =}$} & 
{\small 1} & {\small 2} & {\small 3} & {\small 4} &  & {\small 1} & {\small 2%
} & {\small 3} & {\small 4} &  & {\small 1} & {\small 2} & {\small 3} & 
{\small 4} \\ \hline
\multicolumn{1}{r}{\small 100} &  & {\small 0.85} & {\small 0.87} & {\small %
0.90} & {\small 0.89} &  & {\small 0.82} & {\small 0.89} & {\small 0.88} & 
{\small 0.88} &  & {\small 0.83} & {\small 0.88} & {\small 0.89} & {\small %
0.90} \\ 
\multicolumn{1}{r}{\small 200} &  & {\small 0.87} & {\small 0.91} & {\small %
0.91} & {\small 0.91} &  & {\small 0.87} & \multicolumn{1}{l}{\small 0.90} & 
{\small 0.93} & {\small 0.93} &  & {\small 0.86} & {\small 0.90} & {\small %
0.92} & {\small 0.94} \\ 
\multicolumn{1}{r}{\small 500} &  & {\small 0.89} & {\small 0.92} & {\small %
0.95} & {\small 0.94} &  & {\small 0.87} & \multicolumn{1}{l}{\small 0.93} & 
{\small 0.93} & {\small 0.94} &  & {\small 0.85} & {\small 0.92} & {\small %
0.95} & {\small 0.92} \\ \hline
\end{tabular}

\bigskip

\raggedright {\footnotesize Note: Entries are coverage probabilities of 95\%
confidence intervals for $\beta _{20}$, $\beta _{20}+\delta _{20},$\ and $%
\delta _{20}$. Data are generated from (\ref{sim-model})\ with $\gamma
_{0}\left( s\right) =\sin (s)/2$, where the dependence structure is given in
(\ref{DGP corr})\ with $\rho =0.5$\ and $m=3$.\ The results are based on
1000 simulations.}%
%TCIMACRO{\TeXButton{E}{\end{table}}}%
%BeginExpansion
\end{table}%
%EndExpansion

%TCIMACRO{%
%\TeXButton{B}{\begin{table}[tbp] 
%\centering}}%
%BeginExpansion
\begin{table}[tbp] 
\centering%
%EndExpansion
\caption{Coverage Prob. of the Plug-in Confidence Interval (w/ LRV adj.)}%
\label{tbl bd lrv}

\bigskip

\begin{tabular}{lccccccccccccccc}
\hline\hline
&  & \multicolumn{4}{c}{$\beta _{20}$} &  & \multicolumn{4}{c}{$\beta
_{20}+\delta _{20}$} &  & \multicolumn{4}{c}{$\delta _{20}$} \\ 
\cline{3-6}\cline{8-11}\cline{13-16}
\multicolumn{1}{c}{${\small n}$} & \multicolumn{1}{l}{${\small \delta =}$} & 
{\small 1} & {\small 2} & {\small 3} & {\small 4} &  & {\small 1} & {\small 2%
} & {\small 3} & {\small 4} &  & {\small 1} & {\small 2} & {\small 3} & 
{\small 4} \\ \hline
\multicolumn{1}{r}{\small 100} &  & {\small 0.94} & {\small 0.94} & {\small %
0.94} & {\small 0.94} &  & {\small 0.91} & {\small 0.95} & {\small 0.95} & 
{\small 0.95} &  & {\small 0.92} & {\small 0.94} & {\small 0.96} & {\small %
0.96} \\ 
\multicolumn{1}{r}{\small 200} &  & {\small 0.94} & {\small 0.95} & {\small %
0.96} & {\small 0.96} &  & {\small 0.93} & \multicolumn{1}{l}{\small 0.94} & 
{\small 0.96} & {\small 0.96} &  & {\small 0.92} & {\small 0.96} & {\small %
0.97} & {\small 0.97} \\ 
\multicolumn{1}{r}{\small 500} &  & {\small 0.94} & {\small 0.95} & {\small %
0.98} & {\small 0.97} &  & {\small 0.92} & \multicolumn{1}{l}{\small 0.96} & 
{\small 0.97} & {\small 0.96} &  & {\small 0.91} & {\small 0.97} & {\small %
0.97} & {\small 0.96} \\ \hline
\end{tabular}

\bigskip

\raggedright {\footnotesize Note: Entries are coverage probabilities of 95\%
confidence intervals for $\beta _{20}$, $\beta _{20}$$+\delta _{20},$\ and $%
\delta _{20} $\ with a small sample adjustment of the LRV estimator. Data
are generated from (\ref{sim-model})\ with $\gamma _{0}\left( s\right)
\left. =\right. \sin (s)/2$, where the dependence structure is given in (\ref%
{DGP corr})\ with $\rho =0.5$\ and $m=3$.\ The results are based on 1000
simulations.}%
%TCIMACRO{%
%\TeXButton{E}{
%\end{table}}}%
%BeginExpansion

\end{table}%
%EndExpansion

Second, Table \ref{tbl bd plugin} shows the finite sample coverage
properties of the 95\% confidence intervals for the parametric components $%
\beta _{20}$, $\delta _{20}^{\ast }=\beta _{20}+\delta _{20}$, and $\delta
_{20}$. The results are based on the same simulation design as above with $%
\rho =0.5$ and $m=3$. Regarding the tuning parameters, we use the same
bandwidth choice $b_{n}=0.5n^{-1/2}$ as before and set the truncation
parameter $\pi _{n}=\left( nb_{n}\right) ^{-1/2}$. Unreported results
suggest that choice of the constant in the bandwidth matters particularly
with small samples like $n=100$, but such effect quickly decays as the
sample size gets larger. For the estimator of the LRV, we use the spatial
lag order of $5$ following \cite{Conley07}. Results with other lag choices
are similar and hence omitted. The result suggests that the asymptotic
normality is better approximated with larger samples and larger change
sizes. Table \ref{tbl bd lrv} shows the same results with\ a small sample
adjustment of the LRV estimator for $\Omega ^{\ast }$ by dividing it by the
sample truncation fraction $\sum_{i\in \Lambda _{n}}(\mathbf{1}[q_{i}>%
\widehat{\gamma }(s_{i})+\pi _{n}]+\mathbf{1}[q_{i}<\widehat{\gamma }%
(s_{i})-\pi _{n}])\mathbf{1}[s_{i}\in \mathcal{S}_{0}]/\sum_{i\in \Lambda
_{n}}\mathbf{1}[s_{i}\in \mathcal{S}_{0}]$. This ratio enlarges the LRV
estimator and hence the coverage probabilities, especially when the change
size is small. It only affects the finite sample performance as it
approaches one in probability as $n\rightarrow \infty $.

\section{Applications\label{Section empirics}}

\subsection{Tipping point and social segregation\label{Section tipping}}

The first application is about the tipping point problem in social
segregation, which stimulates a vast literature in labor, public, and
political economics. \cite{Schelling71} initially proposes the tipping point
model to study the fact that the white population decreases substantially
once the minority share exceeds a certain tipping point. \cite{Card08}
empirically estimate this model and find strong evidence for such a tipping
point phenomenon. In particular, they specify the threshold regression model
as%
\begin{equation*}
y_{i}=\beta _{10}+\delta _{10}\mathbf{1}\left[ q_{i}\leq \gamma _{0}\right]
+x_{2i}^{\top }\beta _{20}+u_{i}\text{,}
\end{equation*}%
where for tract $i$ in a certain city, $q_{i}$ is the minority share in
percentage at the beginning of a certain decade, $y_{i}$ is the normalized
white population change in percentage within this decade, and $x_{2i}$ is a
vector of control variables. They apply the least squares method to estimate
the tipping point $\gamma _{0}$. For most cities and for the periods
1970-80, 1980-90, and 1990-2000, they find that white population flows
exhibit the tipping-like behavior, with the estimated tipping points ranging
approximately from 5\% to 20\% across cities.

In Section VII of \cite{Card08}, they also find that the location of the
tipping point substantially depends on white residents' attitudes toward the
minority. Specifically, they first construct a city-level index that
measures white attitudes and regress the estimated tipping point from each
city on this index. The regression coefficient is significantly different
from zero, suggesting that the tipping point is heterogeneous across cities.

We go one step further by considering a more flexible model in the tract
level given as%
\begin{equation*}
y_{i}=\beta _{10}+\delta _{10}\mathbf{1}\left[ q_{i}\leq \gamma _{0}(s_{i})%
\right] +x_{2i}^{\top }\beta _{20}+u_{i}\text{,}
\end{equation*}%
where $\gamma _{0}(\cdot )$ denotes an unknown tipping point function and $%
s_{i}$ denotes the attitude index. The nonparametric function $\gamma
_{0}(\cdot )$ here allows for heterogeneous tipping points across tracts
depending on the level of the attitude index $s_{i}$ in tract $i$.
Unfortunately, the attitude index by \cite{Card08} is only available at the
aggregated city-level, and hence we cannot use it to analyze the census
tract-level observations. For this reason, we instead use the tract-level
unemployment rate as $s_{i}$ to illustrate the nonparametric threshold
function, which is readily available in the original dataset. Such a
compromise is far from being perfect but can be partially justified since
race discrimination has been widely documented to be correlated with
employment (e.g., \cite{DarityMason1998}).

\begin{figure}[tbp] 
\begin{center}
%EndExpansion
\caption{Estimate of the tipping point as a function of the unemployment
rate}\label{fig tipping}%
%TCIMACRO{\TeXButton{more space}{\vspace{+4ex}}}%
%BeginExpansion
\vspace{+4ex}%
%EndExpansion

{\footnotesize Panel A: Estimated tipping point function in Chicago 1980-90}
\includegraphics[width=6.06in,height=2.03in]{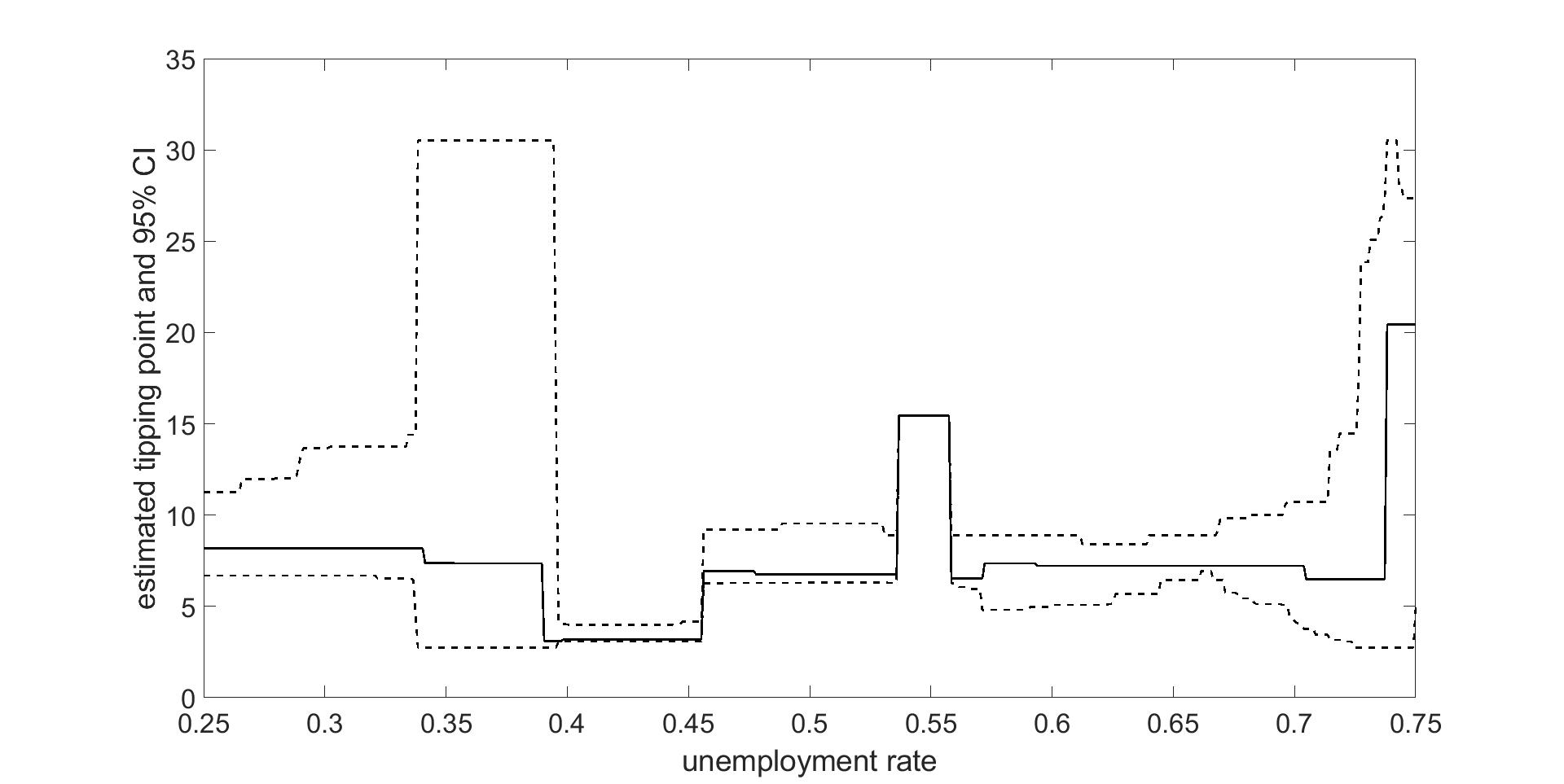}

{\footnotesize Panel B: Estimated tipping point function in Los Angeles 1980-90}
\includegraphics[width=6.06in,height=2.03in]{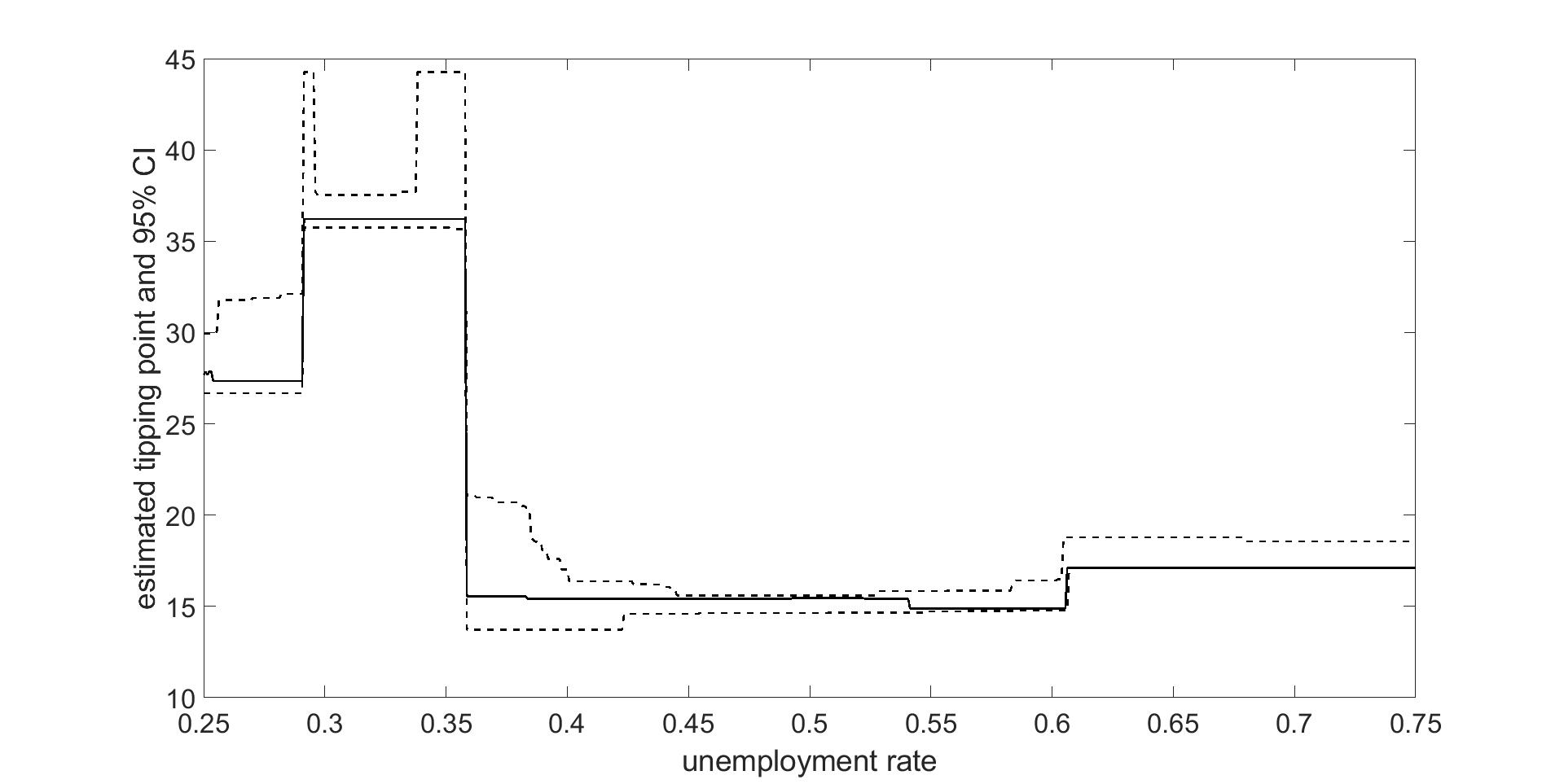}

{\footnotesize Panel C: Estimated tipping point function in New York City 1980-90}
\includegraphics[width=6.06in,height=2.03in]{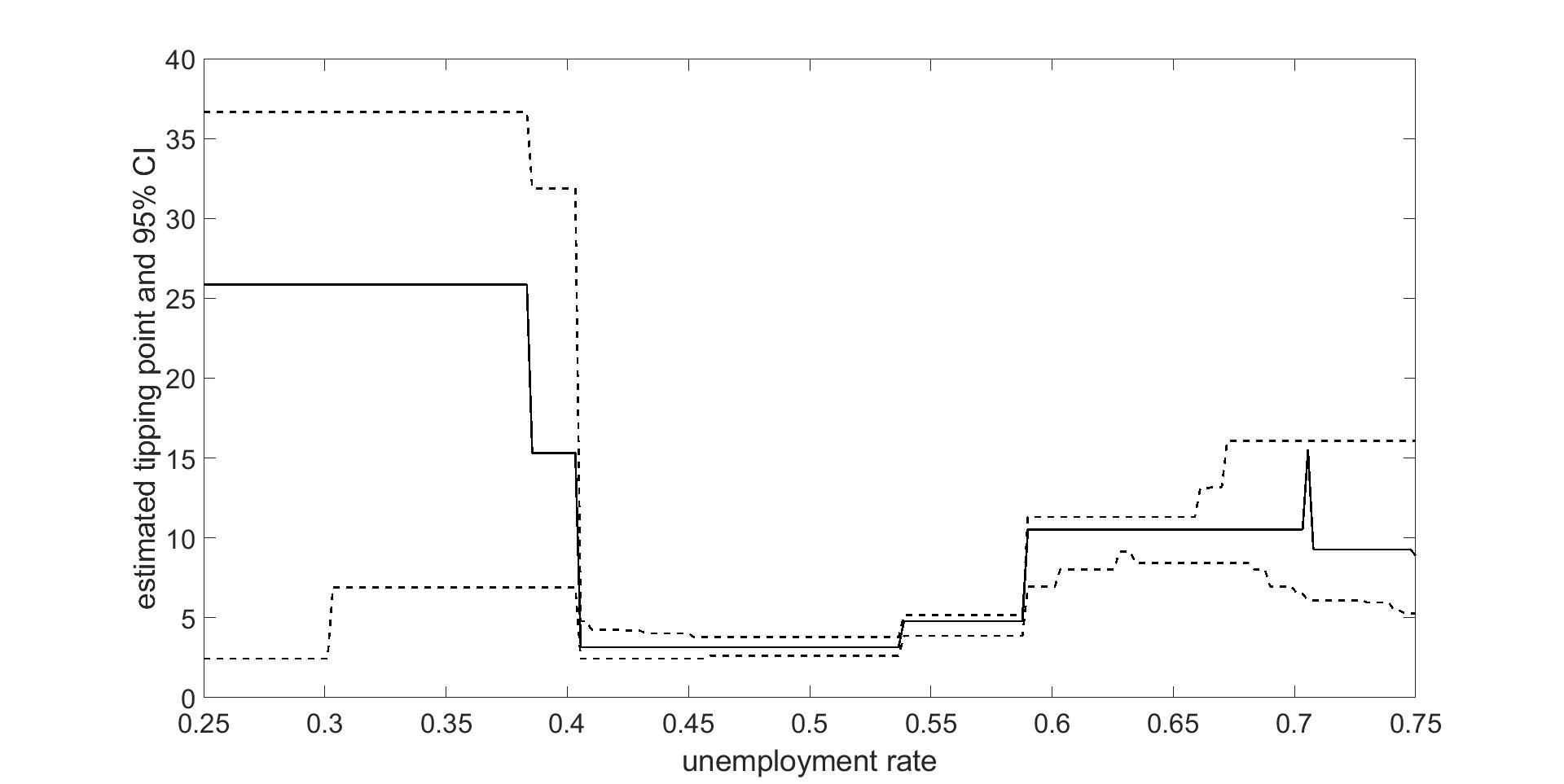}

\end{center}
\begin{small}%
%EndExpansion
Note: The figure depicts the point estimate of the tipping points as a
function of the unemployment rate, using the data in Chicago, Los Angeles, and
New York City in 1980-1990. The vertical axis is the estimated tipping point
in percentage, and the horizontal axis is the tract-level unemployment
normalized into quantile level. Data are available from \cite{Card08}.%
%TCIMACRO{\TeXButton{endsmall}{\end{small}}}%
%BeginExpansion
\end{small}%
%EndExpansion
%TCIMACRO{\TeXButton{E}{\end{figure}}}%
%BeginExpansion
\end{figure}%
%EndExpansion

We use the data provided by \cite{Card08} and estimate the tipping point
function $\gamma _{0}(\cdot )$ over census tracts by the method introduced
in Section \ref{Section estimation}. As in their work, we drop the tracts
where the minority shares are above 60 percentage points and use five
control variables as $x_{2i}$, including the logarithm of mean family
income, the fractions of single-unit, vacant, and renter-occupied housing
units, and the fraction of workers who use public transport to travel to
work. The bandwidth is set as $b_{n}=cn^{-1/2}$ for some $c>0$, so that it
satisfies the technical conditions in the previous sections, where the
constant $c$ is chosen by the leave-one-out cross validation. In particular,
we first construct the leave-one-out estimate, $\widehat{\gamma }_{-i}\left(
s_{i}\right) $, of $\gamma _{0}\left( s_{i}\right) $ as in (\ref{g-hat0})
without using the $i$th observation. Then, leaving the $i$th observation
out, we construct $\widehat{\beta }_{-i}$ and $\widehat{\delta }_{-i}$ as in
(\ref{para-b}) and (\ref{para-d}) with $\pi _{n}=\left( nb_{n}\right)
^{-1/2} $ using the bandwidth $b_{n}$ chosen in the previous step. We choose
the bandwidth that minimizes $\sum_{i\in \Lambda _{n}}(y_{i}-\widehat{\beta }%
_{1,-i}-\widehat{\delta }_{1,-i}\mathbf{1}\left[ q_{i}\leq \widehat{\gamma }%
_{-i}(s_{i})\right] -x_{2i}^{\top }\widehat{\beta }_{2,-i})^{2}\mathbf{1}%
\left[ s_{i}\in \mathcal{S}_{0}\right] $, where $\mathcal{S}_{0}$ again
includes the middle 70\% quantiles of $s_{i}$.

Figure \ref{fig tipping} depicts the estimated tipping points and the 95\%
pointwise confidence intervals by inverting the likelihood ratio test
statistic (\ref{LR}) in the years 1980-90 in Chicago, Los Angeles, and New
York City, whose sample sizes are relatively large. For each city, the
constant $c$ of the bandwidth $b_{n}=cn^{-1/2}$ chosen by the aforementioned
cross validation is $3.20$, $4.87$, and $3.42$, respectively. We make the
following comments. First, the estimates of the tipping points vary
substantially in the unemployment rate within all three cities. Therefore,
the standard constant tipping point model is insufficient to characterize
the segregation fully. Second, the tipping points as functions of the
unemployment rate do not exhibit the same pattern across cities, reinforcing
the heterogeneous tipping points in the city-level as found in \cite{Card08}%
. Finally, the estimated tipping point $\widehat{\gamma }\left( s\right) $
as a function of $s$ can be discontinuous, which does not contrast with
Assumption A-(vi), that is, the true function $\gamma _{0}\left( \cdot
\right) $ is smooth. The discontinuity comes from the fact that $\widehat{%
\gamma }\left( s\right) $ is obtained by grid search and can only take
values among the discrete points $\{q_{1},...,q_{n}\}$ in finite samples.

\subsection{Metropolitan area determination\label{Section boundary}}

The second application is about determining the boundary of a metropolitan
area, which is a fundamental question in urban economics. Recently,
researchers propose to use nighttime light intensity obtained by satellite
imagery to define metropolitan areas. The intuition is straightforward:
metropolitan areas are bright at night while rural areas are dark.

\begin{figure}[tbp] 
\begin{center}%
%EndExpansion
\caption{Nighttime light intensity in Dallas, Texas, in 2010}\label{fig raw}%
%TCIMACRO{\TeXButton{more space}{\vspace{+4ex}}}%
%BeginExpansion
\vspace{+4ex}%
%EndExpansion
\includegraphics[width=3.646in,height=2.508in]{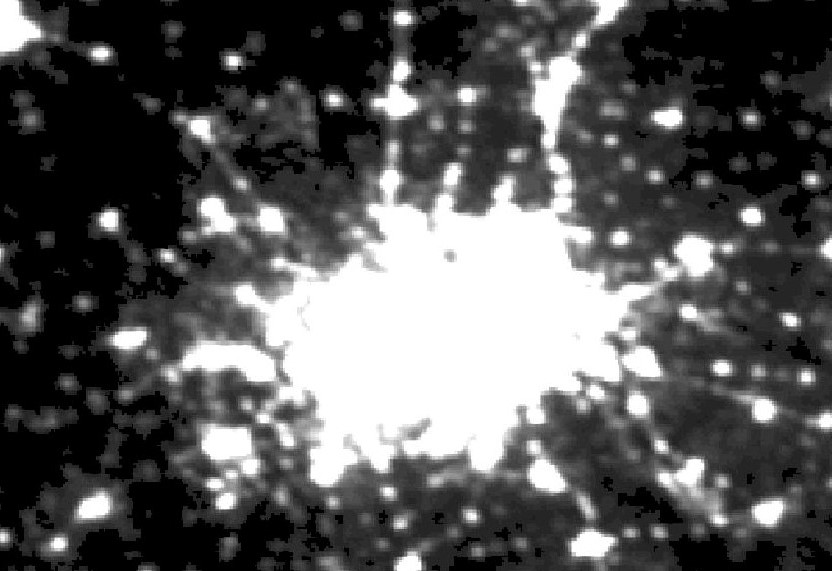}

\end{center}
\begin{small}%
%EndExpansion
Note: The figure depicts the intensity of the stable nighttime light in
Dallas, TX 2010. Data are available from https://www.ncei.noaa.gov/.\ 
%TCIMACRO{\TeXButton{endsmall}{\end{small}}}%
%BeginExpansion
\end{small}%
%EndExpansion
%TCIMACRO{\TeXButton{E}{\end{figure}}}%
%BeginExpansion
\end{figure}%
%EndExpansion

Specifically, the National Oceanic and Atmospheric Administration (NOAA)
collects satellite imagery of nighttime lights at approximately 1-kilometer
resolution since 1992. NOAA further constructs several indices measuring the
annual light intensity. Following the literature (e.g., \cite{Dingel19}), we
choose the \textquotedblleft average visible, stable
lights\textquotedblright\ index that ranges from 0 (dark) to 63 (bright).
For illustration, we focus on Dallas, Texas and use the data from the years
1995, 2000, 2005, and 2010. In each year, the data are recorded as a 240$%
\times $360 grid that covers the latitudes from 32$^{\circ }$N to 34$^{\circ
}$N and the longitudes from 98.5$^{\circ }$W to 95.5$^{\circ }$W. The total
sample size is 240$\times $360=86400 each year. These data are available at
NOAA's website and also provided on the authors' website. Figure \ref{fig
raw} depicts the intensity of the stable nighttime light of the Dallas area
in 2010 as an example.

Let $y_{i}$ be the level of nighttime light intensity and $\left(
q_{i},s_{i}\right) $ be the latitude and longitude of the $i$th pixel, which
is normalized into the equally-spaced grids on $[0,1]^{2}$. To define the
metropolitan area, existing literature in urban economics first chooses an 
\textit{ad hoc} intensity threshold, say 95\% quantile of $y_{i}$, and
categorizes the $i$th pixel as a part of the metropolitan area if $y_{i}$ is
larger than the threshold. See \cite{Dingel19}, \cite{Vogel19}, and
references therein. In particular, on p.3 in \cite{Dingel19}, they note that
\textquotedblleft \lbrack ...] the choice of the light-intensity threshold,
which governs the definitions of the resulting metropolitan areas, is not
pinned down by economic theory or prior empirical
research.\textquotedblright\ Our new approach can provide a data-driven
guidance of choosing the intensity threshold from the econometric
perspective.

\begin{figure}[tbp] 
\begin{center}%
%EndExpansion
\caption{Kernel density estimate of nighttime light intensity, Dallas 2010}%
\label{fig ksdensity}
\includegraphics[width=4.1632in,height=2.9334in]{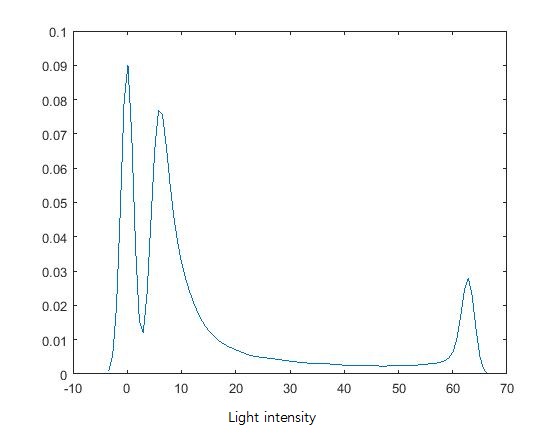}

\end{center}
\begin{small}%
%EndExpansion
Note: The figure depicts the kernel density estimate of the strength of the
stable nighttime light in Dallas, TX 2010. Data are available from
https://www.ncei.noaa.gov/.\ 
%TCIMACRO{\TeXButton{endsmall}{\end{small}}}%
%BeginExpansion
\end{small}%
%EndExpansion
%TCIMACRO{\TeXButton{E}{\end{figure}}}%
%BeginExpansion
\end{figure}%

\begin{figure}[tbp]
\begin{center}%
%EndExpansion
\caption{Metropolitan area determination in Dallas (color online)}
\label{fig cityboundary}
\includegraphics[width=3.0in,height=2.3609in]{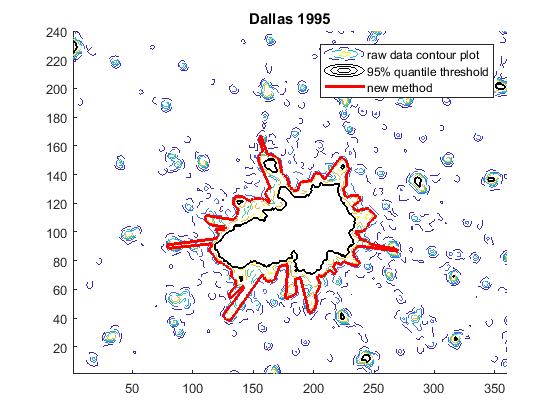}
\includegraphics[width=3.0in,height=2.3609in]{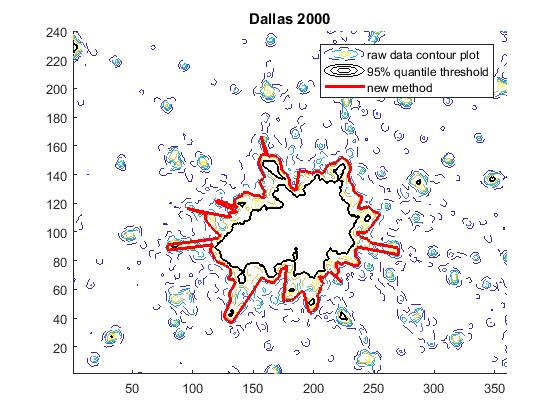}
\includegraphics[width=3.0in,height=2.3609in]{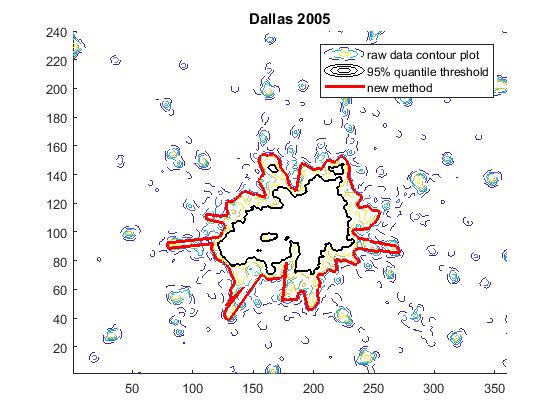}
\includegraphics[width=3.0in,height=2.3609in]{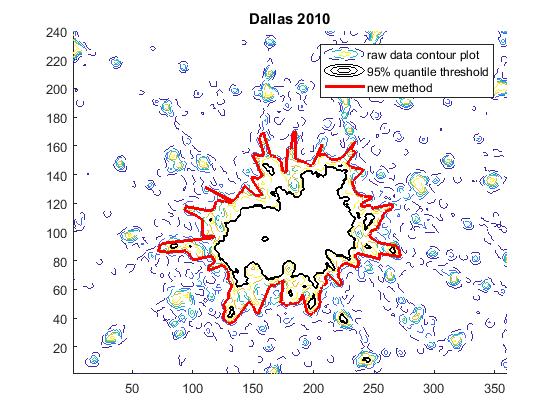}

\end{center}
\begin{small}%
%EndExpansion
Note: The figure depicts the city boundary determined by either the new
method or by taking the 0.95 quantile of nighttime light strength as the
threshold, using the satellite imagery data for Dallas, TX in the years
1995, 2000, 2005, and 2010. Data are available from
https://www.ncei.noaa.gov/.\ 
%TCIMACRO{\TeXButton{endsmall}{\end{small}}}%
%BeginExpansion
\end{small}%
%EndExpansion
%TCIMACRO{\TeXButton{E}{\end{figure}}}%
%BeginExpansion
\end{figure}%
%EndExpansion

To this end, we first examine whether the light intensity data exhibits a
clear threshold pattern. We plot the kernel density estimates of $y_{i}$ in
the year 2010 in Figure \ref{fig ksdensity}. The bandwidth is the standard
rule-of-thumb one. The estimated density exhibits three peaks at around the
intensity levels 0, 8, and 63. They respectively correspond to the rural
area, small towns, and the central metropolitan area. It shows that the
threshold model is appropriate in characterizing such a mean-shift pattern.

Now we implement the rotation and estimation method described in Section \ref%
{Section contour}. In particular, we pick the center point in the bright
middle area as the Dallas metropolitan center, which corresponds to the
pixel point in the 181st column from the left and the 100th row from the
bottom. Then for each $a^{\circ }$ over the 500 equally-spaced grid on $%
[0^{\circ },360^{\circ }]$, we rotate the data by $a^{\circ }$ degrees
counterclockwise and estimate the model (\ref{model3}) with $x_{i}=1$. The
bandwidth is chosen as $cn^{-1/2}$ with $c=1$. Other choices of $c$ lead to almost identical results,
given the large sample size. Figure \ref{fig cityboundary} presents the
estimated metropolitan areas using our nonparametric approach (red) and the
area determined by the \textit{ad hoc} threshold of the 95\% quantile of $%
y_{i}$ (black) in the years 1995, 2000, 2005, and 2010. It clearly shows the
expansion of the Dallas metropolitan area over the 15 years of the sample
period.

Several interesting findings are summarized as follows. First, the estimated
boundary is highly nonlinear as a function of the angle. Therefore, any
parametric threshold model could lead to a substantially misleading result.
Second, our estimated area is larger than that determined by the \textit{ad
hoc} threshold, by 80.31\%, 81.56\%, 106.46\%, and 102.09\% in the years
1995, 2000, 2005, and 2010, respectively. In particular, our nonparametric
estimates tend to include some suburban areas that exhibit strong light
intensity and that are geographically close to the city center. For example,
the very left stretch-out area in the estimated boundary corresponds to Fort
Worth, which is 30 miles from downtown Dallas. Residents can easily commute
by train or driving on the interstate highway 30. It is then reasonable to
include Fort Worth as a part of the metropolitan Dallas area for economic
analysis. Third, given the large sample size, the 95\% confidence intervals
of the boundary are too narrow to be distinguished from the estimates and
therefore omitted from the figure. Such narrow intervals apparently exclude
the boundary determined by the \textit{ad hoc }method. Finally, the
estimated value of $\beta _{0}+\delta _{0}$ is approximately $53$ in these
sample periods, which corresponds to the 89\% quantile of $y_{i}$ in the
sample. This suggests that a more proper choice of the level of light
intensity threshold is the 89\% quantile of $y_{i}$, instead of the 95\%
quantile, if one needs to choose the light-intensity threshold to determine
the Dallas metropolitan area.

\section{Concluding Remarks\label{Section conclusion}}

This paper proposes a novel approach to conduct sample splitting. In
particular, we develop a nonparametric threshold regression model where two
variables can jointly determine the unknown threshold boundary. Our approach
can be easily generalized so that the sample splitting depends on more
numbers of variables, though such an extension is subject to the curse of
dimensionality, as usually observed in the kernel regression literature. The
main interest is in identifying the threshold function that determines how
to split the sample. Thus our model should be distinguished from the
smoothed threshold regression model or the random coefficient regression
model. It instead could be seen as an unsupervised learning tool for
clustering.

This new approach is empirically relevant in broad areas studying sample
splitting (e.g., segregation and group-formation) and heterogeneous effects
over different subsamples. We illustrate some of them with the tipping point
problem in social segregation and metropolitan area determination using
satellite imagery datasets. Though we omit in this paper, we also estimate
the economic border between Brooklyn and Queens boroughs in New York City
using housing prices.\footnote{%
The result is available upon request.} The estimated border is substantially
different from the existing administrative border, which was determined in
1931 and cannot reflect the dramatic city development. Interestingly, the
estimated border coincides with the Jackson Robinson Parkway and the Long
Island Railroad. This finding provides new evidence that local
transportation corridors could increase community segregation (cf.\ \cite%
{Ananat11} and \cite{Heilmann18}).

We list some related works, which could motivate potential theoretical
extensions. First, while we focus on the local constant estimation in this
paper, one could consider the local linear estimation using the threshold
indicator $\mathbf{1}\left[ q_{i}\leq \gamma _{1}+\gamma _{2}(s_{i}-s)\right]
$ in (\ref{sse}). Although grid search is very difficult in determining the
two threshold parameters ($\gamma _{1}$ and $\gamma _{2}$), we could use the
MCMC algorithm developed by \cite{Yu19} and the mixed integer optimization
(MIO) algorithms developed by \cite{LLSS18}. Besides the computational
challenge, however, the asymptotic derivation is more involved since we need
to consider higher-order expansions of the objective function. Second, while
our nonparametric setup is on the threshold function $\gamma _{0}(\cdot )$,
some recent literature studies the nonparametric regression model with a
parametric threshold, such as $y_{i}=m_{1}(x_{i})+m_{2}(x_{i})\mathbf{1}%
[q_{i}\leq \gamma _{0}]+u_{i}$, where $m_{1}\left( \cdot \right) $ and $%
m_{2}\left( \cdot \right) $ are different nonparametric functions. See, for
example, \cite{Henderson17}, \cite{Chiou18}, \cite{Yu18}, \cite%
{YuLiaoPhillips19}, and \cite{DelgadoHidalgo2000}.

{\footnotesize \newpage }

\setcounter{equation}{0}\renewcommand{\theequation}{A.\arabic{equation}}%
\renewcommand{\thelemma}{A.\arabic{lemma}}\renewcommand{%
\baselinestretch}{0.7}\scalefont{0.96}\baselineskip=14pt

\appendix

\section{Appendix}

Throughout the proof, we denote $K_{i}\left( s\right) =K\left(
(s_{i}-s)/b_{n}\right) $ and $\mathbf{1}_{i}\left( \gamma \right) =\mathbf{1}%
\left[ q_{i}\leq \gamma \right] $. We let $C$ and its variants such as $%
C_{1} $ and $C_{1}^{\prime }$ stand for generic positive finite constants
that may vary across lines. We also let $a_{n}=n^{1-2\epsilon }b_{n}$. All
the additional lemmas in the proof assume the conditions in Assumptions ID
and A hold. Omitted proofs for some lemmas are all collected in the
supplementary material.

\subsection{Proof of Theorem \protect\ref{Thm id} (Identification)}

\paragraph{Proof of Theorem \protect\ref{Thm id}}

First, for any $\gamma (s)\in \Gamma $ with given $s\in \mathcal{S}$, we
define an $L_{2}$-loss as%
\begin{eqnarray*}
R(\beta ,\delta ,\gamma ;s) &=&\mathbb{E}\left[ \left. \left(
y_{i}-x_{i}^{\top }\beta -x_{i}^{\top }\delta \mathbf{1}_{i}\left( \gamma
(s_{i})\right) \right) ^{2}\right\vert s_{i}=s\right] \\
&&-\mathbb{E}\left[ \left. \left( y_{i}-x_{i}^{\top }\beta _{0}-x_{i}^{\top
}\delta _{0}\mathbf{1}_{i}\left( \gamma _{0}(s_{i})\right) \right)
^{2}\right\vert s_{i}=s\right] \text{.}
\end{eqnarray*}%
For any $\gamma \in \Gamma \equiv \lbrack \underline{\gamma },\overline{%
\gamma }]$, we have $\mathbb{E}[R(\beta ,\delta ,\gamma ;s_{i})]\geq 0$ from
Assumption ID-(i). Since 
\begin{equation*}
R(\beta ,\delta ,\gamma ;s)=\left\{ 
\begin{array}{ll}
\mathbb{E}\left[ \left. \left( x_{i}^{\top }((\beta +\delta )-\mathbf{(}%
\beta _{0}+\delta _{0}\mathbf{)}\right) ^{2}\right\vert s_{i}=s\right] & 
\text{for }q_{i}\leq \min \{\gamma (s),\gamma _{0}(s)\}\text{;} \\ 
\mathbb{E}\left[ \left. \left( x_{i}^{\top }\left( \beta -\beta _{0}\right)
\right) ^{2}\right\vert s_{i}=s\right] & \text{for }q_{i}>\max \{\gamma
(s),\gamma _{0}(s)\}\text{,}%
\end{array}%
\right.
\end{equation*}%
we have%
\begin{eqnarray*}
\mathbb{E}\left[ R(\beta ,\delta ,\gamma ;s_{i})\right] &\geq &\left\Vert
(\beta -\beta _{0})+\mathbf{(}\delta -\delta _{0}\mathbf{)}\right\Vert ^{2}%
\mathbb{E}\left[ ||x_{i}x_{i}^{\top }||\mathbf{1}\left[ q_{i}\leq \underline{%
\gamma }\right] \right] \\
&&+\left\Vert \beta -\beta _{0}\right\Vert ^{2}\mathbb{E}\left[
||x_{i}x_{i}^{\top }||\mathbf{1}\left[ q_{i}>\overline{\gamma }\right] %
\right] \\
&>&0
\end{eqnarray*}%
when $(\beta ^{\top },\delta ^{\top })^{\top }\neq (\beta _{0}^{\top
},\delta _{0}^{\top })^{\top }$ from Assumption ID-(ii). Therefore, we have $%
\mathbb{E}\left[ R(\beta ,\delta ,\gamma ;s_{i})\right] =0$ only when $%
(\beta ^{\top },\delta ^{\top })^{\top }=(\beta _{0}^{\top },\delta
_{0}^{\top })^{\top }$, which gives that $(\beta _{0}^{\top },\delta
_{0}^{\top })^{\top }$ are identified as the unique minimizer of $\mathbb{E}%
[\left( y_{i}-x_{i}^{\top }\beta -x_{i}^{\top }\delta \mathbf{1}_{i}\left(
\gamma \right) \right) ^{2}]$ for any $\gamma \in \Gamma $.

Second, note that we have $R(\beta _{0},\delta _{0},\gamma ;s)\geq 0$ for
any $\gamma (s)\in \Gamma $ with given $s\in \mathcal{S}$ from Assumption
ID-(i). For any $\gamma (s)\neq \gamma _{0}(s)$ at $s_{i}=s$ and $(\beta
^{\top },\delta ^{\top })^{\top }=(\beta _{0}^{\top },\delta _{0}^{\top
})^{\top }$, however, we have%
\begin{eqnarray*}
&&R(\beta _{0},\delta _{0},\gamma ;s) \\
&=&\delta _{0}^{\top }\mathbb{E}\left[ \left. x_{i}x_{i}^{\top }\left( 
\mathbf{1}_{i}\left( \gamma (s_{i})\right) -\mathbf{1}_{i}\left( \gamma
_{0}(s_{i})\right) \right) ^{2}\right\vert s_{i}=s\right] \delta _{0} \\
&=&\delta _{0}^{\top }\mathbb{E}\left[ \left. x_{i}x_{i}^{\top }\mathbf{1}%
\left[ \min \{\gamma (s_{i}),\gamma _{0}(s_{i})\}<q_{i}\leq \max \{\gamma
(s_{i}),\gamma _{0}(s_{i})\}\right] \right\vert s_{i}=s\right] \delta _{0} \\
&=&\int_{\min \{\gamma (s),\gamma _{0}(s)\}}^{\max \{\gamma (s),\gamma
_{0}(s)\}}\delta _{0}^{\top }\mathbb{E}\left[ \left. x_{i}x_{i}^{\top
}\right\vert q_{i}=q,s_{i}=s\right] \delta _{0}f(q|s)dq \\
&\geq &C(s)\mathbb{P}\left( \left. \min \{\gamma (s_{i}),\gamma
_{0}(s_{i})\}<q_{i}\leq \max \{\gamma (s_{i}),\gamma
_{0}(s_{i})\}\right\vert s_{i}=s\right) \\
&>&0
\end{eqnarray*}%
from Assumptions ID-(i), (iii), and (iv), where $C(s)=\inf_{q\in \mathcal{Q}%
}\delta _{0}^{\top }\mathbb{E}[x_{i}x_{i}^{\top }|q_{i}=q,s_{i}=s]\delta
_{0}>0$. Note that the last probability is strictly positive because we
assume $f(q|s)>0$ for any $(q,s)\in \Gamma \times \mathcal{S}$ and $\gamma
_{0}(s)$ is not located on the boundary of $\mathcal{Q}$ as $\varepsilon (s)<%
\mathbb{P}(q_{i}\leq \gamma _{0}(s_{i})|s_{i}=s)<1-\varepsilon (s)$ for some 
$\varepsilon (s)>0$. Therefore, $R(\beta _{0},\delta _{0},\gamma ;s)=0$ only
when $\gamma (s)=\gamma _{0}(s)$ since $R(\beta _{0},\delta _{0},\gamma ;s)$
is continuous at $\gamma =\gamma _{0}(s)$ and $\gamma $ is in a compact
support from Assumptions ID-(ii) and (iv). It gives that $\gamma _{0}(s)$ is
identified as the unique minimizer of $\mathbb{E}[\left( y_{i}-x_{i}^{\top
}\beta _{0}-x_{i}^{\top }\delta _{0}\mathbf{1}_{i}\left( \gamma
(s_{i})\right) \right) ^{2}|s_{i}=s]$ for each $s\in \mathcal{S}$. $%
\blacksquare $

\subsection{Proof of Theorem \protect\ref{p-roc} (Pointwise Convergence)}

We first present a covariance inequality for strong mixing random field.
Suppose $\Lambda _{1}$ and $\Lambda _{2}$ are finite subsets in $\Lambda
_{n} $ with $\left\vert \Lambda _{1}\right\vert =k_{x},\left\vert \Lambda
_{2}\right\vert =l_{x}$, and let $X_{1}$ and $X_{2}$ be random variables
respectively measurable with respect to the $\sigma $-algebra's generated by 
$\Lambda _{1}$ and $\Lambda _{2}$. If $\mathbb{E}\left[ \left\vert
X_{1}\right\vert ^{p_{x}}\right] <\infty $ and $\mathbb{E}\left[ \left\vert
X_{2}\right\vert ^{q_{x}}\right] <\infty $ with $1/p_{x}+1/q_{x}+1/r_{x}=1$
for some constants $p_{x},q_{x}>1$ and $r_{x}>0$, then 
\begin{equation}
\left\vert Cov\left[ X_{1},X_{2}\right] \right\vert <8\alpha
_{k_{x},l_{x}}(\lambda (\Lambda _{1},\Lambda _{2}))^{1/r_{x}}\mathbb{E}\left[
\left\vert X_{1}\right\vert ^{p_{x}}\right] ^{1/p_{x}}\mathbb{E}\left[
\left\vert X_{2}\right\vert ^{q_{x}}\right] ^{1/q_{x}}  \label{cov ineq}
\end{equation}%
under Assumptions A-(i) and A-(iii). This covariance inequality is presented
as Lemma 1 in the working paper version of \cite{Jenish09}. The proof is
also available in \cite{Hall80}, p.277.

\bigskip

For a given $s\in \mathcal{S}_{0}$, we define%
\begin{eqnarray*}
M_{n}(\gamma ;s) &=&\frac{1}{nb_{n}}\dsum_{i\in \Lambda
_{n}}x_{i}x_{i}^{\top }\mathbf{1}_{i}(\gamma )K_{i}(s) \\
J_{n}(\gamma ;s) &=&\frac{1}{\sqrt{nb_{n}}}\dsum_{i\in \Lambda
_{n}}x_{i}u_{i}\mathbf{1}_{i}(\gamma )K_{i}(s).
\end{eqnarray*}%
The following four lemmas give the asymptotic behavior of $M_{n}(\gamma ;s)$
and $J_{n}(\gamma ;s)$.

\begin{lemma}[Maximal inequality]
\label{max ineq}For any given $s\in \mathcal{S}_{0}$, any $\eta ,\varpi >0$,
and any $\gamma _{1}\in \Gamma $, there exists a constant $C^{\ast }$ such
that 
\begin{equation*}
\mathbb{P}\left( \sup_{\gamma \in \lbrack \gamma _{1},\gamma _{1}+\varpi
]}\left\Vert J_{n}\left( \gamma ;s\right) -J_{n}\left( \gamma _{1};s\right)
\right\Vert >\eta \right) \leq \frac{C^{\ast }\varpi ^{2}}{\eta ^{4}}
\end{equation*}%
if $n$ is sufficiently large.
\end{lemma}

\paragraph*{Proof of Lemma \protect\ref{max ineq}}

For expositional simplicity, we only present the case of scalar $x_{i}$. Let 
$\varpi $ be such that $\varpi \geq (nb_{n})^{-1}$ and $\overline{g}$ be an
integer satisfying $nb_{n}\varpi /2\leq \overline{g}\leq nb_{n}\varpi $,
which always exists since $nb_{n}\varpi \geq 1$. Consider a fine enough grid
over $[\gamma _{1},\gamma _{1}+\varpi ]$ such that $\gamma _{g}=\gamma
_{1}+(g-1)\varpi /\overline{g}$ for $g=1,\ldots ,\overline{g}+1$, where $%
\max_{1\leq g\leq \overline{g}}\left( \gamma _{g}-\gamma _{g-1}\right) \leq
\varpi /\overline{g}$. We define $h_{ig}(s)=x_{i}u_{i}K_{i}\left( s\right) 
\mathbf{1}\left[ \gamma _{g}<q_{i}\leq \gamma _{g+1}\right] $ and $%
H_{ng}(s)=(nb_{n})^{-1}\sum_{i\in \Lambda _{n}}|h_{ig}(s)|$ for $1\leq g\leq 
\overline{g}$. Then for any $\gamma \in \left[ \gamma _{g},\gamma _{g+1}%
\right] $,%
\begin{eqnarray*}
\left\vert J_{n}\left( \gamma ;s\right) -J_{n}\left( \gamma _{g};s\right)
\right\vert &\leq &\sqrt{nb_{n}}H_{ng}(s) \\
&\leq &\sqrt{nb_{n}}\left\vert H_{ng}(s)-\mathbb{E}\left[ H_{ng}(s)\right]
\right\vert +\sqrt{nb_{n}}\mathbb{E}\left[ H_{ng}(s)\right]
\end{eqnarray*}%
and hence%
\begin{eqnarray*}
&&\sup_{\gamma \in \lbrack \gamma _{1},\gamma _{1}+\varpi ]}\left\vert
J_{n}\left( \gamma ;s\right) -J_{n}\left( \gamma _{1};s\right) \right\vert \\
&\leq &\max_{2\leq g\leq \overline{g}+1}\left\vert J_{n}\left( \gamma
_{g};s\right) -J_{n}\left( \gamma _{1};s\right) \right\vert \\
&&+\max_{1\leq g\leq \overline{g}}\sqrt{nb_{n}}\left\vert H_{ng}(s)-\mathbb{E%
}\left[ H_{ng}(s)\right] \right\vert +\max_{1\leq g\leq \overline{g}}\sqrt{%
nb_{n}}\mathbb{E}\left[ H_{ng}(s)\right] \\
&\equiv &\Psi _{1}(s)+\Psi _{2}(s)+\Psi _{3}(s)\text{.}
\end{eqnarray*}%
We let $h_{i}(s)=x_{i}u_{i}K_{i}\left( s\right) \mathbf{1}\left[ \gamma
_{g}<q_{i}\leq \gamma _{k}\right] $ for any given $1\leq g<k\leq \overline{g}
$ and for fixed $s$. First, for $\Psi _{1}(s)$, we have%
\begin{eqnarray*}
&&\mathbb{E}\left[ \left\vert J_{n}\left( \gamma _{g};s\right) -J_{n}\left(
\gamma _{k};s\right) \right\vert ^{4}\right] \\
&=&\frac{1}{n^{2}b_{n}^{2}}\dsum_{i\in \Lambda _{n}}\mathbb{E}\left[
h_{i}^{4}(s)\right] +\frac{1}{n^{2}b_{n}^{2}}\dsum_{\substack{ i,j\in
\Lambda _{n}  \\ i\neq j}}\mathbb{E}\left[ h_{i}^{2}(s)h_{j}^{2}(s)\right] +%
\frac{1}{n^{2}b_{n}^{2}}\dsum_{\substack{ i,j\in \Lambda _{n}  \\ i\neq j}}%
\mathbb{E}\left[ h_{i}^{3}(s)h_{j}(s)\right] \\
&&+\frac{1}{n^{2}b_{n}^{2}}\dsum_{\substack{ i,j,k,l\in \Lambda _{n}  \\ %
i\neq j\neq k\neq l}}\mathbb{E}\left[ h_{i}(s)h_{j}(s)h_{k}(s)h_{l}(s)\right]
+\frac{1}{n^{2}b_{n}^{2}}\dsum_{\substack{ i,j,k\in \Lambda _{n}  \\ i\neq
j\neq k}}\mathbb{E}\left[ h_{i}^{2}(s)h_{j}(s)h_{k}(s)\right] \\
&\equiv &\Psi _{11}(s)+\Psi _{12}(s)+\Psi _{13}(s)+\Psi _{14}(s)+\Psi
_{15}(s)\text{,}
\end{eqnarray*}%
where each term's bound is obtained as follows.

For $\Psi _{11}(s)$, since $\mathbb{E}[|x_{i}u_{i}|^{4}|q_{i},s_{i}]\mathbf{1%
}[\gamma _{g}<q_{i}\leq \gamma _{k}]<C_{1}<\infty $ from Assumption A-(v),
we have 
\begin{eqnarray}
\frac{1}{b_{n}}\mathbb{E}\left[ h_{i}^{4}(s)\right] &=&\frac{1}{b_{n}}\iint 
\mathbb{E}\left[ |x_{i}u_{i}|^{4}|q,v\right] 1\left[ \gamma _{g}<q\leq
\gamma _{k}\right] K^{4}\left( \frac{v-s}{b_{n}}\right) f\left( q,v\right)
dqdv  \label{Eh4} \\
&\leq &\frac{C_{1}}{b_{n}}\iint K^{4}\left( \frac{v-s}{b_{n}}\right) f\left(
q,v\right) dqdv  \notag \\
&=&C_{1}\iint K^{4}\left( t\right) f\left( q,s+b_{n}t\right) dqdt  \notag \\
&=&C_{1}(s)+O\left( b_{n}^{2}\right)  \notag
\end{eqnarray}%
for some $C_{1}(s)<\infty $ for a given $s$ from Assumption A-(x), where we
apply the change of variables $t=(v-s)/b_{n}$. Hence, $\Psi
_{11}(s)=O(n^{-1}b_{n}^{-1})=o(1)$. For $\Psi _{12}(s)$, by the covariance
inequality (\ref{cov ineq}) with $p_{x}=2+\varphi $, $q_{x}=2+\varphi $, $%
r_{x}=(2+\varphi )/\varphi $, and $k_{x}=l_{x}=1$, 
\begin{eqnarray}
&&\frac{1}{n^{2}b_{n}^{2}}\dsum_{\substack{ i,j\in \Lambda _{n}  \\ i\neq j}}%
\left\vert Cov\left[ h_{i}^{2}(s),h_{j}^{2}(s)\right] \right\vert
\label{P12} \\
&\leq &\frac{C_{2}}{n^{2}b_{n}^{2}}\dsum_{\substack{ i,j\in \Lambda _{n}  \\ %
i\neq j}}\alpha _{1,1}(\lambda (i,j))^{\varphi /(2+\varphi )}\mathbb{E}\left[
\left\vert h_{i}^{2}(s)\right\vert ^{2+\varphi }\right] ^{2/\left( 2+\varphi
\right) }  \notag \\
&\leq &\frac{C_{2}}{n^{2}b_{n}^{2}}\mathbb{E}\left[ \left\vert
h_{i}^{2}(s)\right\vert ^{2+\varphi }\right] ^{2/\left( 2+\varphi \right)
}\dsum_{i\in \Lambda _{n}}\dsum_{m=1}^{n-1}\dsum_{\substack{ j\in \Lambda
_{n}  \\ \lambda (i,j)\in \lbrack m,m+1)}}\alpha _{1,1}\left( m\right)
^{\varphi /(2+\varphi )}  \notag \\
&\leq &\frac{C_{2}^{\prime }}{n^{2}b_{n}^{2}}\mathbb{E}\left[ \left\vert
h_{i}^{2}(s)\right\vert ^{2+\varphi }\right] ^{2/\left( 2+\varphi \right)
}\dsum_{i\in \Lambda _{n}}\dsum_{m=1}^{\infty }m\alpha _{1,1}\left( m\right)
^{\varphi /(2+\varphi )}  \notag \\
&\leq &\frac{C_{2}^{\prime \prime }}{n^{2}b_{n}^{(2+2\varphi )/(2+\varphi )}}%
\left( \frac{1}{b_{n}}\mathbb{E}\left[ \left\vert h_{i}^{2}(s)\right\vert
^{2+\varphi }\right] \right) ^{2/\left( 2+\varphi \right)
}n\dsum_{m=1}^{\infty }m\exp \left( -m\varphi /(2+\varphi )\right)  \notag
\end{eqnarray}%
for some $\varphi >0$ and $C_{2},C_{2}^{\prime },C_{2}^{\prime \prime
}<\infty $, where $b_{n}^{-1}\mathbb{E}[\left\vert h_{i}^{2}(s)\right\vert
^{2+\varphi }]<\infty $ as in (\ref{Eh4}). Note that $\alpha _{1,1}(\lambda
(i,j))\leq \alpha _{1,1}(m)$ for $\lambda (i,j)\in \lbrack m,m+1)$ and $%
|\{j\in \Lambda _{n}:\lambda (i,j)\in \lbrack m,m+1)\}|=O(m)$ for any given $%
i\in \Lambda _{n}$ as in Lemma A.1.(iii) of \cite{Jenish09}.\ Furthermore,
by Assumptions A-(vii) and (x),%
\begin{eqnarray}
\frac{1}{b_{n}}\mathbb{E}\left[ h_{i}^{2}(s)\right] &=&\iint V\left(
q,t\right) 1\left[ \gamma _{g}<q\leq \gamma _{k}\right] K^{2}\left( t\right)
f\left( q,s+tb_{n}\right) dqdt  \label{h2} \\
&=&\iint V\left( q,t\right) 1\left[ \gamma _{g}<q\leq \gamma _{k}\right]
K^{2}\left( t\right) f\left( q,s\right) dqdt+O\left( b_{n}^{2}\right)  \notag
\\
&\leq &C_{3}\int 1\left[ \gamma _{g}<q\leq \gamma _{k}\right] f\left(
q,s\right) dq+O\left( b_{n}^{2}\right)  \notag \\
&\leq &C_{3}^{\prime }\left\vert \gamma _{k}-\gamma _{g}\right\vert +O\left(
b_{n}^{2}\right)  \notag
\end{eqnarray}%
for some $C_{3},C_{3}^{\prime }<\infty $, where the inequality is from the
fact that $V\left( q,s\right) $ and $f\left( q,s\right) $ are uniformed
bounded and $\int K^{2}\left( t\right) dt<\infty $. Thus, Assumptions
A-(iii), (v), and (x) yield that%
\begin{eqnarray}
\Psi _{12}(s) &\leq &\frac{1}{n^{2}b_{n}^{2}}\dsum_{\substack{ i,j\in
\Lambda _{n}  \\ i\neq j}}\left( \mathbb{E}\left[ h_{i}^{2}(s)\right] 
\mathbb{E}\left[ h_{j}^{2}(s)\right] +\left\vert Cov\left[
h_{i}^{2}(s),h_{j}^{2}(s)\right] \right\vert \right)  \label{P12b} \\
&\leq &\left( \frac{1}{b_{n}}\mathbb{E}\left[ h_{i}^{2}(s)\right] \right)
^{2}  \notag \\
&&+\frac{C_{2}^{\prime }}{nb_{n}^{(2+2\varphi )/(2+\varphi )}}\left( \frac{1%
}{b_{n}}\mathbb{E}\left[ \left\vert h_{i}^{2}(s)\right\vert ^{2+\varphi }%
\right] \right) ^{2/\left( 2+\varphi \right) }\dsum_{m=1}^{\infty }m\exp
\left( -m\varphi /(2+\varphi )\right)  \notag \\
&\leq &C_{3}^{\prime \prime }\left( \gamma _{k}-\gamma _{g}\right)
^{2}+O(n^{-1}b_{n}^{-(2+2\varphi )/(2+\varphi )})+O\left( b_{n}^{2}\right) 
\notag \\
&=&C_{3}^{\prime \prime }\left( \gamma _{k}-\gamma _{g}\right) ^{2}+o\left(
1\right)  \notag
\end{eqnarray}%
since $\dsum_{m=1}^{\infty }m\exp \left( -m\varphi /(2+\varphi )\right)
<\infty $ for $\varphi >0$ and $n^{-1}b_{n}^{-(2+2\varphi )/(2+\varphi
)}\rightarrow 0$ from Assumption A-(ix). Since $\mathbb{E}\left[ h_{i}(s)%
\right] =0$, using the same argument as (\ref{P12}) and (\ref{P12b}), and
the inequality (\ref{cov ineq}) with $p_{x}=2\left( 2+\varphi \right) /3$, $%
q_{x}=2\left( 2+\varphi \right) $, $r_{x}=(2+\varphi )/\varphi $, and $%
k_{x}=l_{x}=1$, we can also show that 
\begin{eqnarray*}
\Psi _{13}(s) &\leq &\frac{1}{n^{2}b_{n}^{2}}\dsum_{\substack{ i,j\in
\Lambda _{n}  \\ i\neq j}}\left\vert Cov\left[ h_{i}^{3}(s),h_{j}(s)\right]
\right\vert \\
&\leq &\frac{C_{4}}{n^{2}b_{n}^{2}}\dsum_{\substack{ i,j\in \Lambda _{n}  \\ %
i\neq j}}\alpha _{1,1}(\lambda (i,j))^{\varphi /(2+\varphi )}\mathbb{E}\left[
\left\vert h_{i}^{3}(s)\right\vert ^{2(2+\varphi )/3}\right] ^{3/\left(
4+2\varphi \right) }\mathbb{E}\left[ \left\vert h_{j}(s)\right\vert
^{2(2+\varphi )}\right] ^{1/\left( 4+2\varphi \right) } \\
&\leq &\frac{C_{4}^{\prime }}{nb_{n}^{(2+2\varphi )/(2+\varphi )}}\left( 
\frac{1}{b_{n}}\mathbb{E}\left[ \left\vert h_{i}^{2}(s)\right\vert
^{(2+\varphi )}\right] \right) ^{2/\left( 2+\varphi \right)
}\dsum_{m=1}^{\infty }m\exp \left( -m\varphi /(2+\varphi )\right) \\
&=&O(n^{-1}b_{n}^{-(2+2\varphi )/(2+\varphi )})=o\left( 1\right) \text{.}
\end{eqnarray*}%
For $\Psi _{14}(s)$, let $\mathcal{E}=\{(i,j,k,l):i\neq j\neq k\neq l$, $%
0<\lambda (i,j)\leq \lambda (i,k)\leq \lambda (i,l)$, and $\lambda (j,k)\leq
\lambda (j,l)\}$.\footnote{%
In the (one-dimensional) time series case, this set of indices reduces to $%
\{\left( i,j,k,l\right) :1\leq i<j<k<l\leq n\}$.} Then by stationarity, 
\begin{eqnarray*}
\Psi _{14}(s) &\leq &\frac{4!}{n^{2}b_{n}^{2}}\dsum_{i,j,k,l\in \Lambda
_{n}\cap \mathcal{E}}\left\vert \mathbb{E}\left[
h_{i}(s)h_{j}(s)h_{k}(s)h_{l}(s)\right] \right\vert \\
&=&\frac{2\cdot 4!}{n^{2}b_{n}^{2}}\dsum_{\substack{ i,j,k,l\in \Lambda
_{n}\cap \mathcal{E}  \\ \lambda \left( i,j\right) \geq \max \{\lambda
\left( j,k\right) ,\lambda \left( k,l\right) \}}}\left\vert Cov\left[
h_{i}(s),\left\{ h_{j}(s)h_{k}(s)h_{l}(s)\right\} \right] \right\vert \\
&&+\frac{4!}{n^{2}b_{n}^{2}}\dsum_{\substack{ i,j,k,l\in \Lambda _{n}\cap 
\mathcal{E}  \\ \lambda \left( j,k\right) \geq \max \{\lambda \left(
i,j\right) ,\lambda \left( k,l\right) \}}}\left\vert Cov\left[ \left\{
h_{i}(s)h_{j}(s)\right\} ,\left\{ h_{k}(s)h_{l}(s)\right\} \right]
\right\vert \\
&&+\frac{4!}{n^{2}b_{n}^{2}}\dsum_{\substack{ i,j,k,l\in \Lambda _{n}\cap 
\mathcal{E}  \\ \lambda \left( j,k\right) \geq \max \{\lambda \left(
i,j\right) ,\lambda \left( k,l\right) \}}}\left\vert \mathbb{E}\left[
h_{i}(s)h_{j}(s)\right] \mathbb{E}\left[ h_{k}(s)h_{l}(s)\right] \right\vert
\\
&\equiv &\Psi _{14,1}(s)+\Psi _{14,2}(s)+\Psi _{14,3}(s)\text{.}
\end{eqnarray*}%
For $\Psi _{14,1}(s)$, the largest distance among all the pairs is $\lambda
\left( i,j\right) $. Then by the covariance inequality (\ref{cov ineq}) with 
$p_{x}=2\left( 2+\varphi \right) /3$, $q_{x}=2\left( 2+\varphi \right) $, $%
r_{x}=(2+\varphi )/\varphi $, $k_{x}=1$, and $l_{x}=3$, 
\begin{eqnarray*}
&&\Psi _{14,1}(s) \\
&\leq &\frac{C_{5}}{n^{2}b_{n}^{(2+2\varphi )/(2+\varphi )}}\dsum_{\substack{
i,j,k,l\in \Lambda _{n}\cap \mathcal{E}  \\ \lambda \left( i,j\right) \geq
\max \{\lambda \left( j,k\right) ,\lambda \left( k,l\right) \}}}\alpha
_{1,3}(\lambda (i,j))^{\varphi /(2+\varphi )}\left( \frac{1}{b_{n}}\mathbb{E}%
\left[ \left\vert h_{i}(s)\right\vert ^{4+2\varphi }\right] \right)
^{1/\left( 4+2\varphi \right) } \\
&&\ \ \ \ \ \ \ \ \ \ \ \ \ \ \ \ \ \ \ \ \ \ \ \ \ \ \ \ \ \ \ \ \ \ \ \ \
\ \ \ \ \ \times \left( \frac{1}{b_{n}}\mathbb{E}\left[ \left\vert
h_{j}(s)h_{k}(s)h_{l}(s)\right\vert ^{2\left( 2+\varphi \right) /3}\right]
\right) ^{3/\left( 4+2\varphi \right) } \\
&\leq &\frac{C_{5}^{\prime }(s)}{n^{2}b_{n}^{(2+2\varphi )/(2+\varphi )}}%
\dsum_{\substack{ i,j,k,l\in \Lambda _{n}\cap \mathcal{E}  \\ \lambda \left(
i,j\right) \geq \max \{\lambda \left( j,k\right) ,\lambda \left( k,l\right)
\}}}\alpha _{1,3}(\lambda (i,j))^{\varphi /(2+\varphi )} \\
&\leq &\frac{C_{5}^{\prime }(s)}{n^{2}b_{n}^{(2+2\varphi )/(2+\varphi )}}%
\dsum_{i\in \Lambda _{n}}\dsum_{m=1}^{n-1}\dsum_{\substack{ j\in \Lambda
_{n}  \\ \lambda \left( i,j\right) \in \lbrack m,m+1)}}\dsum_{\substack{ %
k\in \Lambda _{n}  \\ \lambda \left( j,k\right) \leq m}}\dsum_{\substack{ %
l\in \Lambda _{n}  \\ \lambda \left( k,l\right) \leq m}}\alpha
_{1,3}(m)^{\varphi /(2+\varphi )} \\
&\leq &\frac{C_{5}^{\prime \prime }(s)}{nb_{n}^{(2+2\varphi )/(2+\varphi )}}%
\dsum_{m=1}^{\infty }m^{5}\exp \left( -m\varphi /(2+\varphi )\right) \\
&=&o(1)
\end{eqnarray*}%
since $|\{k\in \Lambda _{n}:\lambda \left( j,k\right) \leq m\}|=O(m^{2})$
for any given $j\in \Lambda _{n}$. For $\Psi _{14,2}(s)$, the largest
distance among all the pairs is $\lambda \left( j,k\right) $. Similarly as
above, 
\begin{eqnarray*}
&&\Psi _{14,2}(s) \\
&\leq &\frac{C_{6}}{n^{2}b_{n}^{(2+2\varphi )/(2+\varphi )}}\dsum_{\substack{
i,j,k,l\in \Lambda _{n}\cap \mathcal{E}  \\ \lambda \left( j,k\right) \geq
\max \{\lambda \left( i,j\right) ,\lambda \left( k,l\right) \}}}\alpha
_{2,2}(\lambda (j,k))^{\varphi /(2+\varphi )}\left( \frac{1}{b_{n}}\mathbb{E}%
\left[ \left\vert h_{i}(s)h_{j}(s)\right\vert ^{2+\varphi }\right]
^{1/\left( 2+\varphi \right) }\right) \\
&&\ \ \ \ \ \ \ \ \ \ \ \ \ \ \ \ \ \ \ \ \ \ \ \ \ \ \ \ \ \ \ \ \ \ \ \ \
\ \ \ \ \ \times \left( \frac{1}{b_{n}}\mathbb{E}\left[ \left\vert
h_{k}(s)h_{l}(s)\right\vert ^{2+\varphi }\right] ^{1/\left( 2+\varphi
\right) }\right) \\
&\leq &\frac{C_{6}^{\prime }(s)}{n^{2}b_{n}^{(2+2\varphi )/(2+\varphi )}}%
\dsum_{\substack{ i,j,k,l\in \Lambda _{n}\cap \mathcal{E}  \\ \lambda \left(
j,k\right) \geq \max \{\lambda \left( i,j\right) ,\lambda \left( k,l\right)
\}}}\alpha _{2,2}(\lambda (j,k))^{\varphi /(2+\varphi )} \\
&\leq &\frac{C_{6}^{\prime }(s)}{n^{2}b_{n}^{(2+2\varphi )/(2+\varphi )}}%
\dsum_{j\in \Lambda _{n}}\dsum_{m=1}^{n}\dsum_{\substack{ k\in \Lambda _{n} 
\\ \lambda \left( j,k\right) \in \lbrack m,m+1)}}\dsum_{\substack{ i\in
\Lambda _{n}  \\ \lambda \left( i,j\right) \leq m}}\dsum_{\substack{ l\in
\Lambda _{n}  \\ \lambda \left( k,l\right) \leq m}}\alpha _{2,2}\left(
m\right) ^{\varphi /(2+\varphi )} \\
&\leq &\frac{C_{6}^{\prime \prime }(s)}{nb_{n}^{(2+2\varphi )/(2+\varphi )}}%
\dsum_{m=1}^{\infty }m^{5}\exp \left( -m\varphi /(2+\varphi )\right) \\
&=&o(1)\text{.}
\end{eqnarray*}%
For $\Psi _{14,3}(s)$, the largest distance among all the pairs is still $%
\lambda \left( j,k\right) $. We let a sequence of integers $\kappa
_{n}=O\left( b_{n}^{-\ell }\right) $ for some $\ell >0$ such that $\kappa
_{n}\rightarrow \infty $ and $\kappa _{n}^{2}b_{n}\rightarrow 0$ as $%
n\rightarrow \infty $. We decompose $\Psi _{14,3}(s)$ such that 
\begin{eqnarray*}
\Psi _{14,3}(s) &=&\frac{C_{7}}{n^{2}b_{n}^{2}}\dsum_{\substack{ i,j,k,l\in
\Lambda _{n}\cap \mathcal{E}  \\ \lambda \left( j,k\right) \geq \max
\{\lambda \left( i,j\right) ,\lambda \left( k,l\right) \}  \\ \lambda \left(
i,j\right) \leq \kappa _{n},\lambda \left( k,l\right) \leq \kappa _{n}}}%
\left\vert \mathbb{E}\left[ h_{i}(s)h_{j}(s)\right] \mathbb{E}\left[
h_{k}(s)h_{l}(s)\right] \right\vert \\
&&+\frac{C_{7}}{n^{2}b_{n}^{2}}\dsum_{\substack{ i,j,k,l\in \Lambda _{n}\cap 
\mathcal{E}  \\ \lambda \left( j,k\right) \geq \max \{\lambda \left(
i,j\right) ,\lambda \left( k,l\right) \}  \\ \lambda \left( i,j\right)
>\kappa _{n},\lambda \left( k,l\right) >\kappa _{n}}}\left\vert \mathbb{E}%
\left[ h_{i}(s)h_{j}(s)\right] \mathbb{E}\left[ h_{k}(s)h_{l}(s)\right]
\right\vert \\
&&+\frac{2C_{7}}{n^{2}b_{n}^{2}}\dsum_{\substack{ i,j,k,l\in \Lambda
_{n}\cap \mathcal{E}  \\ \lambda \left( j,k\right) \geq \max \{\lambda
\left( i,j\right) ,\lambda \left( k,l\right) \}  \\ \lambda \left(
i,j\right) \leq \kappa _{n},\lambda \left( k,l\right) >\kappa _{n}}}%
\left\vert \mathbb{E}\left[ h_{i}(s)h_{j}(s)\right] \mathbb{E}\left[
h_{k}(s)h_{l}(s)\right] \right\vert \\
&\equiv &\Psi _{14,3}^{\prime }(s)+\Psi _{14,3}^{\prime \prime }(s)+\Psi
_{14,3}^{\prime \prime \prime }(s)\text{.}
\end{eqnarray*}%
For $\Psi _{14,3}^{\prime }(s)$, note that 
\begin{eqnarray}
&&\frac{1}{b_{n}^{2}}\mathbb{E}\left[ h_{i}(s)h_{j}(s)\right]  \notag \\
&=&\int_{\gamma _{g}}^{\gamma _{k}}\int_{\gamma _{g}}^{\gamma _{k}}\mathbb{E}%
\left[ \left( x_{i}u_{i}\right) \left( x_{j}u_{j}\right) |q_{i},q_{j},s,s%
\right] f\left( q_{i},q_{j},s,s\right) dq_{i}dq_{j}+O\left( b_{n}^{2}\right)
\label{Ehh} \\
&=&O\left( 1\right)  \notag
\end{eqnarray}%
from Assumptions A-(v) and (vii). Hence, from the fact that $|\{j\in \Lambda
_{n}:\lambda \left( i,j\right) \leq \kappa _{n}\}|=O(\kappa _{n}^{2})$ for
any fixed $i\in \Lambda _{n}$, we obtain 
\begin{eqnarray*}
\Psi _{14,3}^{\prime }(s) &\leq &\frac{C_{7}}{n^{2}b_{n}^{2}}\dsum 
_{\substack{ i,j\in \Lambda _{n}  \\ 0<\lambda \left( i,j\right) \leq \kappa
_{n}}}\left\vert \mathbb{E}\left[ h_{i}(s)h_{j}(s)\right] \right\vert \dsum 
_{\substack{ k,l\in \Lambda _{n}  \\ 0<\lambda \left( k,l\right) \leq \kappa
_{n}}}\left\vert \mathbb{E}\left[ h_{k}(s)h_{l}(s)\right] \right\vert \\
&\leq &C_{7}^{\prime }(s)b_{n}^{2}\kappa _{n}^{4}=o(1)\text{,}
\end{eqnarray*}%
where the last line follows from the construction of $\kappa _{n}$.

For $\Psi _{14,3}^{\prime \prime }(s)$, since $\mathbb{E}\left[
h_{i}(s)h_{j}(s)\right] =Cov\left[ h_{i}(s),h_{j}(s)\right] $, the
covariance inequality (\ref{cov ineq}) yields 
\begin{eqnarray*}
&&\Psi _{14,3}^{\prime \prime }(s) \\
&\leq &\frac{C_{7}}{n^{2}b_{n}^{2}}\dsum_{\substack{ i,j\in \Lambda _{n}  \\ %
\lambda \left( i,j\right) >\kappa _{n}}}\left\vert \mathbb{E}\left[
h_{i}(s)h_{j}(s)\right] \right\vert \dsum_{\substack{ k,l\in \Lambda _{n} 
\\ \lambda \left( k,l\right) >\kappa _{n}}}\left\vert \mathbb{E}\left[
h_{k}(s)h_{l}(s)\right] \right\vert \\
&\leq &\frac{C_{7}^{\prime }}{n^{2}b_{n}^{2}}\left\{ \dsum_{\substack{ %
i,j\in \Lambda _{n}  \\ \lambda \left( i,j\right) >\kappa _{n}}}\alpha
_{1,1}\left( \lambda \left( i,j\right) \right) ^{\varphi /(2+\varphi )}%
\mathbb{E}\left[ \left\vert h_{i}(s)\right\vert ^{2+\varphi }\right]
^{1/\left( 2+\varphi \right) }\mathbb{E}\left[ \left\vert
h_{j}(s)\right\vert ^{2+\varphi }\right] ^{1/\left( 2+\varphi \right)
}\right\} ^{2} \\
&=&\frac{C_{7}^{\prime }}{b_{n}^{2\varphi /(2+\varphi )}}\left\{ \mathbb{E}%
\left[ \frac{1}{b_{n}}\left\vert h_{i}(s)\right\vert ^{2+\varphi }\right]
^{2/\left( 2+\varphi \right) }\frac{1}{n}\dsum_{\substack{ i,j\in \Lambda
_{n}  \\ \lambda \left( i,j\right) >\kappa _{n}}}\alpha _{1,1}\left( \lambda
\left( i,j\right) \right) ^{\varphi /(2+\varphi )}\right\} ^{2} \\
&\leq &\frac{C_{7}^{\prime \prime }(s)}{b_{n}^{2\varphi /(2+\varphi )}}%
\left\{ \frac{1}{n}\dsum_{i\in \Lambda _{n}}\sum_{m=\kappa
_{n}+1}^{n-1}\dsum _{\substack{ j\in \Lambda _{n}  \\ \lambda \left(
i,j\right) \in \lbrack m,m+1)}}\alpha _{1,1}\left( m\right) ^{\varphi
/(2+\varphi )}\right\} ^{2} \\
&\leq &\frac{C_{7}^{\prime \prime }(s)}{b_{n}^{2\varphi /(2+\varphi )}}%
\left\{ \sum_{m=\kappa _{n}+1}^{\infty }m\exp \left( -m\varphi /(2+\varphi
)\right) \right\} ^{2}\text{.}
\end{eqnarray*}%
However, for some $c>0$, we have 
\begin{equation*}
\dsum_{m=\kappa _{n}+1}^{\infty }m\exp \left( -cm\right) \leq \int_{\kappa
_{n}}^{\infty }t\exp \left( -ct\right) dt=\frac{1}{c}\left( \kappa _{n}+%
\frac{1}{c}\right) \exp \left( -c\kappa _{n}\right) \text{.}
\end{equation*}%
As we set $\kappa _{n}=O\left( b_{n}^{-\ell }\right) $ for some $\ell >0$,
it follows that 
\begin{eqnarray*}
\Psi _{14,3}^{\prime \prime }(s) &=&O\left( \frac{\kappa _{n}^{2}\exp \left(
-(2\varphi /(2+\varphi ))\kappa _{n}\right) }{b_{n}^{2\varphi /(2+\varphi )}}%
\right) \\
&=&O\left( \exp \left( -\frac{2\varphi }{2+\varphi }\times \frac{1}{%
b_{n}^{\ell }}\right) \times \frac{1}{b_{n}^{2(\varphi +\ell )/(2+\varphi )}}%
\right) =o\left( 1\right)
\end{eqnarray*}%
since the exponential term decays faster for any $\ell ,\varphi >0$. For $%
\Psi _{14,3}^{\prime \prime \prime }(s)$, by combining the arguments for
bounding $\Psi _{14,3}^{\prime }(s)$ and $\Psi _{14,3}^{\prime \prime }(s)$,
we obtain that 
\begin{equation*}
\Psi _{14,3}^{\prime \prime \prime }(s)\leq C_{7}^{\prime }(s)\left\{
b_{n}\kappa _{n}^{2}\right\} \left\{ b_{n}^{-\varphi /(2+\varphi )}\kappa
_{n}\exp (-\kappa _{n}\varphi /(2+\varphi ))\right\} =o(1)\text{.}
\end{equation*}

For $\Psi _{15}(s)$, we let $\mathcal{E}^{\prime }=\{(i,j,k):i\neq j\neq k$
and $0<\lambda (i,j)\leq \lambda (i,k)\}$ and decompose it into%
\begin{eqnarray*}
\Psi _{15}(s) &=&\frac{2}{n^{2}b_{n}^{2}}\dsum_{\substack{ i,j,k\in \Lambda
_{n}\cap \mathcal{E}^{\prime }  \\ \lambda \left( i,j\right) <\lambda (j,k)}}%
\mathbb{E}\left[ h_{i}^{2}(s)h_{j}(s)h_{k}(s)\right] +\frac{2}{n^{2}b_{n}^{2}%
}\dsum_{\substack{ i,j,k\in \Lambda _{n}\cap \mathcal{E}^{\prime }  \\ %
\lambda \left( i,j\right) \geq \lambda (j,k)}}\mathbb{E}\left[
h_{i}^{2}(s)h_{j}(s)h_{k}(s)\right] \\
&\leq &\frac{2}{n^{2}b_{n}^{2}}\dsum_{\substack{ i,j,k\in \Lambda _{n}\cap 
\mathcal{E}^{\prime }  \\ \lambda \left( i,j\right) <\lambda (j,k)}}%
\left\vert Cov\left[ \{h_{i}^{2}(s)h_{j}(s)\},h_{k}(s)\right] \right\vert \\
&&+\frac{2}{n^{2}b_{n}^{2}}\dsum_{\substack{ i,j,k\in \Lambda _{n}\cap 
\mathcal{E}^{\prime }  \\ \lambda \left( i,j\right) \geq \lambda (j,k)}}%
\left\vert Cov\left[ h_{i}^{2}(s),\{h_{j}(s)h_{k}(s)\}\right] \right\vert \\
&&+\frac{2}{n^{2}b_{n}^{2}}\dsum_{\substack{ i,j,k\in \Lambda _{n}\cap 
\mathcal{E}^{\prime }  \\ \lambda \left( i,j\right) \geq \lambda (j,k)}}%
\left\vert \mathbb{E}\left[ h_{i}^{2}(s)\right] \mathbb{E}\left[
h_{j}(s)h_{k}(s)\right] \right\vert \\
&\equiv &\Psi _{15,1}(s)+\Psi _{15,2}(s)+\Psi _{15,3}(s)\text{.}
\end{eqnarray*}%
Similarly as $\Psi _{14,1}(s)$, 
\begin{eqnarray*}
&&\Psi _{15,1}(s) \\
&\leq &\frac{C_{8}}{n^{2}b_{n}^{2}}\dsum_{\substack{ i,j,k\in \Lambda
_{n}\cap \mathcal{E}^{\prime }  \\ \lambda \left( i,j\right) <\lambda (j,k)}}%
\alpha _{2,1}\left( \lambda (j,k)\}\right) ^{\varphi /(2+\varphi )}\mathbb{E}%
\left[ \left\vert h_{i}^{2}(s)h_{j}(s)\right\vert ^{2(2+\varphi )/3}\right]
^{3/\left( 4+2\varphi \right) }\mathbb{E}\left[ \left\vert
h_{k}(s)\right\vert ^{2(2+\varphi )}\right] ^{1/\left( 4+2\varphi \right) }
\\
&\leq &\frac{C_{8}^{\prime }}{n^{2}b_{n}^{(2+2\varphi )/(2+\varphi )}}%
\dsum_{j\in \Lambda _{n}}\dsum_{m=1}^{n-1}\dsum_{\substack{ k\in \Lambda
_{n}  \\ \lambda \left( j,k\right) \in \lbrack m,m+1)}}\dsum_{\substack{ %
i\in \Lambda _{n}  \\ \lambda \left( i,j\right) \leq m}}\alpha _{2,1}\left(
m\right) ^{\varphi /(2+\varphi )} \\
&\leq &\frac{C_{8}^{\prime \prime }}{nb_{n}^{(2+2\varphi )/(2+\varphi )}}%
\dsum_{m=1}^{\infty }m^{3}\exp \left( -m\varphi /(2+\varphi )\right) \\
&=&o\left( 1\right) \text{,}
\end{eqnarray*}%
and the same argument implies that $\Psi _{15,2}(s)=o(1)$ as well. For $\Psi
_{15,3}(s)$, we let a sequence of integers $\kappa _{n}^{\prime }=O\left(
b_{n}^{-\ell ^{\prime }}\right) $ for some $\ell ^{\prime }>0$ such that $%
\kappa _{n}^{\prime }\rightarrow \infty $ and $(\kappa _{n}^{\prime
})^{2}b_{n}\rightarrow 0$ as $n\rightarrow \infty $. Then, similarly as $%
\Psi _{14,3}(s)$, we decompose%
\begin{eqnarray*}
&&\Psi _{15,3}(s) \\
&=&\frac{2}{n^{2}b_{n}^{2}}\dsum_{\substack{ i,j,k\in \Lambda _{n}\cap 
\mathcal{E}^{\prime }  \\ \lambda \left( i,j\right) \geq \lambda
(j,k),0<\lambda \left( j,k\right) \leq \kappa _{n}^{\prime }}}\left\vert 
\mathbb{E}\left[ h_{i}^{2}(s)\right] \mathbb{E}\left[ h_{j}(s)h_{k}(s)\right]
\right\vert \\
&&+\frac{2}{n^{2}b_{n}^{2}}\dsum_{\substack{ i,j,k\in \Lambda _{n}\cap 
\mathcal{E}^{\prime }  \\ \lambda \left( i,j\right) \geq \lambda
(j,k),\lambda \left( j,k\right) >\kappa _{n}^{\prime }}}\left\vert \mathbb{E}%
\left[ h_{i}^{2}(s)\right] \mathbb{E}\left[ h_{j}(s)h_{k}(s)\right]
\right\vert \\
&\leq &\frac{2}{nb_{n}}\dsum_{i\in \Lambda _{n}}\left\vert \mathbb{E}\left[
h_{i}^{2}(s)\right] \right\vert \left\{ \frac{1}{nb_{n}}\dsum_{\substack{ %
j,k\in \Lambda _{n}  \\ 0<\lambda \left( j,k\right) \leq \kappa _{n}^{\prime
} }}\left\vert \mathbb{E}\left[ h_{j}(s)h_{k}(s)\right] \right\vert +\frac{1%
}{nb_{n}}\dsum_{\substack{ j,k\in \Lambda _{n}  \\ \lambda \left( j,k\right)
>\kappa _{n}^{\prime }}}\left\vert \mathbb{E}\left[ h_{j}(s)h_{k}(s)\right]
\right\vert \right\} \\
&=&O\left( 1\right) \left\{ O\left( (\kappa _{n}^{\prime })^{2}b_{n}\right)
+O\left( \frac{\kappa _{n}^{\prime }\exp \left( -(\varphi /(2+\varphi
))\kappa _{n}^{\prime }\right) }{b_{n}^{\varphi /(2+\varphi )}}\right)
\right\} \\
&=&o\left( 1\right) \text{.}
\end{eqnarray*}

By combining all the results for $\Psi _{11}$ to $\Psi _{15}$, we thus have 
\begin{equation*}
\mathbb{E}\left[ \left\vert J_{n}\left( \gamma _{g};s\right) -J_{n}\left(
\gamma _{k};s\right) \right\vert ^{4}\right] \leq C\left( \gamma _{k}-\gamma
_{g}\right) ^{2}+o\left( 1\right) \leq C\left( \left\vert k-g\right\vert
\varpi /\overline{g}\right) ^{2}+o\left( 1\right)
\end{equation*}%
for some $C<\infty $, and Theorem 12.2 of \cite{Billingsley68} yields that%
\begin{equation}
\mathbb{P}\left( \Psi _{1}(s)>\eta \right) =\mathbb{P}\left( \max_{2\leq
g\leq \overline{g}+1}\left\vert J_{n}\left( \gamma _{g};s\right)
-J_{n}\left( \gamma _{1};s\right) \right\vert >\eta \right) \leq \frac{%
C\varpi ^{2}}{\eta ^{4}}  \label{P1}
\end{equation}%
for $\eta >0$.

Next, for $\Psi _{2}(s)$, a similar argument yields that 
\begin{equation*}
\mathbb{E}\left[ \left( \sqrt{nb_{n}}\left\vert H_{ng}(s)-\mathbb{E}\left[
H_{ng}(s)\right] \right\vert \right) ^{4}\right] \leq C^{\prime }\varpi ^{2}/%
\overline{g}^{2}
\end{equation*}%
for some $C^{\prime }<\infty $, and hence by Markov's inequality we have%
\begin{equation}
\mathbb{P}\left( \max_{1\leq g\leq \overline{g}}\sqrt{nb_{n}}\left\vert
H_{ng}(s)-\mathbb{E}\left[ H_{ng}(s)\right] \right\vert >\eta \right) \leq 
\overline{g}\frac{C^{\prime }\varpi ^{2}/\overline{g}^{2}}{\eta ^{4}}\leq 
\frac{C^{\prime }\varpi ^{2}}{\eta ^{4}}  \label{P2}
\end{equation}%
since $\overline{g}=O(nb_{n})$ by construction. Finally, to bound $\Psi
_{3}(s)$, note that 
\begin{equation}
\sqrt{nb_{n}}\mathbb{E}\left[ H_{ng}(s)\right] \leq \sqrt{nb_{n}}C\varpi /%
\overline{g}\leq C^{\prime }/\sqrt{nb_{n}}  \label{P3}
\end{equation}%
for some $C,C^{\prime }<\infty $, since $\varpi /\overline{g}%
=O((nb_{n})^{-1})$. So the proof is complete by combining (\ref{P1}), (\ref%
{P2}), and (\ref{P3}). $\blacksquare $

\begin{lemma}
\label{weak J}For any fixed $s\in \mathcal{S}_{0}$,%
\begin{equation*}
J_{n}\left( \gamma ;s\right) \Rightarrow J\left( \gamma ;s\right) \text{,}
\end{equation*}%
where $J\left( \gamma ;s\right) $ is a mean-zero Gaussian process indexed by 
$\gamma $ as $n\rightarrow \infty $.
\end{lemma}

\paragraph{Proof of Lemma \protect\ref{weak J}}

For a fixed $\gamma $, the Theorem of \cite{Bolthausen82} implies that $%
J_{n}\left( \gamma ;s\right) \rightarrow _{d}J\left( \gamma ;s\right) $ as $%
n\rightarrow \infty $ under Assumption A-(iii). Because $\gamma $ is in the
indicator function, such pointwise convergence in $\gamma $ can be
generalized into any finite collection of $\gamma $ to yield the finite
dimensional convergence in distribution. Then the weak convergence follows
from Lemma \ref{max ineq} above and Theorem 15.5 of \cite{Billingsley68}. $%
\blacksquare $

\begin{lemma}
\label{L1}%
\begin{align*}
\sup_{\left( \gamma ,s\right) \in \Gamma \times \mathcal{S}_{0}}\left\Vert
M_{n}\left( \gamma ;s\right) -M\left( \gamma ;s\right) \right\Vert &
\rightarrow _{p}0\text{,} \\
\sup_{\left( \gamma ,s\right) \in \Gamma \times \mathcal{S}%
_{0}}(nb_{n})^{-1/2}\left\Vert J_{n}\left( \gamma ;s\right) \right\Vert &
\rightarrow _{p}0
\end{align*}%
as $n\rightarrow \infty $, where 
\begin{equation}
M\left( \gamma ;s\right) =\int_{-\infty }^{\gamma }D(q,s)f\left( q,s\right)
dq\text{.}  \label{M_gs}
\end{equation}
\end{lemma}

\paragraph{Proof of Lemma \protect\ref{L1}}

For expositional simplicity, we only present the case of scalar $x_{i}$. We
prove the convergence of $M_{n}\left( \gamma ;s\right) $. For $J_{n}\left(
\gamma ;s\right) $, since $\mathbb{E}\left[ u_{i}x_{i}|q_{i},s_{i}\right] =0$%
, the proof is identical as $M_{n}\left( \gamma ;s\right) $ and hence
omitted.

By stationarity, Assumptions A-(vii), (x), and Taylor expansion, we have 
\begin{eqnarray}
\mathbb{E}\left[ M_{n}\left( \gamma ;s\right) \right] &=&\frac{1}{b_{n}}%
\iint \mathbb{E}[x_{i}^{2}|q,v]\mathbf{1}[q\leq \gamma ]K\left( \frac{v-s}{%
b_{n}}\right) f\left( q,v\right) dqdv  \label{EM} \\
&=&\iint D(q,s+b_{n}t)\mathbf{1}[q\leq \gamma ]K\left( t\right) f\left(
q,s+b_{n}t\right) dqdt  \notag \\
&=&M\left( \gamma ;s\right) +b_{n}^{2}\int \widetilde{M}\left( q;s\right) 
\mathbf{1}[q\leq \gamma ]dq\int t^{2}K\left( t\right) dt\text{,}  \notag
\end{eqnarray}%
where $\widetilde{M}\left( q;s\right) =\dot{D}(q,s)\dot{f}\left( q,s\right)
+(\ddot{D}(q,s)+\ddot{f}\left( q,s\right) )/2$. We let $\dot{D}$ and $\dot{f}
$ denote the partial derivatives, and $\ddot{D}$ and $\ddot{f}$ denote the
second-order partial derivatives with respect to $s$. Since $\sup_{s\in 
\mathcal{S}_{0}}||\widetilde{M}\left( q;s\right) ||<\infty $ for any $q$
from Assumption A-(vii), and $K\left( \cdot \right) $ is a second-order
kernel, we have 
\begin{equation}
\sup_{\left( \gamma ,s\right) \in \Gamma \times \mathcal{S}_{0}}\left\Vert 
\mathbb{E}\left[ M_{n}\left( \gamma ;s\right) \right] -M\left( \gamma
;s\right) \right\Vert =O_{p}\left( b_{n}^{2}\right) =o_{p}(1)\text{.}
\label{M1}
\end{equation}

Next, we let $\tau _{n}=(n\log n)^{1/(4+\varphi )}$ and $\varphi $ is given
in Assumption A-(v). By Markov's and H\"{o}lder's inequalities, Assumption
A-(v) gives $\mathbb{P}\left( x_{n}^{2}>\tau _{n}\right) \leq C\tau
_{n}^{-(4+\varphi )}\mathbb{E}[|x_{n}^{2}|^{4+\varphi }]\leq C^{\prime
}\left( n\log n\right) ^{-1}$ for some $C,C^{\prime }<\infty $. Thus%
\begin{equation*}
\dsum_{n\in 
%TCIMACRO{\U{2124} }%
%BeginExpansion
\mathbb{Z}
%EndExpansion
^{2}}\mathbb{P}\left( x_{n}^{2}>\tau _{n}\right) \leq C^{\prime }\dsum_{n\in 
%TCIMACRO{\U{2124} }%
%BeginExpansion
\mathbb{Z}
%EndExpansion
^{2}}\left( n\log n\right) ^{-1}<\infty \text{,}
\end{equation*}%
which yields that $x_{n}^{2}\leq \tau _{n}$ almost surely for sufficiently
large $n$ by the Borel-Cantelli lemma. Since $\tau _{n}\rightarrow \infty $
as $n\rightarrow \infty $, we have $x_{i}^{2}\leq \tau _{n}$ for any $i\in
\Lambda _{n}$ and hence 
\begin{equation*}
\sup_{\left( \gamma ,s\right) \in \Gamma \times \mathcal{S}_{0}}\left\Vert
M_{n}(\gamma ;s)-M_{n}^{\tau }(\gamma ;s)\right\Vert =0\text{ \ and }%
\sup_{\left( \gamma ,s\right) \in \Gamma \times \mathcal{S}_{0}}\left\Vert 
\mathbb{E}\left[ M_{n}(\gamma ;s)\right] -\mathbb{E}\left[ M_{n}^{\tau
}(\gamma ;s)\right] \right\Vert =0
\end{equation*}%
almost surely for sufficiently large $n$, where 
\begin{equation}
M_{n}^{\tau }(\gamma ;s)=\frac{1}{nb_{n}}\sum_{i\in \Lambda _{n}}x_{i}^{2}%
\mathbf{1}_{i}(\gamma )K_{i}(s)\mathbf{1}\left\{ x_{i}^{2}\leq \tau
_{n}\right\} \text{.}  \label{Mnt}
\end{equation}%
It follows that 
\begin{eqnarray}
\sup_{\left( \gamma ,s\right) \in \Gamma \times \mathcal{S}_{0}}\left\Vert
M_{n}(\gamma ;s)-\mathbb{E}\left[ M_{n}(\gamma ;s)\right] \right\Vert &\leq
&\sup_{\left( \gamma ,s\right) \in \Gamma \times \mathcal{S}_{0}}\left\Vert
M_{n}(\gamma ;s)-M_{n}^{\tau }(\gamma ;s)\right\Vert  \label{VM} \\
&&+\sup_{\left( \gamma ,s\right) \in \Gamma \times \mathcal{S}%
_{0}}\left\Vert M_{n}^{\tau }(\gamma ;s)-\mathbb{E}\left[ M_{n}^{\tau
}(\gamma ;s)\right] \right\Vert  \notag \\
&&+\sup_{\left( \gamma ,s\right) \in \Gamma \times \mathcal{S}%
_{0}}\left\Vert \mathbb{E}\left[ M_{n}(\gamma ;s)\right] -\mathbb{E}\left[
M_{n}^{\tau }(\gamma ;s)\right] \right\Vert  \notag
\end{eqnarray}%
and we establish $\sup_{\left( \gamma ,s\right) \in \Gamma \times \mathcal{S}%
_{0}}\left\Vert M_{n}(\gamma ;s)-\mathbb{E}\left[ M_{n}(\gamma ;s)\right]
\right\Vert =o_{p}(1)$ if the second term in (\ref{VM}) is $o_{p}(1)$. Then
we conclude $\sup_{\left( \gamma ,s\right) \in \Gamma \times \mathcal{S}%
_{0}}\left\Vert M_{n}\left( \gamma ;s\right) -M\left( \gamma ;s\right)
\right\Vert \rightarrow _{p}0$ as desired by combining (\ref{M1}) and (\ref%
{VM}).

To this end, we let $m_{n}$ be an integer such that $m_{n}=O(\tau
_{n}(n/(b_{n}^{3}\log n))^{1/2})$ and we cover the compact $\Gamma \times 
\mathcal{S}_{0}$ by small $m_{n}^{2}$ squares centered at $\left( \gamma
_{k_{1}},s_{k_{2}}\right) $, which are defined as $\mathcal{I}_{k}=\{\left(
\gamma ^{\prime },s^{\prime }\right) :\left\vert \gamma ^{\prime }-\gamma
_{k_{1}}\right\vert \leq C/m_{n}$ and $\left\vert s^{\prime
}-s_{k_{2}}\right\vert \leq C/m_{n}\}$ for some $C<\infty $. Note that $\tau
_{n}\left( n^{1-2\epsilon }b_{n}/\log n\right) ^{1/2}\left( n^{2\epsilon
}/b_{n}^{4}\right) ^{1/2}\rightarrow \infty $ as $n\rightarrow \infty $ from
Assumption A-(xi), hence $m_{n}\rightarrow \infty $. We then have 
\begin{eqnarray*}
\sup_{\left( \gamma ,s\right) \in \Gamma \times \mathcal{S}_{0}}\left\Vert
M_{n}^{\tau }(\gamma ;s)-\mathbb{E}\left[ M_{n}^{\tau }(\gamma ;s)\right]
\right\Vert &\leq &\max_{\substack{ 1\leq k_{1}\leq m_{n}  \\ 1\leq
k_{2}\leq m_{n}}}\sup_{\left( \gamma ,s\right) \in \mathcal{I}%
_{k}}\left\Vert M_{n}^{\tau }(\gamma ;s)-\mathbb{E}\left[ M_{n}^{\tau
}(\gamma ;s)\right] \right\Vert \\
&\leq &\max_{\substack{ 1\leq k_{1}\leq m_{n}  \\ 1\leq k_{2}\leq m_{n}}}%
\sup_{\left( \gamma ,s\right) \in \mathcal{I}_{k}}\left\Vert M_{n}^{\tau
}(\gamma ;s)-M_{n}^{\tau }(\gamma _{k_{1}};s_{k_{2}})\right\Vert \\
&&+\max_{\substack{ 1\leq k_{1}\leq m_{n}  \\ 1\leq k_{2}\leq m_{n}}}%
\sup_{\left( \gamma ,s\right) \in \mathcal{I}_{k}}\left\Vert \mathbb{E}\left[
M_{n}^{\tau }(\gamma ;s)\right] -\mathbb{E}\left[ M_{n}^{\tau }(\gamma
_{k_{1}};s_{k_{2}})\right] \right\Vert \\
&&+\max_{\substack{ 1\leq k_{1}\leq m_{n}  \\ 1\leq k_{2}\leq m_{n}}}%
\left\Vert M_{n}^{\tau }(\gamma _{k_{1}};s_{k_{2}})-\mathbb{E}\left[
M_{n}^{\tau }(\gamma _{k_{1}};s_{k_{2}})\right] \right\Vert \\
&\equiv &\Psi _{M1}+\Psi _{M2}+\Psi _{M3}\text{.}
\end{eqnarray*}%
We first decompose $M_{n}^{\tau }(\gamma ;s)-M_{n}^{\tau }(\gamma
_{k_{1}};s_{k_{2}})\leq M_{1n}^{\tau }(\gamma ,\gamma
_{k_{1}};s,s_{k_{2}})+M_{2n}^{\tau }(\gamma ,\gamma _{k_{1}};s,s_{k_{2}})$,
where 
\begin{eqnarray*}
M_{1n}^{\tau }(\gamma ,\gamma _{k_{1}};s) &=&\frac{1}{nb_{n}}\dsum_{i\in
\Lambda _{n}}x_{i}^{2}\left\vert \mathbf{1}\left[ q_{i}\leq \gamma \right] -%
\mathbf{1}\left[ q_{i}\leq \gamma _{k_{1}}\right] \right\vert
K_{i}(s_{k_{2}})\mathbf{1}\left[ x_{i}^{2}\leq \tau _{n}\right] \text{,} \\
M_{2n}^{\tau }(\gamma ;s,s_{k_{2}}) &=&\frac{1}{nb_{n}}\dsum_{i\in \Lambda
_{n}}x_{i}^{2}\mathbf{1}\left[ q_{i}\leq \gamma \right] \left\vert
K_{i}(s)-K_{i}(s_{k_{2}})\right\vert \mathbf{1}\left[ x_{i}^{2}\leq \tau _{n}%
\right] \text{.}
\end{eqnarray*}%
Without loss of generality, we can suppose $\gamma >\gamma _{k_{1}}$ in $%
M_{1n}^{\tau }(\gamma ,\gamma _{k_{1}};s)$. Since $K_{i}(\cdot )$ is bounded
from Assumption A-(x) and we only consider $x_{i}^{2}\leq \tau _{n}$, for
any $\gamma $ such that $\left\vert \gamma -\gamma _{k_{1}}\right\vert \leq
C/m_{n}$, 
\begin{eqnarray}
\left\Vert M_{1n}^{\tau }(\gamma ,\gamma _{k_{1}};s)\right\Vert &\leq &C_{1}%
\frac{\tau _{n}}{nb_{n}}\dsum_{i\in \Lambda _{n}}\mathbf{1}\left[ \gamma
_{k_{1}}<q_{i}\leq \gamma \right]  \label{M1nt} \\
&=&C_{1}\tau _{n}b_{n}^{-1}\mathbb{P}\left( \gamma _{k_{1}}<q_{i}\leq \gamma
\right) \text{ \ almost surely}  \notag \\
&\leq &C_{1}^{\prime }\tau _{n}b_{n}^{-1}m_{n}^{-1}  \notag \\
&=&C_{1}^{\prime \prime }\left( \frac{b_{n}\log n}{n}\right)
^{1/2}=O_{a.s.}\left( \left( \frac{\log n}{nb_{n}}\right) ^{1/2}\right) 
\notag
\end{eqnarray}%
for some $C_{1},C_{1}^{\prime },C_{1}^{\prime \prime }<\infty $, where the
second equality is by the uniform almost sure law of large numbers for
random fields (e.g., \cite{Jenish09}, Theorem 2). This bound holds uniformly
in $\left( \gamma ,s\right) \in \mathcal{I}_{k}$ and $k_{1},k_{2}\in
\{1,\ldots ,m_{n}\}$. Similarly, since $K(\cdot )$ is Lipschitz from
Assumption A-(x), 
\begin{eqnarray}
\left\Vert M_{2n}^{\tau }(\gamma ;s,s_{k_{2}})\right\Vert &\leq &\frac{\tau
_{n}}{nb_{n}}\dsum_{i\in \Lambda _{n}}\left\vert
K_{i}(s)-K_{i}(s_{k_{2}})\right\vert  \label{M2nt} \\
&\leq &C_{2}\frac{\tau _{n}}{b_{n}^{2}}\left\vert s-s_{k_{2}}\right\vert
\leq \frac{C_{2}^{\prime }\tau _{n}}{b_{n}^{2}m_{n}}=O_{a.s.}\left( \left( 
\frac{\log n}{nb_{n}}\right) ^{1/2}\right)  \notag
\end{eqnarray}%
for some $C_{2},C_{2}^{\prime }<\infty $, uniformly in $\gamma $, $s$, $%
k_{1} $ and $k_{2}$. It follows that 
\begin{equation*}
\left\Vert M_{n}^{\tau }(\gamma ;s)-M_{n}^{\tau }(\gamma
_{k_{1}};s_{k_{2}})\right\Vert =O_{a.s.}(\left( \log n/(nb_{n})\right)
^{1/2})
\end{equation*}%
uniformly in $\gamma $, $s$, $k_{1}$ and $k_{2}$, and hence we can readily
verify that both $\Psi _{M1}$ and $\Psi _{M2}$ are $O_{a.s.}((\log
n/(nb_{n}))^{1/2})$. For $\Psi _{M3}$, we follow the same argument for
bounding the $Q_{3n}^{\ast }$ term on pp.794-796 of \cite%
{CarbonFrancqTran2007}. In particular, for any $k_{1}\in \{1,\ldots ,m_{n}\}$%
, $\max_{1\leq k_{2}\leq m_{n}}||M_{n}^{\tau }(\gamma _{k_{1}};s_{k_{2}})-%
\mathbb{E}\left[ M_{n}^{\tau }(\gamma _{k_{1}};s_{k_{2}})\right] ||\leq
C_{3}\left( \log n/(nb_{n})\right) ^{1/2}$ almost surely for some $%
C_{3}<\infty $. Note that $\gamma _{k_{1}}$ shows up in the indicator
function $\mathbf{1}\left[ q_{i}\leq \gamma _{k_{1}}\right] $ only, which is
uniformly bounded by 1. The bound is hence uniform over all $k_{1}\in
\{1,\ldots ,m_{n}\}$ and $\Psi _{M3}=O_{a.s.}((\log n/(nb_{n}))^{1/2})$ as
well. Combining the bounds for $\Psi _{M1},\Psi _{M2}$, and $\Psi _{M3}$, we
have $\sup_{\left( \gamma ,s\right) \in \Gamma \times \mathcal{S}%
_{0}}\left\Vert M_{n}^{\tau }(\gamma ;s)-\mathbb{E}\left[ M_{n}^{\tau
}(\gamma ;s)\right] \right\Vert =o_{a.s.}(1)$ and hence complete the proof
because $\log n/(nb_{n})\rightarrow 0$ from Assumption A-(ix). $\blacksquare 
$

\begin{lemma}
\label{L-int1}Uniformly over $s\in \mathcal{S}_{0}$, 
\begin{equation}
\Delta M_{n}\left( s\right) \equiv \frac{1}{nb_{n}}\dsum_{i\in \Lambda
_{n}}x_{i}x_{i}^{\top }\left\{ \mathbf{1}_{i}\left( \gamma _{0}\left(
s_{i}\right) \right) -\mathbf{1}_{i}\left( \gamma _{0}\left( s\right)
\right) \right\} K_{i}\left( s\right) =O_{a.s.}\left( b_{n}\right) \text{.}
\label{DMn}
\end{equation}
\end{lemma}

\paragraph{Proof of Lemma \protect\ref{L-int1}}

See the supplementary material. $\blacksquare $

\begin{lemma}
\label{L-A3}For a given $s\in \mathcal{S}_{0}$, $\widehat{\gamma }%
(s)\rightarrow _{p}\gamma _{0}(s)$ as $n\rightarrow \infty $.
\end{lemma}

\paragraph{Proof of Lemma \protect\ref{L-A3}}

For given $s\in \mathcal{S}_{0}$, we let $\widetilde{y}%
_{i}(s)=K_{i}(s)^{1/2}y_{i}$, $\widetilde{x}_{i}(s)=K_{i}(s)^{1/2}x_{i}$, $%
\widetilde{u}_{i}(s)=K_{i}(s)^{1/2}u_{i}$, $\widetilde{x}_{i}(\gamma
;s)=K_{i}(s)^{1/2}x_{i}\mathbf{1}_{i}\left( \gamma \right) $, and $%
\widetilde{x}_{i}(\gamma _{0}(s_{i});s)=K_{i}(s)^{1/2}x_{i}\mathbf{1}%
_{i}\left( \gamma _{0}(s_{i})\right) $. We denote $\widetilde{y}(s)$, $%
\widetilde{X}(s)$, $\widetilde{u}(s)$, $\widetilde{X}(\gamma ;s)$, and $%
\widetilde{X}(\gamma _{0}(s_{i});s)$ as their corresponding matrices of $n$%
-stacks. Then $\widehat{\theta }(\gamma ;s)=(\widehat{\beta }(\gamma
;s)^{\top },\widehat{\delta }(\gamma ;s)^{\top })^{\top }$ in (\ref{first})
is given as 
\begin{equation}
\widehat{\theta }(\gamma ;s)=(\widetilde{Z}(\gamma ;s)^{\top }\widetilde{Z}%
(\gamma ;s))^{-1}\widetilde{Z}(\gamma ;s)^{\top }\widetilde{y}(s)\text{,}
\label{theta}
\end{equation}%
where $\widetilde{Z}(\gamma ;s)=[\widetilde{X}(s),\widetilde{X}(\gamma ;s)]$%
. Therefore, since $\widetilde{y}(s)=\widetilde{X}(s)\beta _{0}+\widetilde{X}%
(\gamma _{0}(s_{i});s)\delta _{0}+\widetilde{u}(s)$ and $\widetilde{X}(s)$
lies in the space spanned by $\widetilde{Z}(\gamma ;s)$, we have%
\begin{eqnarray*}
Q_{n}\left( \gamma ;s\right) -\widetilde{u}(s)^{\top }\widetilde{u}(s) &=&%
\widetilde{y}(s)^{\top }\left( I_{n}-P_{\widetilde{Z}}(\gamma ;s)\right) 
\widetilde{y}(s)-\widetilde{u}(s)^{\top }\widetilde{u}(s) \\
&=&-\widetilde{u}(s)^{\top }P_{\widetilde{Z}}(\gamma ;s)\widetilde{u}%
(s)+2\delta _{0}^{\top }\widetilde{X}(\gamma _{0}(s_{i});s)^{\top }\left(
I_{n}-P_{\widetilde{Z}}(\gamma ;s)\right) \widetilde{u}(s) \\
&&+\delta _{0}^{\top }\widetilde{X}(\gamma _{0}(s_{i});s)^{\top }\left(
I_{n}-P_{\widetilde{Z}}(\gamma ;s)\right) \widetilde{X}(\gamma
_{0}(s_{i});s)\delta _{0}\text{,}
\end{eqnarray*}%
where $P_{\widetilde{Z}}(\gamma ;s)=\widetilde{Z}(\gamma ;s)(\widetilde{Z}%
(\gamma ;s)^{\top }\widetilde{Z}(\gamma ;s))^{-1}\widetilde{Z}(\gamma
;s)^{\top }$ and $I_{n}$ is the identity matrix of rank $n$. Note that $P_{%
\widetilde{Z}}(\gamma ;s)$ is the same as the projection onto $[\widetilde{X}%
(s)-\widetilde{X}(\gamma ;s),\widetilde{X}(\gamma ;s)]$, where $\widetilde{X}%
(\gamma ;s)^{\top }(\widetilde{X}(s)-\widetilde{X}(\gamma ;s))=0$.
Furthermore, for $\gamma \geq \gamma _{0}(s_{i})$, $\widetilde{x}_{i}(\gamma
_{0}(s_{i});s)^{\top }(\widetilde{x}_{i}(s)-\widetilde{x}_{i}(\gamma ;s))=0$
and hence $\widetilde{X}(\gamma _{0}(s_{i});s)^{\top }\widetilde{X}(\gamma
;s)=\widetilde{X}(\gamma _{0}(s_{i});s)^{\top }\widetilde{X}(\gamma
_{0}(s_{i});s)$. Since we can rewrite 
\begin{eqnarray*}
M_{n}(\gamma ;s) &=&\frac{1}{nb_{n}}\dsum_{i\in \Lambda _{n}}\widetilde{x}%
_{i}(\gamma ;s)\widetilde{x}_{i}(\gamma ;s)^{\top }\text{ \ and} \\
J_{n}(\gamma ;s) &=&\frac{1}{\sqrt{nb_{n}}}\dsum_{i\in \Lambda _{n}}%
\widetilde{x}_{i}(\gamma ;s)\widetilde{u}_{i}(s)\text{,}
\end{eqnarray*}%
Lemma \ref{L1} yields that%
\begin{eqnarray*}
\widetilde{Z}(\gamma ;s)^{\top }\widetilde{u}(s) &=&[\widetilde{X}(s)^{\top }%
\widetilde{u}(s),\widetilde{X}(\gamma ;s)^{\top }\widetilde{u}%
(s)]=O_{p}\left( (nb_{n})^{1/2}\right) \\
\widetilde{Z}(\gamma ;s)^{\top }\widetilde{X}(\gamma _{0}(s_{i});s) &=&[%
\widetilde{X}(s)^{\top }\widetilde{X}(\gamma _{0}(s_{i});s),\widetilde{X}%
(\gamma ;s)^{\top }\widetilde{X}(\gamma _{0}(s_{i});s)] \\
&=&[\widetilde{X}(s)^{\top }\widetilde{X}(\gamma _{0}(s_{i});s),\widetilde{X}%
(\gamma _{0}(s_{i});s)^{\top }\widetilde{X}(\gamma
_{0}(s_{i});s)]=O_{p}\left( nb_{n}\right)
\end{eqnarray*}%
for given $s$. It follows that 
\begin{eqnarray}
\Upsilon _{n}\left( \gamma ;s\right) &\equiv &\frac{1}{a_{n}}\left(
Q_{n}\left( \gamma ;s\right) -\widetilde{u}(s)^{\top }\widetilde{u}(s)\right)
\label{pf1} \\
&=&O_{p}\left( \frac{1}{a_{n}}\right) +O_{p}\left( \frac{1}{a_{n}^{1/2}}%
\right) +\frac{1}{nb_{n}}c_{0}^{\top }\widetilde{X}(\gamma
_{0}(s_{i});s)^{\top }\left( I_{n}-P_{\widetilde{Z}}(\gamma ;s)\right) 
\widetilde{X}(\gamma _{0}(s_{i});s)c_{0}  \notag \\
&=&\frac{1}{nb_{n}}c_{0}^{\top }\widetilde{X}(\gamma _{0}(s_{i});s)^{\top
}\left( I-P_{\widetilde{Z}}(\gamma ;s)\right) \widetilde{X}(\gamma
_{0}(s_{i});s)c_{0}+o_{p}(1)  \notag
\end{eqnarray}%
for $a_{n}=n^{1-2\epsilon }b_{n}\rightarrow \infty $ as $n\rightarrow \infty 
$. Moreover, we have 
\begin{eqnarray}
M_{n}\left( \gamma _{0}(s_{i});s\right) &=&\frac{1}{nb_{n}}\dsum_{i\in
\Lambda _{n}}\widetilde{x}_{i}(\gamma _{0}(s_{i});s)\widetilde{x}_{i}(\gamma
_{0}(s_{i});s)^{\top }  \label{m0} \\
&=&M_{n}\left( \gamma _{0}(s);s\right) +\Delta M_{n}\left( s\right)  \notag
\\
&=&M_{n}\left( \gamma _{0}(s);s\right) +o_{a.s.}\left( 1\right)  \notag
\end{eqnarray}%
from Lemma \ref{L-int1}, where $\Delta M_{n}\left( s\right) $ is defined in (%
\ref{DMn}). It follows that 
\begin{align}
& \frac{1}{nb_{n}}c_{0}^{\top }\widetilde{X}(\gamma _{0}(s_{i});s)^{\top
}\left( I_{n}-P_{\widetilde{Z}}(\gamma ;s)\right) \widetilde{X}(\gamma
_{0}(s_{i});s)c_{0}  \label{pf2} \\
& \rightarrow _{p}c_{0}^{\top }M(\gamma _{0}(s);s)c_{0}-c_{0}^{\top
}M(\gamma _{0}(s);s)^{\top }M(\gamma ;s)^{-1}M(\gamma _{0}(s);s)c_{0}\equiv
\Upsilon _{0}(\gamma ;s)<\infty  \notag
\end{align}%
uniformly over $\gamma \in \Gamma \cap \lbrack \gamma _{0}(s),\infty )$ as $%
n\rightarrow \infty $, from Lemma \ref{L1} and Assumptions ID-(ii) and
A-(viii). However, 
\begin{equation*}
\partial \Upsilon _{0}(\gamma ;s)/\partial \gamma =c_{0}^{\top }M(\gamma
_{0}(s);s)^{\top }M(\gamma ;s)^{-1}D(\gamma ,s)f(\gamma ,s)M(\gamma
;s)^{-1}M(\gamma _{0}(s);s)c_{0}\geq 0
\end{equation*}%
and 
\begin{equation}
\partial \Upsilon _{0}(\gamma _{0}(s);s)/\partial \gamma =c_{0}^{\top
}D(\gamma _{0}(s),s)f(\gamma _{0}(s),s)c_{0}>0  \label{dY0}
\end{equation}%
from Assumption A-(viii), which implies that $\Upsilon _{0}(\gamma ;s)$ is
continuous, non-decreasing, and uniquely minimized at $\gamma _{0}(s)$ given 
$s\in \mathcal{S}_{0}$.

We can symmetrically show that, uniformly over $\gamma \in \Gamma \cap
(-\infty ,\gamma _{0}(s)]$, $\Upsilon _{0}(\gamma ;s)$ in (\ref{pf2}) is
continuous, non-increasing, and uniquely minimized at $\gamma _{0}(s)$ as
well. Therefore, given $s\in \mathcal{S}_{0}$, $\sup_{\gamma \in \Gamma
}|\Upsilon _{n}(\gamma ;s)-\Upsilon _{0}(\gamma ;s)|=o_{p}(1)$; $\Upsilon
_{0}(\gamma ;s)$ is continuous and uniquely minimized at $\gamma _{0}(s)$.
Since $\Gamma $ is compact and $\widehat{\gamma }(s)$ is the minimizer of $%
\Upsilon _{n}(\gamma ;s)$, the pointwise consistency follows as Theorem 2.1
of \cite{NeweyMcFadden1994}. $\blacksquare $

\bigskip

We let $\phi _{1n}=a_{n}^{-1}$, where $a_{n}=n^{1-2\epsilon }b_{n}$ and $%
\epsilon $ is given in Assumption A-(ii). For a given $s\in \mathcal{S}_{0}$
and any $\gamma :\mathcal{S}_{0}\mapsto \Gamma $, we define%
\begin{eqnarray}
T_{n}\left( \gamma ;s\right) &=&\frac{1}{nb_{n}}\dsum_{i\in \Lambda
_{n}}\left( c_{0}^{\top }x_{i}\right) ^{2}\left\vert \mathbf{1}_{i}\left(
\gamma \left( s\right) \right) -\mathbf{1}_{i}\left( \gamma _{0}\left(
s\right) \right) \right\vert K_{i}\left( s\right) \text{,}  \label{Tn_def} \\
\overline{T}_{n}(\gamma ,s) &=&\frac{1}{nb_{n}}\dsum_{i\in \Lambda
_{n}}\left\Vert x_{i}\right\Vert ^{2}\left\vert \mathbf{1}_{i}\left( \gamma
\left( s\right) \right) -\mathbf{1}_{i}\left( \gamma _{0}\left( s\right)
\right) \right\vert K_{i}\left( s\right) \text{,}  \label{Tnb_def} \\
L_{nj}\left( \gamma ;s\right) &=&\frac{1}{\sqrt{nb_{n}}}\dsum_{i\in \Lambda
_{n}}x_{ij}u_{i}\left\{ \mathbf{1}_{i}\left( \gamma \left( s\right) \right) -%
\mathbf{1}_{i}\left( \gamma _{0}\left( s\right) \right) \right\} K_{i}\left(
s\right)  \label{Lnj_def}
\end{eqnarray}%
for $j=1,\ldots ,\dim (x)$, where $x_{ij}$ denotes the $j$th element of $%
x_{i}$.

\begin{lemma}
\label{infTsupL pointwise}For a given $s\in \mathcal{S}_{0}$, for any $%
\gamma \left( \cdot \right) :\mathcal{S}_{0}\mapsto \Gamma $, $\eta (s)>0$,
and $\varepsilon (s)>0$, there exist constants $0<C_{T}(s),C_{\overline{T}%
}(s),\overline{C}(s),\overline{r}(s)<\infty $ such that if $n$ is
sufficiently large and $n^{1-2\epsilon }b_{n}^{2}\rightarrow \varrho <\infty 
$,%
\begin{eqnarray}
\mathbb{P}\left( \inf_{\overline{r}(s)\phi _{1n}<\left\vert \gamma
(s)-\gamma _{0}\left( s\right) \right\vert <\overline{C}(s)}\frac{%
T_{n}\left( \gamma ;s\right) }{\left\vert \gamma \left( s\right) -\gamma
_{0}\left( s\right) \right\vert }<C_{T}(1-\eta (s))\right) &\leq
&\varepsilon (s)\text{,}  \label{infsupL pointwise 1} \\
\mathbb{P}\left( \sup_{\overline{r}(s)\phi _{1n}<\left\vert \gamma
(s)-\gamma _{0}\left( s\right) \right\vert <\overline{C}(s)}\frac{\overline{T%
}_{n}\left( \gamma ;s\right) }{\left\vert \gamma \left( s\right) -\gamma
_{0}\left( s\right) \right\vert }>C_{\overline{T}}(1+\eta (s))\right) &\leq
&\varepsilon (s)\text{,}  \label{infsupL pointwise 2} \\
\mathbb{P}\left( \sup_{\overline{r}(s)\phi _{1n}<\left\vert \gamma
(s)-\gamma _{0}\left( s\right) \right\vert <\overline{C}(s)}\frac{\left\vert
L_{nj}\left( \gamma ;s\right) \right\vert }{\sqrt{a_{n}}\left\vert \gamma
\left( s\right) -\gamma _{0}\left( s\right) \right\vert }>\eta (s)\right)
&\leq &\varepsilon (s)  \label{infsupL pointwise 3}
\end{eqnarray}%
for $j=1,\ldots ,\dim (x)$.
\end{lemma}

\paragraph{Proof of Lemma \protect\ref{infTsupL pointwise}}

See the supplementary material. $\blacksquare $

\bigskip

For a given $s\in \mathcal{S}_{0}$, we let $\widehat{\theta }(\widehat{%
\gamma }(s))=(\widehat{\beta }(\widehat{\gamma }(s))^{\top },\widehat{\delta 
}(\widehat{\gamma }(s))^{\top })^{\top }$ and $\theta _{0}=(\beta _{0}^{\top
},\delta _{0}^{\top })^{\top }$.

\begin{lemma}
\label{neg pointwise}For a given $s\in \mathcal{S}_{0}$, $n^{\epsilon }(%
\widehat{\theta }(\widehat{\gamma }(s))-\theta _{0})=o_{p}(1)$.
\end{lemma}

\paragraph{Proof of Lemma \protect\ref{neg pointwise}}

See the supplementary material. $\blacksquare $

\paragraph{Proof of Theorem \protect\ref{p-roc}}

The consistency is proved in Lemma \ref{L-A3} above. For given $s\in 
\mathcal{S}_{0}$, we let 
\begin{eqnarray}
Q_{n}^{\ast }(\gamma (s);s) &=&Q_{n}(\widehat{\beta }\left( \widehat{\gamma }%
\left( s\right) \right) ,\widehat{\delta }\left( \widehat{\gamma }\left(
s\right) \right) ,\gamma (s);s)  \label{Q*} \\
&=&\dsum_{i\in \Lambda _{n}}\left\{ y_{i}-x_{i}^{\top }\widehat{\beta }%
\left( \widehat{\gamma }\left( s\right) \right) -x_{i}^{\top }\widehat{%
\delta }\left( \widehat{\gamma }\left( s\right) \right) \mathbf{1}%
_{i}(\gamma (s))\right\} ^{2}K_{i}\left( s\right)  \notag
\end{eqnarray}%
for any $\gamma (\cdot )$, where $Q_{n}(\beta ,\delta ,\gamma ;s)$ is the
sum of squared errors function in (\ref{sse}). Consider $\gamma (s)$ such
that $\gamma \left( s\right) \in \left[ \gamma _{0}\left( s\right) +%
\overline{r}(s)\phi _{1n},\gamma _{0}\left( s\right) +\overline{C}(s)\right] 
$ for some $0<\overline{r}(s),\overline{C}(s)<\infty $ that are chosen in
Lemma \ref{infTsupL pointwise}. We let $\Delta _{i}(\gamma ;s)=\mathbf{1}%
_{i}\left( \gamma \left( s\right) \right) -\mathbf{1}_{i}\left( \gamma
_{0}\left( s\right) \right) $; $\widehat{c}_{j}(\widehat{\gamma }\left(
s\right) )$ and $c_{0j}$ be the $j$th element of $\widehat{c}(\widehat{%
\gamma }\left( s\right) )\in \mathbb{%
%TCIMACRO{\U{211d} }%
%BeginExpansion
\mathbb{R}
%EndExpansion
}^{\dim (x)}$ and $c_{0}\in 
%TCIMACRO{\U{211d} }%
%BeginExpansion
\mathbb{R}
%EndExpansion
^{\dim (x)}$, respectively. Then, since $y_{i}=\beta _{0}^{\top
}x_{i}+\delta _{0}^{\top }x_{i}1_{i}\left( \gamma _{0}\left( s_{i}\right)
\right) +u_{i}$,%
\begin{eqnarray}
&&Q_{n}^{\ast }(\gamma (s);s)-Q_{n}^{\ast }(\gamma _{0}(s);s)  \notag \\
&=&\dsum_{i\in \Lambda _{n}}\left( \widehat{\delta }\left( \widehat{\gamma }%
\left( s\right) \right) ^{\top }x_{i}\right) ^{2}\Delta _{i}(\gamma
;s)K_{i}\left( s\right)  \notag \\
&&-2\dsum_{i\in \Lambda _{n}}\left( y_{i}-\widehat{\beta }\left( \widehat{%
\gamma }\left( s\right) \right) ^{\top }x_{i}-\widehat{\delta }\left( 
\widehat{\gamma }\left( s\right) \right) ^{\top }x_{i}\mathbf{1}_{i}\left(
\gamma _{0}\left( s\right) \right) \right) \left( \widehat{\delta }\left( 
\widehat{\gamma }\left( s\right) \right) ^{\top }x_{i}\right) \Delta
_{i}(\gamma ;s)K_{i}\left( s\right)  \notag \\
&=&\dsum_{i\in \Lambda _{n}}\left( \delta _{0}^{\top }x_{i}\right)
^{2}\Delta _{i}(\gamma ;s)K_{i}\left( s\right) +\dsum_{i\in \Lambda
_{n}}\left\{ \left( \widehat{\delta }\left( \widehat{\gamma }\left( s\right)
\right) ^{\top }x_{i}\right) ^{2}-\left( \delta _{0}^{\top }x_{i}\right)
^{2}\right\} \Delta _{i}(\gamma ;s)K_{i}\left( s\right)  \notag \\
&&-2\dsum_{i\in \Lambda _{n}}\delta _{0}^{\top }x_{i}u_{i}\Delta _{i}(\gamma
;s)K_{i}\left( s\right) -2\dsum_{i\in \Lambda _{n}}\left( \widehat{\delta }%
\left( \widehat{\gamma }\left( s\right) \right) -\delta _{0}\right) ^{\top
}x_{i}u_{i}\Delta _{i}(\gamma ;s)K_{i}\left( s\right)  \notag \\
&&-2\dsum_{i\in \Lambda _{n}}\left( \widehat{\beta }\left( \widehat{\gamma }%
\left( s\right) \right) -\beta _{0}\right) ^{\top }x_{i}x_{i}^{\top }%
\widehat{\delta }\left( \widehat{\gamma }\left( s\right) \right) \Delta
_{i}(\gamma ;s)K_{i}\left( s\right)  \notag \\
&&-2\dsum_{i\in \Lambda _{n}}\delta _{0}^{\top }x_{i}x_{i}^{\top }\delta
_{0}\left\{ \mathbf{1}_{i}\left( \gamma _{0}\left( s_{i}\right) \right) -%
\mathbf{1}_{i}\left( \gamma _{0}\left( s\right) \right) \right\} \Delta
_{i}(\gamma ;s)K_{i}\left( s\right)  \label{aa1} \\
&&-2\dsum_{i\in \Lambda _{n}}\delta _{0}^{\top }x_{i}x_{i}^{\top }\left( 
\widehat{\delta }\left( \widehat{\gamma }\left( s\right) \right) -\delta
_{0}\right) \left\{ \mathbf{1}_{i}\left( \gamma _{0}\left( s_{i}\right)
\right) -\mathbf{1}_{i}\left( \gamma _{0}\left( s\right) \right) \right\}
\Delta _{i}(\gamma ;s)K_{i}\left( s\right)  \label{aa2} \\
&&-2\dsum_{i\in \Lambda _{n}}\left( \widehat{\delta }\left( \widehat{\gamma }%
\left( s\right) \right) -\delta _{0}\right) ^{\top }x_{i}x_{i}^{\top }%
\widehat{\delta }\left( \widehat{\gamma }\left( s\right) \right) \mathbf{1}%
_{i}\left( \gamma _{0}\left( s\right) \right) \Delta _{i}(\gamma
;s)K_{i}\left( s\right) \text{,}  \label{aa3}
\end{eqnarray}%
where the absolute values of the last two terms in lines (\ref{aa2}) and (%
\ref{aa3}) are bounded by 
\begin{eqnarray*}
&&2\left\vert \dsum_{i\in \Lambda _{n}}\delta _{0}^{\top }x_{i}x_{i}^{\top
}\left( \widehat{\delta }\left( \widehat{\gamma }\left( s\right) \right)
-\delta _{0}\right) \left\vert \Delta _{i}(\gamma ;s)\right\vert K_{i}\left(
s\right) \right\vert \text{ \ and} \\
&&2\left\vert \dsum_{i\in \Lambda _{n}}\left( \widehat{\delta }\left( 
\widehat{\gamma }\left( s\right) \right) -\delta _{0}\right) ^{\top
}x_{i}x_{i}^{\top }\widehat{\delta }\left( \widehat{\gamma }\left( s\right)
\right) \left\vert \Delta _{i}(\gamma ;s)\right\vert K_{i}\left( s\right)
\right\vert \text{,}
\end{eqnarray*}%
respectively, since $\left\vert \mathbf{1}_{i}\left( \gamma _{0}\left(
s\right) \right) \right\vert \leq 1$ and $\left\vert \mathbf{1}_{i}\left(
\gamma _{0}\left( s_{i}\right) \right) -\mathbf{1}_{i}\left( \gamma
_{0}\left( s\right) \right) \right\vert \leq 1$. Moreover, for the term in
line (\ref{aa1}), we have 
\begin{eqnarray}
&&\frac{1}{a_{n}}\dsum_{i\in \Lambda _{n}}\delta _{0}^{\top
}x_{i}x_{i}^{\top }\delta _{0}\left\{ \mathbf{1}_{i}\left( \gamma _{0}\left(
s_{i}\right) \right) -\mathbf{1}_{i}\left( \gamma _{0}\left( s\right)
\right) \right\} \Delta _{i}(\gamma ;s)K_{i}\left( s\right)  \notag \\
&\leq &\frac{1}{a_{n}}\dsum_{i\in \Lambda _{n}}\delta _{0}^{\top
}x_{i}x_{i}^{\top }\delta _{0}\left\vert \mathbf{1}_{i}\left( \gamma
_{0}\left( s_{i}\right) \right) -\mathbf{1}_{i}\left( \gamma _{0}\left(
s\right) \right) \right\vert K_{i}\left( s\right) =C_{n}^{\ast }(s)b_{n}
\label{C*term}
\end{eqnarray}%
for some $C_{n}^{\ast }(s)=O_{a.s.}(1)$ as in (\ref{m0}).

For any vector $v=(v_{1},\ldots ,v_{\dim (v)})^{\top }$, we let $\left\Vert
v\right\Vert _{\infty }=\max_{1\leq j\leq \dim (x)}|v_{j}|$. From Lemma \ref%
{neg pointwise}, we also let a sufficiently small $\kappa _{n}(s)$ such that 
$n^{\epsilon }||\widehat{\theta }(\widehat{\gamma }(s))-\theta _{0}||\leq
\kappa _{n}(s)$ and $\kappa _{n}(s)\rightarrow 0$ as $n\rightarrow \infty $
for any $s$. Then, $\left\Vert \widehat{c}\left( \widehat{\gamma }\left(
s\right) \right) -c_{0}\right\Vert \leq \kappa _{n}(s)$, $\left\Vert 
\widehat{c}(\widehat{\gamma }\left( s\right) )\right\Vert \leq \left\Vert
c_{0}\right\Vert +\kappa _{n}(s)$, and $\left\Vert \widehat{c}\left( 
\widehat{\gamma }\left( s\right) \right) +c_{0}\right\Vert \leq 2\left\Vert
c_{0}\right\Vert +\kappa _{n}(s)$. In addition, given Lemma \ref{infTsupL
pointwise}, there exist $0<C(s),\overline{C}(s),\overline{r}(s),\eta
(s),\varepsilon (s)<\infty $ such that 
\begin{eqnarray*}
\mathbb{P}\left( \inf_{\overline{r}(s)\phi _{1n}<\left\vert \gamma \left(
s\right) -\gamma _{0}\left( s\right) \right\vert <\overline{C}(s)}\frac{%
T_{n}\left( \gamma ;s\right) }{\left\vert \gamma \left( s\right) -\gamma
_{0}\left( s\right) \right\vert }<C(s)\left( 1-\eta (s)\right) \right) &\leq
&\frac{\varepsilon (s)}{3}\text{,} \\
\mathbb{P}\left( \sup_{\overline{r}(s)\phi _{1n}<\left\vert \gamma \left(
s\right) -\gamma _{0}\left( s\right) \right\vert <\overline{C}(s)}\frac{%
\overline{T}_{n}\left( \gamma ;s\right) }{\left\vert \gamma \left( s\right)
-\gamma _{0}\left( s\right) \right\vert }>C_{\overline{T}}(1+\eta
(s))\right) &\leq &\frac{\varepsilon (s)}{3}\text{,} \\
\mathbb{P}\left( \sup_{\overline{r}(s)\phi _{1n}<\left\vert \gamma \left(
s\right) -\gamma _{0}\left( s\right) \right\vert <\overline{C}(s)}\frac{%
2\dim (x)\left\Vert c_{0}\right\Vert _{\infty }\left\Vert L_{n}\left( \gamma
;s\right) \right\Vert _{\infty }}{\sqrt{a_{n}}\left\vert \gamma \left(
s\right) -\gamma _{0}\left( s\right) \right\vert }>\eta (s)\right) &\leq &%
\frac{\varepsilon (s)}{3}
\end{eqnarray*}%
for $\left\Vert c_{0}\right\Vert _{\infty }<\infty $. For $\gamma \left(
s\right) \in \left[ \gamma _{0}\left( s\right) +\overline{r}(s)\phi
_{1n},\gamma _{0}\left( s\right) +\overline{C}(s)\right] $, we also have 
\begin{equation*}
\mathbb{P}\left( \sup_{\overline{r}(s)\phi _{1n}<\left\vert \gamma
(s)-\gamma _{0}\left( s\right) \right\vert <\overline{C}(s)}\frac{%
2C_{n}^{\ast }(s)b_{n}}{\left\vert \gamma \left( s\right) -\gamma
_{0}(s)\right\vert }>\eta (s)\right) \leq \frac{\varepsilon (s)}{3}
\end{equation*}%
by choosing $\overline{r}(s)$ large enough, since 
\begin{equation*}
\sup_{\overline{r}(s)\phi _{1n}<\left\vert \gamma (s)-\gamma _{0}\left(
s\right) \right\vert <\overline{C}(s)}\frac{C_{n}^{\ast }(s)b_{n}}{\gamma
\left( s\right) -\gamma _{0}(s)}\leq \frac{C_{n}^{\ast }(s)b_{n}}{\overline{r%
}(s)\phi _{1n}}=a_{n}b_{n}\frac{C_{n}^{\ast }(s)}{\overline{r}(s)}<\infty
\end{equation*}%
almost surely provided $n^{1-2\epsilon }b_{n}^{2}\rightarrow \varrho <\infty 
$.

It follows that, with probability approaching to one, 
\begin{eqnarray}
&&\frac{Q_{n}^{\ast }(\gamma (s);s)-Q_{n}^{\ast }(\gamma _{0}(s);s)}{%
a_{n}(\gamma (s)-\gamma _{0}(s))}  \label{Q decomposition} \\
&\geq &\frac{T_{n}\left( \gamma ;s\right) }{\gamma (s)-\gamma _{0}(s)}%
-\left\Vert \widehat{c}\left( \widehat{\gamma }\left( s\right) \right)
-c_{0}\right\Vert \left\Vert \widehat{c}\left( \widehat{\gamma }\left(
s\right) \right) +c_{0}\right\Vert \frac{\overline{T}_{n}(\gamma ,s)}{\gamma
(s)-\gamma _{0}(s)}  \notag \\
&&-2\dim (x)\left\Vert c_{0}\right\Vert _{\infty }\frac{\left\Vert
L_{n}\left( \gamma ;s\right) \right\Vert _{\infty }}{\sqrt{a_{n}}(\gamma
(s)-\gamma _{0}(s))}-2\dim (x)\left\Vert \widehat{c}(\widehat{\gamma }\left(
s\right) )-c_{0}\right\Vert _{\infty }\frac{\left\Vert L_{n}\left( \gamma
;s\right) \right\Vert _{\infty }}{\sqrt{a_{n}}\left( \gamma (s)-\gamma
_{0}(s)\right) }  \notag \\
&&-2\left\Vert n^{\epsilon }(\widehat{\beta }\left( \widehat{\gamma }\left(
s\right) \right) -\beta _{0})\right\Vert \left\Vert \widehat{c}(\widehat{%
\gamma }\left( s\right) )\right\Vert \frac{\overline{T}_{n}(\gamma ,s)}{%
\gamma (s)-\gamma _{0}(s)}  \notag \\
&&-2\frac{C_{n}^{\ast }(s)b_{n}}{\gamma (s)-\gamma _{0}(s)}  \notag \\
&&-2\left\Vert c_{0}\right\Vert \left\Vert \widehat{c}\left( \widehat{\gamma 
}\left( s\right) \right) -c_{0}\right\Vert \frac{\overline{T}_{n}(\gamma ,s)%
}{\gamma (s)-\gamma _{0}(s)}  \notag \\
&&-2\left\Vert n^{\epsilon }(\widehat{\delta }\left( \widehat{\gamma }\left(
s\right) \right) -\delta _{0})\right\Vert \left\Vert \widehat{c}(\widehat{%
\gamma }\left( s\right) )\right\Vert \frac{\overline{T}_{n}(\gamma ,s)}{%
\gamma (s)-\gamma _{0}(s)}  \notag \\
&\geq &C_{T}\left( s\right) \left( 1-\eta \left( s\right) \right) -\kappa
_{n}(s)\left\{ 2||c_{0}||+\kappa _{n}(s)\right\} C_{\overline{T}}\left(
s\right) \left( 1+\eta \left( s\right) \right)  \notag \\
&&-2\dim (x)\left\Vert c_{0}\right\Vert _{\infty }\eta \left( s\right)
-2\dim (x)\kappa _{n}(s)\eta \left( s\right)  \notag \\
&&-2\kappa _{n}(s)\left\{ ||c_{0}||+\kappa _{n}(s)\right\} C_{\overline{T}%
}\left( s\right) \left( 1+\eta \left( s\right) \right)  \notag \\
&&-2\eta \left( s\right)  \notag \\
&&-2\left\Vert c_{0}\right\Vert \kappa _{n}(s)C_{\overline{T}}\left(
s\right) \left( 1+\eta \left( s\right) \right)  \notag \\
&&-2\kappa _{n}(s)\left\{ ||c_{0}||+\kappa _{n}(s)\right\} C_{\overline{T}%
}\left( s\right) \left( 1+\eta \left( s\right) \right)  \notag \\
&>&0  \notag
\end{eqnarray}%
by choosing sufficiently small $\kappa _{n}(s)$ and $\eta (s)$.

Since we suppose $a_{n}(\gamma (s)-\gamma _{0}(s))>0$, it implies that, for
any $\varepsilon (s)\in (0,1)$ and $\eta (s)>0$, 
\begin{equation*}
\mathbb{P}\left( \inf_{\overline{r}(s)\phi _{1n}<\left\vert \gamma
(s)-\gamma _{0}\left( s\right) \right\vert <\overline{C}(s)}\left\{
Q_{n}^{\ast }(\gamma (s);s)-Q_{n}^{\ast }(\gamma _{0}(s);s)\right\} >\eta
(s)\right) \geq 1-\varepsilon (s)\text{,}
\end{equation*}%
which yields $\mathbb{P}\left( Q_{n}^{\ast }(\gamma (s);s)-Q_{n}^{\ast
}(\gamma _{0}(s);s)>0\right) \rightarrow 1$ as $n\rightarrow \infty $ for
given $s\in \mathcal{S}_{0}$. We similarly show the same result when $\gamma
\left( s\right) \in \left[ \gamma _{0}\left( s\right) -\overline{C}%
(s),\gamma _{0}\left( s\right) -\overline{r}(s)\phi _{1n}\right] $.
Therefore, because $Q_{n}^{\ast }(\widehat{\gamma }(s);s)-Q_{n}^{\ast
}(\gamma _{0}(s);s)\leq 0$ for any $s\in \mathcal{S}_{0}$ by construction,
it should hold that $|\widehat{\gamma }\left( s\right) -\gamma _{0}\left(
s\right) |\leq \overline{r}(s)\phi _{1n}$ with probability approaching to
one; or for any $\varepsilon (s)>0$ and $s\in \mathcal{S}_{0}$, there exists 
$\overline{r}(s)>0$ such that%
\begin{equation*}
\mathbb{P}\left( a_{n}|\widehat{\gamma }\left( s\right) -\gamma _{0}\left(
s\right) |>\overline{r}(s)\right) <\varepsilon (s)
\end{equation*}%
for sufficiently large $n$, since $\phi _{1n}=a_{n}^{-1}$. $\blacksquare $

\subsection{Proof of Theorem \protect\ref{g-an} and Corollary \protect\ref%
{g-test} (Asymptotic Distribution)}

For a given $s\in \mathcal{S}_{0}$, we let $\gamma _{n}\left( s\right)
=\gamma _{0}\left( s\right) +r/a_{n}$ with some $|r|<\infty $, where $%
a_{n}=n^{1-2\epsilon }b_{n}$ and $\epsilon $ is given in Assumption A-(ii).
We define 
\begin{eqnarray*}
A_{n}^{\ast }\left( r,s\right) &=&\dsum_{i\in \Lambda _{n}}\left( \delta
_{0}^{\top }x_{i}\right) ^{2}\left\vert \mathbf{1}_{i}\left( \gamma
_{n}\left( s\right) \right) -\mathbf{1}_{i}\left( \gamma _{0}\left( s\right)
\right) \right\vert K_{i}\left( s\right) \text{,} \\
B_{n}^{\ast }\left( r,s\right) &=&\dsum_{i\in \Lambda _{n}}\delta _{0}^{\top
}x_{i}u_{i}\left\{ \mathbf{1}_{i}\left( \gamma _{n}\left( s\right) \right) -%
\mathbf{1}_{i}\left( \gamma _{0}\left( s\right) \right) \right\} K_{i}\left(
s\right) \text{.}
\end{eqnarray*}

\begin{lemma}
\label{L2}Suppose $n^{1-2\epsilon }b_{n}^{2}\rightarrow \varrho <\infty $.
Then, for fixed $s\in \mathcal{S}_{0}$, uniformly over $r$ in any compact
set,%
\begin{equation*}
A_{n}^{\ast }\left( r,s\right) \rightarrow _{p}\left\vert r\right\vert
c_{0}^{\top }D\left( \gamma _{0}\left( s\right) ,s\right) c_{0}f\left(
\gamma _{0}\left( s\right) ,s\right)
\end{equation*}%
and 
\begin{equation*}
B_{n}^{\ast }\left( r,s\right) \Rightarrow W\left( r\right) \sqrt{%
c_{0}^{\top }V\left( \gamma _{0}\left( s\right) ,s\right) c_{0}f\left(
\gamma _{0}\left( s\right) ,s\right) \kappa _{2}}
\end{equation*}%
as $n\rightarrow \infty $, where $\kappa _{2}=\int K^{2}(v)dv$ and $W\left(
r\right) $ is the two-sided Brownian Motion defined in (\ref{2BM}).
\end{lemma}

\paragraph{Proof of Lemma \protect\ref{L2}}

First, for $A_{n}^{\ast }\left( r,s\right) $, we consider the case with $r>0$%
. We let $\Delta _{i}(\gamma _{n};s)=\mathbf{1}_{i}\left( \gamma _{n}\left(
s\right) \right) -\mathbf{1}_{i}\left( \gamma _{0}\left( s\right) \right) $, 
$h_{i}(r,s)=\left( c_{0}^{\top }x_{i}\right) ^{2}\Delta _{i}(\gamma
_{n};s)K_{i}\left( s\right) $, and recall that $\delta
_{0}=c_{0}n^{-\epsilon }=c_{0}(a_{n}/\left( nb_{n}\right) )^{1/2}$. By
change of variables and Taylor expansion, Assumptions A-(v), (viii), and (x)
imply that 
\begin{eqnarray}
\mathbb{E}\left[ A_{n}^{\ast }\left( r,s\right) \right] &=&\frac{a_{n}}{%
nb_{n}}\dsum_{i\in \Lambda _{n}}\mathbb{E}\left[ h_{i}(r,s)\right]
\label{EA*} \\
&=&a_{n}\iint_{\gamma _{0}\left( s\right) }^{\gamma _{0}\left( s\right)
+r/a_{n}}\mathbb{E}\left[ \left( c_{0}^{\top }x_{i}\right) ^{2}|q,s+b_{n}t%
\right] K\left( t\right) f\left( q,s+b_{n}t\right) dqdt  \notag \\
&=&rc_{0}^{\top }D\left( \gamma _{0}\left( s\right) ,s\right) c_{0}f\left(
\gamma _{0}\left( s\right) ,s\right) +O\left( \frac{1}{a_{n}}%
+b_{n}^{2}\right) \text{,}  \notag
\end{eqnarray}%
where the third equality holds under Assumption A-(vi). Next, we have 
\begin{eqnarray}
Var\left[ A_{n}^{\ast }\left( r,s\right) \right] &=&\frac{a_{n}^{2}}{%
n^{2}b_{n}^{2}}Var\left[ \dsum_{i\in \Lambda _{n}}h_{i}(r,s)\right]
\label{VA*} \\
&=&\frac{a_{n}^{2}}{nb_{n}^{2}}Var\left[ h_{i}(r,s)\right] +\frac{a_{n}^{2}}{%
n^{2}b_{n}^{2}}\dsum_{\substack{ i,j\in \Lambda _{n}  \\ i\neq j}}Cov\left[
h_{i}(r,s),h_{j}(r,s)\right]  \notag \\
&\equiv &\Psi _{A1}(r,s)+\Psi _{A2}(r,s)\text{.}  \notag
\end{eqnarray}%
Taylor expansion and Assumptions A-(vii), (viii), and (x) lead to 
\begin{eqnarray*}
\Psi _{A1}(r,s) &=&\frac{a_{n}}{nb_{n}}\left( \frac{a_{n}}{b_{n}}\mathbb{E}%
\left[ \left( c_{0}^{\top }x_{i}\right) ^{4}\Delta _{i}(\gamma
_{n};s)K_{i}^{2}\left( s\right) \right] \right) -\frac{1}{n}\left( \frac{%
a_{n}}{b_{n}}\mathbb{E}\left[ \left( c_{0}^{\top }x_{i}\right) ^{2}\Delta
_{i}(\gamma _{n};s)K_{i}\left( s\right) \right] \right) ^{2} \\
&=&O\left( \frac{a_{n}}{nb_{n}}+\frac{1}{n}\right) =O\left( n^{-2\epsilon }+%
\frac{1}{n}\right)
\end{eqnarray*}%
since $\Delta _{i}(\gamma _{n};s)^{2}=\Delta _{i}(\gamma _{n};s)$ for $r>0$,
where each moment term is bounded as in (\ref{EA*}). For $\Psi _{A2}$, we
define a sequence of integers $\kappa _{n}=O\left( n^{\ell }\right) $ for
some $\ell >0$ such that $\kappa _{n}\rightarrow \infty $ and $\kappa
_{n}^{2}/n\rightarrow 0$, and decompose 
\begin{eqnarray*}
\Psi _{A2}(r,s) &=&\frac{a_{n}^{2}}{n^{2}b_{n}^{2}}\dsum_{\substack{ i,j\in
\Lambda _{n}  \\ 0<\lambda \left( i,j\right) \leq \kappa _{n}}}Cov\left[
h_{i}(r,s),h_{j}(r,s)\right] +\frac{a_{n}^{2}}{n^{2}b_{n}^{2}}\dsum 
_{\substack{ i,j\in \Lambda _{n}  \\ \lambda \left( i,j\right) >\kappa _{n}}}%
Cov\left[ h_{i}(r,s),h_{j}(r,s)\right] \\
&=&\Psi _{A2}^{\prime }(r,s)+\Psi _{A2}^{\prime \prime }(r,s)\text{.}
\end{eqnarray*}%
Then, since 
\begin{eqnarray*}
&&Cov\left[ \frac{a_{n}}{b_{n}}h_{i}(r,s),\frac{a_{n}}{b_{n}}h_{j}(r,s)%
\right] \\
&\leq &r^{2}\mathbb{E}\left[ \left( c_{0}^{\top }x_{i}\right) ^{2}\left(
c_{0}^{\top }x_{j}\right) ^{2}|\gamma _{0}\left( s\right) ,\gamma _{0}\left(
s\right) ,s,s\right] f\left( \gamma _{0}\left( s\right) ,\gamma _{0}\left(
s\right) ,s,s\right) +o\left( 1\right)
\end{eqnarray*}%
using a similar argument as in (\ref{Ehh}) and (\ref{EA*}), similarly as the
proof of $\Psi _{14,3}^{\prime }(s)$ in Lemma \ref{max ineq}, we have%
\begin{equation*}
\Psi _{A2}^{\prime }(r,s)\leq Cr^{2}\kappa _{n}^{2}/n=o\left( 1\right)
\end{equation*}%
for some $C<\infty $. Furthermore, by the covariance inequality (\ref{cov
ineq}) and Assumption A-(iii), we have%
\begin{eqnarray*}
\left\vert \Psi _{A2}^{\prime \prime }(r,s)\right\vert &\leq &\frac{%
C^{\prime }}{n^{2}}\left( \frac{a_{n}}{b_{n}}\right) ^{\frac{2+2\varphi }{%
2+\varphi }}\dsum_{\substack{ i,j\in \Lambda _{n}  \\ \lambda \left(
i,j\right) >\kappa _{n}}}\alpha _{1,1}\left( \lambda \left( i,j\right)
\right) ^{\varphi /(2+\varphi )}\mathbb{E}\left[ \frac{a_{n}}{b_{n}}%
\left\vert h_{i}(r,s)\right\vert ^{2+\varphi }\right] ^{2/\left( 2+\varphi
\right) } \\
&\leq &\frac{C^{\prime \prime }}{n}\left( \frac{a_{n}}{b_{n}}\right) ^{\frac{%
2+2\varphi }{2+\varphi }}\dsum_{i\in \Lambda _{n}}\sum_{m=\kappa
_{n}+1}^{n-1}\dsum_{\substack{ j\in \Lambda _{n}  \\ \lambda \left(
i,j\right) \in \lbrack m,m+1)}}\alpha _{1,1}\left( m\right) ^{\varphi
/(2+\varphi )} \\
&\leq &\frac{C^{\prime \prime }}{n}\left( \frac{a_{n}}{b_{n}}\right) ^{\frac{%
2+2\varphi }{2+\varphi }}\sum_{m=\kappa _{n}+1}^{\infty }m\exp \left(
-m\varphi /(2+\varphi )\right) \\
&=&O\left( n^{(\left( 1-2\epsilon \right) (2+2\varphi )/(2+\varphi
))-1}\kappa _{n}\exp (-\kappa _{n}\varphi /(2+\varphi ))\right) \\
&=&o\left( 1\right) \text{,}
\end{eqnarray*}%
similarly as the proof of $\Psi _{14,3}^{\prime \prime }(s)$ in Lemma \ref%
{max ineq}, because $\mathbb{E}[\left( a_{n}/b_{n}\right)
|h_{i}(r,s)|^{2+\varphi }]$ is bounded as in (\ref{EA*}) and we set $\kappa
_{n}$ such that $\kappa _{n}=O(n^{\ell })$ for $\ell >0$. Hence, the
pointwise convergence of $A_{n}^{\ast }\left( r,s\right) $ is obtained.
Furthermore, since $A_{n}^{\ast }(r,s)$ is monotonically increasing in $r$
and the limit function $rc_{0}^{\top }D\left( \gamma _{0}\left( s\right)
,s\right) c_{0}f\left( \gamma _{0}\left( s\right) ,s\right) $ is continuous
in $r$, the convergence holds uniformly on any compact set. Symmetrically,
we can show that $\mathbb{E}\left[ A_{n}^{\ast }\left( r,s\right) \right]
=-rc_{0}^{\top }D\left( \gamma _{0}\left( s\right) ,s\right) c_{0}f\left(
\gamma _{0}\left( s\right) ,s\right) +O\left( a_{n}^{-1}+b_{n}^{2}\right) $
when $r<0$. The uniform convergence also holds in this case using the same
argument as above, which completes the proof for $A_{n}^{\ast }\left(
r,s\right) $.

Next, for $B_{n}^{\ast }\left( r,s\right) $, Assumption ID-(i) leads to $%
\mathbb{E}\left[ B_{n}^{\ast }\left( r,s\right) \right] =0$. We let $%
\widetilde{h}_{i}(r,s)=c_{0}^{\top }x_{i}u_{i}\Delta _{i}(\gamma
_{n};s)K_{i}\left( s\right) $ and write 
\begin{eqnarray*}
Var[B_{n}^{\ast }\left( r,s\right) ] &=&\frac{a_{n}}{b_{n}}Var[\widetilde{h}%
_{i}(r,s)]+\frac{a_{n}}{nb_{n}}\dsum_{\substack{ i,j\in \Lambda _{n}  \\ %
i\neq j}}Cov[\widetilde{h}_{i}(r,s),\widetilde{h}_{j}(r,s)] \\
&\equiv &\Psi _{B1}(r,s)+\Psi _{B2}(r,s)\text{.}
\end{eqnarray*}%
As in (\ref{EA*}), we have 
\begin{equation*}
\Psi _{B1}(r,s)=\left\vert r\right\vert c_{0}^{\top }V\left( \gamma
_{0}\left( s\right) ,s\right) c_{0}f\left( \gamma _{0}(s),s\right) \int
K^{2}(v)dv+O\left( \frac{1}{a_{n}}+b_{n}^{2}\right) \text{,}
\end{equation*}%
which is nonsingular for $|r|>0$ from Assumption A-(viii). For $\Psi
_{B2}(r,s)$, we define a sequence of integers $\kappa _{n}^{\prime }=O\left(
n^{\ell ^{\prime }}\right) $ for some $\ell ^{\prime }>0$ such that $\kappa
_{n}^{\prime }\rightarrow \infty $ and $(\kappa _{n}^{\prime
})^{2}/n^{1-2\epsilon }\rightarrow 0$, and decompose%
\begin{eqnarray*}
\Psi _{B2}(r,s) &=&\frac{a_{n}}{nb_{n}}\dsum_{\substack{ i,j\in \Lambda _{n} 
\\ 0<\lambda \left( i,j\right) \leq \kappa _{n}}}Cov[\widetilde{h}_{i}(r,s),%
\widetilde{h}_{j}(r,s)]+\frac{a_{n}}{nb_{n}}\dsum_{\substack{ i,j\in \Lambda
_{n}  \\ \lambda \left( i,j\right) >\kappa _{n}}}Cov[\widetilde{h}_{i}(r,s),%
\widetilde{h}_{j}(r,s)] \\
&\equiv &\Psi _{B2}^{\prime }(r,s)+\Psi _{B2}^{\prime \prime }(r,s)\text{.}
\end{eqnarray*}%
Then similarly as $\Psi _{A2}^{\prime }$ and $\Psi _{A2}^{\prime \prime }$
above, we have 
\begin{eqnarray*}
\left\vert \Psi _{B2}^{\prime }(r,s)\right\vert &\leq &Cr^{2}(\kappa
_{n}^{\prime })^{2}\times \frac{b_{n}}{a_{n}}=O\left( \frac{(\kappa
_{n}^{\prime })^{2}}{n^{1-2\epsilon }}\right) =o(1)\text{,} \\
\left\vert \Psi _{B2}^{\prime \prime }(r,s)\right\vert &\leq &C^{\prime
}\left( \frac{a_{n}}{b_{n}}\right) ^{\varphi /(2+\varphi )}\sum_{m=\kappa
_{n}+1}^{\infty }m\exp \left( -m\varphi /(2+\varphi )\right) \\
&=&C^{\prime }n^{\left( 1-2\epsilon \right) \varphi /(2+\varphi )}\kappa
_{n}^{\prime }\exp (-\kappa _{n}^{\prime }\varphi /(2+\varphi ))=o(1)
\end{eqnarray*}%
for some $C,C^{\prime }<\infty $. By combining these results, we have 
\begin{equation*}
Var[B_{n}^{\ast }\left( r,s\right) ]=\left\vert r\right\vert c_{0}^{\top
}V\left( \gamma _{0}\left( s\right) ,s\right) c_{0}f\left( \gamma
_{0}(s),s\right) \kappa _{2}+o\left( 1\right)
\end{equation*}%
with $\kappa _{2}=\int K^{2}(v)dv$, and by the CLT for stationary and mixing
random field (e.g., \cite{Bolthausen82} and \cite{Jenish09}), we have%
\begin{equation*}
B_{n}^{\ast }\left( r,s\right) \Rightarrow W\left( r\right) \sqrt{%
c_{0}^{\top }V\left( \gamma _{0}\left( s\right) ,s\right) c_{0}f\left(
\gamma _{0}\left( s\right) ,s\right) \kappa _{2}}
\end{equation*}%
as $n\rightarrow \infty $, where $W\left( r\right) $ is the two-sided
Brownian Motion defined in (\ref{2BM}). This pointwise convergence in $r$
can be extended to any finite-dimensional convergence in $r$ by the fact
that for any $r_{1}<r_{2}$, $Cov\left[ B_{n}^{\ast }\left( r_{1},s\right)
,B_{n}^{\ast }\left( r_{2},s\right) \right] =Var\left[ B_{n}^{\ast }\left(
r_{1},s\right) \right] +o\left( 1\right) $, which is because $\left( \mathbf{%
1}_{i}\left( \gamma _{0}+r_{2}/a_{n}\right) -\mathbf{1}_{i}\left( \gamma
_{0}+r_{1}/a_{n}\right) \right) \mathbf{1}_{i}\left( \gamma
_{0}+r_{1}/a_{n}\right) =0$. The tightness follows from a similar argument
as $J_{n}(\gamma ;s)$ in Lemma \ref{max ineq} and the desired result follows
by Theorem 15.5 in \cite{Billingsley68}. $\blacksquare $

\bigskip

For a given $s\in \mathcal{S}_{0}$, we let $\widehat{\theta }\left( \gamma
_{0}\left( s\right) \right) =(\widehat{\beta }\left( \gamma _{0}\left(
s\right) \right) ^{\top },\widehat{\delta }\left( \gamma _{0}\left( s\right)
\right) ^{\top })^{\top }$. Recall that $\theta _{0}=(\beta _{0}^{\top
},\delta _{0}^{\top })^{\top }$ and $\widehat{\theta }\left( \widehat{\gamma 
}\left( s\right) \right) =(\widehat{\beta }\left( \widehat{\gamma }\left(
s\right) \right) ^{\top },\widehat{\delta }\left( \widehat{\gamma }\left(
s\right) \right) ^{\top })^{\top }$.

\begin{lemma}
\label{L3}For a given $s\in \mathcal{S}_{0}$, $\sqrt{nb_{n}}(\widehat{\theta 
}\left( \widehat{\gamma }\left( s\right) \right) -\theta _{0})=O_{p}(1)$ and 
$\sqrt{nb_{n}}(\widehat{\theta }\left( \widehat{\gamma }\left( s\right)
\right) -\widehat{\theta }\left( \gamma _{0}\left( s\right) \right)
)=o_{p}(1)$, if $n^{1-2\epsilon }b_{n}^{2}\rightarrow \varrho <\infty $ as $%
n\rightarrow \infty $.
\end{lemma}

\paragraph{Proof of Lemma \protect\ref{L3}}

See the supplementary material. $\blacksquare $

\paragraph*{Proof of Theorem \protect\ref{g-an}}

From Theorem \ref{p-roc}, we define a random variable $r^{\ast }(s)$ such
that 
\begin{equation*}
r^{\ast }(s)=a_{n}(\widehat{\gamma }\left( s\right) -\gamma _{0}\left(
s\right) )=\arg \max_{r\in 
%TCIMACRO{\U{211d} }%
%BeginExpansion
\mathbb{R}
%EndExpansion
}\left\{ Q_{n}^{\ast }(\gamma _{0}(s);s)-Q_{n}^{\ast }\left( \gamma _{0}(s)+%
\frac{r}{a_{n}};s\right) \right\} \text{,}
\end{equation*}%
where $Q_{n}^{\ast }(\gamma (s);s)$ is defined in (\ref{Q*}). We let $\Delta
_{i}(s)=\mathbf{1}_{i}\left( \gamma _{0}\left( s\right) +r/a_{n}\right) -%
\mathbf{1}_{i}\left( \gamma _{0}\left( s\right) \right) $. We then have 
\begin{eqnarray}
&&\Delta Q_{n}^{\ast }(r;s)  \label{diff} \\
&=&Q_{n}^{\ast }(\gamma _{0}(s);s)-Q_{n}^{\ast }\left( \gamma _{0}(s)+\frac{r%
}{a_{n}};s\right)  \notag \\
&=&-\dsum_{i\in \Lambda _{n}}\left( \widehat{\delta }\left( \widehat{\gamma }%
\left( s\right) \right) ^{\top }x_{i}\right) ^{2}\left\vert \Delta
_{i}(s)\right\vert K_{i}\left( s\right)  \notag \\
&&+2\dsum_{i\in \Lambda _{n}}\left( y_{i}-\widehat{\beta }\left( \widehat{%
\gamma }\left( s\right) \right) ^{\top }x_{i}-\widehat{\delta }\left( 
\widehat{\gamma }\left( s\right) \right) ^{\top }x_{i}\mathbf{1}_{i}\left(
\gamma _{0}\left( s\right) \right) \right) \left( \widehat{\delta }\left( 
\widehat{\gamma }\left( s\right) \right) ^{\top }x_{i}\right) \Delta
_{i}(s)K_{i}\left( s\right)  \notag \\
&\equiv &-A_{n}(r;s)+2B_{n}(r;s)\text{.}  \notag
\end{eqnarray}%
For $A_{n}(r;s)$, Lemmas \ref{L2} and \ref{L3} yield%
\begin{equation}
A_{n}(r;s)=A_{n}^{\ast }\left( r,s\right) +o_{p}\left( 1\right)
\label{Aterm}
\end{equation}%
since $\widehat{\delta }\left( \widehat{\gamma }\left( s\right) \right)
-\delta _{0}=O_{p}((nb_{n})^{-1/2})$. Similarly, for $B_{n}(r;s)$, since $%
y_{i}=\beta _{0}^{\top }x_{i}+\delta _{0}^{\top }x_{i}1_{i}\left( \gamma
_{0}(s_{i})\right) +u_{i}$, $\widehat{\delta }\left( \widehat{\gamma }\left(
s\right) \right) -\delta _{0}=O_{p}((nb_{n})^{-1/2})$, and $\widehat{\beta }%
\left( \widehat{\gamma }\left( s\right) \right) -\beta
_{0}=O_{p}((nb_{n})^{-1/2})$, we have 
\begin{eqnarray}
&&B_{n}(r;s)  \label{Bterm} \\
&=&\dsum_{i\in \Lambda _{n}}\left( u_{i}+\delta _{0}^{\top }x_{i}\left\{ 
\mathbf{1}_{i}\left( \gamma _{0}\left( s_{i}\right) \right) -\mathbf{1}%
_{i}\left( \gamma _{0}\left( s\right) \right) \right\} -\left( \widehat{%
\beta }\left( \widehat{\gamma }\left( s\right) \right) -\beta _{0}\right)
^{\top }x_{i}\right.  \notag \\
&&\left. \ \ \ \ \ \ \ \ -\left( \widehat{\delta }\left( \widehat{\gamma }%
\left( s\right) \right) -\delta _{0}\right) ^{\top }x_{i}\mathbf{1}%
_{i}\left( \gamma _{0}\left( s\right) \right) \right) \widehat{\delta }%
\left( \widehat{\gamma }\left( s\right) \right) ^{\top }x_{i}\Delta
_{i}(s)K_{i}\left( s\right)  \notag \\
&=&\dsum_{i\in \Lambda _{n}}u_{i}\widehat{\delta }\left( \widehat{\gamma }%
\left( s\right) \right) ^{\top }x_{i}\Delta _{i}(s)K_{i}\left( s\right) 
\notag \\
&&+\dsum_{i\in \Lambda _{n}}\delta _{0}^{\top }x_{i}\left\{ \mathbf{1}%
_{i}\left( \gamma _{0}\left( s_{i}\right) \right) -\mathbf{1}_{i}\left(
\gamma _{0}\left( s\right) \right) \right\} \widehat{\delta }\left( \widehat{%
\gamma }\left( s\right) \right) ^{\top }x_{i}\Delta _{i}(s)K_{i}\left(
s\right)  \notag \\
&&-\dsum_{i\in \Lambda _{n}}\left\{ \left( \widehat{\beta }\left( \widehat{%
\gamma }\left( s\right) \right) -\beta _{0}\right) ^{\top }x_{i}+\left( 
\widehat{\delta }\left( \widehat{\gamma }\left( s\right) \right) -\delta
_{0}\right) ^{\top }x_{i}\mathbf{1}_{i}\left( \gamma _{0}\left( s\right)
\right) \right\} \widehat{\delta }\left( \widehat{\gamma }\left( s\right)
\right) ^{\top }x_{i}\Delta _{i}(s)K_{i}\left( s\right)  \notag \\
&=&\dsum_{i\in \Lambda _{n}}u_{i}\delta _{0}^{\top }x_{i}\Delta
_{i}(s)K_{i}\left( s\right) +\dsum_{i\in \Lambda _{n}}\delta _{0}^{\top
}x_{i}\left\{ \mathbf{1}_{i}\left( \gamma _{0}\left( s_{i}\right) \right) -%
\mathbf{1}_{i}\left( \gamma _{0}\left( s\right) \right) \right\} \delta
_{0}^{\top }x_{i}\Delta _{i}(s)K_{i}\left( s\right) +o_{p}(1)  \notag \\
&=&B_{n}^{\ast }\left( r,s\right) +B_{n}^{\ast \ast }(r,s)+o_{p}\left(
1\right) \text{,}  \notag
\end{eqnarray}%
where we let 
\begin{equation*}
B_{n}^{\ast \ast }(r,s)\equiv \dsum_{i\in \Lambda _{n}}\delta _{0}^{\top
}x_{i}\left\{ \mathbf{1}_{i}\left( \gamma _{0}\left( s_{i}\right) \right) -%
\mathbf{1}_{i}\left( \gamma _{0}\left( s\right) \right) \right\} \delta
_{0}^{\top }x_{i}\Delta _{i}(s)K_{i}\left( s\right) \text{.}
\end{equation*}%
In Lemma \ref{Bn3} below, we show that, if $n^{1-2\epsilon
}b_{n}^{2}\rightarrow \varrho \in (0,\infty )$, 
\begin{align*}
B_{n}^{\ast \ast }(r,s)& \rightarrow _{p}\left\vert r\right\vert c_{0}^{\top
}D\left( \gamma _{0}\left( s\right) ,s\right) c_{0}f\left( \gamma _{0}\left(
s\right) ,s\right) \left\{ \frac{1}{2}-\mathcal{K}_{0}\left( r,\varrho
;s\right) \right\} \\
& \ \ \ \ \ \ \ +\varrho c_{0}^{\top }D\left( \gamma _{0}\left( s\right)
,s\right) c_{0}f\left( \gamma _{0}\left( s\right) ,s\right) \left\vert \dot{%
\gamma}_{0}(s)\right\vert \mathcal{K}_{1}\left( r,\varrho ;s\right)
\end{align*}%
as $n\rightarrow \infty $, where $\dot{\gamma}_{0}\left( \cdot \right) $ is
the first derivative of $\gamma _{0}(\cdot )$ and $\mathcal{K}_{j}\left(
r,\varrho ;s\right) =\int_{0}^{|r|/(\varrho |\dot{\gamma}_{0}(s)|)}t^{j}K%
\left( t\right) dt$ for $j=0,1$.

From Lemma \ref{L2}, it follows that%
\begin{eqnarray}
\Delta Q_{n}^{\ast }(r;s) &=&-A_{n}^{\ast }\left( r,s\right) +2B_{n}^{\ast
\ast }(r,s)+2B_{n}^{\ast }\left( r,s\right)  \notag \\
&\Rightarrow &-\left\vert r\right\vert c_{0}^{\top }D\left( \gamma
_{0}\left( s\right) ,s\right) c_{0}f\left( \gamma _{0}\left( s\right)
,s\right)  \notag \\
&&+\left\vert r\right\vert c_{0}^{\top }D\left( \gamma _{0}\left( s\right)
,s\right) c_{0}f\left( \gamma _{0}\left( s\right) ,s\right) \left\{ 1-2%
\mathcal{K}_{0}\left( r,\varrho ;s\right) \right\}  \notag \\
&&+2\varrho c_{0}^{\top }D\left( \gamma _{0}\left( s\right) ,s\right)
c_{0}f\left( \gamma _{0}\left( s\right) ,s\right) \left\vert \dot{\gamma}%
_{0}(s)\right\vert \mathcal{K}_{1}\left( r,\varrho ;s\right)  \notag \\
&&+2W\left( r\right) \sqrt{c_{0}^{\top }V\left( \gamma _{0}\left( s\right)
,s\right) c_{0}f\left( \gamma _{0}\left( s\right) ,s\right) \kappa _{2}} 
\notag \\
&=&-2\left\vert r\right\vert \ell _{D}(s)\mathcal{K}_{0}\left( r,\varrho
;s\right) +2\varrho \ell _{D}(s)\left\vert \dot{\gamma}_{0}(s)\right\vert 
\mathcal{K}_{1}\left( r,\varrho ;s\right)  \label{DQ*n} \\
&&+2W\left( r\right) \sqrt{\ell _{V}(s)}\text{,}  \notag
\end{eqnarray}%
where%
\begin{eqnarray*}
\ell _{D}(s) &=&c_{0}^{\top }D\left( \gamma _{0}\left( s\right) ,s\right)
c_{0}f\left( \gamma _{0}\left( s\right) ,s\right) \text{,} \\
\ell _{V}(s) &=&c_{0}^{\top }V\left( \gamma _{0}\left( s\right) ,s\right)
c_{0}f\left( \gamma _{0}\left( s\right) ,s\right) \kappa _{2}\text{.}
\end{eqnarray*}%
However, if we let $\xi (s)=\ell _{V}(s)/\ell _{D}^{2}(s)>0$ and $r=\xi
(s)\nu $, we have 
\begin{eqnarray*}
&&\arg \max_{r\in 
%TCIMACRO{\U{211d} }%
%BeginExpansion
\mathbb{R}
%EndExpansion
}\left( 2W\left( r\right) \sqrt{\ell _{V}(s)}-2\left\vert r\right\vert \ell
_{D}(s)\mathcal{K}_{0}\left( r,\varrho ;s\right) +2\varrho \ell
_{D}(s)\left\vert \dot{\gamma}_{0}(s)\right\vert \mathcal{K}_{1}\left(
r,\varrho ;s\right) \right) \\
&=&\xi (s)\arg \max_{\nu \in 
%TCIMACRO{\U{211d} }%
%BeginExpansion
\mathbb{R}
%EndExpansion
}\left( W\left( \xi (s)\nu \right) \sqrt{\ell _{V}(s)}-\left\vert \xi (s)\nu
\right\vert \ell _{D}(s)\mathcal{K}_{0}\left( \xi (s)\nu ,\varrho ;s\right)
+\varrho \ell _{D}(s)\left\vert \dot{\gamma}_{0}(s)\right\vert \mathcal{K}%
_{1}\left( \xi (s)\nu ,\varrho ;s\right) \right) \\
&=&\xi (s)\arg \max_{\nu \in 
%TCIMACRO{\U{211d} }%
%BeginExpansion
\mathbb{R}
%EndExpansion
}\left( W\left( \nu \right) \frac{\ell _{V}(s)}{\ell _{D}(s)}-\left\vert \nu
\right\vert \frac{\ell _{V}(s)}{\ell _{D}(s)}\mathcal{K}_{0}\left( \xi
(s)\nu ,\varrho ;s\right) +\varrho \frac{\ell _{V}(s)}{\ell _{D}(s)}\cdot 
\frac{\left\vert \dot{\gamma}_{0}(s)\right\vert }{\xi (s)}\mathcal{K}%
_{1}\left( \xi (s)\nu ,\varrho ;s\right) \right) \\
&=&\xi (s)\arg \max_{\nu \in 
%TCIMACRO{\U{211d} }%
%BeginExpansion
\mathbb{R}
%EndExpansion
}\left( W\left( \nu \right) -\left\vert \nu \right\vert \mathcal{K}%
_{0}\left( \xi (s)\nu ,\varrho ;s\right) +\varrho \frac{\left\vert \dot{%
\gamma}_{0}(s)\right\vert }{\xi (s)}\mathcal{K}_{1}\left( \xi (s)\nu
,\varrho ;s\right) \right)
\end{eqnarray*}%
similar to the proof of Theorem 1 in \cite{Hansen00a}. By Theorem 2.7 of 
\cite{Kim90}, it follows that (rewriting $\nu $ as $r$)%
\begin{equation*}
n^{1-2\epsilon }b_{n}\left( \widehat{\gamma }\left( s\right) -\gamma
_{0}\left( s\right) \right) \rightarrow _{d}\xi \left( s\right) \arg
\max_{r\in 
%TCIMACRO{\U{211d} }%
%BeginExpansion
\mathbb{R}
%EndExpansion
}\left( W\left( r\right) -\left\vert r\right\vert \psi _{0}\left( r,\varrho
;s\right) +\varrho \frac{\left\vert \dot{\gamma}_{0}(s)\right\vert }{\xi (s)}%
\psi _{1}\left( r,\varrho ;s\right) \right)
\end{equation*}%
as $n\rightarrow \infty $, where%
\begin{equation*}
\psi _{j}\left( r,\varrho ;s\right) =\int_{0}^{|r|\xi (s)/\left( \varrho |%
\dot{\gamma}_{0}(s)|\right) }t^{j}K\left( t\right) dt
\end{equation*}%
for $j=0,1$. Finally, letting 
\begin{equation}
\mu \left( r,\varrho ;s\right) =-\left\vert r\right\vert \psi _{0}\left(
r,\varrho ;s\right) +\varrho \frac{\left\vert \dot{\gamma}_{0}(s)\right\vert 
}{\xi (s)}\psi _{1}\left( r,\varrho ;s\right) \text{,}  \label{Th3-mu}
\end{equation}%
$\mathbb{E}\left[ \arg \max_{r\in 
%TCIMACRO{\U{211d} }%
%BeginExpansion
\mathbb{R}
%EndExpansion
}\left( W\left( r\right) +\mu \left( r,\varrho ;s\right) \right) \right] =0$
follows from Lemmas \ref{lemma-drift} and \ref{lemma-mu} below. $%
\blacksquare $

\begin{lemma}
\label{Bn3}For a given $s\in \mathcal{S}_{0}$, let $r$ be the same term used
in Lemma \ref{L2}. If $n^{1-2\epsilon }b_{n}^{2}\rightarrow \varrho \in
(0,\infty )$, uniformly over $r$ in any compact set,%
\begin{align*}
B_{n}^{\ast \ast }(r,s)& \equiv \dsum_{i\in \Lambda _{n}}\delta _{0}^{\top
}x_{i}\left\{ \mathbf{1}_{i}\left( \gamma _{0}\left( s_{i}\right) \right) -%
\mathbf{1}_{i}\left( \gamma _{0}\left( s\right) \right) \right\} \delta
_{0}^{\top }x_{i}\Delta _{i}(s)K_{i}\left( s\right) \\
& \rightarrow _{p}\left\vert r\right\vert c_{0}^{\top }D\left( \gamma
_{0}\left( s\right) ,s\right) c_{0}f\left( \gamma _{0}\left( s\right)
,s\right) \left\{ \frac{1}{2}-\mathcal{K}_{0}\left( r,\varrho ;s\right)
\right\} \\
& \ \ \ \ \ +\varrho c_{0}^{\top }D\left( \gamma _{0}\left( s\right)
,s\right) c_{0}f\left( \gamma _{0}\left( s\right) ,s\right) \left\vert \dot{%
\gamma}_{0}(s)\right\vert \mathcal{K}_{1}\left( r,\varrho ;s\right)
\end{align*}%
as $n\rightarrow \infty $, where $\dot{\gamma}_{0}\left( \cdot \right) $ is
the first derivatives of $\gamma _{0}(\cdot )$ and 
\begin{equation*}
\mathcal{K}_{j}\left( r,\varrho ;s\right) =\int_{0}^{|r|/\left( \varrho |%
\dot{\gamma}_{0}(s)|\right) }t^{j}K\left( t\right) dt
\end{equation*}%
for $j=0,1$.
\end{lemma}

\paragraph{Proof of Lemma \protect\ref{Bn3}}

See the supplementary material. $\blacksquare $

\begin{lemma}
\label{lemma-drift}Let $\tau =\arg \max_{r\in 
%TCIMACRO{\U{211d} }%
%BeginExpansion
\mathbb{R}
%EndExpansion
}\left( W(r)+\mu (r)\right) $, where $W(r)$ is a two-sided Brownian motion
in (\ref{2BM}) and $\mu (r)$ is a continuous and symmetric function
satisfying: $\mu (0)=0$, $\mu (-r)=\mu (r)$, $\mu (r)/r^{1/2+\varepsilon }$
is monotonically decreasing to $-\infty $ on $[\underline{r},\infty )$ for
some $\underline{r}>0$ and $\varepsilon >0$. Then, $\mathbb{E}[\tau ]=0$.
\end{lemma}

\paragraph{Proof of Lemma \protect\ref{lemma-drift}}

See the supplementary material. $\blacksquare $

\begin{lemma}
\label{lemma-mu}For any given $\varrho <\infty $ and $s\in \mathcal{S}_{0}$, 
$\mu \left( r,\varrho ;s\right) $ in (\ref{Th3-mu}) satisfies conditions in
Lemma \ref{lemma-drift}.
\end{lemma}

\paragraph{Proof of Lemma \protect\ref{lemma-mu}}

See the supplementary material. $\blacksquare $

\paragraph{Proof of Corollary \protect\ref{g-test}}

Under $H_{0}:\gamma _{0}\left( s\right) =\gamma _{\ast }\left( s\right) $,
we write%
\begin{equation*}
LR_{n}(s)=\frac{1}{nb_{n}}\dsum_{i\in \Lambda _{n}}K\left( \frac{s_{i}-s}{%
b_{n}}\right) \times \frac{\left\{ Q_{n}\left( \gamma _{\ast }\left(
s\right) ,s\right) -Q_{n}\left( \widehat{\gamma }\left( s\right) ,s\right)
\right\} }{(nb_{n})^{-1}Q_{n}\left( \widehat{\gamma }\left( s\right)
,s\right) }\text{.}
\end{equation*}%
From (\ref{pf1}) and (\ref{pf2}), we have 
\begin{equation*}
\frac{1}{nb_{n}}Q_{n}\left( \widehat{\gamma }\left( s\right) ,s\right) =%
\frac{1}{nb_{n}}\dsum_{i\in \Lambda _{n}}u_{i}^{2}K_{i}\left( s\right)
+o_{p}(1)\rightarrow _{p}\mathbb{E}\left[ u_{i}^{2}|s_{i}=s\right]
f_{s}\left( s\right)
\end{equation*}%
as $n\rightarrow \infty $, where $f_{s}\left( s\right) $ is the marginal
density of $s_{i}$. In addition, from Theorem \ref{g-an} and Lemmas \ref{L1}
and \ref{L3}, we have%
\begin{eqnarray*}
&&Q_{n}\left( \gamma _{0}\left( s\right) ,s\right) -Q_{n}\left( \widehat{%
\gamma }\left( s\right) ,s\right) \\
&=&Q_{n}^{\ast }\left( \gamma _{0}\left( s\right) ,s\right) -Q_{n}^{\ast
}\left( \widehat{\gamma }\left( s\right) ,s\right) \\
&&+\left( \widehat{\theta }\left( \widehat{\gamma }\left( s\right) \right) -%
\widehat{\theta }\left( \gamma _{0}\left( s\right) \right) \right) ^{\top }%
\widetilde{Z}(\gamma _{0}(s);s)\widetilde{Z}(\gamma _{0}(s);s)^{\top }\left( 
\widehat{\theta }\left( \widehat{\gamma }\left( s\right) \right) -\widehat{%
\theta }\left( \gamma _{0}\left( s\right) \right) \right) \\
&=&Q_{n}^{\ast }\left( \gamma _{0}\left( s\right) ,s\right) -Q_{n}^{\ast
}\left( \widehat{\gamma }\left( s\right) ,s\right) +o_{p}(1)\text{,}
\end{eqnarray*}%
where $\widetilde{Z}(\gamma ;s)$ is defined in Lemma \ref{L-A3}. Similar to
Theorem 2 of \cite{Hansen00a}, the rest of the proof follows from the change
of variables and the continuous mapping theorem because the limiting
expression in (\ref{DQ*n}) and $(nb_{n})^{-1}\sum_{i\in \Lambda
_{n}}K_{i}\left( s\right) \rightarrow _{p}f_{s}\left( s\right) $ by the
standard result of the kernel density estimator. $\blacksquare $

\subsection{Proof of Theorem \protect\ref{u-roc} (Uniform Convergence)}

We let $\phi _{2n}=\log n/a_{n}$, where $a_{n}=n^{1-2\epsilon }b_{n}$ and $%
\epsilon $ is given in Assumption A-(ii). We also define $\mathcal{G}_{n}(%
\mathcal{S}_{0};\Gamma )$ as a class of cadlag and piecewise constant
functions $\mathcal{S}_{0}\mapsto \Gamma $ with at most $n$ discontinuity
points. Recall that $T_{n}\left( \gamma ;s\right) $, $\overline{T}_{n}\left(
\gamma ;s\right) $, and $L_{nj}\left( \gamma ;s\right) $ are defined in (\ref%
{Tn_def}), (\ref{Tnb_def}), and (\ref{Lnj_def}), respectively; $\sup_{s\in 
\mathcal{S}_{0}}\left\vert \gamma \left( s\right) -\gamma _{0}\left(
s\right) \right\vert $ is bounded since $\gamma \left( s\right) \in \Gamma $%
, a compact set, for any $s\in \mathcal{S}_{0}$.

\begin{lemma}
\label{T uniform}There exist constants $C^{\ast }$ and $\bar{C}^{\ast }$
such that for any $\gamma \left( \cdot \right) \in \mathcal{G}_{n}(\mathcal{S%
}_{0};\Gamma )$ 
\begin{eqnarray*}
\sup_{s\in \mathcal{S}_{0}}\left\vert T_{n}\left( \gamma ;s\right) -\mathbb{E%
}\left[ T_{n}\left( \gamma ;s\right) \right] \right\vert &\leq &C^{\ast
}\left( \sup_{s\in \mathcal{S}_{0}}\left\vert \gamma \left( s\right) -\gamma
_{0}\left( s\right) \right\vert \frac{\log n}{nb_{n}}\right) ^{1/2} \\
\sup_{s\in \mathcal{S}_{0}}\left\vert \overline{T}_{n}\left( \gamma
;s\right) -\mathbb{E}\left[ \overline{T}_{n}\left( \gamma ;s\right) \right]
\right\vert &\leq &\bar{C}^{\ast }\left( \sup_{s\in \mathcal{S}%
_{0}}\left\vert \gamma \left( s\right) -\gamma _{0}\left( s\right)
\right\vert \frac{\log n}{nb_{n}}\right) ^{1/2}
\end{eqnarray*}%
almost surely if $n^{1-2\epsilon }b_{n}^{2}\rightarrow \varrho <\infty $.
\end{lemma}

\paragraph{Proof of Lemma \protect\ref{T uniform}}

See the supplementary material. $\blacksquare $

\begin{lemma}
\label{L uniform}There exists some constant $C_{L}$ such that for any $%
\gamma \left( \cdot \right) \in \mathcal{G}_{n}(\mathcal{S}_{0};\Gamma )$
and any $j=1,\ldots ,\dim (x)$ 
\begin{equation*}
\sup_{s\in \mathcal{S}_{0}}\left\vert L_{nj}\left( \gamma ;s\right)
\right\vert =C_{L}\left( \sup_{s\in \mathcal{S}_{0}}\left\vert \gamma \left(
s\right) -\gamma _{0}\left( s\right) \right\vert \log n\right) ^{1/2}
\end{equation*}%
almost surely if $n^{1-2\epsilon }b_{n}^{2}\rightarrow \varrho <\infty $.
\end{lemma}

\paragraph{Proof of Lemma \protect\ref{L uniform}}

See the supplementary material. $\blacksquare $

\begin{lemma}
\label{L-A2}For any $\gamma \left( \cdot \right) \in \mathcal{G}_{n}(%
\mathcal{S}_{0};\Gamma )$, $\eta >0$, and $\varepsilon >0$, there exist
constants $\overline{C}$, $\overline{r}$, $C_{T}$, and $C_{\overline{T}}$
such that if $n^{1-2\epsilon }b_{n}^{2}\rightarrow \varrho <\infty $ and $n$
is sufficiently large, 
\begin{eqnarray}
\mathbb{P}\left( \inf_{\overline{r}\phi _{2n}<\sup_{s\in \mathcal{S}%
_{0}}\left\vert \gamma \left( s\right) -\gamma _{0}\left( s\right)
\right\vert <\overline{C}}\frac{\sup_{s\in \mathcal{S}_{0}}T_{n}\left(
\gamma ;s\right) }{\sup_{s\in \mathcal{S}_{0}}\left\vert \gamma \left(
s\right) -\gamma _{0}\left( s\right) \right\vert }<C_{T}(1-\eta )\right)
&\leq &\varepsilon \text{,}  \label{infTsupL uniform 1} \\
\mathbb{P}\left( \sup_{\overline{r}\phi _{2n}<\sup_{s\in \mathcal{S}%
_{0}}\left\vert \gamma \left( s\right) -\gamma _{0}\left( s\right)
\right\vert <\overline{C}}\frac{\sup_{s\in \mathcal{S}_{0}}\overline{T}%
_{n}\left( \gamma ;s\right) }{\sup_{s\in \mathcal{S}_{0}}\left\vert \gamma
\left( s\right) -\gamma _{0}\left( s\right) \right\vert }>C_{\overline{T}%
}(1+\eta )\right) &\leq &\varepsilon \text{,}  \label{infTsupL uniform 2}
\end{eqnarray}%
and for $j=1,\ldots ,\dim (x)$ 
\begin{equation}
\mathbb{P}\left( \sup_{\overline{r}\phi _{2n}<\sup_{s\in \mathcal{S}%
_{0}}\left\vert \gamma \left( s\right) -\gamma _{0}\left( s\right)
\right\vert <\overline{C}}\frac{\sup_{s\in \mathcal{S}_{0}}\left\vert
L_{nj}\left( \gamma ;s\right) \right\vert }{\sqrt{a_{n}}\sup_{s\in \mathcal{S%
}_{0}}\left\vert \gamma \left( s\right) -\gamma _{0}\left( s\right)
\right\vert }>\eta \right) \leq \varepsilon \text{.}
\label{infTsupL uniform 3}
\end{equation}
\end{lemma}

\paragraph{Proof of Lemma \protect\ref{L-A2}}

See the supplementary material. $\blacksquare $

\begin{lemma}
\label{neg uniform} $n^{\epsilon }\sup_{s\in \mathcal{S}_{0}}||\widehat{%
\theta }(\widehat{\gamma }(s))-\theta _{0}||=o_{p}(1)$.
\end{lemma}

\paragraph{Proof of Lemma \protect\ref{neg uniform}}

See the supplementary material. $\blacksquare $

\paragraph{Proof of Theorem \protect\ref{u-roc}}

Note that $\widehat{\gamma }(\cdot )$ belongs to $\mathcal{G}_{n}(\mathcal{S}%
_{0};\Gamma )$. For $Q_{n}^{\ast }(\cdot ;\cdot )$ defined in (\ref{Q*}),
since $\sup_{s\in \mathcal{S}_{0}}\left( Q_{n}^{\ast }(\widehat{\gamma }%
(s);s)-Q_{n}^{\ast }(\gamma _{0}(s);s)\right) \leq 0$ by construction,\ it
suffices to show that as $n\rightarrow \infty $, 
\begin{equation*}
\mathbb{P}\left( \sup_{s\in \mathcal{S}_{0}}\left\{ Q_{n}^{\ast }(\gamma
(s);s)-Q_{n}^{\ast }(\gamma _{0}(s);s)\right\} >0\right) \rightarrow 1
\end{equation*}%
for any $\gamma \left( \cdot \right) \in \mathcal{G}_{n}(\mathcal{S}%
_{0};\Gamma )$ such that $\sup_{s\in \mathcal{S}_{0}}\left\vert \gamma
\left( s\right) -\gamma _{0}\left( s\right) \right\vert >\overline{r}\phi
_{2n}$ where $\overline{r}$ is chosen in Lemma \ref{L-A2}.

To this end, consider $\gamma \left( \cdot \right) $ such that $\overline{r}%
\phi _{2n}\leq \sup_{s\in \mathcal{S}_{0}}\left\vert \gamma \left( s\right)
-\gamma _{0}\left( s\right) \right\vert \leq \overline{C}$ for some $0<%
\overline{r},\overline{C}<\infty $. Then, similarly as (\ref{Q decomposition}%
) and using Lemmas \ref{L-A2} and \ref{neg uniform}, we have%
\begin{eqnarray*}
&&\frac{Q_{n}^{\ast }(\gamma (s);s)-Q_{n}^{\ast }(\gamma _{0}(s);s)}{%
a_{n}\sup_{s\in \mathcal{S}_{0}}\left\vert \gamma (s)-\gamma
_{0}(s)\right\vert } \\
&\geq &\frac{T_{n}\left( \gamma ;s\right) }{\sup_{s\in \mathcal{S}%
_{0}}\left\vert \gamma (s)-\gamma _{0}(s)\right\vert }-\frac{2\dim
(x)\left\Vert c_{0}\right\Vert _{\infty }\left\Vert L_{n}\left( \gamma
;s\right) \right\Vert _{\infty }}{\sqrt{a_{n}}\sup_{s\in \mathcal{S}%
_{0}}\left\vert \gamma (s)-\gamma _{0}(s)\right\vert }-\frac{2C_{n}^{\ast
}(s)b_{n}}{\sup_{s\in \mathcal{S}_{0}}\left\vert \gamma (s)-\gamma
_{0}(s)\right\vert }+o_{p}(1) \\
&>&0
\end{eqnarray*}%
for sufficiently large $n$ and small $\eta (s)$, where all the notations are
the same as in (\ref{Q decomposition}). Note that the $C_{n}^{\ast }(s)$
term in (\ref{C*term}) satisfies $\sup_{s\in \mathcal{S}_{0}}C_{n}^{\ast
}\left( s\right) =O_{a.s.}(1)$ from \ref{L-int1}, and 
\begin{eqnarray*}
\sup_{\overline{r}\phi _{2n}<\left\vert \gamma \left( s\right) -\gamma
_{0}\left( s\right) \right\vert <\overline{C}}\frac{\sup_{s\in \mathcal{S}%
_{0}}C_{n}^{\ast }\left( s\right) b_{n}}{\sup_{s\in \mathcal{S}%
_{0}}\left\vert \gamma (s)-\gamma _{0}(s)\right\vert } &<&\frac{\sup_{s\in 
\mathcal{S}_{0}}C_{n}^{\ast }(s)b_{n}}{\overline{r}\phi _{2n}} \\
&=&\frac{\sup_{s\in \mathcal{S}_{0}}C_{n}^{\ast }(s)}{\overline{r}}\left( 
\frac{a_{n}b_{n}}{\log n}\right) \\
&=&o_{a.s.}(1)
\end{eqnarray*}%
given $a_{n}b_{n}\rightarrow \varrho <\infty $. Thus, we have 
\begin{equation*}
\mathbb{P}\left( \sup_{\overline{r}\phi _{2n}<\left\vert \gamma \left(
s\right) -\gamma _{0}\left( s\right) \right\vert <\overline{C}}\frac{%
2\sup_{s\in \mathcal{S}_{0}}C^{\ast }(s)b_{n}}{\sup_{s\in \mathcal{S}%
_{0}}\left\vert \gamma \left( s\right) -\gamma _{0}\left( s\right)
\right\vert }>\eta \right) \leq \frac{\varepsilon }{3}
\end{equation*}%
when $n$ is sufficiently large. Therefore, for any $\varepsilon \in (0,1)$
and $\eta >0$, 
\begin{equation*}
\mathbb{P}\left( \inf_{\overline{r}\phi _{2n}<\sup_{s\in \mathcal{S}%
_{0}}\left\vert \gamma \left( s\right) -\gamma _{0}\left( s\right)
\right\vert <\overline{C}}\sup_{s\in \mathcal{S}_{0}}\left\{ Q_{n}^{\ast
}(\gamma (s);s)-Q_{n}^{\ast }(\gamma _{0}(s);s)\right\} >\eta \right) \geq
1-\varepsilon \text{,}
\end{equation*}%
which completes the proof by the same argument as Theorem \ref{p-roc}. $%
\blacksquare $

\subsection{Proof of Theorem \protect\ref{bd} (Asymptotic Normality of $%
\protect\widehat{\protect\theta }$)}

\paragraph{Proof of Theorem \protect\ref{bd}}

We let $\mathbf{1}_{\mathcal{S}_{0}}=\mathbf{1}[s_{i}\in \mathcal{S}_{0}]$
and consider a sequence of positive constants $\pi _{n}\rightarrow 0$ as $%
n\rightarrow \infty $. Then, 
\begin{eqnarray}
\sqrt{n}\left( \widehat{\beta }-\beta _{0}\right) &=&\left( \frac{1}{n}%
\dsum_{i\in \Lambda _{n}}x_{i}x_{i}^{\top }\mathbf{1}\left[ q_{i}>\widehat{%
\gamma }\left( s_{i}\right) +\pi _{n}\right] \mathbf{1}_{\mathcal{S}%
_{0}}\right) ^{-1}  \notag \\
&&\times \left\{ \frac{1}{\sqrt{n}}\dsum_{i\in \Lambda _{n}}x_{i}u_{i}%
\mathbf{1}\left[ q_{i}>\gamma _{0}\left( s_{i}\right) +\pi _{n}\right] 
\mathbf{1}_{\mathcal{S}_{0}}\right.  \notag \\
&&\ \ \ +\frac{1}{\sqrt{n}}\dsum_{i\in \Lambda _{n}}x_{i}u_{i}\left\{ 
\mathbf{1}\left[ q_{i}>\widehat{\gamma }\left( s_{i}\right) +\pi _{n}\right]
-\mathbf{1}\left[ q_{i}>\gamma _{0}\left( s_{i}\right) +\pi _{n}\right]
\right\} \mathbf{1}_{\mathcal{S}_{0}}  \notag \\
&&\ \ \ \left. +\frac{1}{\sqrt{n}}\dsum_{i\in \Lambda _{n}}x_{i}x_{i}^{\top
}\delta _{0}\mathbf{1}\left[ q_{i}\leq \gamma _{0}\left( s_{i}\right) \right]
\mathbf{1}\left[ q_{i}>\widehat{\gamma }\left( s_{i}\right) +\pi _{n}\right] 
\mathbf{1}_{\mathcal{S}_{0}}\right\}  \notag \\
&\equiv &\Xi _{\beta 0}^{-1}\left\{ \Xi _{\beta 1}+\Xi _{\beta 2}+\Xi
_{\beta 3}\right\}  \label{beta*}
\end{eqnarray}%
and 
\begin{eqnarray}
\sqrt{n}\left( \widehat{\delta }^{\ast }-\delta _{0}^{\ast }\right)
&=&\left( \frac{1}{n}\dsum_{i\in \Lambda _{n}}x_{i}x_{i}^{\top }\mathbf{1}%
\left[ q_{i}<\widehat{\gamma }\left( s_{i}\right) -\pi _{n}\right] \mathbf{1}%
_{\mathcal{S}_{0}}\right) ^{-1}  \notag \\
&&\times \left\{ \frac{1}{\sqrt{n}}\dsum_{i\in \Lambda _{n}}x_{i}u_{i}%
\mathbf{1}\left[ q_{i}<\gamma _{0}\left( s_{i}\right) -\pi _{n}\right] 
\mathbf{1}_{\mathcal{S}_{0}}\right.  \notag \\
&&\ \ \ +\frac{1}{\sqrt{n}}\dsum_{i\in \Lambda _{n}}x_{i}u_{i}\left\{ 
\mathbf{1}\left[ q_{i}<\widehat{\gamma }\left( s_{i}\right) -\pi _{n}\right]
-\mathbf{1}\left[ q_{i}<\gamma _{0}\left( s_{i}\right) -\pi _{n}\right]
\right\} \mathbf{1}_{\mathcal{S}_{0}}  \notag \\
&&\ \ \ \left. +\frac{1}{\sqrt{n}}\dsum_{i\in \Lambda _{n}}x_{i}x_{i}^{\top
}\delta _{0}^{\ast }\mathbf{1}\left[ q_{i}>\gamma _{0}\left( s_{i}\right) %
\right] \mathbf{1}\left[ q_{i}<\widehat{\gamma }\left( s_{i}\right) -\pi _{n}%
\right] \mathbf{1}_{\mathcal{S}_{0}}\right\}  \notag \\
&\equiv &\Xi _{\delta 0}^{-1}\left\{ \Xi _{\delta 1}+\Xi _{\delta 2}+\Xi
_{\delta 3}\right\} \text{,}  \label{delta*}
\end{eqnarray}%
where $\Xi _{\beta 2}$, $\Xi _{\beta 3}$, $\Xi _{\delta 2}$, and $\Xi
_{\delta 3}$ are all $o_{p}(1)$ from Lemma \ref{bias1} below, provided $\phi
_{2n}/\pi _{n}\rightarrow 0$ as $n\rightarrow \infty $. Therefore, 
\begin{equation*}
\sqrt{n}\left( \widehat{\theta }^{\ast }-\theta _{0}^{\ast }\right) =\left( 
\begin{array}{cc}
\Xi _{\beta 0} & 0 \\ 
0 & \Xi _{\delta 0}%
\end{array}%
\right) ^{-1}\left( 
\begin{array}{c}
\Xi _{\beta 1} \\ 
\Xi _{\delta 1}%
\end{array}%
\right) +o_{p}\left( 1\right)
\end{equation*}%
and the desired result follows once we establish that 
\begin{equation}
\Xi _{\beta 0}\rightarrow _{p}\mathbb{E}\left[ x_{i}x_{i}^{\top }\mathbf{1}%
\left[ q_{i}>\gamma _{0}\left( s_{i}\right) \right] \mathbf{1}_{\mathcal{S}%
_{0}}\right] \text{,}  \label{th00}
\end{equation}%
\begin{equation}
\Xi _{\delta 0}\rightarrow _{p}\mathbb{E}\left[ x_{i}x_{i}^{\top }\mathbf{1}%
\left[ q_{i}<\gamma _{0}\left( s_{i}\right) \right] \mathbf{1}_{\mathcal{S}%
_{0}}\right] \text{,}  \label{th10}
\end{equation}%
and 
\begin{equation}
\left( 
\begin{array}{c}
\Xi _{\beta 1} \\ 
\Xi _{\delta 1}%
\end{array}%
\right) \rightarrow _{d}\mathcal{N}\left( 0,\lim_{n\rightarrow \infty }\frac{%
1}{n}Var\left[ \left( 
\begin{array}{c}
\sum_{i\in \Lambda _{n}}x_{i}u_{i}\mathbf{1}\left[ q_{i}>\gamma _{0}\left(
s_{i}\right) \right] \mathbf{1}_{\mathcal{S}_{0}} \\ 
\sum_{i\in \Lambda _{n}}x_{i}u_{i}\mathbf{1}\left[ q_{i}<\gamma _{0}\left(
s_{i}\right) \right] \mathbf{1}_{\mathcal{S}_{0}}%
\end{array}%
\right) \right] \right)  \label{th2}
\end{equation}%
as $n\rightarrow \infty $.

First, by Assumptions A-(v) and (ix), (\ref{th00}) can be readily verified
since we have 
\begin{eqnarray*}
&&\frac{1}{n}\dsum_{i\in \Lambda _{n}}x_{i}x_{i}^{\top }\mathbf{1}\left[
q_{i}>\widehat{\gamma }\left( s_{i}\right) +\pi _{n}\right] \mathbf{1}_{%
\mathcal{S}_{0}} \\
&=&\frac{1}{n}\dsum_{i\in \Lambda _{n}}x_{i}x_{i}^{\top }\mathbf{1}\left[
q_{i}>\gamma _{0}\left( s_{i}\right) +\pi _{n}\right] \mathbf{1}_{\mathcal{S}%
_{0}} \\
&&+\frac{1}{n}\dsum_{i\in \Lambda _{n}}x_{i}x_{i}^{\top }\left\{ \mathbf{1}%
\left[ q_{i}>\widehat{\gamma }\left( s_{i}\right) +\pi _{n}\right] -\mathbf{1%
}\left[ q_{i}>\gamma _{0}\left( s_{i}\right) +\pi _{n}\right] \right\} 
\mathbf{1}_{\mathcal{S}_{0}} \\
&=&\frac{1}{n}\dsum_{i\in \Lambda _{n}}x_{i}x_{i}^{\top }\mathbf{1}\left[
q_{i}>\gamma _{0}\left( s_{i}\right) +\pi _{n}\right] \mathbf{1}_{\mathcal{S}%
_{0}}+O_{p}\left( \phi _{2n}\right)
\end{eqnarray*}%
with $\pi _{n}\rightarrow 0$ as $n\rightarrow \infty $. More precisely,
given Theorem \ref{u-roc}, we consider $\widehat{\gamma }\left( s\right) $
in a neighborhood of $\gamma _{0}\left( s\right) $ with uniform distance at
most $\overline{r}\phi _{2n}$ for some large enough constant $\overline{r}$.
We define a non-random function $\widetilde{\gamma }\left( s\right) =\gamma
_{0}\left( s\right) +\overline{r}\phi _{2n}$. Then, on the event $%
E_{n}^{\ast }=\{\sup_{s\in \mathcal{S}_{0}}\left\vert \widehat{\gamma }%
\left( s\right) -\gamma _{0}\left( s\right) \right\vert \leq \overline{r}%
\phi _{2n}\}$, 
\begin{eqnarray}
&&\mathbb{E}\left[ x_{i}x_{i}^{\top }\left\{ \mathbf{1}\left[ q_{i}>\widehat{%
\gamma }\left( s_{i}\right) +\pi _{n}\right] -\mathbf{1}\left[ q_{i}>\gamma
_{0}\left( s_{i}\right) +\pi _{n}\right] \right\} \mathbf{1}_{\mathcal{S}%
_{0}}\right]  \label{xxD} \\
&\leq &\mathbb{E}\left[ x_{i}x_{i}^{\top }\left\{ \mathbf{1}\left[ q_{i}>%
\widetilde{\gamma }\left( s_{i}\right) +\pi _{n}\right] -\mathbf{1}\left[
q_{i}>\gamma _{0}\left( s_{i}\right) +\pi _{n}\right] \right\} \mathbf{1}_{%
\mathcal{S}_{0}}\right]  \notag \\
&=&\int_{\mathcal{S}_{0}}\int_{\gamma _{0}\left( v\right) +\pi _{n}}^{%
\widetilde{\gamma }\left( v\right) +\pi _{n}}D\left( q,v\right) f\left(
q,v\right) dqdv  \notag \\
&=&\int_{\mathcal{S}_{0}}\left\{ D\left( \gamma _{0}\left( v\right)
,v\right) f\left( \gamma _{0}\left( v\right) ,v\right) \left( \widetilde{%
\gamma }\left( v\right) -\gamma _{0}\left( v\right) \right) +o_{p}\left(
\phi _{2n}\right) \right\} dv  \notag \\
&\leq &\overline{r}\phi _{2n}\int D\left( \gamma _{0}\left( v\right)
,v\right) f\left( \gamma _{0}\left( v\right) ,v\right) dv  \notag \\
&=&O_{p}\left( \phi _{2n}\right) =o_{p}\left( 1\right)  \notag
\end{eqnarray}%
from Theorem \ref{u-roc}, Assumptions A-(v), (vii), and (ix). (\ref{th10})
can be verified symmetrically. Using a similar argument, since $\mathbb{E}%
\left[ x_{i}u_{i}\mathbf{1}\left[ q_{i}>\gamma _{0}\left( s_{i}\right) %
\right] \mathbf{1}_{\mathcal{S}_{0}}\right] =\mathbb{E}\left[ x_{i}u_{i}%
\mathbf{1}\left[ q_{i}<\gamma _{0}\left( s_{i}\right) \right] \mathbf{1}_{%
\mathcal{S}_{0}}\right] =0$ from Assumption ID-(i), the asymptotic normality
in (\ref{th2}) follows by the Theorem of \cite{Bolthausen82} under
Assumption A-(iii), which completes the proof. $\blacksquare $

\begin{lemma}
\label{bias1}When $\phi _{2n}\rightarrow 0$ as $n\rightarrow \infty $, if we
let $\pi _{n}>0$ such that $\pi _{n}\rightarrow 0$ and $\phi _{2n}/\pi
_{n}\rightarrow 0$ as $n\rightarrow \infty $, then it holds that $\Xi
_{\beta 2}$, $\Xi _{\beta 3}$, $\Xi _{\delta 2}$, and $\Xi _{\delta 3}$ in (%
\ref{beta*}) and (\ref{delta*}) are all $o_{p}(1)$.
\end{lemma}

\paragraph{Proof of Lemma \protect\ref{bias1}}

See the supplementary material. $\blacksquare $

\bibliographystyle{econometrica}
\bibliography{diss}

\newpage

\setcounter{page}{1}\setcounter{equation}{0}\renewcommand{\theequation}{B.%
\arabic{equation}}\renewcommand{\thelemma}{A.\arabic{lemma}}

\appendix

\begin{center}
{\Large Supplementary Material for \textquotedblleft Threshold Regression with Nonparametric Sample Splitting\textquotedblright }

\bigskip

\bigskip

By Yoonseok Lee and Yulong Wang

\bigskip

January 2021

\bigskip
\end{center}

\begin{quotation}
\noindent This supplementary material contains omitted proofs of some
technical lemmas.

\bigskip
\end{quotation}

\paragraph{Proof of Lemma \protect\ref{L-int1}}

For expositional simplicity, we only present the case of scalar $x_{i}$.
Similarly as (\ref{EM}), we have%
\begin{eqnarray}
&&\mathbb{E}\left[ \Delta M_{n}\left( s\right) \right]  \label{EDM} \\
&=&\iint \mathcal{D}(q,s+b_{n}t)\left\{ \mathbf{1}\left[ q<\gamma _{0}\left(
s+b_{n}t\right) \right] -\mathbf{1}\left[ q<\gamma _{0}\left( s\right) %
\right] \right\} K(t)dqdt  \notag \\
&=&\int_{\mathcal{T}^{+}(s)}\int_{\gamma _{0}\left( s\right) }^{\gamma
_{0}\left( s+b_{n}t\right) }\mathcal{D}(q,s+b_{n}t)K(t)dqdt+\int_{\mathcal{T}%
^{-}(s)}\int_{\gamma _{0}\left( s+b_{n}t\right) }^{\gamma _{0}\left(
s\right) }\mathcal{D}(q,s+b_{n}t)K(t)dqdt  \notag \\
&\equiv &\Psi _{M}^{+}(s)+\Psi _{M}^{-}(s)\text{,}  \notag
\end{eqnarray}%
where $\mathcal{D}(q,s+b_{n}t)=D(q,s+b_{n}t)f(q,s+b_{n}t)$ and we denote $%
\mathcal{T}^{+}(s)=\{t:\gamma _{0}(s)\leq \gamma _{0}\left( s+b_{n}t\right)
\}$ and $\mathcal{T}^{-}(s)=\{t:\gamma _{0}(s)>\gamma _{0}\left(
s+b_{n}t\right) \}$. We consider three cases of $\dot{\gamma}%
_{0}(s)=\partial \gamma _{0}\left( s\right) /\partial s>0$, $\dot{\gamma}%
_{0}(s)<0$, and $\dot{\gamma}_{0}(s)=0$ separately, which are well-defined
from Assumption A-(vi).

First, we suppose $\dot{\gamma}_{0}(s)>0$. We choose a positive sequence $%
t_{n}\rightarrow \infty $ such that $t_{n}b_{n}\rightarrow 0$ as $%
n\rightarrow \infty $. It follows that for any fixed $\varepsilon >0$, $%
t_{n}b_{n}\leq \varepsilon $ if $n$ is sufficiently large and hence $%
\mathcal{T}^{+}(s)\cap \{t:\left\vert t\right\vert \leq t_{n}\}$ becomes $%
\left[ 0,t_{n}\right] $ since $\dot{\gamma}_{0}\left( \cdot \right) $ is
continuous. The mean value theorem gives 
\begin{eqnarray}
&&\Psi _{M}^{+}(s)  \label{PsiM} \\
&=&\int_{0}^{t_{n}}\int_{\gamma _{0}\left( s\right) }^{\gamma _{0}\left(
s+b_{n}t\right) }\mathcal{D}(q,s+b_{n}t)K(t)dqdt+\int_{\left\vert
t\right\vert >t_{n}}\int_{\gamma _{0}\left( s\right) }^{\gamma _{0}\left(
s+b_{n}t\right) }\mathcal{D}(q,s+b_{n}t)K(t)dqdt  \notag \\
&\leq &\left\{ b_{n}\mathcal{D}(\gamma _{0}\left( s\right) ,s)\dot{\gamma}%
_{0}\left( s\right) \int_{0}^{t_{n}}tK(t)dt+O(b_{n}^{2})\right\} +O\left(
b_{n}\right) \int_{t_{n}}^{\infty }tK(t)dt  \notag \\
&=&b_{n}\mathcal{D}(\gamma _{0}\left( s\right) ,s)\dot{\gamma}_{0}\left(
s\right) \int_{0}^{\infty }tK(t)dt+O(b_{n}^{2})\text{,}  \notag
\end{eqnarray}%
where $\mathcal{D}(q,s+b_{n}t)<\infty $ and $\dot{\gamma}_{0}\left( s\right)
<\infty $ from Assumptions A-(vi) and (vii). Note that Assumption A-(x)
implies $K\left( t\right) t^{-(2+\eta )}\rightarrow 0$ for some $\eta >0$ as 
$t\rightarrow \infty $ and hence $\int_{t_{n}}^{\infty }tK(t)dt\rightarrow 0$
as $t_{n}\rightarrow \infty $. Similarly, 
\begin{eqnarray*}
\Psi _{M}^{-}(s) &=&\int_{-t_{n}}^{0}\int_{\gamma _{0}\left( s+b_{n}t\right)
}^{\gamma _{0}\left( s\right) }\mathcal{D}(q,s+b_{n}t)K(t)dqdt+o(b_{n}) \\
&=&-\mathcal{D}(\gamma _{0}\left( s\right) ,s)\dot{\gamma}_{0}\left(
s\right) b_{n}\int_{-\infty }^{0}tK(t)dt+O(b_{n}^{2})\text{,}
\end{eqnarray*}%
which yields $\mathbb{E}\left[ \Delta M_{n}\left( s\right) \right] =\mathcal{%
D}(\gamma _{0}\left( s\right) ,s)\dot{\gamma}_{0}\left( s\right)
b_{n}+O(b_{n}^{2})$. When $\dot{\gamma}_{0}(s)<0$, we can symmetrically show
that $\mathbb{E}\left[ \Delta M_{n}\left( s\right) \right] =-\mathcal{D}%
(\gamma _{0}\left( s\right) ,s)\dot{\gamma}_{0}\left( s\right)
b_{n}+O(b_{n}^{2})$.

Second, we suppose $\dot{\gamma}_{0}(s)=0$ and $s$ is the local maximizer.
Then, $\mathcal{T}^{+}(s)\cap \{t:\left\vert t\right\vert \leq t_{n}\}$
becomes empty and hence%
\begin{equation*}
\Psi _{M}^{+}(s)=0+\int_{\left\vert t\right\vert >t_{n}}\int_{\gamma
_{0}\left( s\right) }^{\gamma _{0}\left( s+b_{n}t\right) }\mathcal{D}%
(q,s+b_{n}t)K(t)dqdt=o(b_{n})\text{.}
\end{equation*}%
However, $\mathcal{T}^{-}(s)\cap \{t:\left\vert t\right\vert \leq t_{n}\}$
becomes $\{t:\left\vert t\right\vert \leq t_{n}\}$ in this case and hence%
\begin{eqnarray*}
\Psi _{M}^{-}(s) &=&\int_{-\infty }^{\infty }\int_{\gamma _{0}\left(
s+b_{n}t\right) }^{\gamma _{0}\left( s\right) }\mathcal{D}%
(q,s+b_{n}t)K(t)dqdt \\
&=&-\mathcal{D}(\gamma _{0}\left( s\right) ,s)\dot{\gamma}_{0}\left(
s\right) b_{n}\int_{-\infty }^{\infty }tK(t)dt+O(b_{n}^{2})=O(b_{n}^{2})
\end{eqnarray*}%
since $\int_{-\infty }^{\infty }tK(t)dt=0$. When $\dot{\gamma}_{0}(s)=0$ and 
$s$ is the local minimizer, we can similarly show that 
\begin{eqnarray}
\Psi _{M}^{+}(s) &=&\int_{-\infty }^{\infty }\int_{\gamma _{0}\left(
s\right) }^{\gamma _{0}\left( s+b_{n}t\right) }\mathcal{D}%
(q,s+b_{n}t)K(t)dqdt  \label{PsiM0} \\
&=&\mathcal{D}(\gamma _{0}\left( s\right) ,s)\dot{\gamma}_{0}\left( s\right)
b_{n}\int_{-\infty }^{\infty }tK(t)dt+O(b_{n}^{2})=O(b_{n}^{2})  \notag
\end{eqnarray}%
but $\Psi _{M}^{-}(s)=o(b_{n})$. By combining these results, we have $%
\mathbb{E}\left[ \Delta M_{n}\left( s\right) \right] =\mathcal{D}(\gamma
_{0}\left( s\right) ,s)|\dot{\gamma}_{0}\left( s\right) |b_{n}+O(b_{n}^{2})$
for a given $s\in \mathcal{S}_{0}$, and hence%
\begin{equation*}
\sup_{s\in \mathcal{S}_{0}}\mathbb{E}\left[ \Delta M_{n}\left( s\right) %
\right] =O(b_{n})
\end{equation*}%
since $\sup_{s\in \mathcal{S}_{0}}\mathcal{D}(\gamma _{0}\left( s\right) ,s)|%
\dot{\gamma}_{0}\left( s\right) |<\infty $ from Assumptions A-(vi) and (vii).

The desired uniform convergence result then follows if $\sup_{s\in \mathcal{S%
}_{0}}||\Delta M_{n}\left( s\right) -\mathbb{E}\left[ \Delta M_{n}\left(
s\right) \right] ||=o(b_{n})$ almost surely, which can be shown as Theorem
2.2 in Carbon, Francq, and Tran (2007) (see also Section 3 in Tran (1990)
and Section 5 in Carbon, Tran, and Wu (1997)). Similarly as the proof of (%
\ref{VM}) in Lemma \ref{L1}, we let $\tau _{n}=\left( n\log n\right)
^{1/\left( 4+\varphi \right) }$ and define 
\begin{equation*}
\Delta M_{n}^{\tau }(s)=\frac{1}{nb_{n}}\sum_{i\in \Lambda
_{n}}x_{i}^{2}\Delta _{i}(s_{i},s)K_{i}(s)\mathbf{1}_{\tau _{n}}
\end{equation*}%
as in (\ref{Mnt}), where $\Delta _{i}(s_{i},s)=\mathbf{1}_{i}\left( \gamma
_{0}\left( s_{i}\right) \right) -\mathbf{1}_{i}\left( \gamma _{0}\left(
s\right) \right) $ and $\mathbf{1}_{\tau _{n}}=\mathbf{1}\left\{
x_{i}^{2}\leq \tau _{n}\right\} $. We also let $m_{n}$ be an integer such
that $m_{n}=O(\tau _{n}n^{1-2\epsilon }/b_{n}^{2})$ and we cover the compact 
$\mathcal{S}_{0}$ by small $m_{n}$ intervals centered at $s_{k}$, which are
defined as $\mathcal{I}_{k}=\{s^{\prime }:\left\vert s^{\prime
}-s_{k}\right\vert \leq C/m_{n}\}$ for some $C<\infty $. Then, 
\begin{eqnarray}
\sup_{s\in \mathcal{S}_{0}}\left\Vert \Delta M_{n}^{\tau }(s)-\mathbb{E}%
\left[ M\Delta _{n}^{\tau }(s)\right] \right\Vert &\leq &\max_{1\leq k\leq
m_{n}}\sup_{s\in \mathcal{I}_{k}}\left\Vert \Delta M_{n}^{\tau }(s)-\Delta
M_{n}^{\tau }(s_{k})\right\Vert  \notag \\
&&+\max_{1\leq k\leq m_{n}}\sup_{s\in \mathcal{I}_{k}}\left\Vert \mathbb{E}%
\left[ \Delta M_{n}^{\tau }(s)\right] -\mathbb{E}\left[ \Delta M_{n}^{\tau
}(s_{k})\right] \right\Vert  \notag \\
&&+\max_{1\leq k\leq m_{n}}\left\Vert \Delta M_{n}^{\tau }(s_{k})-\mathbb{E}%
\left[ \Delta M_{n}^{\tau }(s_{k})\right] \right\Vert  \notag \\
&\equiv &\Psi _{\Delta M1}+\Psi _{\Delta M2}+\Psi _{\Delta M3}\text{.}
\label{PsiDM}
\end{eqnarray}%
However, 
\begin{eqnarray*}
\left\Vert \Delta M_{n}^{\tau }(s)-\Delta M_{n}^{\tau }(s_{k})\right\Vert
&\leq &\frac{1}{nb_{n}}\dsum_{i\in \Lambda _{n}}x_{i}^{2}\left\vert \Delta
_{i}(s_{i},s)-\Delta _{i}(s_{i},s_{k})\right\vert K_{i}(s_{k})\mathbf{1}%
_{\tau _{n}} \\
&&+\frac{1}{nb_{n}}\dsum_{i\in \Lambda _{n}}x_{i}^{2}\left\vert \Delta
_{i}(s_{i},s)\right\vert \left\vert K_{i}(s)-K_{i}(s_{k})\right\vert \mathbf{%
1}_{\tau _{n}} \\
&=&\frac{1}{nb_{n}}\dsum_{i\in \Lambda _{n}}x_{i}^{2}\left\vert \mathbf{1}%
_{i}\left( \gamma _{0}\left( s_{k}\right) \right) -\mathbf{1}_{i}\left(
\gamma _{0}\left( s\right) \right) \right\vert K_{i}(s_{k})\mathbf{1}_{\tau
_{n}} \\
&&+\frac{1}{nb_{n}}\dsum_{i\in \Lambda _{n}}x_{i}^{2}\left\vert \mathbf{1}%
_{i}\left( \gamma _{0}\left( s_{i}\right) \right) -\mathbf{1}_{i}\left(
\gamma _{0}\left( s\right) \right) \right\vert \left\vert
K_{i}(s)-K_{i}(s_{k})\right\vert \mathbf{1}_{\tau _{n}} \\
&=&O_{a.s.}\left( \frac{\tau _{n}}{b_{n}^{2}m_{n}}\right) =O_{a.s.}\left( 
\frac{1}{n^{1-2\epsilon }}\right)
\end{eqnarray*}%
as in (\ref{M1nt}) and (\ref{M2nt}), and hence $\Psi _{\Delta M1}=\Psi
_{\Delta M2}=o_{a.s.}(b_{n})$ as $n^{1-2\epsilon }b_{n}\rightarrow \infty $.
We also have $\Psi _{\Delta M3}=o_{a.s.}(b_{n})$ as proved below,\footnote{%
Unlike the Lemma \ref{L1}, We cannot directly use the results for $%
Q_{3n}^{\ast }$ in Carbon, Francq, and Tran (2007) here. This is because $%
O((\log n/(nb_{n}))^{1/2})$ is not necassarily $o(b_{n})$ without further
restrictions.} which completes the proof. $\blacksquare $

\paragraph{Proof of $\Psi _{\Delta M3}=o_{a.s.}(b_{n})$:}

We let%
\begin{equation*}
Z_{i}^{\tau }(s)=(nb_{n})^{-1}\left\{ (c_{0}^{\top }x_{i})^{2}\Delta
_{i}(s_{i},s)K_{i}\left( s\right) \mathbf{1}_{\tau _{n}}-\mathbb{E}%
[(c_{0}^{\top }x_{i})^{2}\Delta _{i}(s_{i},s)K_{i}\left( s\right) \mathbf{1}%
_{\tau _{n}}]\right\}
\end{equation*}%
and apply the blocking technique as in Carbon, Francq, and Tran (2007),
p.788. For $i=(i_{1},i_{2})\in \Lambda _{n}\subset \mathbb{R}^{2}$, let $%
n_{1}$ and $n_{2}$ are the numbers of grids in two dimensions, then $%
\left\vert \Lambda _{n}\right\vert =n=n_{1}n_{2}$. Without loss of
generality, we assume $n_{\ell }=2wr_{\ell }$ for $\ell =1,2$, where $w$ and 
$r_{\ell }$ are constants to be specified later. For $j=(j_{1},j_{2})$,
define%
\begin{eqnarray}
U^{[1]}(j;s)
&=&\dsum_{i_{1}=2j_{1}w+1}^{(2j_{1}+1)w}%
\dsum_{i_{2}=2j_{2}w+1}^{(2j_{2}+1)w}Z_{i}^{\tau }(s)\text{,}  \label{u[1]}
\\
U^{[2]}(j;s)
&=&\dsum_{i_{1}=2j_{1}w+1}^{(2j_{1}+1)w}%
\dsum_{i_{2}=(2j_{2}+1)w+1}^{2(j_{2}+1)w}Z_{i}^{\tau }(s)\text{,}  \notag \\
U^{[3]}(j;s) &=&\dsum_{i_{1}=\left( 2j_{1}+1\right)
w+1}^{2(j_{1}+1)w}\dsum_{i_{2}=2j_{2}w+1}^{\left( 2j_{2}+1\right)
w}Z_{i}^{\tau }(s)\text{,}  \notag \\
U^{[4]}(j;s) &=&\dsum_{i_{1}=\left( 2j_{1}+1\right)
w+1}^{2(j_{1}+1)w}\dsum_{i_{2}=(2j_{2}+1)w+1}^{2\left( j_{2}+1\right)
w}Z_{i}^{\tau }(s)\text{,}  \notag
\end{eqnarray}%
and define four blocks as 
\begin{equation*}
\mathcal{B}^{[h]}(s)=\dsum_{j_{1}=0}^{r_{1}-1}%
\dsum_{j_{2}=0}^{r_{2}-1}U^{[h]}(j;s)\text{ \ for }h=1,2,3,4,
\end{equation*}%
so that $\sum_{i\in \Lambda _{n}}Z_{i}^{\tau }(s)=\sum_{h=1}^{4}\mathcal{B}%
^{[h]}(s)$ and $\Psi _{\Delta M3}=\max_{1\leq k\leq m_{n}}\left\vert
\sum_{h=1}^{4}\mathcal{B}^{[h]}(s_{k})\right\vert $. Since these four blocks
have the same number of summands, it suffices to show $\max_{1\leq k\leq
m_{n}}\left\vert \mathcal{B}^{[1]}(s_{k})\right\vert =o_{a.s.}(b_{n})$. To
this end, we show that for some $\varepsilon _{n}=o(b_{n})$,%
\begin{eqnarray}
\mathbb{P}\left( \max_{1\leq k\leq m_{n}}\left\vert \mathcal{B}%
^{[1]}(s_{k})\right\vert >\varepsilon _{n}\right) &\leq &\dsum_{k=1}^{m_{n}}%
\mathbb{P}\left( \left\vert \mathcal{B}^{[1]}(s_{k})\right\vert >\varepsilon
_{n}\right)  \label{PBlock} \\
&\leq &m_{n}\sup_{s\in \mathcal{S}_{0}}\mathbb{P}\left( \left\vert \mathcal{B%
}^{[1]}(s)\right\vert >\varepsilon _{n}\right)  \notag \\
&=&O(n^{-c})  \notag
\end{eqnarray}%
for some $c>1$ and hence $\sum_{n=1}^{\infty }\mathbb{P}\left( \max_{1\leq
k\leq m_{n}}\left\vert \mathcal{B}^{[1]}(s_{k})\right\vert >\varepsilon
_{n}\right) <\infty $. Then the almost sure convergence is obtained by the
Borel-Cantelli lemma.

For any $s\in \mathcal{S}_{0}$, $\mathcal{B}^{[1]}(s)$ is the sum of $%
r=r_{1}r_{2}=n/\left( 2w^{2}\right) $ of $U^{[1]}(j;s)$'s. In addition, $%
U^{[1]}(j;s)$ is measurable with the $\sigma $-field generated by $%
Z_{i}^{\tau }(s)$ with $i$ belonging to the set%
\begin{equation*}
\{i=(i_{1},i_{2}):2j_{\ell }w+1\leq i_{\ell }\leq (2j_{\ell }+1)w\text{ for }%
\ell =1,2\}\text{.}
\end{equation*}%
These sets are separated by a distance of at least $w$. We enumerate the
random variables $U^{[1]}(j;s)$ and the corresponding $\sigma $-fields with
which they are measurable in an arbitrary manner, and refer to those $%
U^{[1]}(j;s)$'s as $U_{1}(s),U_{2}(s),\ldots ,U_{r}(s)$. By the uniform
almost sure law of large numbers in random fields (e.g., Theorem 2 in Jenish
and Prucha (2009)) and the fact that $\mathbb{E}\left[ K_{i}\left( s\right)
b_{n}^{-1}\right] \leq C$, we have that for any $t=1,\ldots ,r$ and $s\in 
\mathcal{S}_{0}$,%
\begin{eqnarray}
\left\vert U_{t}(s)\right\vert &\leq &\frac{Cw^{2}\tau _{n}}{n}\left( \frac{1%
}{w^{2}b_{n}}\dsum_{i_{1}=2j_{1}w+1}^{(2j_{1}+1)w}%
\dsum_{i_{2}=2j_{2}w+1}^{(2j_{2}+1)w}\left\vert \Delta _{i}(\gamma
;s)\right\vert K_{i}\left( s\right) \right)  \label{Utbd} \\
&\leq &\frac{Cw^{2}\tau _{n}}{n}\left( \frac{1}{w^{2}}%
\dsum_{i_{1}=2j_{1}w+1}^{(2j_{1}+1)w}%
\dsum_{i_{2}=2j_{2}w+1}^{(2j_{2}+1)w}K_{i}\left( s\right) b_{n}^{-1}\right) 
\notag \\
&=&\frac{C^{\prime }w^{2}\tau _{n}}{n}  \notag
\end{eqnarray}%
almost surely from (\ref{u[1]}), for some $C,C^{\prime }<\infty $, where the
last equality is obtained similarly as (\ref{PsiM}). From Lemma 3.6 in
Carbon, Francq, and Tran (2007), we can approximate\footnote{%
This approximation is reminiscient of the Berbee's lemma (Berbee (1987)) and
is based on Rio (1995), who studies the time series case. It can also be
found as Lemma 4.5 in Carbon, Tran, and Wu (1997).} $\{U_{t}(s)\}_{t=1}^{r}$
by another sequence of random variables $\{U_{t}^{\ast }(s)\}_{t=1}^{r}$
that satisfies (i)\ elements of $\{U_{t}^{\ast }(s)\}_{t=1}^{r}$ are
independent, (ii) $U_{t}^{\ast }(s)$ has the same distribution as $U_{t}(s)$
for all $t=1,\ldots ,r$, and (iii)%
\begin{equation}
\dsum_{t=1}^{r}\mathbb{E}\left[ \left\vert U_{t}^{\ast
}(s)-U_{t}(s)\right\vert \right] \leq rC^{\prime \prime }n^{-1}w^{2}\tau
_{n}\alpha _{w^{2},w^{2}}(w)  \label{indep approx}
\end{equation}%
for some $C^{\prime \prime }<\infty $. Recall that $\alpha _{w^{2},w^{2}}(w)$
is the $\alpha $-mixing coefficient defined in (\ref{mix}). Then, it follows
that%
\begin{equation}
\mathbb{P}\left( \mathcal{B}^{[1]}(s)>\varepsilon _{n}\right) \leq \mathbb{P}%
\left( \dsum_{t=1}^{r}\left\vert U_{t}^{\ast }(s)-U_{t}(s)\right\vert
>\varepsilon _{n}\right) +\mathbb{P}\left( \left\vert
\dsum_{t=1}^{r}U_{t}^{\ast }(s)\right\vert >\varepsilon _{n}\right)
\label{block decom}
\end{equation}%
for any given $s\in \mathcal{S}_{0}$, and hence in view of (\ref{PBlock})
and (\ref{block decom}) 
\begin{eqnarray}
\mathbb{P}\left( \max_{1\leq k\leq m_{n}}\left\vert \mathcal{B}%
^{[1]}(s_{k})\right\vert >\varepsilon _{n}\right) &\leq &m_{n}\sup_{s\in 
\mathcal{S}_{0}}\mathbb{P}\left( \dsum_{t=1}^{r}\left\vert U_{t}^{\ast
}(s)-U_{t}(s)\right\vert >\varepsilon _{n}\right)  \notag \\
&&+m_{n}\sup_{s\in \mathcal{S}_{0}}\mathbb{P}\left( \left\vert
\dsum_{t=1}^{r}U_{t}^{\ast }(s)\right\vert >\varepsilon _{n}\right)  \notag
\\
&\equiv &P_{U1}+P_{U2}\text{.}  \label{Pu1Pu2}
\end{eqnarray}

First, we let $\varepsilon _{n}=O((\log n/n)^{1/2})$. By Markov's
inequality, (\ref{indep approx}), and Assumption A-(iii), we have%
\begin{equation*}
P_{U1}\leq m_{n}\frac{rC^{\prime \prime }n^{-1}w^{2}\tau _{n}\alpha
_{w^{2},w^{2}}(w)}{\varepsilon _{n}}\leq C_{1}n^{\kappa _{1}}(\log
n)^{\kappa _{2}}\times \frac{\exp (-C_{1}^{\prime }n^{\kappa _{3}})}{\left(
n^{1-2\epsilon }b_{n}\right) ^{2}}
\end{equation*}%
for some $\kappa _{1},\kappa _{2},\kappa _{3}>0$ and $C_{1},C_{1}^{\prime
}<\infty $. Recall that we chose $m_{n}=O(\tau _{n}n^{1-2\epsilon
}/b_{n}^{2})$, $n=4w^{2}r$, and $\tau _{n}=\left( n\log n\right) ^{1/\left(
4+\varphi \right) }$. Hence $P_{U1}=O(\exp (-n^{\kappa _{3}}))\rightarrow 0$
as $n\rightarrow \infty $, since the second term in the last inequality
diminishes faster than the polynomial order.

Second, we now choose an integer $w$ such that%
\begin{eqnarray*}
w &=&\left( n/\left( C_{w}\tau _{n}\lambda _{n}\right) \right) ^{1/2}\text{,}
\\
\lambda _{n} &=&(n\log n)^{1/2}
\end{eqnarray*}%
for some large positive constant $C_{w}$. Note that, substituting $\lambda
_{n}$ and $\tau _{n}$ into $w$ gives%
\begin{equation*}
w=O\left( \left[ \frac{n^{\frac{1}{2}-\frac{1}{4+\varphi }}}{\left( \log
n\right) ^{\frac{1}{4+\varphi }+\frac{1}{2}}}\right] ^{1/2}\right) \text{,}
\end{equation*}%
which diverges as $n\rightarrow \infty $ for $\varphi >0$. Since $%
U_{t}^{\ast }(s)$ has the same distribution as $U_{t}(s)$, $\left\vert
U_{t}^{\ast }(s)\right\vert $ is also uniformly bounded by $C^{\prime
}n^{-1}\tau _{n}w^{2}$ almost surely for all $t=1,\ldots ,r$ from (\ref{Utbd}%
). Therefore, $\left\vert \lambda _{n}U_{t}^{\ast }(s)\right\vert \leq 1/2$
for all $t$ if $C_{w}$ is chosen to be large enough. Using the inequality $%
\exp (v)\leq 1+v+v^{2}$ for $\left\vert v\right\vert \leq 1/2$, we have $%
\exp (\lambda _{n}U_{t}^{\ast }(s))\leq 1+\lambda _{n}U_{t}^{\ast
}(s)+\lambda _{n}^{2}U_{t}^{\ast }(s)^{2}$. Hence%
\begin{equation}
\mathbb{E}[\exp (\lambda _{n}U_{t}^{\ast }(s))]\leq 1+\lambda _{n}^{2}%
\mathbb{E}\left[ U_{t}^{\ast }(s)^{2}\right] \leq \exp \left( \lambda
_{n}^{2}\mathbb{E}\left[ U_{t}^{\ast }(s)^{2}\right] \right)
\label{exp ineq}
\end{equation}%
since $\mathbb{E}\left[ U_{t}^{\ast }(s)\right] =0$ and $1+v\leq \exp (v)$
for $v\geq 0$. Using the fact that $\mathbb{P}(X>c)\leq \mathbb{E}[\exp
(Xa)]/\exp (ac)$ for any random variable $X$ and nonrandom constants $a$ and 
$c$, and that $\{U_{t}^{\ast }(s)\}_{t=1}^{r}$ are independent, we have%
\begin{eqnarray}
&&\mathbb{P}\left( \left\vert \dsum_{t=1}^{r}U_{t}^{\ast }(s)\right\vert
>\varepsilon _{n}\right)  \notag \\
&=&\mathbb{P}\left( \dsum_{t=1}^{r}\lambda _{n}U_{t}^{\ast }(s)>\lambda
_{n}\varepsilon _{n}\right) +\mathbb{P}\left( -\dsum_{t=1}^{r}\lambda
_{n}U_{t}^{\ast }(s)>\lambda _{n}\varepsilon _{n}\right)  \notag \\
&\leq &\frac{\mathbb{E}\left[ \exp \left( \lambda
_{n}\dsum_{t=1}^{r}U_{t}^{\ast }(s)\right) \right] +\mathbb{E}\left[ \exp
\left( -\lambda _{n}\dsum_{t=1}^{r}U_{t}^{\ast }(s)\right) \right] }{\exp
(\lambda _{n}\varepsilon _{n})}  \notag \\
&\leq &2\exp (-\lambda _{n}\varepsilon _{n})\exp \left( \lambda
_{n}^{2}\dsum_{t=1}^{r}\mathbb{E}\left[ U_{t}^{\ast }(s)^{2}\right] \right)
\label{PUbd}
\end{eqnarray}%
by (\ref{exp ineq}). However, using the same argument as in (\ref{EDM})
above, we can show that%
\begin{equation*}
\mathbb{E}\left[ U_{t}^{\ast }(s)^{2}\right] \leq \dsum_{\substack{ 1\leq
i_{1}\leq w  \\ 1\leq i_{2}\leq w}}\mathbb{E}\left[ Z_{i}^{\tau }(s)^{2}%
\right] +\dsum_{\substack{ i\neq j  \\ 1\leq i_{1},i_{2}\leq w  \\ 1\leq
j_{1},j_{2}\leq w}}Cov\left[ Z_{i}^{\tau }(s),Z_{j}^{\tau }(s)\right] \leq 
\frac{C_{2}w^{2}}{n^{2}}
\end{equation*}%
for some $C_{2}<\infty $, which does not depend on $s$ given Assumptions
A-(v) and (x). It follows that (\ref{PUbd}) satisfies 
\begin{eqnarray}
\sup_{s\in \mathcal{S}_{0}}\mathbb{P}\left( \left\vert
\dsum_{t=1}^{r}U_{t}^{\ast }\right\vert >\varepsilon _{n}\right) &\leq
&2\exp \left( -\lambda _{n}\varepsilon _{n}+\frac{C_{2}\lambda _{n}^{2}rw^{2}%
}{n^{2}}\right)  \label{T3n bnd} \\
&=&2\exp \left( -\lambda _{n}\varepsilon _{n}+C_{2}\lambda
_{n}^{2}n^{-1}\right) \text{.}  \notag
\end{eqnarray}%
We choose $\varepsilon _{n}=C^{\ast }\lambda _{n}^{-1}\log n=C^{\ast }(\log
n/n)^{1/2}$ for some $C^{\ast }>0$ and have%
\begin{equation*}
-\lambda _{n}\varepsilon _{n}+C_{2}\lambda _{n}^{2}n^{-1}=-C^{\ast }\log
n+C_{2}\log n=-\left( C^{\ast }-C_{2}\right) \log n\text{.}
\end{equation*}%
Therefore, in view of (\ref{T3n bnd}), we have%
\begin{equation*}
m_{n}\sup_{s\in \mathcal{S}_{0}}\mathbb{P}\left( \left\vert
\dsum_{t=1}^{r}U_{t}^{\ast }\right\vert >C^{\ast }\sqrt{\frac{\log n}{n}}%
\right) \leq \frac{2m_{n}}{n^{C^{\ast }-C_{2}}}=\frac{C_{3}(\log n)^{\kappa
_{4}}}{n^{\kappa _{5}}(n^{1-2\epsilon }b_{n})^{2}}
\end{equation*}%
for some $C_{3}<\infty $, $\kappa _{4}>0$, and $\kappa _{5}>1$ by choosing $%
C^{\ast }$ sufficiently large. Since $n^{1-2\epsilon }b_{n}\rightarrow
\infty $, we have $P_{U2}=O(n^{-\kappa _{5}})\rightarrow 0$ as $n\rightarrow
\infty $. Therefore, the desired result follows since $\varepsilon
_{n}=O((\log n/n)^{1/2})=o(b_{n})$ from Assumption A-(ix) and $%
P_{U1}+P_{U2}=O(n^{-c})$ for some $c>1$. $\blacksquare $

\paragraph{Proof of Lemma \protect\ref{infTsupL pointwise}}

For a given $s\in \mathcal{S}_{0}$, we first show (\ref{infsupL pointwise 1}%
). We consider the case with $\gamma (s)>\gamma _{0}(s)$, and the other
direction can be shown symmetrically. Since $c_{0}^{\top }D(\cdot
,s)c_{0}f\left( \cdot ,s\right) $ is continuous at $\gamma _{0}(s)$ and $%
c_{0}^{\top }D(\gamma _{0}(s),s)c_{0}f\left( \gamma _{0}(s),s\right) >0$
from Assumptions A-(vii) and (viii), there exists a sufficiently small $%
\overline{C}(s)>0$ such that 
\begin{equation}
\underline{\ell }_{D}(s)=\inf_{\left\vert \gamma (s)-\gamma _{0}\left(
s\right) \right\vert <\overline{C}(s)}c_{0}^{\top }D(\gamma
(s),s)c_{0}f\left( \gamma (s),s\right) >0\text{.}  \label{L_D}
\end{equation}%
By the mean value expansion and the fact that $T_{n}\left( \gamma ;s\right)
=c_{0}^{\top }(M_{n}\left( \gamma (s);s\right) -M_{n}\left( \gamma
_{0}(s);s\right) )c_{0}$, we have%
\begin{eqnarray*}
\mathbb{E}\left[ T_{n}\left( \gamma ;s\right) \right] &=&\int \int_{\gamma
_{0}(s)}^{\gamma (s)}\mathbb{E}\left[ \left( c_{0}^{\top }x_{i}\right)
^{2}|q,s+b_{n}t\right] f(q,s+b_{n}t)K\left( t\right) dqdt \\
&=&\left( \gamma (s)-\gamma _{0}(s)\right) c_{0}^{\top }D(\widetilde{\gamma }%
(s),s)c_{0}f\left( \widetilde{\gamma }(s),s\right)
\end{eqnarray*}%
for some $\widetilde{\gamma }(s)\in (\gamma _{0}(s),\gamma (s))$, which
yields%
\begin{equation}
\mathbb{E}\left[ T_{n}\left( \gamma ;s\right) \right] \geq \left( \gamma
\left( s\right) -\gamma _{0}\left( s\right) \right) \underline{\ell }_{D}(s)%
\text{.}  \label{ET bnd}
\end{equation}%
Furthermore, if we let $\Delta _{i}(\gamma ;s)=\mathbf{1}_{i}\left( \gamma
\left( s\right) \right) -\mathbf{1}_{i}\left( \gamma _{0}\left( s\right)
\right) $ and $Z_{n,i}(s)=\left( c_{0}^{\top }x_{i}\right) ^{2}\Delta
_{i}(\gamma ;s)K_{i}\left( s\right) -\mathbb{E}[\left( c_{0}^{\top
}x_{i}\right) ^{2}\Delta _{i}(\gamma ;s)K_{i}\left( s\right) ]$, using a
similar argument as (\ref{P12}), we have 
\begin{eqnarray}
&&\mathbb{E}\left[ \left( T_{n}\left( \gamma ;s\right) -\mathbb{E}\left[
T_{n}\left( \gamma ;s\right) \right] \right) ^{2}\right]  \label{ETn} \\
&=&\frac{1}{n^{2}b_{n}^{2}}\dsum_{i\in \Lambda _{n}}\mathbb{E}\left[
Z_{n,i}^{2}(s)\right] +\frac{1}{n^{2}b_{n}^{2}}\dsum_{i,j\in \Lambda
_{n},i\neq j}Cov[Z_{n,i}(s),Z_{n,j}(s)]  \notag \\
&\leq &\frac{C_{1}\left( s\right) }{nb_{n}}\left( \gamma \left( s\right)
-\gamma _{0}\left( s\right) \right)  \notag
\end{eqnarray}%
for some $C_{1}(s)<\infty $.

We suppose $n$ is large enough so that $\overline{r}(s)\phi _{1n}\leq 
\overline{C}(s)$. Similarly as Lemma A.7 in Hansen (2000), we set $\gamma
_{g}$ for $g=1,2,...,\overline{g}+1$ such that, for any $s\in \mathcal{S}%
_{0} $, $\gamma _{g}\left( s\right) =\gamma _{0}\left( s\right) +2^{g-1}%
\overline{r}(s)\phi _{1n}$, where $\overline{g}$ is an integer satisfying $%
\gamma _{\overline{g}}\left( s\right) -\gamma _{0}\left( s\right) =2^{%
\overline{g}-1}\overline{r}(s)\phi _{1n}\leq \overline{C}(s)$ and $\gamma _{%
\overline{g}+1}\left( s\right) -\gamma _{0}\left( s\right) >\overline{C}(s)$%
. Then Markov's inequality and (\ref{ETn}) yield that for any fixed $\eta
(s)>0$,%
\begin{eqnarray}
&&\mathbb{P}\left( \max_{1\leq g\leq \overline{g}}\left\vert \frac{%
T_{n}\left( \gamma _{g};s\right) }{\mathbb{E}\left[ T_{n}\left( \gamma
_{g};s\right) \right] }-1\right\vert >\eta (s)\right)  \label{PT bnd} \\
&\leq &\mathbb{P}\left( \max_{1\leq g\leq \overline{g}}\left\vert \frac{%
T_{n}\left( \gamma _{g};s\right) -\mathbb{E}\left[ T_{n}\left( \gamma
_{g};s\right) \right] }{\mathbb{E}\left[ T_{n}\left( \gamma _{g};s\right) %
\right] }\right\vert >\eta (s)\right)  \notag \\
&\leq &\frac{1}{\eta ^{2}(s)}\dsum_{g=1}^{\overline{g}}\frac{\mathbb{E}\left[
\left( T_{n}\left( \gamma _{g};s\right) -\mathbb{E}\left[ T_{n}\left( \gamma
_{g};s\right) \right] \right) ^{2}\right] }{\left\vert \mathbb{E}\left[
T_{n}\left( \gamma _{g};s\right) \right] \right\vert ^{2}}  \notag \\
&\leq &\frac{1}{\eta ^{2}(s)}\dsum_{g=1}^{\overline{g}}\frac{C_{1}\left(
s\right) \left( nb_{n}\right) ^{-1}\left( \gamma \left( s\right) -\gamma
_{0}\left( s\right) \right) }{\left\vert \left( \gamma \left( s\right)
-\gamma _{0}\left( s\right) \right) \underline{\ell }_{D}(s)\right\vert ^{2}}
\notag \\
&\leq &\frac{1}{\eta ^{2}(s)}\dsum_{g=1}^{\overline{g}}\frac{C_{1}\left(
s\right) \left( nb_{n}\right) ^{-1}}{2^{g-1}\overline{r}(s)\phi _{1n}%
\underline{\ell }_{D}^{2}(s)}  \notag \\
&\leq &\frac{C_{1}\left( s\right) }{\eta ^{2}(s)\overline{r}(s)\underline{%
\ell }_{D}^{2}(s)}\dsum_{g=1}^{\infty }\frac{1}{2^{g-1}}\times \frac{1}{%
n^{2\epsilon }}  \notag \\
&\leq &\varepsilon (s)  \notag
\end{eqnarray}%
for any $\varepsilon (s)>0$. From eq.\ (33) of \cite{Hansen00a}, for any $%
\gamma \left( s\right) $ such that $\overline{r}(s)\phi _{1n}\leq \gamma
\left( s\right) -\gamma _{0}\left( s\right) \leq \overline{C}(s)$, there
exists some $g^{\ast }$ satisfying $\gamma _{g^{\ast }}\left( s\right)
-\gamma _{0}\left( s\right) <\gamma \left( s\right) -\gamma _{0}\left(
s\right) <\gamma _{g^{\ast }+1}\left( s\right) -\gamma _{0}\left( s\right) $%
, and then 
\begin{eqnarray}
\frac{T_{n}\left( \gamma ;s\right) }{\left\vert \gamma \left( s\right)
-\gamma _{0}\left( s\right) \right\vert } &\geq &\frac{T_{n}\left( \gamma
_{g^{\ast }};s\right) }{\mathbb{E}\left[ T_{n}\left( \gamma _{g^{\ast
}};s\right) \right] }\times \frac{\mathbb{E}\left[ T_{n}\left( \gamma
_{g^{\ast }};s\right) \right] }{\left\vert \gamma _{g^{\ast }+1}\left(
s\right) -\gamma _{0}\left( s\right) \right\vert }  \label{Tn_Lbound} \\
&\geq &\left\{ 1-\max_{1\leq g\leq \overline{g}}\left\vert \frac{T_{n}\left(
\gamma _{g};s\right) }{\mathbb{E}\left[ T_{n}\left( \gamma _{g};s\right) %
\right] }-1\right\vert \right\} \frac{\mathbb{E}\left[ T_{n}\left( \gamma
_{g^{\ast }};s\right) \right] }{\left\vert \gamma _{g^{\ast }+1}\left(
s\right) -\gamma _{0}\left( s\right) \right\vert }  \notag \\
&\geq &\left( 1-\eta \left( s\right) \right) \frac{\left\vert \gamma
_{g^{\ast }}\left( s\right) -\gamma _{0}\left( s\right) \right\vert 
\underline{\ell }_{D}(s)}{\left\vert \gamma _{g^{\ast }+1}\left( s\right)
-\gamma _{0}\left( s\right) \right\vert }  \notag
\end{eqnarray}%
from (\ref{ET bnd}), where $|\gamma _{g^{\ast }}\left( s\right) -\gamma
_{0}\left( s\right) |/|\gamma _{g^{\ast }+1}\left( s\right) -\gamma
_{0}\left( s\right) |\underline{\ell }_{D}(s)$ is some finite non-zero
constant by construction. Hence, in view of (\ref{Tn_Lbound}), we can find $%
C_{T}(s)<\infty $ such that 
\begin{equation*}
\mathbb{P}\left( \inf_{\overline{r}(s)\phi _{1n}<\left\vert \gamma
(s)-\gamma _{0}\left( s\right) \right\vert <\overline{C}(s)}\frac{%
T_{n}\left( \gamma ;s\right) }{\left\vert \gamma (s)-\gamma _{0}\left(
s\right) \right\vert }<C_{T}(s)(1-\eta (s))\right) \leq \varepsilon (s)\text{%
.}
\end{equation*}%
The proof for (\ref{infsupL pointwise 2}) is similar to that for (\ref%
{infsupL pointwise 1}) and hence omitted.

We next show (\ref{infsupL pointwise 3}). Without loss of generality we
assume $x_{i}$ is a scalar, and so is $L_{n}(\gamma ;s)$. Similarly as (\ref%
{ETn}), we have 
\begin{equation}
\mathbb{E}\left[ \left\vert L_{n}\left( \gamma ;s\right) \right\vert ^{2}%
\right] \leq C_{2}\left( s\right) \left\vert \gamma (s)-\gamma
_{0}(s)\right\vert  \label{Lbound}
\end{equation}%
for some $C_{2}(s)<\infty $. By defining $\gamma _{g}$ in the same way as
above, Markov's inequality and (\ref{Lbound}) yields that for any fixed $%
\eta (s)>0$,%
\begin{eqnarray}
&&\mathbb{P}\left( \max_{1\leq g\leq \overline{g}}\frac{\left\vert
L_{n}\left( \gamma _{g};s\right) \right\vert }{\sqrt{a_{n}}\left( \gamma
_{g}\left( s\right) -\gamma _{0}\left( s\right) \right) }>\frac{\eta (s)}{4}%
\right)  \label{L 1} \\
&\leq &\frac{16}{\eta ^{2}(s)}\dsum_{g=1}^{\infty }\frac{\mathbb{E}\left[
L_{n}\left( \gamma _{g},s\right) ^{2}\right] }{a_{n}\left\vert \gamma
_{g}\left( s\right) -\gamma _{0}\left( s\right) \right\vert ^{2}}  \notag \\
&\leq &\frac{16}{\eta ^{2}(s)}\dsum_{g=1}^{\infty }\frac{C_{2}\left(
s\right) }{a_{n}\left\vert \gamma _{g}\left( s\right) -\gamma _{0}\left(
s\right) \right\vert }  \notag \\
&\leq &\frac{16C_{2}\left( s\right) }{\eta ^{2}(s)\overline{r}(s)}%
\dsum_{g=1}^{\infty }\frac{1}{2^{g-1}}  \notag
\end{eqnarray}%
since $a_{n}=\phi _{1n}^{-1}$. This probability is arbitrarily close to zero
if $\overline{r}(s)$ is chosen large enough. It is worth to note that (\ref%
{L 1}) provides the maximal (or sharp) rate of $\phi _{1n}$ as $a_{n}^{-1}$
because we need $a_{n}\left\vert \gamma _{g}\left( s\right) -\gamma
_{0}\left( s\right) \right\vert =O(\phi _{1n}a_{n})=O(1)$ as $n\rightarrow
\infty $. This $\phi _{1n}a_{n}=O(1)$ condition also satisfies (\ref{PT bnd}%
).

Similarly, from Lemma \ref{max ineq}, we have 
\begin{eqnarray}
&&\mathbb{P}\left( \max_{1\leq g\leq \overline{g}}\sup_{\gamma _{g}(s)\leq
\gamma (s)\leq \gamma _{g+1}(s)}\frac{\left\vert L_{n}\left( \gamma
;s\right) -L_{n}\left( \gamma _{g};s\right) \right\vert }{\sqrt{a_{n}}\left(
\gamma _{g}(s)-\gamma _{0}\left( s\right) \right) }>\frac{\eta \left(
s\right) }{4}\right)  \label{L 2} \\
&\leq &\dsum_{g=1}^{\bar{g}}\mathbb{P}\left( \sup_{\gamma _{g}(s)\leq \gamma
(s)\leq \gamma _{g+1}(s)}\left\vert L_{n}\left( \gamma ;s\right)
-L_{n}\left( \gamma _{g};s\right) \right\vert >\sqrt{a_{n}}\left( \gamma
_{g}\left( s\right) -\gamma _{0}\left( s\right) \right) \frac{\eta \left(
s\right) }{4}\right)  \notag \\
&\leq &\dsum_{g=1}^{\infty }\frac{C_{3}\left( s\right) \left\vert \gamma
_{g+1}(s)-\gamma _{g}(s)\right\vert ^{2}}{\eta ^{4}\left( s\right)
a_{n}^{2}\left\vert \gamma _{g}\left( s\right) -\gamma _{0}\left( s\right)
\right\vert ^{4}}  \notag \\
&\leq &\frac{C_{3}^{\prime }\left( s\right) }{\eta ^{4}\left( s\right) 
\overline{r}(s)^{2}}  \notag
\end{eqnarray}%
for some $C_{3}(s),C_{3}^{\prime }(s)<\infty $, where $\gamma _{g}\left(
s\right) =\gamma _{0}\left( s\right) +2^{g-1}\overline{r}(s)\phi _{1n}$.
This probability is also arbitrarily close to zero if $\overline{r}(s)$ is
chosen large enough. Since 
\begin{eqnarray}
&&\sup_{\overline{r}(s)\phi _{1n}<\left\vert \gamma (s)-\gamma _{0}\left(
s\right) \right\vert <\overline{C}(s)}\frac{\left\vert L_{n}\left( \gamma
;s\right) \right\vert }{\sqrt{a_{n}}\left( \gamma (s)-\gamma _{0}\left(
s\right) \right) }  \label{Ln_bd} \\
&\leq &2\max_{1\leq g\leq \overline{g}}\frac{\left\vert L_{n}\left( \gamma
_{g};s\right) \right\vert }{\sqrt{a_{n}}\left( \gamma _{g}\left( s\right)
-\gamma _{0}\left( s\right) \right) }  \notag \\
&&+2\max_{1\leq g\leq \overline{g}}\sup_{\gamma _{g}(s)\leq \gamma (s)\leq
\gamma _{g+1}(s)}\frac{\left\vert L_{n}\left( \gamma ;s\right) -L_{n}\left(
\gamma _{g};s\right) \right\vert }{\sqrt{a_{n}}\left( \gamma _{g}(s)-\gamma
_{0}\left( s\right) \right) }\text{,}  \notag
\end{eqnarray}%
(\ref{L 1}) and (\ref{L 2}) yield 
\begin{eqnarray*}
&&\mathbb{P}\left( \sup_{\overline{r}(s)\phi _{1n}<\left\vert \gamma
(s)-\gamma _{0}\left( s\right) \right\vert <\overline{C}(s)}\frac{\left\vert
L_{n}\left( \gamma ;s\right) \right\vert }{\sqrt{a_{n}}\left( \gamma
(s)-\gamma _{0}\left( s\right) \right) }>\eta \left( s\right) \right) \\
&\leq &\mathbb{P}\left( 2\max_{1\leq g\leq \overline{g}}\frac{\left\vert
L_{n}\left( \gamma _{g};s\right) \right\vert }{\sqrt{a_{n}}\left( \gamma
_{g}\left( s\right) -\gamma _{0}\left( s\right) \right) }>\frac{\eta \left(
s\right) }{2}\right) \\
&&+\mathbb{P}\left( 2\max_{1\leq g\leq \overline{g}}\sup_{\gamma _{g}(s)\leq
\gamma (s)\leq \gamma _{g+1}(s)}\frac{\left\vert L_{n}\left( \gamma
;s\right) -L_{n}\left( \gamma _{g};s\right) \right\vert }{\sqrt{a_{n}}\left(
\gamma _{g}(s)-\gamma _{0}\left( s\right) \right) }>\frac{\eta \left(
s\right) }{2}\right) \\
&\leq &\varepsilon (s)
\end{eqnarray*}%
for any $\varepsilon (s)>0$ if we pick $\overline{r}(s)$ sufficiently large. 
$\blacksquare $

\paragraph{Proof of Lemma \protect\ref{neg pointwise}}

Using the same notations in Lemma \ref{L-A3}, (\ref{theta}) yields%
\begin{eqnarray}
&&n^{\epsilon }\left( \widehat{\theta }(\widehat{\gamma }(s))-\theta
_{0}\right)  \label{para1} \\
&=&\left\{ \frac{1}{nb_{n}}\widetilde{Z}(\widehat{\gamma }(s);s)^{\top }%
\widetilde{Z}(\widehat{\gamma }(s);s)\right\} ^{-1}  \notag \\
&&\times \left\{ \frac{n^{\epsilon }}{nb_{n}}\widetilde{Z}(\widehat{\gamma }%
(s);s)^{\top }\widetilde{u}(s)-\frac{n^{\epsilon }}{nb_{n}}\widetilde{Z}(%
\widehat{\gamma }(s);s)^{\top }\left( \widetilde{Z}(\widehat{\gamma }(s);s)-%
\widetilde{Z}(\gamma _{0}(s_{i});s)\right) \theta _{0}\right\}  \notag \\
&\equiv &\Theta _{A1}^{-1}(s)\left\{ \Theta _{A2}(s)-\Theta _{A3}(s)\right\} 
\text{.}  \notag
\end{eqnarray}%
We let $M(s)\equiv \int_{-\infty }^{\infty }D(q,s)f\left( q,s\right)
dq<\infty $. For the denominator $\Theta _{A1}(s)$, we have 
\begin{align}
\Theta _{A1}(s)& =\left( 
\begin{array}{cc}
(nb_{n})^{-1}\sum_{i\in \Lambda _{n}}x_{i}x_{i}^{\top }K_{i}(s) & 
M_{n}\left( \widehat{\gamma }(s);s\right) \\ 
M_{n}\left( \widehat{\gamma }(s);s\right) & M_{n}\left( \widehat{\gamma }%
(s);s\right)%
\end{array}%
\right)  \label{DEN} \\
& \rightarrow _{p}\left( 
\begin{array}{cc}
M(s) & M\left( \gamma _{0}(s);s\right) \\ 
M\left( \gamma _{0}(s);s\right) & M\left( \gamma _{0}(s);s\right)%
\end{array}%
\right) \text{,}  \notag
\end{align}%
where $M\left( \gamma ;s\right) <\infty $ is defined in (\ref{M_gs}), which
is continuously differentiable in $\gamma $. Note that $|M_{n}\left( 
\widehat{\gamma }(s);s\right) -M\left( \gamma _{0}(s);s\right) |\leq
|M_{n}\left( \widehat{\gamma }(s);s\right) -M\left( \widehat{\gamma }%
(s);s\right) |+|M\left( \widehat{\gamma }(s);s\right) -M\left( \gamma
_{0}(s);s\right) |=o_{p}(1)$ from Lemma \ref{L1} and the pointwise
consistency of $\widehat{\gamma }(s)$ in Lemma \ref{L-A3}. In addition, $%
(nb_{n})^{-1}\dsum_{i\in \Lambda _{n}}x_{i}x_{i}^{\top }K_{i}\left( s\right)
\rightarrow _{p}M(s)$ from the standard kernel estimation result. Note that
the probability limit of $\Theta _{A1}(s)$ is positive definite since both $%
M(s)$ and $M\left( \gamma _{0}(s);s\right) $ are positive definite and 
\begin{equation*}
M(s)-M\left( \gamma _{0}(s);s\right) =\int_{\gamma _{0}(s)}^{\infty
}D(q,s)f\left( q,s\right) dq>0
\end{equation*}%
for any $\gamma _{0}(s)\in \Gamma $ from Assumption A-(viii).

For the numerator part $\Theta _{A2}(s)$, we have $\Theta
_{A2}(s)=O_{p}(a_{n}^{-1/2})=o_{p}(1)$ because 
\begin{equation}
\frac{1}{\sqrt{nb_{n}}}\widetilde{Z}(\widehat{\gamma }(s);s)^{\top }%
\widetilde{u}(s)=\left( 
\begin{array}{c}
(nb_{n})^{-1/2}\sum_{i\in \Lambda _{n}}x_{i}u_{i}K_{i}(s) \\ 
J_{n}\left( \widehat{\gamma }(s);s\right)%
\end{array}%
\right) =O_{p}\left( 1\right)  \label{NUM1}
\end{equation}%
from Lemma \ref{L1} and the pointwise consistency of $\widehat{\gamma }(s)$
in Lemma \ref{L-A3}. Note that the standard kernel estimation result gives $%
(nb_{n})^{-1/2}\sum_{i\in \Lambda _{n}}x_{i}u_{i}K_{i}\left( s\right)
=O_{p}(1)$. Moreover, we have%
\begin{equation}
\Theta _{A3}(s)=\left( 
\begin{array}{c}
(nb_{n})^{-1}\sum_{i\in \Lambda _{n}}x_{i}x_{i}^{\top }c_{0}\left\{ \mathbf{1%
}_{i}\left( \widehat{\gamma }(s)\right) -\mathbf{1}_{i}\left( \gamma
_{0}\left( s_{i}\right) \right) \right\} K_{i}\left( s\right) \\ 
(nb_{n})^{-1}\sum_{i\in \Lambda _{n}}x_{i}x_{i}^{\top }c_{0}\mathbf{1}%
_{i}\left( \widehat{\gamma }(s)\right) \left\{ \mathbf{1}_{i}\left( \widehat{%
\gamma }(s)\right) -\mathbf{1}_{i}\left( \gamma _{0}\left( s_{i}\right)
\right) \right\} K_{i}\left( s\right)%
\end{array}%
\right)  \label{NUM2}
\end{equation}%
and 
\begin{eqnarray}
&&\frac{1}{nb_{n}}\dsum_{i\in \Lambda _{n}}c_{0}^{\top }x_{i}x_{i}^{\top
}\left\{ \mathbf{1}_{i}\left( \widehat{\gamma }(s)\right) -\mathbf{1}%
_{i}\left( \gamma _{0}\left( s_{i}\right) \right) \right\} K_{i}\left(
s\right)  \label{num*} \\
&\leq &\left\Vert c_{0}\right\Vert \left\Vert M_{n}\left( \widehat{\gamma }%
(s);s\right) -M_{n}\left( \gamma _{0}(s_{i});s\right) \right\Vert  \notag \\
&\leq &\left\Vert c_{0}\right\Vert \left\{ \left\Vert M_{n}\left( \widehat{%
\gamma }(s);s\right) -M_{n}\left( \gamma _{0}(s);s\right) \right\Vert
+O_{p}(b_{n})\right\}  \notag \\
&=&o_{p}(1)\text{,}  \notag
\end{eqnarray}%
where the second inequality is from (\ref{m0}) and the last equality is
because $M_{n}\left( \gamma ;s\right) \rightarrow _{p}M\left( \gamma
;s\right) $ from Lemma \ref{L1}, which is continuous in $\gamma $ and $%
\widehat{\gamma }(s)\rightarrow _{p}\gamma _{0}(s)$ in Lemma \ref{L-A3}.
Since 
\begin{eqnarray}
&&\frac{1}{nb_{n}}\dsum_{i\in \Lambda _{n}}x_{i}x_{i}^{\top }c_{0}\mathbf{1}%
_{i}\left( \widehat{\gamma }(s)\right) \left\{ \mathbf{1}_{i}\left( \widehat{%
\gamma }(s)\right) -\mathbf{1}_{i}\left( \gamma _{0}\left( s_{i}\right)
\right) \right\} K_{i}\left( s\right)  \label{NUM3} \\
&\leq &\left\Vert c_{0}\right\Vert \left\Vert M_{n}\left( \widehat{\gamma }%
(s);s\right) -M_{n}\left( \gamma _{0}(s_{i});s\right) \right\Vert =o_{p}(1) 
\notag
\end{eqnarray}%
from (\ref{num*}), we have $\Theta _{A3}(s)=o_{p}(1)$ as well, which
completes the proof. $\blacksquare $

\paragraph*{Proof of Lemma \protect\ref{L3}}

For the first result, using the same notations in Lemma \ref{L-A3}, we write%
\begin{eqnarray*}
&&\sqrt{nb_{n}}\left( \widehat{\theta }\left( \widehat{\gamma }\left(
s\right) \right) -\theta _{0}\right) \\
&=&\left\{ \frac{1}{nb_{n}}\widetilde{Z}(\widehat{\gamma }(s);s)^{\top }%
\widetilde{Z}(\widehat{\gamma }(s);s)\right\} ^{-1} \\
&&\times \left\{ \frac{1}{\sqrt{nb_{n}}}\widetilde{Z}(\widehat{\gamma }%
(s);s)^{\top }\widetilde{u}(s)-\frac{1}{\sqrt{nb_{n}}}\widetilde{Z}(\widehat{%
\gamma }(s);s)^{\top }\left( \widetilde{Z}(\widehat{\gamma }(s);s)-%
\widetilde{Z}(\gamma _{0}(s_{i});s)\right) \theta _{0}\right\} \\
&\equiv &\Theta _{B1}^{-1}(s)\left\{ \Theta _{B2}(s)-\Theta _{B3}(s)\right\}
\end{eqnarray*}%
similarly as (\ref{para1}). For the denominator, since $\Theta
_{B1}(s)=\Theta _{A1}(s)$ in (\ref{para1}), then $\Theta
_{B1}^{-1}(s)=O_{p}(1)$ from (\ref{DEN}). For the numerator, we first have $%
\Theta _{B2}(s)=O_{p}(1)$ from (\ref{NUM1}). For $\Theta _{B3}(s)$,
similarly as (\ref{NUM2}), 
\begin{equation*}
\Theta _{B3}(s)=\left( 
\begin{array}{l}
a_{n}^{-1/2}\sum_{i\in \Lambda _{n}}n^{-\epsilon }x_{i}x_{i}^{\top }\delta
_{0}\left\{ \mathbf{1}_{i}\left( \widehat{\gamma }(s)\right) -\mathbf{1}%
_{i}\left( \gamma _{0}\left( s_{i}\right) \right) \right\} K_{i}\left(
s\right) \\ 
a_{n}^{-1/2}\sum_{i\in \Lambda _{n}}n^{-\epsilon }x_{i}x_{i}^{\top }\delta
_{0}\mathbf{1}_{i}\left( \widehat{\gamma }(s)\right) \left\{ \mathbf{1}%
_{i}\left( \widehat{\gamma }(s)\right) -\mathbf{1}_{i}\left( \gamma
_{0}\left( s_{i}\right) \right) \right\} K_{i}\left( s\right)%
\end{array}%
\right) \text{.}
\end{equation*}%
However, since $\widehat{\gamma }(s)=\gamma _{0}(s)+r(s)\phi _{1n}$ for some 
$r(s)$ bounded in probability from Theorem \ref{p-roc}, similarly as (\ref%
{EA*}), we have%
\begin{eqnarray*}
&&\mathbb{E}\left[ \dsum_{i\in \Lambda _{n}}n^{-\epsilon }\delta _{0}^{\top
}x_{i}x_{i}^{\top }\left\{ \mathbf{1}_{i}\left( \widehat{\gamma }(s)\right) -%
\mathbf{1}_{i}\left( \gamma _{0}\left( s_{i}\right) \right) \right\}
K_{i}\left( s\right) \right] \\
&\leq &a_{n}\left\vert \iint_{\min \{\gamma _{0}\left( s+b_{n}t\right)
,\gamma _{0}\left( s\right) +r(s)\phi _{1n}\}}^{\max \{\gamma _{0}\left(
s+b_{n}t\right) ,\gamma _{0}\left( s\right) +r(s)\phi _{1n}\}}\mathbb{E}%
\left[ x_{i}x_{i}^{\top }c_{0}|q,s+b_{n}t\right] K\left( t\right) f\left(
q,s+b_{n}t\right) dqdt\right\vert \\
&\leq &a_{n}\left\vert \iint_{\min \{\gamma _{0}\left( s\right) +r(s)\phi
_{1n},\gamma _{0}(s)\}}^{\max \{\gamma _{0}\left( s\right) +r(s)\phi
_{1n},\gamma _{0}(s)\}}\mathbb{E}\left[ x_{i}x_{i}^{\top }c_{0}|q,s+b_{n}t%
\right] K\left( t\right) f\left( q,s+b_{n}t\right) dqdt\right\vert \\
&&+a_{n}\left\vert \iint_{\min \{\gamma _{0}\left( s+b_{n}t\right) ,\gamma
_{0}(s)\}}^{\max \{\gamma _{0}\left( s+b_{n}t\right) ,\gamma _{0}(s)\}}%
\mathbb{E}\left[ x_{i}x_{i}^{\top }c_{0}|q,s+b_{n}t\right] K\left( t\right)
f\left( q,s+b_{n}t\right) dqdt\right\vert \\
&=&a_{n}\phi _{1n}\left\vert r(s)\right\vert \left\vert D\left( \gamma
_{0}\left( s\right) ,s\right) c_{0}\right\vert f\left( \gamma _{0}\left(
s\right) ,s\right) +O(a_{n}b_{n}) \\
&=&O(1)
\end{eqnarray*}%
as $a_{n}\phi _{1n}=1$ and $a_{n}b_{n}=n^{1-2\epsilon }b_{n}^{2}\rightarrow
\varrho <\infty $. We also have 
\begin{equation*}
Var\left[ \dsum_{i\in \Lambda _{n}}n^{-\epsilon }x_{i}x_{i}^{\top }\delta
_{0}\left\{ \mathbf{1}_{i}\left( \widehat{\gamma }(s)\right) -\mathbf{1}%
_{i}\left( \gamma _{0}\left( s_{i}\right) \right) \right\} K_{i}\left(
s\right) \right] =O(n^{-2\epsilon })=o(1),
\end{equation*}%
similarly as (\ref{VA*}). Therefore, from the same reason as (\ref{NUM3}),
we have $\Theta _{B3}(s)=O_{p}(a_{n}^{-1/2})=o_{p}(1)$, which completes the
proof.

For the second result, given the same derivation for $\Theta _{B1}^{-1}(s)$
and $\Theta _{B3}(s)$ above, it suffices to show that%
\begin{equation*}
\frac{1}{\sqrt{nb_{n}}}\widetilde{Z}(\widehat{\gamma }(s);s)^{\top }%
\widetilde{u}(s)-\frac{1}{\sqrt{nb_{n}}}\widetilde{Z}(\gamma
_{0}(s);s)^{\top }\widetilde{u}(s)=o_{p}(1)\text{,}
\end{equation*}%
which is implied by Lemma \ref{max ineq}. $\blacksquare $

\paragraph*{Proof of Lemma \protect\ref{Bn3}}

First, we consider the case with $r>0$. For a fixed $s\in \mathcal{S}_{0}$,
we have%
\begin{eqnarray*}
&&\left\{ \mathbf{1}[q\leq \gamma _{0}\left( s\right) +r/a_{n}]-\mathbf{1}%
[q\leq \gamma _{0}\left( s\right) ]\right\} \left\{ \mathbf{1}[q\leq \gamma
_{0}\left( s+b_{n}t\right) ]-\mathbf{1}[q\leq \gamma _{0}\left( s\right)
]\right\} \\
&=&\left\{ 
\begin{array}{ll}
\mathbf{1}\left[ \gamma _{0}\left( s\right) <q\leq \gamma _{0}\left(
s+b_{n}t\right) \right] & \text{if }\gamma _{0}\left( s+b_{n}t\right) \leq
\gamma _{0}\left( s\right) +r/a_{n}\text{,} \\ 
\mathbf{1}\left[ \gamma _{0}\left( s\right) <q\leq \gamma _{0}\left(
s\right) +r/a_{n}\right] & \text{otherwise.}%
\end{array}%
\right.
\end{eqnarray*}%
Denote $\mathcal{D}_{c_{0}}(q,s)=c_{0}^{\top }D(q,s)c_{0}f\left( q,s\right) $%
. Then 
\begin{eqnarray*}
&&\mathbb{E}\left[ B_{n}^{\ast \ast }(r,s)\right] \\
&=&a_{n}\iint c_{0}^{\top }D(q,s+b_{n}t)c_{0}\left\{ \mathbf{1}[q\leq \gamma
_{0}\left( s\right) +r/a_{n}]-\mathbf{1}[q\leq \gamma _{0}\left( s\right)
]\right\} \\
&&\ \ \ \ \ \ \ \ \ \ \ \ \ \ \ \ \times \left\{ \mathbf{1}[q\leq \gamma
_{0}\left( s+b_{n}t\right) ]-\mathbf{1}[q\leq \gamma _{0}\left( s\right)
]\right\} K\left( t\right) f\left( q,s+b_{n}t\right) dqdt \\
&=&a_{n}\int_{\mathcal{T}_{1}^{\ast }(r;s)}\int_{\gamma _{0}(s)}^{\gamma
_{0}(s+b_{n}t)}\mathcal{D}_{c_{0}}(q,s+b_{n}t)K\left( t\right) dqdt \\
&&+a_{n}\int_{\mathcal{T}_{2}^{\ast }(r;s)}\int_{\gamma _{0}(s)}^{\gamma
_{0}(s)+r/a_{n}}\mathcal{D}_{c_{0}}(q,s+b_{n}t)K\left( t\right) dqdt \\
&\equiv &B_{n1}^{\ast \ast }(r,s)+B_{n2}^{\ast \ast }(r,s)\text{,}
\end{eqnarray*}%
where 
\begin{eqnarray*}
\mathcal{T}_{1}^{\ast }(r;s) &=&\left\{ t:\gamma _{0}\left( s\right) <\gamma
_{0}\left( s+b_{n}t\right) \right\} \cap \left\{ t:\gamma _{0}\left(
s+b_{n}t\right) \leq \gamma _{0}\left( s\right) +r/a_{n}\right\} \text{,} \\
\mathcal{T}_{2}^{\ast }(r;s) &=&\left\{ t:\gamma _{0}\left( s\right) <\gamma
_{0}\left( s+b_{n}t\right) \right\} \cap \left\{ t:\gamma _{0}\left(
s\right) +r/a_{n}<\gamma _{0}\left( s+b_{n}t\right) \right\} \text{.}
\end{eqnarray*}%
Note that $\gamma _{0}\left( s\right) <\gamma _{0}\left( s\right) +r/a_{n}$
always holds for $r>0$. Similarly as in the proof of Lemma \ref{L-int1}, we
let a positive sequence $t_{n}\rightarrow \infty $ such that $%
t_{n}b_{n}\rightarrow 0$ as $n\rightarrow \infty $. Since $%
\int_{t_{n}}^{\infty }tK(t)dt\rightarrow 0$ by Assumption A-(x) with $%
t_{n}\rightarrow \infty $, both $\mathcal{T}_{1}^{\ast }(r;s)\cap
\{t:\left\vert t\right\vert >t_{n}\}$ and $\mathcal{T}_{2}^{\ast }(r;s)\cap
\{t:\left\vert t\right\vert >t_{n}\}$ becomes negligible as $%
t_{n}\rightarrow \infty $ using the same argument in (\ref{PsiM}). It
follows that 
\begin{eqnarray*}
B_{n1}^{\ast \ast }(r,s) &=&a_{n}\int_{\mathcal{T}_{1}(r;s)}\int_{\gamma
_{0}(s)}^{\gamma _{0}(s+b_{n}t)}\mathcal{D}_{c_{0}}(q,s+b_{n}t)K\left(
t\right) dqdt+o\left( a_{n}b_{n}\right) \text{,} \\
B_{n2}^{\ast \ast }(r,s) &=&a_{n}\int_{\mathcal{T}_{2}(r;s)}\int_{\gamma
_{0}(s)}^{\gamma _{0}(s)+r/a_{n}}\mathcal{D}_{c_{0}}(q,s+b_{n}t)K\left(
t\right) dqdt+o\left( a_{n}b_{n}\right) \text{,}
\end{eqnarray*}%
where 
\begin{eqnarray*}
\mathcal{T}_{1}(r;s) &=&\mathcal{T}_{1}^{\ast }(r;s)\cap \{t:\left\vert
t\right\vert \leq t_{n}\}\text{,} \\
\mathcal{T}_{2}(r;s) &=&\mathcal{T}_{2}^{\ast }(r;s)\cap \{t:\left\vert
t\right\vert \leq t_{n}\}\text{.}
\end{eqnarray*}%
Recall that $a_{n}b_{n}=n^{1-2\epsilon }b_{n}^{2}\rightarrow \varrho <\infty 
$ and hence $o\left( a_{n}b_{n}\right) =o(1)$. We consider three cases of $%
\dot{\gamma}_{0}(s)>0$, $\dot{\gamma}_{0}(s)<0$, and $\dot{\gamma}_{0}(s)=0$
separately.

First, we suppose $\dot{\gamma}_{0}(s)>0$. For any fixed $\varepsilon >0$,
it holds $t_{n}b_{n}\leq \varepsilon $ if $n$ is sufficiently large.
Therefore, for both $\mathcal{T}_{1}(r;s)$ and $\mathcal{T}_{2}(r;s)$, $%
\gamma _{0}\left( s\right) <\gamma _{0}\left( s+b_{n}t\right) $ requires
that $t>0$ for sufficiently large $n$. Furthermore, $\gamma _{0}\left(
s+b_{n}t\right) <\gamma _{0}\left( s\right) +r/a_{n}$ implies that $%
t<r/\left( a_{n}b_{n}\dot{\gamma}_{0}(\widetilde{s})\right) $ for some $%
\widetilde{s}\in \left[ s,s+b_{n}t\right] $, where $0<r/\left( a_{n}b_{n}%
\dot{\gamma}_{0}(\widetilde{s})\right) <\infty $. Therefore, $\mathcal{T}%
_{1}(r;s)=\{t:t>0$ and $t<r/\left( a_{n}b_{n}\dot{\gamma}_{0}(\widetilde{s}%
)\right) \}$ for sufficiently large $n$. It follows that, by Taylor
expansion,%
\begin{eqnarray*}
B_{n1}^{\ast \ast }(r,s) &=&a_{n}\int_{0}^{r/\left( a_{n}b_{n}\dot{\gamma}%
_{0}(\widetilde{s})\right) }\int_{\gamma _{0}(s)}^{\gamma _{0}(s+b_{n}t)}%
\mathcal{D}_{c_{0}}(q,s+b_{n}t)K\left( t\right) dqdt \\
&=&a_{n}b_{n}\mathcal{D}_{c_{0}}(\gamma _{0}(s),s)\dot{\gamma}%
_{0}(s)\int_{0}^{r/\left( a_{n}b_{n}\dot{\gamma}_{0}(\widetilde{s})\right)
}tK\left( t\right) dt+a_{n}b_{n}O\left( b_{n}\right) \\
&=&\varrho \mathcal{D}_{c_{0}}(\gamma _{0}(s),s)\dot{\gamma}_{0}(s)\mathcal{K%
}_{1}\left( r,\varrho ;s\right) +o(1)
\end{eqnarray*}%
for sufficiently large $n$, since $a_{n}b_{n}=n^{1-2\epsilon
}b_{n}^{2}\rightarrow \varrho <\infty $ and $\widetilde{s}\rightarrow s$ as $%
n\rightarrow \infty $. Similarly, since $\gamma _{0}\left( s\right)
+r/a_{n}<\gamma _{0}\left( s+b_{n}t\right) $ implies $t>r/\left( a_{n}b_{n}%
\dot{\gamma}_{0}(\widetilde{s})\right) $ for some $\widetilde{s}\in \left[
s,s+b_{n}t\right] $, we have $\mathcal{T}_{2}(r;s)=\{t:t>0$ and $t>r/\left(
a_{n}b_{n}\dot{\gamma}_{0}(\widetilde{s})\right) \}$. Hence, 
\begin{eqnarray*}
B_{n2}^{\ast \ast }(r,s) &=&a_{n}\int_{r/(a_{n}b_{n}\dot{\gamma}_{0}(%
\widetilde{s}))}^{t_{n}}\int_{\gamma _{0}(s)}^{\gamma _{0}(s)+r/a_{n}}%
\mathcal{D}_{c_{0}}(q,s+b_{n}t)K\left( t\right) dqdt \\
&=&r\mathcal{D}_{c_{0}}(\gamma _{0}(s),s)\int_{r/\left( a_{n}b_{n}\dot{\gamma%
}_{0}(\widetilde{s})\right) }^{t_{n}}K\left( t\right) dt+O(b_{n}) \\
&=&r\mathcal{D}_{c_{0}}(\gamma _{0}(s),s)\left\{ \frac{1}{2}-\mathcal{K}%
_{0}\left( r,\varrho ;s\right) \right\} +o(1)
\end{eqnarray*}%
for sufficiently large $n$. Recall that $\left\vert \mathcal{K}_{0}\left(
r,\varrho ;s\right) \right\vert \leq 1/2$ and $\left\vert \mathcal{K}%
_{1}\left( r,\varrho ;s\right) \right\vert \leq 1/2$.

When $\dot{\gamma}_{0}(s)<0$, $-\infty <r/\left( a_{n}b_{n}\dot{\gamma}%
_{0}(s)\right) <0$ and we can similarly derive 
\begin{eqnarray*}
B_{n1}^{\ast \ast }(r,s) &=&a_{n}\int_{r/\left( a_{n}b_{n}\dot{\gamma}_{0}(%
\widetilde{s})\right) }^{0}\int_{\gamma _{0}(s)}^{\gamma _{0}(s+b_{n}t)}%
\mathcal{D}_{c_{0}}(q,s+b_{n}t)K\left( t\right) dqdt \\
&=&-\varrho \mathcal{D}_{c_{0}}(\gamma _{0}(s),s)\dot{\gamma}_{0}(s)\mathcal{%
K}_{1}\left( r,\varrho ;s\right) +o\left( 1\right) \text{,} \\
B_{n2}^{\ast \ast }(r,s) &=&a_{n}\int_{-t_{n}}^{r/\left( a_{n}b_{n}\dot{%
\gamma}_{0}(\widetilde{s})\right) }\int_{\gamma _{0}(s)}^{\gamma
_{0}(s)+r/a_{n}}\mathcal{D}_{c_{0}}(q,s+b_{n}t)K\left( t\right) dqdt \\
&=&r\mathcal{D}_{c_{0}}(\gamma _{0}(s),s)\left\{ \frac{1}{2}-\mathcal{K}%
_{0}\left( r,\varrho ;s\right) \right\} +o\left( 1\right) \text{.}
\end{eqnarray*}

When $\dot{\gamma}_{0}(s)=0$, it suffices to consider $\gamma _{0}(s)$ as
the local minimum, so that $\dot{\gamma}_{0}(t)\leq 0$ for $t\in \lbrack
s-\varepsilon ,s]$ and $\dot{\gamma}_{0}(t)\geq 0$ for $t\in \lbrack
s,s+\varepsilon ]$ for some small $\varepsilon $. In this case, based on the
same argument as (\ref{PsiM0}), 
\begin{eqnarray*}
\mathcal{T}_{1}(r;s) &=&\{t:\gamma _{0}\left( s+b_{n}t\right) \leq \gamma
_{0}\left( s\right) +r/a_{n}\}\cap \{t:\left\vert t\right\vert \leq t_{n}\}%
\text{,} \\
\mathcal{T}_{2}(r;s) &=&\{t:\gamma _{0}\left( s\right) +r/a_{n}<\gamma
_{0}\left( s+b_{n}t\right) \}\cap \{t:\left\vert t\right\vert \leq t_{n}\}%
\text{.}
\end{eqnarray*}%
Therefore, for sufficiently large $n$, 
\begin{eqnarray*}
B_{n1}^{\ast \ast }(r,s) &=&a_{n}\int_{0}^{t_{n}}\int_{\gamma
_{0}(s)}^{\gamma _{0}(s+b_{n}t)}\mathcal{D}_{c_{0}}(q,s+b_{n}t)K\left(
t\right) dqdt \\
&=&-\varrho \mathcal{D}_{c_{0}}(\gamma _{0}(s),s)\dot{\gamma}%
_{0}(s)\int_{0}^{\infty }tK\left( t\right) dt+o\left( 1\right) =o\left(
1\right) \text{,} \\
B_{n2}^{\ast \ast }(r,s) &=&a_{n}\int_{-t_{n}}^{r/\left( a_{n}b_{n}\dot{%
\gamma}_{0}(\widetilde{s})\right) }\int_{\gamma _{0}(s)}^{\gamma
_{0}(s)+r/a_{n}}\mathcal{D}_{c_{0}}(q,s+b_{n}t)K\left( t\right) dqdt \\
&=&r\mathcal{D}_{c_{0}}(\gamma _{0}(s),s)\left\{ \frac{1}{2}-\mathcal{K}%
_{0}\left( r,\varrho ;s\right) \right\} +o\left( 1\right) =o\left( 1\right)
\end{eqnarray*}%
since $\mathcal{K}_{0}\left( r,\varrho ;s\right) =1/2$ when $\dot{\gamma}%
_{0}(s)=0$.

By combining all three cases and the symmetric argument for $r<0$, we have 
\begin{equation*}
\mathbb{E}\left[ B_{n}^{\ast \ast }(r,s)\right] =\left\vert r\right\vert 
\mathcal{D}_{c_{0}}(\gamma _{0}(s),s)\left\{ \frac{1}{2}-\mathcal{K}%
_{0}\left( r,\varrho ;s\right) \right\} +\varrho \mathcal{D}_{c_{0}}(\gamma
_{0}(s),s)\left\vert \dot{\gamma}_{0}(s)\right\vert \mathcal{K}_{1}\left(
r,\varrho ;s\right) +o\left( 1\right) \text{.}
\end{equation*}%
Furthermore, since $\left\vert B_{n}^{\ast \ast }(r,s)\right\vert \leq
\sum_{i\in \Lambda _{n}}(\delta _{0}^{\top }x_{i})^{2}\left\vert \mathbf{1}%
_{i}\left( \gamma _{0}\left( s\right) +r/a_{n}\right) -\mathbf{1}_{i}\left(
\gamma _{0}\left( s\right) \right) \right\vert K_{i}\left( s\right) $, we
have $Var\left[ B_{n}^{\ast \ast }(r,s)\right] =O(n^{-2\epsilon })=o(1)$
from (\ref{VA*}), which establishes the pointwise convergence for each $r$.
The tightness follows from a similar argument as in Lemma \ref{max ineq} and
the desired result follows by Theorem 15.5 in \cite{Billingsley68}. $%
\blacksquare $

\paragraph{Proof of Lemma \protect\ref{lemma-drift}}

Define $W_{\mu }(r)=W(r)+\mu (r)$, $\tau ^{+}=\arg \max_{r\in 
%TCIMACRO{\U{211d} }%
%BeginExpansion
\mathbb{R}
%EndExpansion
^{+}}W_{\mu }(r)$, and $\tau ^{-}=\arg \max_{r\in 
%TCIMACRO{\U{211d} }%
%BeginExpansion
\mathbb{R}
%EndExpansion
^{-}}W_{\mu }(r)$.\ The process $W_{\mu }(\cdot )$ is a Gaussian process,
and hence Lemma 2.6 of \cite{Kim90} implies that $\tau ^{+}$ and $\tau ^{-}$
are unique almost surely. Recall that we define $%
W(r)=W_{1}(-r)1[r<0]+W_{2}(r)1[r>0]$, where $W_{1}(\cdot )$ and $W_{2}(\cdot
)$ are two independent standard Wiener processes defined on $%
%TCIMACRO{\U{211d} }%
%BeginExpansion
\mathbb{R}
%EndExpansion
^{+}$. We claim that 
\begin{equation}
\mathbb{E}[\tau ^{+}]=-\mathbb{E}[\tau ^{-}]<\infty \text{,}  \label{claim}
\end{equation}%
which gives the desired result.

The equality in (\ref{claim}) follows directly from the symmetry (i.e., $%
\mathbb{P}(\tau ^{+}\leq t)=\mathbb{P}(\tau ^{-}\geq -t)$ for any $t>0$) and
the fact that $W_{1}$ is independent of $W_{2}$. Now, we focus on $r>0$ and
show that $\mathbb{E}[\tau ^{+}]<\infty $. First, for any $r>0$,%
\begin{equation*}
\mathbb{P}\left( W_{\mu }(r)\geq 0\right) =\mathbb{P}\left( W_{2}(r)\geq
-\mu (r)\right) =\mathbb{P}\left( \frac{W_{2}(r)}{\sqrt{r}}\geq -\frac{\mu
(r)}{\sqrt{r}}\right) =1-\Phi \left( -\frac{\mu (r)}{\sqrt{r}}\right) \text{,%
}
\end{equation*}%
where $\Phi (\cdot )$ denotes the standard normal distribution function.
Since the sample path of $W_{\mu }(\cdot )$ is continuous, for some $%
\underline{r}>0$, we then have%
\begin{eqnarray}
\mathbb{E}[\tau ^{+}] &=&\int_{0}^{\infty }\left\{ 1-\mathbb{P}\left( \tau
^{+}\leq r\right) \right\} dr  \notag \\
&=&\int_{0}^{\underline{r}}\mathbb{P}\left( \tau ^{+}>r\right) dr+\int_{%
\underline{r}}^{\infty }\mathbb{P}\left( \tau ^{+}>r\right) dr  \notag \\
&\leq &C_{1}+\int_{\underline{r}}^{\infty }\mathbb{P}\left( W_{\mu }(\tau
^{+})\geq 0\text{ \textit{and} }\tau ^{+}>r\right) dr  \notag \\
&\leq &C_{1}+\int_{\underline{r}}^{\infty }\mathbb{P}\left( W_{\mu }(r)\geq
0\right) dr  \notag \\
&=&C_{1}+\int_{\underline{r}}^{\infty }\left( 1-\Phi \left( -\frac{\mu (r)}{%
\sqrt{r}}\right) \right) dr  \label{Etau}
\end{eqnarray}%
for some $C_{1}<\infty $, where the first inequality is because $W_{\mu
}(\tau ^{+})=\max_{r\in 
%TCIMACRO{\U{211d} }%
%BeginExpansion
\mathbb{R}
%EndExpansion
^{+}}W_{\mu }(r)\geq 0$ given $W_{\mu }(0)=0$, and the second inequality is
because $\mathbb{P}\left( W_{\mu }(r)\geq 0\right) $ is monotonically
decreasing to zero on $[\underline{r},\infty )$ by assumption. The second
term in (\ref{Etau}) can be bounded as follows. Using the change of
variables $t=r^{\varepsilon }$, integral by parts, and the condition that $%
r^{-(1/2+\varepsilon )}\mu (r)$ monotonically decreases to $-\infty $ on $[%
\underline{r},\infty )$ for some $\varepsilon >0$, we have%
\begin{eqnarray*}
\int_{\underline{r}}^{\infty }\left( 1-\Phi \left( -\frac{\mu (r)}{\sqrt{r}}%
\right) \right) dr &\leq &C_{2}\int_{\underline{r}}^{\infty }\left( 1-\Phi
\left( r^{\varepsilon }\right) \right) dr \\
&=&C_{3}\int_{\underline{r}^{1/\varepsilon }}^{\infty }\left( 1-\Phi \left(
t\right) \right) dt^{1/\varepsilon } \\
&=&C_{4}+C_{5}\int_{\underline{r}^{1/\varepsilon }}^{\infty
}t^{1/\varepsilon }\phi (t)dt<\infty
\end{eqnarray*}%
for some $C_{j}<\infty $ for $j=2,3,4,5$, where $\phi (\cdot )$ denotes the
standard normal density function and we use $\lim_{t\rightarrow \infty
}t^{1/\varepsilon }\left( 1-\Phi \left( t\right) \right) =0$. The same
result can be obtained for $r<0$ symmetrically, which completes the proof. $%
\blacksquare $

\paragraph{Proof of Lemma \protect\ref{lemma-mu}}

For given $(\varrho ,s)$, we simply denote $\mu (r)=\mu \left( r,\varrho
;s\right) $. Then, for the kernel functions satisfying Assumption A-(x), it
is readily verified that $\mu (0)=0$, $\mu (r)$ is continuous in $r$, and $%
\mu (r)$ is symmetric about zero. To check the monotonically decreasing
condition, for $r>0$, we write 
\begin{equation*}
\mu (r)=-r\int_{0}^{rC_{1}}K(t)dt+C_{2}\int_{0}^{rC_{1}}tK(t)dt\text{,}
\end{equation*}%
where $C_{1}$ and $C_{2}$ are some positive constants depending on $(\varrho
,\left\vert \dot{\gamma}_{0}(s)\right\vert ,\xi (s))$ from (\ref{Th3-mu}).
We consider the two possible cases.

First, if $K(\cdot )$ has a bounded support, say $[-\underline{r},\underline{%
r}]$ for some $0<\underline{r}<\infty $, then $\mu (r)=-rC_{3}+C_{4}$ for $r>%
\underline{r}$ and some $0<C_{3},C_{4}<\infty $. Thus, $\mu
(r)r^{-(1/2+\varepsilon )}$ is monotonically decreasing to $-\infty $ on $[%
\underline{r},\infty )$ for any $\varepsilon \in (0,1/2)$.

Second, if $K(\cdot )$ has an unbounded support, 
\begin{equation*}
\mu (r)r^{-((1/2)+\varepsilon )}=-r^{1/2-\varepsilon
}\int_{0}^{rC_{1}}K(t)dt+r^{-(1/2+\varepsilon
)}C_{2}\int_{0}^{rC_{1}}tK(t)dt,
\end{equation*}%
which goes to $-\infty $ as $r\rightarrow \infty $ since $%
\int_{0}^{rC_{1}}tK(t)dt\leq \int_{0}^{\infty }tK(t)dt<\infty $ and $%
\int_{0}^{rC_{1}}K(t)dt>0$. We can verify the monotonicity since 
\begin{eqnarray*}
\frac{\partial }{\partial r}\left\{ \mu (r)r^{-((1/2)+\varepsilon )}\right\}
&=&-\left( \frac{1}{2}-\varepsilon \right) r^{-\left( 1/2+\varepsilon
\right) }\int_{0}^{rC_{1}}K(t)dt-r^{1/2-\varepsilon }C_{1}K(C_{1}r) \\
&&-\left( \frac{1}{2}+\varepsilon \right) r^{-\left( 3/2+\varepsilon \right)
}C_{2}\int_{0}^{rC_{1}}tK(t)dt+r^{1/2-\varepsilon }C_{1}^{2}C_{2}K(C_{1}r) \\
&=&-r^{-\left( 1/2+\varepsilon \right) }\left\{ \left( \frac{1}{2}%
-\varepsilon \right) \int_{0}^{rC_{1}}K(t)dt+rK(C_{1}r)\left(
C_{1}-C_{1}^{2}C_{2}\right) \right\} \\
&&-\left( \frac{1}{2}+\varepsilon \right) r^{-3/2-\varepsilon
}C_{2}\int_{0}^{rC_{1}}tK(t)dt
\end{eqnarray*}%
by the Leibniz integral rule. For $r>\underline{r}$ for some large enough $%
\underline{r}$ and $\varepsilon \in \left( 0,1/2\right) $, this derivative
is strictly negative because $(1/2-\varepsilon )\int_{0}^{rC_{1}}K(t)dt>0$
and $\lim_{r\rightarrow \infty }rK(r)=0$, which proves $\mu
(r)r^{-((1/2)+\varepsilon )}$ is monotonically decreasing on $[\underline{r}%
,\infty )$. The case with $r<0$ follows symmetrically. $\blacksquare $

\paragraph{Proof of Lemma \protect\ref{T uniform}}

We only prove the first results for $T_{n}\left( \gamma ;s\right) $ because
the proof for $\overline{T}_{n}\left( \gamma ;s\right) $ is identical. We
define 
\begin{equation*}
\phi _{3n}=\left\Vert \gamma -\gamma _{0}\right\Vert _{\infty }\frac{\log n}{%
nb_{n}}\text{,}
\end{equation*}%
where $\left\Vert \gamma -\gamma _{0}\right\Vert _{\infty }=\sup_{s\in 
\mathcal{S}_{0}}\left\vert \gamma \left( s\right) -\gamma _{0}\left(
s\right) \right\vert $, which is bounded since $\gamma \left( s\right) \in
\Gamma $, a compact set, for any $s$. In addition, when $\left\Vert \gamma
-\gamma _{0}\right\Vert _{\infty }=0$, $T_{n}\left( \gamma ;s\right) =0$ and
hence the result trivially holds. So we suppose $\left\Vert \gamma -\gamma
_{0}\right\Vert _{\infty }>0$ without loss of generality. Similar to the
proof of Lemma \ref{L1}, we let $\tau _{n}=(n\log n)^{1/(4+\varphi )}$ with $%
\varphi \ $given in Assumption A-(v) and 
\begin{equation}
T_{n}^{\tau }(\gamma ,s)=\frac{1}{nb_{n}}\dsum_{i\in \Lambda _{n}}\left(
c_{0}^{\top }x_{i}\right) ^{2}\left\vert \Delta _{i}(\gamma ;s)\right\vert
K_{i}\left( s\right) \mathbf{1}_{\tau _{n}}\text{,}  \label{T-tau}
\end{equation}%
where $\Delta _{i}(\gamma ;s)=\mathbf{1}_{i}\left( \gamma \left( s\right)
\right) -\mathbf{1}_{i}\left( \gamma _{0}\left( s\right) \right) $ and $%
\mathbf{1}_{\tau _{n}}=\mathbf{1}[\left( c_{0}^{\top }x_{i}\right) ^{2}\leq
\tau _{n}]$. The triangular inequality gives that 
\begin{eqnarray}
\sup_{s\in \mathcal{S}_{0}}\left\vert T_{n}\left( \gamma ;s\right) -\mathbb{E%
}\left[ T_{n}\left( \gamma ;s\right) \right] \right\vert &\leq &\sup_{s\in 
\mathcal{S}_{0}}\left\vert T_{n}^{\tau }\left( \gamma ;s\right)
-T_{n}(\gamma ;s)\right\vert  \label{truncate} \\
&&+\sup_{s\in \mathcal{S}_{0}}\left\vert \mathbb{E}\left[ T_{n}^{\tau
}\left( \gamma ;s\right) \right] -\mathbb{E}\left[ T_{n}\left( \gamma
;s\right) \right] \right\vert  \notag \\
&&+\sup_{s\in \mathcal{S}_{0}}\left\vert T_{n}^{\tau }\left( \gamma
;s\right) -\mathbb{E}\left[ T_{n}^{\tau }\left( \gamma ;s\right) \right]
\right\vert  \notag \\
&\equiv &P_{T1}+P_{T2}+P_{T3}\text{,}  \notag
\end{eqnarray}%
and we bound each of the three terms as follows.

First, we show $P_{T1}=0$ almost surely if $n$ is sufficiently large. By
Markov's and H\"{o}lder's inequalities,%
\begin{equation*}
\mathbb{P}\left( \left( c_{0}^{\top }x_{i}\right) ^{2}\left\vert \Delta
_{i}(\gamma ;s)\right\vert >\tau _{n}\right) \leq C\tau _{n}^{-(4+\varphi )}%
\mathbb{E}\left[ \left\Vert x_{i}^{2}\right\Vert ^{4+\varphi }\right] \leq
C^{\prime }\left( n\log n\right) ^{-1}
\end{equation*}%
for some $C,C^{\prime }<\infty $ from Assumption A-(v) and the fact that $%
\left\vert \Delta _{i}(\gamma ;s)\right\vert \leq 1$. Then, as in the proof
of Lemma \ref{L1}, the Borel-Cantelli lemma implies that $\left( c_{0}^{\top
}x_{n}\right) ^{2}\left\vert \Delta _{n}(\gamma ;s)\right\vert \leq \tau
_{n} $ almost surely for sufficiently large $n$. Since $\tau _{n}\rightarrow
\infty $, we have $\left( c_{0}^{\top }x_{i}\right) ^{2}\left\vert \Delta
_{i}(\gamma ;s)\right\vert \leq \tau _{n}$ almost surely for all $i\in
\Lambda _{n}$ with sufficiently large $n$. The desired results hence follows.

Second, we show $P_{T2}\leq C^{\ast }\phi _{3n}^{1/2}$ almost surely for
some $C^{\ast }<\infty $ if $n$ is sufficiently large. For any $s\in 
\mathcal{S}_{0}$, 
\begin{eqnarray}
&&\left\vert \mathbb{E}\left[ T_{n}^{\tau }\left( \gamma ;s\right) \right] -%
\mathbb{E}\left[ T_{n}\left( \gamma ;s\right) \right] \right\vert
\label{PT2n} \\
&\leq &b_{n}^{-1}\mathbb{E}\left[ \left\vert \left( c_{0}^{\top
}x_{i}\right) ^{2}\mathbf{1}\left[ \min \{\gamma _{0}(s),\gamma
(s)\}<q_{i}\leq \max \{\gamma _{0}(s),\gamma (s)\}\right] K_{i}\left(
s\right) \left( 1-\mathbf{1}_{\tau _{n}}\right) \right\vert \right]  \notag
\\
&\leq &\int \int_{\min \{\gamma _{0}(s),\gamma (s)\}}^{\max \{\gamma
_{0}(s),\gamma (s)\}}\mathbb{E}\left[ \left( c_{0}^{\top }x_{i}\right)
^{2}\left( 1-\mathbf{1}_{\tau _{n}}\right) |q,s+b_{n}t\right]
f(q,s+b_{n}t)K(t)dqdt  \notag \\
&\leq &\tau _{n}^{-(3+\varphi )}\int \int_{\min \{\gamma _{0}(s),\gamma
(s)\}}^{\max \{\gamma _{0}(s),\gamma (s)\}}\mathbb{E}\left[ \left(
c_{0}^{\top }x_{i}\right) ^{2(4+\varphi )}|q,s+b_{n}t\right]
f(q,s+b_{n}t)K(t)dqdt  \notag \\
&\leq &C\tau _{n}^{-(3+\varphi )}\left\Vert \gamma -\gamma _{0}\right\Vert
_{\infty }  \notag
\end{eqnarray}%
for some $C<\infty $, where $\mathbb{E}[\left( c_{0}^{\top }x_{i}\right)
^{2(4+\varphi )}|q,s]f(q,s)\ $is uniformly bounded over $\left( q,s\right) $
by Assumptions A-(v) and (vii); and we use the inequality%
\begin{equation*}
\int_{\left\vert a\right\vert >\tau _{n}}af_{A}\left( a\right) da\leq \tau
_{n}^{-(3+\varphi )}\int_{\left\vert a\right\vert >\tau _{n}}\left\vert
a\right\vert ^{4+\varphi }f_{A}\left( a\right) da\leq \tau _{n}^{-(3+\varphi
)}\mathbb{E}\left[ A^{4+\varphi }\right]
\end{equation*}%
for a generic random variable $A$. Hence, the desired result follows since 
\begin{equation*}
\frac{\tau _{n}^{-(3+\varphi )}\left\Vert \gamma -\gamma _{0}\right\Vert
_{\infty }}{\phi _{3n}^{1/2}}=\left\Vert \gamma -\gamma _{0}\right\Vert
_{\infty }^{1/2}\frac{b_{n}^{1/2}}{n^{\frac{3+\varphi }{4+\varphi }-\frac{1}{%
2}}\left( \log n\right) ^{\frac{3+\varphi }{4+\varphi }+\frac{1}{2}}}%
=o\left( 1\right) \text{,}
\end{equation*}%
where $\left\Vert \gamma -\gamma _{0}\right\Vert _{\infty }$ is bounded.

Finally, we show $P_{T3}\leq C^{\ast }\phi _{3n}^{1/2}$ almost surely for
some $C^{\ast }<\infty $ if $n$ is sufficiently large, which follows
similarly as the proof of Lemma \ref{L-int1}. To this end, we partition the
compact $\mathcal{S}_{0}$\ into $m_{n}$-number of intervals $\mathcal{I}%
_{k}=[s_{k},s_{k+1})$\ for $k=1,\ldots ,m_{n}$. We choose the integer $%
m_{n}>n$\ such that $\left\vert s_{k+1}-s_{k}\right\vert \leq
C/m_{n}=C^{\prime }b_{n}^{2}\tau _{n}^{-1}\phi _{3n}^{1/2}$\ for all $k$ and
for some $C,C^{\prime }<\infty $. In addition, since we let $\gamma (\cdot )$%
\ be a cadlag and piecewise constant function with at most $n$ discontinuity
points, which is less than $m_{n}$\textbf{, }Theorem 28.2 in Davidson (1994)
entails that we can choose these finite partitions such that%
\begin{equation}
\sup_{s\in \mathcal{I}_{k}}\left\vert \gamma (s)-\gamma (s_{k})\right\vert =0
\label{cadlag}
\end{equation}%
for each $k$. Then we have 
\begin{eqnarray}
\sup_{s\in \mathcal{S}_{0}}\left\vert T_{n}^{\tau }\left( \gamma ;s\right) -%
\mathbb{E}\left[ T_{n}^{\tau }\left( \gamma ;s\right) \right] \right\vert
&\leq &\max_{1\leq k\leq m_{n}}\sup_{s\in \mathcal{I}_{k}}\left\vert
T_{n}^{\tau }\left( \gamma ;s\right) -T_{n}^{\tau }\left( \gamma
;s_{k}\right) \right\vert  \label{part3} \\
&&+\max_{1\leq k\leq m_{n}}\sup_{s\in \mathcal{I}_{k}}\left\vert \mathbb{E}%
\left[ T_{n}^{\tau }\left( \gamma ;s\right) \right] -\mathbb{E}\left[
T_{n}^{\tau }\left( \gamma ;s_{k}\right) \right] \right\vert  \notag \\
&&+\max_{1\leq k\leq m_{n}}\left\vert T_{n}^{\tau }\left( \gamma
;s_{k}\right) -\mathbb{E}\left[ T_{n}^{\tau }\left( \gamma ;s_{k}\right) %
\right] \right\vert  \notag \\
&\equiv &\Psi _{T1}+\Psi _{T2}+\Psi _{T3}\text{.}  \notag
\end{eqnarray}%
Below we show $\Psi _{T1}$, $\Psi _{T2}$, and $\Psi _{T3}$ are all $%
O_{a.s.}(\phi _{3n}^{1/2})$.

\textit{Part 1: }$\Psi _{T1}$\textit{\ and }$\Psi _{T2}$\textsl{\ }\textit{%
are both }$O_{a.s.}(\phi _{3n}^{1/2})$. Similarly as $\Psi _{M1}$ term in
Lemma \ref{L1}, we first decompose $T_{n}^{\tau }\left( \gamma ;s\right)
-T_{n}^{\tau }\left( \gamma ;s_{k}\right) \leq T_{1n}^{\tau }\left( \gamma
;s,s_{k}\right) +T_{2n}^{\tau }\left( \gamma ;s,s_{k}\right) $, where 
\begin{eqnarray*}
T_{1n}^{\tau }\left( \gamma ;s,s_{k}\right) &=&\frac{1}{nb_{n}}\dsum_{i\in
\Lambda _{n}}\left( c_{0}^{\top }x_{i}\right) ^{2}\left\vert \Delta
_{i}(\gamma ;s)-\Delta _{i}(\gamma ;s_{k})\right\vert K_{i}\left(
s_{k}\right) \mathbf{1}_{\tau _{n}}\text{,} \\
T_{2n}^{\tau }\left( \gamma ;s,s_{k}\right) &=&\frac{1}{nb_{n}}\dsum_{i\in
\Lambda _{n}}\left( c_{0}^{\top }x_{i}\right) ^{2}\left\vert \Delta
_{i}(\gamma ;s)\right\vert \left\vert K_{i}\left( s\right) -K_{i}\left(
s_{k}\right) \right\vert \mathbf{1}_{\tau _{n}}\text{.}
\end{eqnarray*}%
Without loss of generality, we can suppose $\gamma (s)>\gamma (s_{k})$ and $%
\gamma _{0}(s)>\gamma _{0}(s_{k})$ in $T_{1n}^{\tau }\left( \gamma
;s,s_{k}\right) $. Since $K_{i}(\cdot )$ is bounded from Assumption A-(x)
and we only consider $x_{i}^{2}\leq \tau _{n}$, 
\begin{eqnarray*}
\left\vert T_{1n}^{\tau }\left( \gamma ;s,s_{k}\right) \right\vert &\leq &%
\frac{1}{nb_{n}}\dsum_{i\in \Lambda _{n}}\left( c_{0}^{\top }x_{i}\right)
^{2}\mathbf{1}\left[ \gamma _{0}(s)<q_{i}\leq \gamma _{0}(s_{k})\right]
K_{i}\left( s_{k}\right) \mathbf{1}_{\tau _{n}} \\
&&+\frac{1}{nb_{n}}\dsum_{i\in \Lambda _{n}}\left( c_{0}^{\top }x_{i}\right)
^{2}\mathbf{1}\left[ \gamma (s)<q_{i}\leq \gamma (s_{k})\right] K_{i}\left(
s_{k}\right) \mathbf{1}_{\tau _{n}} \\
&\leq &\frac{C_{1}\tau _{n}}{b_{n}}\mathbb{P}\left( \gamma _{0}(s)<q_{i}\leq
\gamma _{0}(s_{k})\right) +\frac{C_{1}\tau _{n}}{b_{n}}\mathbb{P}\left(
\gamma (s)<q_{i}\leq \gamma (s_{k})\right) \text{ \ almost surely} \\
&\leq &C_{1}^{\prime }\tau _{n}b_{n}^{-1}\sup_{s\in \mathcal{I}%
_{k}}\left\vert s-s_{k}\right\vert +0 \\
&\leq &C_{1}^{\prime }\tau _{n}b_{n}^{-1}m_{n}^{-1} \\
&=&C_{1}^{\prime \prime }b_{n}\phi _{3n}^{1/2}
\end{eqnarray*}%
for $C_{1},C_{1}^{\prime },C_{1}^{\prime \prime }<\infty $, where the second
second equality is by the uniform almost sure law of large numbers for
random fields (e.g., Jenish and Prucha (2009), Theorem 2); the third
inequality is since $\gamma _{0}(\cdot )$ is continuously differentiable, $%
q_{i}$ is continuous, and (\ref{cadlag}). Hence, $T_{1n}^{\tau }\left(
\gamma ;s,s_{k}\right) =O_{a.s.}(\phi _{3n}^{1/2})$, which holds uniformly
in $s\in \mathcal{I}_{k}$ and $k\in \{1,\ldots ,m_{n}\}$. Similarly, since $%
K(\cdot )$ is Lipschitz from Assumption A-(x) and $\left\vert \Delta
_{i}(\gamma ;s)\right\vert \leq 1$, 
\begin{eqnarray*}
\left\vert T_{2n}^{\tau }\left( \gamma ;s,s_{k}\right) \right\vert &\leq &%
\frac{\tau _{n}}{nb_{n}}\dsum_{i\in \Lambda _{n}}\left\vert
K_{i}(s)-K_{i}(s_{k_{2}})\right\vert \\
&\leq &C_{2}\frac{\tau _{n}}{b_{n}^{2}}\left\vert s-s_{k_{2}}\right\vert
\leq \frac{C_{2}^{\prime }\tau _{n}}{b_{n}^{2}m_{n}}=O_{a.s.}\left( \phi
_{3n}^{1/2}\right)
\end{eqnarray*}%
for some $C_{2},C_{2}^{\prime }<\infty $, uniformly in $s$ and $k$. Hence, $%
\Psi _{T1}=O_{a.s.}(\phi _{3n}^{1/2})$ and we can readily verify that $\Psi
_{T2}=O_{a.s.}(\phi _{3n}^{1/2})$ similarly.

\textit{Part 2: }$\Psi _{T3}=O_{a.s.}(\phi _{3n}^{1/2})$.\textsl{\ }We let%
\begin{equation*}
Z_{i}^{\tau }(s)=(nb_{n})^{-1}\left\{ (c_{0}^{\top }x_{i})^{2}\Delta
_{i}(\gamma ;s)K_{i}\left( s\right) \mathbf{1}_{\tau _{n}}-\mathbb{E}%
[(c_{0}^{\top }x_{i})^{2}\Delta _{i}(\gamma ;s)K_{i}\left( s\right) \mathbf{1%
}_{\tau _{n}}]\right\}
\end{equation*}%
and apply the similar proof as $\Psi _{\Delta M3}$ in Lemma \ref{L-int1}
above. In particular, construct the block $\mathcal{B}^{[1]}(s_{k})$ in the
same fashion as (\ref{PBlock}). Then, it suffices to show $\max_{1\leq k\leq
m_{n}}\left\vert \mathcal{B}^{[1]}(s_{k})\right\vert =O_{a.s.}(\phi
_{3n}^{1/2})$ as $n\rightarrow \infty $. Using the same notations as in
Lemma \ref{L-int1}, by the uniform almost sure law of large numbers for
random fields, we have that for any $t=1,\ldots ,r$ and $s\in \mathcal{S}%
_{0} $,%
\begin{equation}
\left\vert U_{t}(s)\right\vert \leq \frac{C_{3}w^{2}\tau _{n}}{nb_{n}}\left( 
\frac{1}{w^{2}}\dsum_{i_{1}=2j_{1}w+1}^{(2j_{1}+1)w}%
\dsum_{i_{2}=2j_{2}w+1}^{(2j_{2}+1)w}\left\vert \Delta _{i}(\gamma
;s)\right\vert \right) \leq \frac{C_{3}w^{2}\tau _{n}\left\Vert \gamma
-\gamma _{0}\right\Vert _{\infty }}{nb_{n}}  \label{Utbd2}
\end{equation}%
almost surely from (\ref{u[1]}), for some $C_{3}<\infty $. We also
approximate $\{U_{t}(s)\}_{t=1}^{r}$ by a version of independent random
variables $\{U_{t}^{\ast }(s)\}_{t=1}^{r}$ that satisfies 
\begin{equation}
\dsum_{t=1}^{r}\mathbb{E}\left[ \left\vert U_{t}^{\ast
}(s)-U_{t}(s)\right\vert \right] \leq rC_{3}\left( nb_{n}\right)
^{-1}w^{2}\tau _{n}\left\Vert \gamma -\gamma _{0}\right\Vert _{\infty
}\alpha _{w^{2},w^{2}}(w)\text{.}  \label{indep approx2}
\end{equation}%
Then, similar to (\ref{Pu1Pu2}), for some positive $C^{\ast }<\infty $,%
\begin{eqnarray*}
\mathbb{P}\left( \max_{1\leq k\leq m_{n}}\left\vert \mathcal{B}%
^{[1]}(s_{k})\right\vert >C^{\ast }\phi _{3n}^{1/2}\right) &\leq
&m_{n}\sup_{s\in \mathcal{S}_{0}}\mathbb{P}\left( \dsum_{t=1}^{r}\left\vert
U_{t}^{\ast }(s)-U_{t}(s)\right\vert >C^{\ast }\phi _{3n}^{1/2}\right) \\
&&+m_{n}\sup_{s\in \mathcal{S}_{0}}\mathbb{P}\left( \left\vert
\dsum_{t=1}^{r}U_{t}^{\ast }(s)\right\vert >C^{\ast }\phi _{3n}^{1/2}\right)
\\
&\equiv &\widetilde{P}_{U1}+\widetilde{P}_{U2}\text{.}
\end{eqnarray*}

For $\widetilde{P}_{U1}$, 
\begin{eqnarray*}
\widetilde{P}_{U1} &\leq &m_{n}\frac{rC_{3}\left( nb_{n}\right)
^{-1}w^{2}\tau _{n}\left\Vert \gamma -\gamma _{0}\right\Vert _{\infty
}\alpha _{w^{2},w^{2}}(w)}{C^{\ast }\phi _{3n}^{1/2}} \\
&\leq &C_{3}^{\prime }\frac{\tau _{n}^{2}}{b_{n}^{3}\phi _{3n}}\left\Vert
\gamma -\gamma _{0}\right\Vert _{\infty }\exp (-C_{3}^{\prime \prime }w) \\
&\leq &\frac{C_{3}^{\prime \prime \prime }n^{\kappa _{1}}(\log n)^{\kappa
_{2}}\exp (-C_{3}^{\prime \prime \prime \prime }n^{\kappa _{3}})}{%
(n^{1-2\epsilon }b_{n})^{2}}
\end{eqnarray*}%
for some $\kappa _{1},\kappa _{2},\kappa _{3}>0$ and $C_{3}^{\prime
},C_{3}^{\prime \prime },C_{3}^{\prime \prime \prime },C_{3}^{\prime \prime
\prime \prime }<\infty $. Hence $\widetilde{P}_{U1}=O(\exp (-n^{\kappa
_{3}}))\rightarrow 0$ as $n\rightarrow \infty $. Recall that we chose $%
m_{n}=O(\tau _{n}/(b_{n}^{2}\phi _{3n}^{1/2}))$, $n=4w^{2}r$, and $\tau
_{n}=\left( n\log n\right) ^{1/\left( 4+\varphi \right) }$.

For $\widetilde{P}_{U2}$, using the same argument as (\ref{P12b}) in Lemma %
\ref{max ineq}, we can show that 
\begin{equation*}
\mathbb{E}\left[ U_{t}^{\ast }(s)^{2}\right] =\dsum_{\substack{ 1\leq
i_{1}\leq w  \\ 1\leq i_{2}\leq w}}\mathbb{E}\left[ Z_{i}^{\tau }(s)^{2}%
\right] +\dsum_{\substack{ i\neq j  \\ 1\leq i_{1},i_{2}\leq w  \\ 1\leq
j_{1},j_{2}\leq w}}Cov\left[ Z_{i}^{\tau }(s),Z_{j}^{\tau }(s)\right] \leq 
\frac{C_{4}w^{2}}{n^{2}b_{n}}\left\Vert \gamma -\gamma _{0}\right\Vert
_{\infty }
\end{equation*}%
for some $C_{4}<\infty $, which does not depend on $s$ given Assumptions
A-(v) and (x). We now choose an integer $w$ such that%
\begin{eqnarray*}
w &=&\left( nb_{n}/\left( C_{w}\tau _{n}\lambda _{n}\right) \right) ^{1/2}%
\text{,} \\
\lambda _{n} &=&(nb_{n}\log n)^{1/2}
\end{eqnarray*}%
for some large positive constant $C_{w}$. Note that, substituting $\lambda
_{n}$ and $\tau _{n}$ into $w$ gives%
\begin{equation*}
w=O\left( \left[ \frac{n^{\frac{1}{4}-\frac{1}{4+\varphi }}}{\left( \log
n\right) ^{\frac{3}{4}+\frac{1}{4+\varphi }}}\times \left( \frac{nb_{n}^{2}}{%
\log n}\right) ^{1/4}\right] ^{1/2}\right) \text{,}
\end{equation*}%
which diverges as $n\rightarrow \infty $ for $\varphi >0$ and from
Assumption A-(ix). From (\ref{Utbd2}), we have $|\lambda _{n}U_{t}^{\ast
}(s)/\left\Vert \gamma -\gamma _{0}\right\Vert _{\infty }^{1/2}|<1/2$ by
choosing $C_{w}$ large enough, and hence 
\begin{eqnarray*}
\sup_{s\in \mathcal{S}_{0}}\mathbb{P}\left( \left\vert
\dsum_{t=1}^{r}U_{t}^{\ast }(s)\right\vert >C^{\ast }\phi _{3n}^{1/2}\right)
&=&\sup_{s\in \mathcal{S}_{0}}\mathbb{P}\left( \left\vert \dsum_{t=1}^{r}%
\frac{\lambda _{n}U_{t}^{\ast }(s)}{\left\Vert \gamma -\gamma
_{0}\right\Vert _{\infty }^{1/2}}\right\vert >C^{\ast }\left( \frac{\log n}{%
nb_{n}}\right) ^{1/2}\right) \\
&\leq &2\exp \left( -C^{\ast }\lambda _{n}\left( \frac{\log n}{nb_{n}}%
\right) ^{1/2}+\frac{C_{4}\lambda _{n}^{2}rw^{2}}{n^{2}b_{n}}\right) \\
&=&2\exp \left( -C^{\ast }\lambda _{n}\left( \frac{\log n}{nb_{n}}\right)
^{1/2}+C_{4}\lambda _{n}^{2}(nb_{n})^{-1}\right) \\
&=&2\exp \left( -C^{\ast }\log n+C_{4}^{\prime }\log n\right)
\end{eqnarray*}%
for some $C_{4},C_{4}^{\prime }<\infty $ as in (\ref{PUbd}) and (\ref{T3n
bnd}). It follows that 
\begin{equation*}
m_{n}\sup_{s\in \mathcal{S}_{0}}\mathbb{P}\left( \left\vert
\dsum_{t=1}^{r}U_{t}^{\ast }(s)\right\vert >C^{\ast }\phi _{3n}^{1/2}\right)
\leq \frac{2m_{n}}{n^{C^{\ast }-C_{4}^{\prime }}}=\frac{C_{5}(\log
n)^{\kappa _{4}}}{n^{\kappa _{5}}(n^{1-2\epsilon }b_{n})^{3/2}}
\end{equation*}%
for some $C_{5}<\infty $, $\kappa _{4}>0$, and $\kappa _{5}>1$ by choosing $%
C^{\ast }$ sufficiently large. Therefore, $\widetilde{P}_{U2}=O(n^{-\kappa
_{5}})\rightarrow 0$ as $n\rightarrow \infty $. Since $\widetilde{P}_{U1}+%
\widetilde{P}_{U2}=O(n^{-c})$ for some $c>1$, we have $\sum_{n=1}^{\infty }%
\mathbb{P}(\max_{1\leq k\leq m_{n}}\left\vert \mathcal{B}^{[1]}(s_{k})\right%
\vert >C^{\ast }\phi _{3n}^{1/2})<\infty $ and hence we obtain the desired
result by the Borel-Cantelli lemma. $\blacksquare $

\paragraph{Proof of Lemma \protect\ref{L uniform}}

Since the proof is similar to that in Lemma \ref{T uniform}, we only
highlight the different parts. Without loss of generality we assume $x_{i}$
is a scalar, so as $L_{n}(\gamma ;s)$. As in (\ref{T-tau}), we let 
\begin{equation*}
L_{n}^{\tau }\left( \gamma ;s\right) =\frac{1}{\sqrt{nb_{n}}}\dsum_{i\in
\Lambda _{n}}x_{i}u_{i}\Delta _{i}(\gamma ;s)K_{i}\left( s\right) \mathbf{1}%
_{\tau _{n}}\text{,}
\end{equation*}%
where $\Delta _{i}(\gamma ;s)=\mathbf{1}_{i}\left( \gamma \left( s\right)
\right) -\mathbf{1}_{i}\left( \gamma _{0}\left( s\right) \right) $ and $%
\mathbf{1}_{\tau _{n}}=\mathbf{1}[|x_{i}u_{i}|\leq \tau _{n}]$ with $\tau
_{n}=\left( n\log n\right) ^{1/(4+\varphi )}$. Since $\mathbb{E}[L_{n}^{\tau
}(\gamma ,s)]=0$, we write 
\begin{eqnarray*}
&&\sup_{s\in \mathcal{S}_{0}}\left\vert L_{n}\left( \gamma ;s\right)
\right\vert \\
&\leq &\sup_{s\in \mathcal{S}_{0}}\left\vert L_{n}^{\tau }\left( \gamma
;s\right) -L_{n}(\gamma ;s)\right\vert +\sup_{s\in \mathcal{S}%
_{0}}\left\vert L_{n}^{\tau }\left( \gamma ;s\right) \right\vert \\
&\equiv &P_{L1}+P_{L2}\text{.}
\end{eqnarray*}%
Using the same argument as $P_{T1}$ in the proof of Lemma \ref{T uniform},
we have 
\begin{equation*}
\mathbb{P}\left( \left\vert x_{i}u_{i}\right\vert \left\vert \Delta
_{i}(\gamma ;s)\right\vert >\tau _{n}\right) \leq C\tau _{n}^{-(4+\varphi )}%
\mathbb{E}\left[ \left\Vert x_{i}u_{i}\right\Vert ^{4+\varphi }\right] \leq
C^{\prime }\left( n\log n\right) ^{-1}
\end{equation*}%
for some $C,C^{\prime }<\infty $. Then the Borel-Cantelli lemma implies that 
$\left\vert x_{i}u_{i}\right\vert \left\vert \Delta _{n}(\gamma
;s)\right\vert \leq \tau _{n}$ almost surely for sufficiently large $n$.
Since $\tau _{n}\rightarrow \infty $, we have $\left\vert
x_{i}u_{i}\right\vert \left\vert \Delta _{i}(\gamma ;s)\right\vert \leq \tau
_{n}$ almost surely for all $i\in \Lambda _{n}$ with sufficiently large $n$,
which yields $P_{L1}=0$ almost surely for a sufficiently large $n$.

For $P_{L2}$, we let $\widetilde{\phi }_{3n}=\left\Vert \gamma -\gamma
_{0}\right\Vert _{\infty }\log n$ and write 
\begin{eqnarray*}
\sup_{s\in \mathcal{S}_{0}}\left\vert L_{n}^{\tau }\left( \gamma ;s\right)
\right\vert &\leq &\max_{1\leq k\leq m_{n}}\sup_{s\in \mathcal{I}%
_{k}}\left\vert L_{n}^{\tau }\left( \gamma ;s\right) -L_{n}^{\tau }\left(
\gamma ;s_{k}\right) \right\vert +\max_{1\leq k\leq m_{n}}\left\vert
L_{n}^{\tau }\left( \gamma ;s_{k}\right) \right\vert \\
&\equiv &\Psi _{L1}+\Psi _{L2}\text{,}
\end{eqnarray*}%
for some integer $m_{n}=O(\tau _{n}/(b_{n}^{2}\widetilde{\phi }_{3n}^{1/2}))$%
. We let $Z_{i}^{\tau }(s)=(nb_{n})^{-1/2}x_{i}u_{i}\Delta _{i}(\gamma
;s)K_{i}\left( s\right) \mathbf{1}_{\tau _{n}}$, and we choose $w=(\left(
nb_{n}\right) /\left( C_{w}\tau _{n}\lambda _{n}\right) )^{1/2}$ for some
large positive constant $C_{w}$ and $\lambda _{n}=(\log n)^{1/2}$. Then, the
rest of the proof follows identically as $P_{T3}$ in the proof of Lemma \ref%
{T uniform}. $\blacksquare $

\paragraph{Proof of Lemma \protect\ref{L-A2}}

We first show (\ref{infTsupL uniform 1}). Similarly as the proof of Lemma %
\ref{infTsupL pointwise}, we consider the case with $\gamma (s)>\gamma
_{0}(s)$, and the other direction can be shown symmetrically. We suppose $n$
is large enough so that $\overline{r}\phi _{2n}\leq \overline{C}$ for some $%
\overline{r},\overline{C}\in (0,\infty )$ and $\sup_{s\in \mathcal{S}%
_{0}}\left( \gamma \left( s\right) -\gamma _{0}\left( s\right) \right) \in %
\left[ \overline{r}\phi _{2n},\overline{C}\right] $. We also let 
\begin{equation*}
\underline{\ell }=\inf_{s\in \mathcal{S}_{0}}\underline{\ell }_{D}(s)>0
\end{equation*}%
where $\underline{\ell }_{D}(s)$ is defined in (\ref{L_D}). Then, from (\ref%
{ET bnd}), we have 
\begin{equation}
\sup_{s\in \mathcal{S}_{0}}\mathbb{E}\left[ T_{n}\left( \gamma ;s\right) %
\right] \geq \underline{\ell }\sup_{s\in \mathcal{S}_{0}}\left( \gamma
\left( s\right) -\gamma _{0}\left( s\right) \right) \text{.}  \label{ineq1}
\end{equation}%
For any $\varepsilon >0$ and for any $\gamma \left( \cdot \right) $ such
that $\sup_{s\in \mathcal{S}_{0}}\left( \gamma \left( s\right) -\gamma
_{0}\left( s\right) \right) \in \left[ \overline{r}\phi _{2n},\overline{C}%
\right] $, Lemma \ref{T uniform} and (\ref{ineq1}) implies that with
probability approaching to one%
\begin{eqnarray*}
&&\frac{\sup_{s\in \mathcal{S}_{0}}T_{n}\left( \gamma ;s\right) }{\sup_{s\in 
\mathcal{S}_{0}}\left\vert \gamma \left( s\right) -\gamma _{0}\left(
s\right) \right\vert } \\
&\geq &\frac{\sup_{s\in \mathcal{S}_{0}}\mathbb{E}\left[ T_{n}\left( \gamma
;s\right) \right] -\sup_{s\in \mathcal{S}_{0}}\left\vert T_{n}\left( \gamma
;s\right) -\mathbb{E}\left[ T_{n}\left( \gamma ;s\right) \right] \right\vert 
}{\sup_{s\in \mathcal{S}_{0}}\left\vert \gamma \left( s\right) -\gamma
_{0}\left( s\right) \right\vert } \\
&\geq &\frac{\sup_{s\in \mathcal{S}_{0}}\mathbb{E}\left[ T_{n}\left( \gamma
;s\right) \right] }{\sup_{s\in \mathcal{S}_{0}}\left\vert \gamma \left(
s\right) -\gamma _{0}\left( s\right) \right\vert }-\frac{\left( \sup_{s\in 
\mathcal{S}_{0}}\left\vert \gamma \left( s\right) -\gamma _{0}\left(
s\right) \right\vert (\log n/n)\right) ^{1/2}}{\sup_{s\in \mathcal{S}%
_{0}}\left\vert \gamma \left( s\right) -\gamma _{0}\left( s\right)
\right\vert } \\
&\geq &\underline{\ell }-\frac{(\log n/n)^{1/2}}{\overline{r}\phi _{2n}^{1/2}%
}\geq \underline{\ell }-\overline{r}^{-1}n^{-\epsilon }\text{.}
\end{eqnarray*}%
Since $\underline{\ell }>0$ and $\overline{r}^{-1}n^{-\epsilon }\rightarrow
0 $ as $n\rightarrow \infty $, we thus can find $C_{T}<\infty $ such that 
\begin{equation*}
\mathbb{P}\left( \inf_{\overline{r}\phi _{2n}<\sup_{s\in \mathcal{S}%
_{0}}\left\vert \gamma \left( s\right) -\gamma _{0}\left( s\right)
\right\vert <\overline{C}}\frac{\sup_{s\in \mathcal{S}_{0}}T_{n}\left(
\gamma ;s\right) }{\sup_{s\in \mathcal{S}_{0}}\left\vert \gamma \left(
s\right) -\gamma _{0}\left( s\right) \right\vert }<C_{T}(1-\eta )\right)
\leq \varepsilon \text{.}
\end{equation*}%
for any $\varepsilon ,\eta >0$. The proof for (\ref{infTsupL uniform 2}) is
similar and hence omitted.

For (\ref{infTsupL uniform 3}), without loss of generality we assume $x_{i}$
is a scalar, and so is $L_{n}(\gamma ;s)$. We set $\gamma _{g}$ for $%
g=1,2,...,\overline{g}+1$ such that, for any $s\in \mathcal{S}_{0}$, $\gamma
_{g}\left( s\right) =\gamma _{0}\left( s\right) +2^{g-1}\overline{r}\phi
_{2n}$, where $\overline{g}$ is an integer satisfying $\sup_{s\in \mathcal{S}%
_{0}}(\gamma _{\overline{g}}\left( s\right) -\gamma _{0}\left( s\right) )=2^{%
\overline{g}-1}\overline{r}\phi _{2n}\leq \overline{C}$ and $\sup_{s\in 
\mathcal{S}_{0}}(\gamma _{\overline{g}+1}\left( s\right) -\gamma _{0}\left(
s\right) )>\overline{C}$. Then Lemma \ref{L uniform} yield that for any $%
\eta >0$,%
\begin{eqnarray}
&&\mathbb{P}\left( \max_{1\leq g\leq \overline{g}}\frac{\sup_{s\in \mathcal{S%
}_{0}}\left\vert L_{n}\left( \gamma _{g};s\right) \right\vert }{\sqrt{a_{n}}%
\sup_{s\in \mathcal{S}_{0}}\left( \gamma _{g}\left( s\right) -\gamma
_{0}\left( s\right) \right) }>\frac{\eta }{4}\right)  \label{Lbound1} \\
&\leq &\dsum_{g=1}^{\overline{g}}\mathbb{P}\left( \frac{\sup_{s\in \mathcal{S%
}_{0}}\left\vert L_{n}\left( \gamma _{g};s\right) \right\vert }{\sqrt{a_{n}}%
\sup_{s\in \mathcal{S}_{0}}\left( \gamma _{g}\left( s\right) -\gamma
_{0}\left( s\right) \right) }>\frac{\eta }{4}\right)  \notag \\
&\leq &\frac{4}{\eta }\dsum_{g=1}^{\overline{g}}\frac{C_{L}\left( \phi
_{2n}\log n\right) ^{1/2}}{\sqrt{a_{n}}2^{g-1}\overline{r}\phi _{2n}}  \notag
\\
&\leq &\frac{C_{L}^{\prime }}{\eta \overline{r}}\dsum_{g=1}^{\infty }\frac{1%
}{2^{(g-1)}}  \notag
\end{eqnarray}%
for some $C_{L},C_{L}^{\prime }<\infty $. This probability is arbitrarily
close to zero if $\overline{r}$ is chosen large enough. Following a similar
discussion after (\ref{L 1}), this result also provides the maximal (or
sharp) rate of $\phi _{2n}$ as $\log n/a_{n}$ because we need $(\log
n/a_{n})/\phi _{2n}=O(1)$ but $\phi _{2n}\rightarrow 0$ as $\log
n/a_{n}\rightarrow 0$ with $n\rightarrow \infty $. For a given $g$, we
define $\Gamma _{g}$ as the collection of $\gamma \left( s\right) $
satisfying $\overline{r}2^{g-1}\phi _{2n}<\sup_{s\in \mathcal{S}%
_{0}}\left\vert \gamma \left( s\right) -\gamma _{0}\left( s\right)
\right\vert <\overline{r}2^{g}\phi _{2n}$. By a similar argument as (\ref%
{Lbound1}), we have 
\begin{equation}
\mathbb{P}\left( \max_{1\leq g\leq \overline{g}}\sup_{\gamma \in \Gamma _{g}}%
\frac{\sup_{s\in \mathcal{S}_{0}}\left\vert L_{n}\left( \gamma ;s\right)
-L_{n}\left( \gamma _{g};s\right) \right\vert }{\sqrt{a_{n}}\sup_{s\in 
\mathcal{S}_{0}}\left( \gamma _{g}(s)-\gamma _{0}\left( s\right) \right) }>%
\frac{\eta }{4}\right) \leq \frac{C_{L}^{\prime \prime }}{\eta \bar{r}}
\label{Lbound2}
\end{equation}%
for some $C_{L}^{\prime \prime }<\infty $, which is arbitrarily close to
zero if $\overline{r}$ is chosen large enough. From (\ref{Ln_bd}), and by
combining (\ref{Lbound1}) and (\ref{Lbound2}), we thus have%
\begin{eqnarray*}
&&\mathbb{P}\left( \sup_{\overline{r}\phi _{2n}<\sup_{s\in \mathcal{S}%
_{0}}\left\vert \gamma \left( s\right) -\gamma _{0}\left( s\right)
\right\vert <\overline{C}}\frac{\sup_{s\in \mathcal{S}_{0}}\left\vert
L_{n}\left( \gamma ;s\right) \right\vert }{\sqrt{a_{n}}\sup_{s\in \mathcal{S}%
_{0}}\left( \gamma \left( s\right) -\gamma _{0}\left( s\right) \right) }%
>\eta \right) \\
&\leq &\mathbb{P}\left( 2\max_{1\leq g\leq \overline{g}}\frac{\sup_{s\in 
\mathcal{S}_{0}}\left\vert L_{n}\left( \gamma _{g};s\right) \right\vert }{%
\sqrt{a_{n}}\sup_{s\in \mathcal{S}_{0}}\left( \gamma _{g}\left( s\right)
-\gamma _{0}\left( s\right) \right) }>\frac{\eta }{2}\right) \\
&&+\mathbb{P}\left( 2\max_{1\leq g\leq \overline{g}}\sup_{\gamma \in \Gamma
_{g}}\frac{\sup_{s\in \mathcal{S}_{0}}\left\vert L_{n}\left( \gamma
;s\right) -L_{n}\left( \gamma _{g};s\right) \right\vert }{\sqrt{a_{n}}%
\sup_{s\in \mathcal{S}_{0}}\left( \gamma _{g}(s)-\gamma _{0}\left( s\right)
\right) }>\frac{\eta }{2}\right) \\
&\leq &\varepsilon
\end{eqnarray*}%
for any $\varepsilon ,\eta >0$ if $\overline{r}$ is chosen sufficiently
large. $\blacksquare $

\paragraph{Proof of Lemma \protect\ref{neg uniform}}

Given Lemma \ref{L1} and the uniform convergence of the standard kernel
estimators, the desired result can be obtained similarly as the proof of
Lemma \ref{neg pointwise}, provided that we have $\sup_{s\in \mathcal{S}%
_{0}}\left\vert \widehat{\gamma }(s)-\gamma _{0}(s)\right\vert \rightarrow
_{p}0$ as $n\rightarrow \infty $. To this end, recall that $\widehat{\gamma }%
(s)$ is the minimizer of $\Upsilon _{n}(\gamma ;s)$ in (\ref{pf1}) and $%
\gamma _{0}(s)$ is the minimizer of $\Upsilon _{0}(\gamma ;s)$ in (\ref{pf2}%
) for any given $s\in \mathcal{S}_{0}$. See Lemma \ref{L-A3} for the
definitions of $\Upsilon _{n}(\gamma ;s)$ and $\Upsilon _{0}(\gamma ;s)$.

Suppose $\widehat{\gamma }(s)$ is not uniformly consistent, implying that
there exist $\eta >0$ and $\varepsilon >0$ such that for any $N\in 
%TCIMACRO{\U{2115} }%
%BeginExpansion
\mathbb{N}
%EndExpansion
$, there exists $n>N$ satisfying 
\begin{eqnarray*}
&&\mathbb{P}\left( \sup_{s\in \mathcal{S}_{0}}\left\vert \widehat{\gamma }%
(s)-\gamma _{0}(s)\right\vert >\eta \right) \\
&=&\mathbb{P}\left( \sup_{s\in \mathcal{S}_{0}}\left( \widehat{\gamma }%
(s)-\gamma _{0}(s)\right) >\eta \right) +\mathbb{P}\left( \sup_{s\in 
\mathcal{S}_{0}}\left( \widehat{\gamma }(s)-\gamma _{0}(s)\right) <-\eta
\right) >\varepsilon
\end{eqnarray*}%
or simply 
\begin{equation}
\mathbb{P}\left( \sup_{s\in \mathcal{S}_{0}}\left( \widehat{\gamma }%
(s)-\gamma _{0}(s)\right) >\eta \right) >\varepsilon  \label{contra}
\end{equation}%
without loss of generality. From (\ref{dY0}), we can define $C\in (0,\infty
) $ such that%
\begin{equation*}
\inf_{s\in \mathcal{S}_{0}}\frac{\partial \Upsilon _{0}(\gamma _{0}(s);s)}{%
\partial \gamma }>C>0\text{,}
\end{equation*}%
and hence the mean value theorem yields 
\begin{eqnarray*}
\Upsilon _{0}(\widehat{\gamma }(s),s)-\Upsilon _{0}(\gamma _{0}(s),s) &=&%
\frac{\partial \Upsilon _{0}(\widetilde{\gamma }(s),s)}{\partial \gamma }%
\left( \widehat{\gamma }\left( s\right) -\gamma _{0}(s)\right) \\
&>&C\left( \widehat{\gamma }\left( s\right) -\gamma _{0}(s)\right)
\end{eqnarray*}%
for sufficiently large $n$, where $\widetilde{\gamma }(s)$ is between $%
\widehat{\gamma }(s)$ and $\gamma _{0}(s)$. Therefore, 
\begin{eqnarray}
&&\mathbb{P}\left( \sup_{s\in \mathcal{S}_{0}}\left\{ \Upsilon _{0}(\widehat{%
\gamma }(s),s)-\Upsilon _{0}(\gamma _{0}(s),s)\right\} >C\eta \right)
\label{contra2} \\
&>&\mathbb{P}\left( \inf_{s\in \mathcal{S}_{0}}\frac{\partial \Upsilon
_{0}(\gamma _{0}(s);s)}{\partial \gamma }\sup_{s\in \mathcal{S}_{0}}\left( 
\widehat{\gamma }(s)-\gamma _{0}(s)\right) >C\eta \right)  \notag \\
&=&\mathbb{P}\left( \sup_{s\in \mathcal{S}_{0}}\left( \widehat{\gamma }%
(s)-\gamma _{0}(s)\right) >\eta \right) >\varepsilon  \notag
\end{eqnarray}%
from (\ref{contra}).

However, by construction, $\Upsilon _{n}(\widehat{\gamma }(s),s)-\Upsilon
_{n}(\gamma _{0}(s),s)\leq 0$ for every $s\in \mathcal{S}_{0}$, which
implies 
\begin{equation}
\sup_{s\in \mathcal{S}_{0}}\left\{ \Upsilon _{n}(\widehat{\gamma }%
(s),s)-\Upsilon _{n}(\gamma _{0}(s),s)\right\} \leq 0\text{ \ almost surely.}
\label{fact1}
\end{equation}%
Furthermore, using the triangular inequality and the uniform convergence
result in Lemma \ref{L1}, we can verify that 
\begin{equation}
\sup_{\left( r,s\right) \in \Gamma \times \mathcal{S}_{0}}\left\vert
\Upsilon _{n}(r,s)-\Upsilon _{0}(r,s)\right\vert \rightarrow _{p}0
\label{fact2}
\end{equation}%
as $n\rightarrow \infty $ from the proof of Lemma \ref{L-A3}. From (\ref%
{fact1}) and (\ref{fact2}), we thus have 
\begin{eqnarray*}
&&\mathbb{P}\left( \sup_{s\in \mathcal{S}_{0}}\left\{ \Upsilon _{0}(\widehat{%
\gamma }(s),s)-\Upsilon _{0}(\gamma _{0}(s),s)\right\} >C\eta \right) \\
&\leq &\mathbb{P}\left( \sup_{s\in \mathcal{S}_{0}}\left\{ \Upsilon _{0}(%
\widehat{\gamma }(s),s)-\Upsilon _{n}(\widehat{\gamma }(s),s)\right\} >C\eta
/3\right) \\
&&+\mathbb{P}\left( \sup_{s\in \mathcal{S}_{0}}\left\{ \Upsilon _{n}(%
\widehat{\gamma }(s),s)-\Upsilon _{n}(\gamma _{0}(s),s)\right\} >C\eta
/3\right) \\
&&+\mathbb{P}\left( \sup_{s\in \mathcal{S}_{0}}\left\{ \Upsilon _{n}(\gamma
_{0}(s),s)-\Upsilon _{0}(\gamma _{0}(s),s)\right\} >C\eta /3\right) \\
&\leq &(\varepsilon ^{\ast }/3)+(\varepsilon ^{\ast }/3)+(\varepsilon ^{\ast
}/3)=\varepsilon ^{\ast }
\end{eqnarray*}%
for any $\varepsilon ^{\ast }>0$ if $n$ is sufficiently large. It
contradicts to (\ref{contra2}) by choosing $\varepsilon ^{\ast }\leq
\varepsilon $, hence the uniform consistency should hold. $\blacksquare $

\paragraph*{Proof of Lemma \protect\ref{bias1}}

We prove $\Xi _{\beta 2}=o_{p}(1)$ and $\Xi _{\beta 3}=o_{p}(1)$. The
results for $\Xi _{\delta 2}$ and $\Xi _{\delta 3}$ can be shown
symmetrically. For expositional simplicity, we present the case of scalar $%
x_{i}$.

\textsl{For }$\Xi _{\beta 2}$\textsl{:} Note that $\widehat{\gamma }(\cdot )$
belongs to $\mathcal{G}_{n}(\mathcal{S}_{0};\Gamma )$. We define intervals $%
\mathcal{I}_{k}$ for $k=1,\ldots ,n$, which are centered at the
discontinuity points of $\widehat{\gamma }(s)$ with length $\ell _{n}$ such
that $\ell _{n}\rightarrow 0$ as $n\rightarrow \infty $. Without loss of
generality, we choose $\ell _{n}=O(n^{-3})$. Then, we can interpolate on
each $\mathcal{I}_{k}$ and define $\widetilde{\gamma }(s)$ as a smooth
version of $\widehat{\gamma }(s)$, which satisfies 
\begin{equation}
\mathbb{P}\left( \sup_{s\in \mathcal{S}_{0}}\left\vert \widehat{\gamma }%
\left( s\right) -\widetilde{\gamma }\left( s\right) \right\vert >\varepsilon
\right) \leq \mathbb{P}\left( \max_{1\leq k\leq n}\sup_{s\in \mathcal{I}%
_{k}}\left\vert \widehat{\gamma }\left( s\right) -\widetilde{\gamma }\left(
s\right) \right\vert >\varepsilon \right) \leq \varepsilon  \label{P_gamma}
\end{equation}%
for any $\varepsilon >0$, if $n$ is sufficiently large. Since $\sup_{s%
\mathcal{\in S}_{0}}\left\vert \widehat{\gamma }\left( s\right) -\gamma
_{0}(s)\right\vert =o_{p}(1)$ in the proof of Lemma \ref{neg uniform}, we
have 
\begin{equation}
\sup_{s\in \mathcal{S}_{0}}\left\vert \widetilde{\gamma }\left( s\right)
-\gamma _{0}\left( s\right) \right\vert \leq \sup_{s\in \mathcal{S}%
_{0}}\left\vert \widehat{\gamma }\left( s\right) -\widetilde{\gamma }\left(
s\right) \right\vert +\sup_{s\in \mathcal{S}_{0}}\left\vert \widehat{\gamma }%
\left( s\right) -\gamma _{0}\left( s\right) \right\vert =o_{p}(1)
\label{g_tild}
\end{equation}%
from (\ref{P_gamma}).

Now we define 
\begin{equation*}
G_{n}(\gamma )=\frac{1}{\sqrt{n}}\dsum_{i\in \Lambda _{n}}x_{i}u_{i}\mathbf{1%
}\left[ q_{i}>\gamma (s_{i})+\pi _{n}\right] \mathbf{1}_{\mathcal{S}_{0}}%
\text{,}
\end{equation*}%
and then 
\begin{eqnarray*}
\Xi _{\beta 2} &=&G_{n}(\widehat{\gamma })-G_{n}(\gamma _{0}) \\
&=&\left\{ G_{n}(\widehat{\gamma })-G_{n}(\widetilde{\gamma })\right\}
+\left\{ G_{n}(\widetilde{\gamma })-G_{n}(\gamma _{0})\right\} \\
&\equiv &\Psi _{G1}+\Psi _{G2}\text{.}
\end{eqnarray*}

First, for $\Psi _{G1}$, let $\Delta _{i}^{\pi }(\widehat{\gamma },%
\widetilde{\gamma })=\mathbf{1}\left[ q_{i}>\widehat{\gamma }(s_{i})+\pi _{n}%
\right] -\mathbf{1}\left[ q_{i}>\widetilde{\gamma }(s_{i})+\pi _{n}\right] $%
. By construction, $|\Delta _{i}^{\pi }(\widehat{\gamma },\widetilde{\gamma }%
)|\leq \mathbf{1[}s_{i}\in \mathcal{I}_{k}$ for some $k]$. Therefore, by the
Cauchy-Schwarz inequality and Assumptions A-(v) and A-(viii), 
\begin{eqnarray*}
\mathbb{E}\left[ \left\vert \Psi _{G1}\right\vert \right] &\leq &n^{1/2}%
\mathbb{E}\left[ \left\vert x_{i}u_{i}\right\vert \left\vert \Delta
_{i}^{\pi }(\widehat{\gamma },\widetilde{\gamma })\right\vert \mathbf{1}_{%
\mathcal{S}_{0}}\right] \\
&\leq &n^{1/2}\mathbb{E}\left[ (x_{i}u_{i})^{2}\right] ^{1/2}\mathbb{E}\left[
\left( \mathbf{1[}s_{i}\in \mathcal{I}_{k}\text{ for some }k]\mathbf{1}_{%
\mathcal{S}_{0}}\right) ^{2}\right] ^{1/2} \\
&\leq &C_{1}n^{1/2}\left( \mathbb{P}\left[ s_{i}\in \mathcal{I}_{k}\cap 
\mathcal{S}_{0}\text{ for some }k\right] \right) ^{1/2} \\
&\leq &C_{1}^{\prime }n^{1/2}n^{-3/2}=o(1)
\end{eqnarray*}%
for some $C_{1},C_{1}^{\prime }<\infty $. Hence, $\Psi _{G1}=o_{p}(1)$.

Second, for $\Psi _{G2}$, we let $\mathbf{1}_{\tau }=\mathbf{1}[\left\vert
x_{i}u_{i}\right\vert \leq \tau ]$ for some $\tau <\infty $. Then, for any $%
\varepsilon ^{\prime }>0$ and $\gamma :\mathcal{S}_{0}\mapsto \Gamma $,%
\begin{eqnarray*}
&&\mathbb{P}\left( \frac{1}{\sqrt{n}}\dsum_{i\in \Lambda _{n}}x_{i}u_{i}%
\mathbf{1}\left[ q_{i}>\gamma (s_{i})+\pi _{n}\right] \left( 1-\mathbf{1}%
_{\tau }\right) \mathbf{1}_{\mathcal{S}_{0}}>\varepsilon ^{\prime }\right) \\
&\leq &\varepsilon ^{-2}\frac{1}{n}\mathbb{E}\left[ \left( \dsum_{i\in
\Lambda _{n}}x_{i}u_{i}\mathbf{1}\left[ q_{i}>\gamma (s_{i})+\pi _{n}\right]
\left( 1-\mathbf{1}_{\tau }\right) \mathbf{1}_{\mathcal{S}_{0}}\right) ^{2}%
\right] \\
&\leq &C\varepsilon ^{-2}\mathbb{E}\left[ \left( x_{i}u_{i}\right) ^{2}%
\mathbf{1}\left[ \left\vert x_{i}u_{i}\right\vert >\tau \right] \right] \\
&\leq &C\varepsilon ^{-2}\mathbb{E}\left[ \left( x_{i}u_{i}\right) ^{4}%
\right] ^{1/2}\left( \mathbb{P}\left[ \left\vert x_{i}u_{i}\right\vert >\tau %
\right] \right) ^{1/2} \\
&\leq &C\varepsilon ^{-2}\tau ^{-2}\mathbb{E}\left[ \left( x_{i}u_{i}\right)
^{4}\right]
\end{eqnarray*}%
for some $C<\infty $, where we apply the Markov's and the Cauchy-Schwarz
inequalities. From Assumption A-(v), by choosing $\tau $ sufficiently large,
this probability can be arbitrarily small. Hence, 
\begin{equation*}
G_{n}(\gamma )=\frac{1}{\sqrt{n}}\dsum_{i\in \Lambda _{n}}x_{i}u_{i}\mathbf{1%
}\left[ q_{i}>\gamma (s_{i})+\pi _{n}\right] \mathbf{1}_{\tau }\mathbf{1}_{%
\mathcal{S}_{0}}+o_{p}(1)
\end{equation*}%
for sufficiently large $n$ and we simply consider $\left\vert
x_{i}u_{i}\right\vert \leq \tau $ almost surely in what follows.

We let $\mathcal{F}^{\ast }$ be the class of functions $\{xu\mathbf{1}\left[
q>\gamma (s)+\pi _{n}\right] $ for $\gamma \in \mathcal{C}^{2}[\mathcal{S}%
_{0}]\}$, where $\mathcal{C}^{2}[\mathcal{S}_{0}]$ denotes the family of
twice-continuously differentiable functions defined on $\mathcal{S}_{0}$.
Using Theorem 2.5.6 in der Vaart and Wellner (1996), we establish that $%
\mathcal{F}^{\ast }$ is P-Donsker, which requires three elements: an entropy
bound, a maximal inequality, and the chaining argument. For the entropy
bound, by Corollaries 2.7.2 and 2.7.3 in der Vaart and Wellner (1996) (with
their $r=d=1$ and $\alpha =2$), $\mathcal{F}^{\ast }$ has the same
bracketing number (up to a constant) as that for the collection of subgraphs
of $\mathcal{C}^{2}[\mathcal{S}_{0}]$, so that $\log N_{[]}\left(
\varepsilon ,\mathcal{F}^{\ast },||\cdot ||_{\infty }\right) \leq
C\varepsilon ^{-1/2}$, where $||\cdot ||_{\infty }$ denotes the uniform
norm. For the maximal inequality, since we consider $\left\vert
xu\right\vert \leq \tau $, Corollary 3.3 in Valenzuela-Dom\'{\i}guez, Krebs,
and Franke (2017) gives the Bernstein inequality for spatial lattice
processes with exponentially decaying $\alpha $-mixing coefficients. This
satisfies the conditions in Lemma 2.2.10 in der Vaart and Wellner (1996),
which implies that for any finite collection of functions $\gamma
_{1},\ldots ,\gamma _{m}\in \mathcal{C}^{2}[\mathcal{S}_{0}]$,%
\begin{equation}
\mathbb{E}\left[ \max_{1\leq k\leq m}G_{n}(\gamma _{k})\right] \leq
C^{\prime }\left( \log (1+m)+\sqrt{\log (1+m)}\right)  \label{maximal ineq}
\end{equation}%
for $C^{\prime }<\infty $. For the chaining argument, the same analysis in
der Vaart and Wellner (1996), pp.131-132 applies with the following two
changes: their envelope function $F$ is $\left\vert xu\right\vert $, which
satisfies $\mathbb{E}\left[ F^{2}\right] <\infty $; and their inequality
(2.5.5) is implied by (\ref{maximal ineq}) with $m=\log N_{[]}\left(
\varepsilon ,\mathcal{F}^{\ast },||\cdot ||_{\infty }\right) $. Note that
the spatial dependence only shows up in deriving the maximal inequality but
not the entropy or the chaining argument.

Since Donsker implies stochastic equicontinuity, it follows that $%
G_{n}(\cdot )$ satisfies, for every positive $\eta _{n}\rightarrow 0$, 
\begin{equation*}
\sup_{\sup_{s\in \mathcal{S}_{0}}\left\vert \gamma (s)-\gamma ^{\prime
}(s)\right\vert \leq \eta _{n}}\left\vert G_{n}(\gamma )-G_{n}(\gamma
^{\prime })\right\vert \rightarrow _{p}0
\end{equation*}%
as $n\rightarrow \infty $. Therefore, $\Psi _{G2}=o_{p}(1)$ since $%
\sup_{s\in \mathcal{S}_{0}}\left\vert \widetilde{\gamma }(s)-\gamma
_{0}(s)\right\vert =o_{p}(1)$ from (\ref{g_tild}).

\textsl{For }$\Xi _{\beta 3}$\textsl{:} On the event $E_{n}^{\ast }$ that $%
\sup_{s\in \mathcal{S}_{0}}\left\vert \widehat{\gamma }(s)-\gamma
_{0}(s)\right\vert \leq \phi _{2n}$, we have%
\begin{eqnarray*}
\mathbb{E}\left[ \left\vert \Xi _{\beta 3}\right\vert \right] &=&\frac{1}{%
\sqrt{n}}\dsum_{i\in \Lambda _{n}}\mathbb{E}\left[ \left\vert
x_{i}^{2}\delta _{0}\right\vert \mathbf{1}\left[ q_{i}\leq \gamma _{0}(s_{i})%
\right] \mathbf{1}\left[ q_{i}>\widehat{\gamma }(s_{i})+\pi _{n}\right] 
\mathbf{1}_{\mathcal{S}_{0}}\right] \\
&\leq &n^{1/2-\epsilon }C\mathbb{E}\left[ \mathbf{1}\left[ q_{i}\leq \gamma
_{0}(s_{i})\right] \mathbf{1}\left[ q_{i}>\widehat{\gamma }(s_{i})+\pi _{n}%
\right] \mathbf{1}_{\mathcal{S}_{0}}\right] \\
&\leq &n^{1/2-\epsilon }C\mathbb{E}\left[ \mathbf{1}\left[ q_{i}\leq \gamma
_{0}(s_{i})\right] \mathbf{1}\left[ q_{i}>\gamma _{0}(s_{i})-\phi _{2n}+\pi
_{n}\right] \mathbf{1}_{\mathcal{S}_{0}}\right] \\
&=&n^{1/2-\epsilon }C\int_{\mathcal{S}_{0}}\int_{\mathcal{I}(q;s)}f(q,s)dqds
\end{eqnarray*}%
for some $0<C<\infty $, where $\mathcal{I}(q;s)=\{q:q\leq \gamma _{0}(s)$
and $q>\gamma _{0}(s)-\phi _{2n}+\pi _{n}\}$. Since we define $\pi _{n}>0$
such that $\phi _{2n}/\pi _{n}\rightarrow 0$, it holds that $\pi _{n}-\phi
_{2n}>0$ for sufficiently large $n$. Therefore, $\mathcal{I}(q;s)$ becomes
empty for all $s$ when $n$ is sufficiently large. The desired result follows
from Markov's inequality and the fact that $\mathbb{P}\left( E_{n}^{\ast
}\right) >1-\varepsilon $ for any $\varepsilon >0$. $\blacksquare $

\paragraph{References}

\begin{description}
\item \textsc{Berbee, H.} (1987): "Convergence Rates in the Strong Law for
Bounded Mixing Sequences", \textit{Probability Theory and Related Fields},
74, 255-270

\item \textsc{Carbon, M., L. T. Tran,} and \textsc{B. Wu} (1997): "Kernel
Density Estimation for Random Fields", \textit{Statistics and Probability
Letters}, 36, 115-125

\item \textsc{Carbon, M., C. Francq,} and \textsc{L. T. Tran} (2007):
"Kernel Regression Estimation for Random Fields," \textit{Journal of
Statistical Planning and Inference}, 137, 778-798

\item \textsc{Davidson, J.} (1994): "Stochastic Limit Theory", Oxford
University Press

\item \textsc{Hansen, B. E.} (2000): "Sample Splitting and Threshold
Estimation,"\ \textit{Econometrica}, 68, 575--603.

\item \textsc{Jenish, N. and I. R. Prucha} (2009): "Central Limit Theorems
and Uniform Laws of Large Numbers for Arrays of Random Fields," \textit{%
Journal of Econometrics}, 150, 86-98.

\item \textsc{Rio, E.} (1995): "The Functional Law of the Iterated Logarithm
for Stationary Strongly Mixing Sequences," \textit{Annals of Probability},
23(3), 1188-1203.

\item \textsc{Tran, L. T.} (1990): "Kernel Density Estimation on Random
Fields," \textit{Journal of Multivariate Analysis}, 34, 37-53.

\item \textsc{Valenzuela-Dom\'{\i}nguez, E., J. T. N. Krebs,} and \textsc{J.
E. Franke} (2017): A Bernstein Inequality for Spatial Lattice Processes, 
\textit{Working Paper}

\item \textsc{van der Vaart, A. W.} and \textsc{J. A. Wellner (1996)}: "Weak
Convergence and Empirical Processes with Applications to Statistics,"\
Springer
\end{description}

\end{document}